\documentclass[11pt,reqno,a4paper]{article}
\usepackage{lmodern} 
\usepackage[T1]{fontenc} 
\usepackage[margin=2.30cm]{geometry}
\usepackage{float}
\usepackage{comment}
\usepackage{adjustbox}
\usepackage{indentfirst}
\usepackage{enumerate}
\usepackage{url}
\usepackage{cite}
\usepackage{ltablex}
\keepXColumns
\usepackage{booktabs}
\usepackage{array}
\usepackage{hyperref}
\hypersetup{
pdftitle={},
pdfauthor={},
pdfsubject={},
pdfkeywords={},
colorlinks=true,
linkcolor=black,
urlcolor=black,
linktocpage={false},
pageanchor={},
citecolor={black},
}
\usepackage{amsmath,amssymb,amsthm,amsfonts,amscd,authblk,dsfont,cancel,diagbox,
enumerate,epsfig,eucal,fancyhdr,latexsym,lmodern,mathrsfs,mathtools,multirow,
musicography,skull,simplewick,subcaption,tikz-cd}
\usetikzlibrary{matrix,arrows,decorations.pathmorphing}
\usepackage{relsize}
\usepackage[normalem]{ulem} 
\usepackage[english]{babel}
\usepackage[utf8]{inputenc}
\usepackage{titlesec}
\titleformat{\section}{\large\bfseries\filcenter}{\thesection}{1em}{}
\titleformat{\subsection}{\bfseries}{\thesubsection}{1em}{}
\let\oldbibliography\thebibliography
\renewcommand{\thebibliography}[1]{\oldbibliography{#1}
\setlength{\itemsep}{-0.15\baselineskip}}
\makeatletter
\renewcommand\@makefntext[1]{\leftskip=2em\hskip-0.5em\@makefnmark#1}
\makeatother
\newtheorem*{maintheorem}{Main Theorem}
\newtheorem{theorem}{Theorem}[section]
\newtheorem{theorem-intro}{Theorem}[]
\newtheorem{corollary}[theorem]{Corollary}
\newtheorem{lemma}[theorem]{Lemma}
\newtheorem{proposition}[theorem]{Proposition}
\theoremstyle{definition}
\newtheorem{remark}[theorem]{Remark}

\newtheorem{assumption}[theorem]{Assumption}
\newtheorem{definition}[theorem]{Definition}
\newtheorem{example}[theorem]{Example}
\newtheorem{examples}[theorem]{Examples}
\newtheorem{notation}[theorem]{Notation}

\numberwithin{equation}{section}
\allowdisplaybreaks[1]
\catcode`@=12
\renewcommand\thanks[1]{%
  \begingroup
  \renewcommand\thefootnote{}\footnote{#1}%
  \addtocounter{footnote}{-1}%
  \endgroup
}
\renewcommand{\:}{\colon} 
\newcommand{\defeq}{\vcentcolon=}

\renewcommand{\i}{\mathrm{i}}

\renewcommand{\epsilon}{\varepsilon}

\newcommand{\A}{A}
\newcommand{\B}{\mathsf{B}}

\newcommand{\E}{\mathsf{E}}
\renewcommand{\L}{\mathsf{L}}
\newcommand{\F}{\mathsf{F}}
\newcommand{\G}{\mathsf{G}}
\renewcommand{\H}{\mathsf{H}}
\renewcommand{\P}{\mathsf{P}}
\newcommand{\K}{\mathsf{K}}

\newcommand{\N}{\mathsf{N}}
\newcommand{\R}{\mathbb{R}}
\newcommand{\V}{\mathsf{V}}
\newcommand{\X}{\mathsf{X}}

\renewcommand{\d}{\mathrm{d}}

\newcommand{\Ad}{\mathrm{Ad}}

\newcommand{\g}{\mathfrak{g}}
\DeclareMathOperator{\im}{\textnormal{ran}}

\newcommand{\D}{\square}

\newcommand{\ran}{\textnormal{ran}}
\renewcommand{\ker}{\textnormal{ker}}
\newcommand{\supp}{{\textnormal{supp}}}
\newcommand{\WF}{\textnormal{WF}}
 
\newcommand{\BS}{\mathrm{B}\mathcal{S}_{\mathrm{cl}}}
\newcommand{\tc}{\mathrm{tc}}

\renewcommand{\c}{\mathrm{c}}

\newcommand{\M}{\mathsf{M}}	
\renewcommand{\g}{\mathsf{g}}
\newcommand{\h}{\mathsf{h}}
\newcommand{\T}{\mathsf{T}}

\newcommand{\bb}[1]{\mathbb{#1}}
\renewcommand{\sf}[1]{\mathsf{#1}}

\usepackage{tikz-cd}
\tikzset{%
  symbol/.style={
    draw=none,
    every to/.append style={
      edge node={node [sloped, allow upside down, auto=false]{$#1$}}
    },
  },
}
\begin{document}
%
%
%
%
\begin{center}
\vspace*{-1cm}

\noindent\hrulefill
\begin{center}
\begin{Large}
	{\textbf{Hadamard States for Yang-Mills Theory Linearised Around\\[0.4em] Flat Connections on Globally Hyperbolic Spacetimes}}\par
	\end{Large}
\end{center}
\hrulefill

\vspace{5mm}

{\bf by}

\vspace{3mm}
\noindent
{  \bf  Beatrice Costeri$\,^1$, Simone Murro$\,^2$, Gabriel Schmid$\,^3$, Raman Deep Singh$\,^1$}\\[2mm]
\noindent   $^1${\it Dipartimento di Fisica,  Universit\`a di Pavia \& INdAM \& INFN, Italy}\\[2mm]
\noindent   $^2${\it  Dipartimento di Matematica,  Universit\`a di Genova \& INdAM \& INFN, Italy}\\[2mm]
\noindent   $^3${\it  Dipartimento di Matematica,  Universit\`a di Trento \& INdAM, Italy}\\[2mm]

Emails: \ {\tt  beatrice.costeri01@universitadipavia.it, simone.murro@unige.it, gabriel.schmid@unitn.it, ramandeep.singh01@universitadipavia.it}
\\[10mm]
\end{center}
\begin{abstract}
The aim of this paper is to prove the existence of Hadamard states for Yang-Mills theory linearised around a flat background connection on (possibly non-trivial) principal bundles over arbitrary globally hyperbolic spacetimes. The core of our approach combines a complete gauge fixing with a microlocal factorisation of the time-dependent, twisted de Rham-Hodge d'Alembertian in ultrastatic globally hyperbolic spacetimes. Moreover, we also characterise the classical phase space of the theory via a Poincaré duality on flat vector bundles, analysing its weak non-degeneracy and its provision of a separating, non-redundant set of gauge-invariant observables.
\end{abstract}
\vspace*{0.5cm}
\tableofcontents
\vspace*{0.5cm}
\section{Introduction}
The quantisation of a classical field propagating on curved spacetimes has been, and remains, at the forefront of scientific research, with \emph{gauge theories} providing a particularly relevant class of examples. Among these, Yang–Mills theory occupies a particularly prominent position. First formulated in 1954 in \cite{YangMills,ShawPhD} to generalise Abelian gauge theories to the non-Abelian regime, on the one hand, it represents a fundamental ingredient of the celebrated \emph{Standard Model of Particle Physics}. On the other hand, its modern formulation using the geometry of principal fibre bundles, see \emph{e.g.}, \cite{WuYang1,AtiyahYM1,AtiyahYM2} for early works on that formalism, has also sparked developments in pure mathematics, yielding deep insights into the topology of smooth 4-manifolds through the study of instantons, see \emph{e.g.}, \cite{Donaldson,Scorpan,FreedUhlenbeck}. 

In the present work, we address the problem of quantisation of Yang-Mills fields \emph{linearised around flat connections} on  globally hyperbolic spacetimes. Formulated as a linear gauge theory, its classical phase space can be systematically described within the axiomatic framework introduced by Hack-Schenkel \cite{HackSchenkel} and further refined by Gérard-Wrochna \cite{GerardWrochna} (see also \cite[Chap.~2]{SchmidPhD} for a detailed review). In this formalism, the dynamics of a linear gauge theory is parametrised by two linear differential operators
\begin{align*}
	\sf{P}\:\Gamma(\sf{E}_{1})\to\Gamma(\sf{E}_{1})\,,\qquad \sf{K}\:\Gamma(\sf{E}_{0})\to\Gamma(\sf{E}_{1})\,,\qquad \sf{P}\circ\sf{K}=0\,,
\end{align*}
acting on smooth sections of the Hermitian vector bundles $(\sf{E}_{0},\langle\cdot,\cdot\rangle_{\sf{E}_{0}})$ and $(\sf{E}_{1},\langle\cdot,\cdot\rangle_{\sf{E}_{1}})$ defined over a globally hyperbolic Lorentzian manifold $(\sf{M},\sf{g})$. The operator $\sf{P}$ parametrises the equations of motion, while the operator $\sf{K}$ encodes linear gauge transformations $\psi\mapsto\psi+\sf{K}\omega$. Although $\sf{P}$ is in general non-hyperbolic, one can introduce two operators 
\begin{align*}
\sf{D}_{0}:=\sf{K}^{\ast}\sf{K}\:\Gamma(\sf{E}_{0})\to\Gamma
(\sf{E}_{0})\,,\qquad\sf{D}_{1}:=\sf{P}+\sf{K}\sf{K}^{\ast}\:\Gamma(\sf{E}_{1})\to\Gamma(\sf{E}_{1}), 
\end{align*}
assumed to be (Green) hyperbolic, and reduce the problem to the analysis of the \emph{gauge-fixed} operator $\sf{D}_{1}$, while $\sf{D}_{0}$ parametrises the \emph{remaining gauge freedom}. This is motivated by the fact that from a physics perspective, we are interested in the space of \emph{gauge equivalent classes}, which can be identified with a suitable quotient space of solutions to the hyperbolic problem. In this formalism, the \emph{classical phase space} associated with a linear gauge theory, namely, the space of \emph{linear, on-shell, gauge-invariant observables}, is given by
\begin{align*}
	\mathcal{V}_{\mathrm{P}}:=\cfrac{\mathrm{ker}(\sf{K}^{\ast}\vert_{\Gamma_{\mathrm{c}}})}{\mathrm{ran}(\sf{P}\vert_{\Gamma_{\mathrm{c}}})}\,,\qquad\sigma\:\mathcal{V}_{\mathrm{P}}\times\mathcal{V}_{\mathrm{P}}\to\mathbb{C}\,,\qquad\sigma([\psi],[\phi]):=i\int_{\sf{M}}\langle\psi,\sf{G}_{1}\phi\rangle_{\sf{E}_{1}}\,\d\mu_{\sf{g}}\,,
\end{align*}
where $\sf{G}_{1}\:\Gamma_{\mathrm{c}}(\sf{E}_{1})\to \Gamma_{\mathrm{sc}}(\sf{E}_{1})$ denotes the \emph{causal propagator} (or \emph{Pauli-Jordan distribution}) associated to $\sf{D}_{1}$. Physically, every element $[\psi]\in\mathcal{V}_{\mathrm{P}}$ determines a corresponding linear on-shell observable, which is gauge invariant, see Section \ref{Subsec:Linearised Yang-Mills Theory in the Hack-Schenkel Formalism} and the discussion therein.  

Having the classical theory under control, the next step consists of its \emph{quantisation}. Within this context, we employ the \emph{algebraic approach to quantum field theory} (see \emph{e.g.}~\cite{AQFT,GerardBook}), which offers a mathematically rigorous framework, ideally suited for field theories on curved Lorentzian backgrounds. In this formalism, quantisation can be formulated as a two-step procedure:
\begin{itemize}\itemsep0em
\item[1.] The first step consists in assigning to the classical phase space $(\mathcal{V}_{\mathrm{P}},\sigma)$ a suitable $*$-algebra of observables encoding structural properties such as causality and dynamics, together with the canonical commutation relations in the Bosonic case.
\item[2.] The second step calls for the identification of a physical state, which is a positive,
linear and normalised functional on the algebra of observables, satisfying the so-called {\it Hadamard condition}. 
\end{itemize}

The Hadamard condition prescribes the short-distance singularity structure of the two-point function and plays a central role in perturbative quantum field theory~\cite{Kasia}. It allows for the construction of Wick polynomials~\cite{HW01} and other nonlinear observables~\cite{VI}, including the stress-energy tensor, and ensures finite fluctuations of suitably smeared Wick polynomials~\cite{FV}. For a complex scalar field~\cite{GerardOulghaziWrochna}, whose dynamics is ruled by a normally hyperbolic operator~$\D$, finding a quasifree Hadamard state amounts to constructing a pair of linear operators $\lambda^\pm\:\,C^\infty_\c(\M)\to C^\infty(\M)$ that satisfies the following properties:
\begin{itemize}
\item[(i)] \quad $(\lambda^{\pm})^{\ast}=\lambda^{\pm}$  with respect to a positive Hermitian form $(\cdot,\cdot)_{\M}$ on $C^{\infty}_{\c}(\M)$\,;
\item[(ii)]\quad $\D\lambda^{\pm}=\lambda^{\pm}\D=0$   and $\i(\lambda^{+}-\lambda^{-})$ is the causal propagator of $\D$\,;
\item[(iii)]\quad $\lambda^{\pm}\geq 0$ w.r.t.~$(\cdot,\cdot)_{\M}$\,;
\item[(iv)]\quad  $\WF^{\prime}(\lambda^{\pm})\subset\mathcal{N}^{\pm}\times\mathcal{N}^{\pm}\, $.\end{itemize}
Here, $\WF^{\prime}(\lambda^{\pm})$ denotes the primed wavefront set of the Schwartz kernel of $\lambda^\pm$ and 
$\mathcal{N}^{\pm}$ are the two connected components of the lightcone $\mathcal{N}\defeq \{(p,\xi)\in \T^*\M \,|\, g^{-1}(\xi,\xi)=0 \}$.  

The construction of Hadamard states for linear field theories that possess a gauge symmetry, such as linearised Yang-Mills theory, however, presents much more delicate challenges, mainly due to the following reasons:
\begin{itemize}\itemsep0em 
	\item[1.] The presence of gauge symmetry implies that the differential operator $\P$ governing the dynamics of a gauge theory is in general not hyperbolic.
    \item[2.] The operator $\P$ is formally self-adjoint w.r.t.~a Hermitian product that is typically non-positive on the fibres of the vector bundle;
	\item[3.] Linear gauge theories also exhibit infrared obstructions. In ultrastatic spacetime, for example,  the construction of positive and negative frequency projectors for a massless field may fail. For ordinary fields, this difficulty can often be removed by a smoothing perturbation, which leaves the Hadamard property unchanged. In gauge theories, however, such a modification must also preserve gauge invariance, requiring the resulting projectors to remain compatible with the gauge operator $\sf K$.
\end{itemize}

The first issue is well understood and can be addressed by means of a suitable gauge-fixing procedure, as discussed above. The second, concerning the non-positivity of the fibre metric, is considerably more subtle, as it gives rise to a tension between the Hadamard condition and the positivity of states. In addition, infrared problems introduce further complications that must be taken into account. Although it is well-known how to construct bidistributions satisfying the Hadamard condition, ensuring the positivity and gauge-invariance of states in linear gauge theories remains a highly non-trivial task at the forefront of current research. 

Among the strategies developed to address these difficulties, a particularly effective approach is based on \emph{complete gauge fixing}, recently implemented for Maxwell theory in \cite{MurroSchmid} and for linearised gravity in \cite{Gerard,GerardMolodykWrochna}. Extending this framework to linearised Yang-Mills theory is, however, not immediate. In the present setting, one must accommodate possibly non-trivial principal bundles, twisted differential operators and spatial background connections that may retain a non-trivial time dependence even on ultrastatic spacetimes. The purpose of this work is to develop a complete gauge-fixing construction adapted to these features and to use it to establish the existence of quasifree Hadamard states for linearised Yang-Mills theory around flat background connections on general globally hyperbolic spacetimes.

\subsection{Main Results and Structure of the Paper}
Our analysis is carried out in a genuinely bundle-theoretic setting. Unlike the framework considered in \cite{GerardWrochna}, we allow $\P\to\M$ to be a possibly non-trivial principal $\G$-bundle over a globally hyperbolic spacetime $(\M,\g)$, and we linearise the theory around a flat background connection $A \in \mathcal{C}(\P)$. The restriction to flat backgrounds is motivated by our aim of developing a \emph{complete gauge-fixing} approach to the construction of Hadamard states, a strategy that has recently proved effective for other gauge theories, notably Maxwell theory \cite{MurroSchmid} and linearised gravity \cite{Gerard,GerardMolodykWrochna}. At the same time, we stress that flatness does not force the underlying gauge configuration to be trivial. As a matter of fact, whenever non-trivial flat connections exist, they may retain non-trivial global and topological information, for instance through their holonomy. The present setting therefore provides a natural geometric framework in which to develop the complete gauge-fixing construction, while at the same time paving the way to future extensions to more general Yang-Mills backgrounds.

In contrast with the trivial-bundle setting briefly reviewed in Section~\ref{Subsec:Flat Yang-Mills Connections}, linearised Yang-Mills fields cannot in general be represented globally by $\mathfrak{g}$-valued one-forms. Rather, they are naturally described as elements of $\Omega^1(\M,\operatorname{Ad}(\P))$, namely, as $\operatorname{Ad}(\P)$-valued one-forms on $\M$, while infinitesimal gauge transformations are parametrised by smooth sections of the adjoint bundle $\operatorname{Ad}(\P)=\P\times_{\mathrm{Ad}}\mathfrak{g}$. Flatness of the background connection implies that the corresponding linear gauge theory is governed by the operators
$$\P=\delta^A \d^A \qquad \text{and} \qquad \K=\d^A\,,$$
with the gauge-fixed dynamics described by the twisted de Rham-Hodge d'Alembertian $\square^A$, see Section \ref{Sec:Geometric Preliminaries on Yang-Mills Theory}. Since every globally hyperbolic spacetime admits a splitting $\M\simeq\mathbb R\times\Sigma$, where $\Sigma$ is a fixed Cauchy hypersurface, the principal bundle $\P$ may be identified with the pullback of a principal bundle $\P_\Sigma\to\Sigma$, see Lemma \ref{Lemma:BundleSigma}. After fixing such an identification, the background connection can be decomposed into temporal and spatial components, \textit{i.e.}, 
$$A=A_{\mathrm T}\,\d t+A_\Sigma(t), \qquad A_{\mathrm{T}} \doteq \partial_t \lrcorner A \in \Gamma(\M, \Ad(\P)), \quad A_\Sigma \in C^\infty(\mathbb{R}, \mathcal{C}(\P_\Sigma)),$$
where $A_\Sigma(t)$ is a time-dependent family of flat connections on $\P_\Sigma$, \textit{cf.}, Proposition \ref{Prop:DecompConn}. Thus, even when the spacetime metric is ultrastatic, this gives rise to time-dependent twisted differential operators on the Cauchy hypersurface and constitutes one of the main analytical differences from the corresponding Maxwell construction.

Our principal result is the following existence theorem. 

\begin{maintheorem} \label{Thm:GlobalHadamardYM}
Let $(\M,\g)$ be a globally hyperbolic spacetime and $\P\to\M$ be a principal $\G$-bundle with the following two properties:
\begin{itemize}
    \item[(i)]its structure group $\G$ admits bi-invariant Riemannian metrics;
    \item[(ii)]there exist flat connections on $\P$.
\end{itemize}
Then, for a given flat connection $A\in\mathcal{C}(\P)$, the CCR-algebra associated with the physical phase space $(\mathcal V_{\mathrm P},\sigma)$ of Yang-Mills theory linearised around $A$ 
admits a quasifree Hadamard state.
\end{maintheorem}

Assumption (i) is necessary to guarantee the existence of a suitable \emph{positive-definite} inner product on the corresponding Lie algebra $\mathfrak{g}$ that is compatible with the gauge symmetries. This property holds generally for any \emph{compact} Lie group, notably including the gauge group of the \emph{Standard Model of Particle Physics}. Regarding Assumption (ii), we emphasise that any manifold $(\M,\g)$ can always be equipped with the trivial principal $\G$-bundle and its corresponding trivial flat connection. If $(\M,\g)$ accommodates non-trivial principal bundles, a property contingent on both the topology of $\M$ and $\G$, further constraints on the existence of flat connections on a given non-trivial $\G$-bundle $\P$ may arise. In particular, a necessary condition is that $(\M,\g)$ must admit non-simply-connected Cauchy hypersurfaces. More details are provided in Section~\ref{Subsec:Flat Yang-Mills Connections}.

The proof rests on three main ingredients. As a preliminary step, we establish several structural properties of the classical phase space which, in the present Yang-Mills setting, were not previously available in this generality. Proposition~\ref{Prop:DegPS} characterises its weak non-degeneracy and, in particular, implies non-degeneracy whenever the Cauchy hypersurface is compact. Moreover, Proposition~\ref{Prop:SepObs} shows that the gauge-invariant observables form a separating and non-redundant family: they distinguish gauge-equivalence classes of on-shell configurations and contain no observables that vanish identically on the solution space.

Next, we implement the \emph{complete Cauchy radiation gauge} (first proposed for Maxwell's theory in~\cite{MurroSchmid}, see also the detailed review in \cite[Sec.~3.1.4]{SchmidPhD}) in the twisted bundle-valued setting. Proposition~\ref{Prop:AchCauchyRadGauge} reduces its construction, in the spatially compact case, to the solvability of a covariant Poisson equation on the Cauchy hypersurface. To remove the compactness assumption, Section~\ref{Subsec:Hodge Decomposition in Flat Vector Bundles on Complete Riemannian Manifolds} develops the corresponding Hodge theory for $\operatorname{Ad}(\P_\Sigma)$-valued forms on complete Riemannian manifolds. In particular, one obtains the sought decomposition and Poisson solvability required to construct the radiation-gauge projection also in the non-compact case. This culminates in the explicit radiation-gauge projector of Proposition~\ref{Prop:TSigmaYM}.

The last, and key technical, ingredient is the microlocal construction of Hadamard states on ultrastatic spacetimes with Cauchy hypersurfaces of bounded geometry. The residual time dependence of $A_\Sigma(t)$ prevents a direct reduction to the static pseudodifferential problem familiar from the scalar case. We overcome this by means of covariant parallel transport, see Lemma~\ref{Lem:ParallelTransportMicrolocalYM}, which identifies the time-dependent spatial problem with a family of operators acting on a fixed reference Cauchy hypersurface. This makes it possible to develop a microlocal factorisation adapted to the time-dependent twisted evolution, \textit{cf.},~Proposition~\ref{Prop:MicrolocalFactorizationYM}, and to construct the Cauchy Hadamard projectors of Theorem~\ref{Thm:HadamardProjectorsYM}. The resulting pseudo-covariances are compatible with the gauge equivalence relation, positive on the physical phase space and satisfy the Hadamard wavefront set condition recalled above.
Finally, the restriction to the ultrastatic setting is removed by a paracausal deformation argument. The essential observation, established in Lemma~\ref{Lem:FlatBackgroundMetricDeformation}, is that a flat connection remains a Yang-Mills background under arbitrary deformations of the spacetime metric. Consequently, the ultrastatic Hadamard state can be transported to an arbitrary globally hyperbolic spacetime, yielding the Main Theorem.

\paragraph{Structure of the Paper.} A synopsis of the paper is as follows. Section~\ref{Sec:Geometric Preliminaries on Yang-Mills Theory} recalls some relevant geometric preliminaries on \emph{mathematical gauge theory}, that is, the geometry of principal fibre bundles and the geometry of non-linear Yang-Mills theory. Section~\ref{Sec:The Quantisation of Linearised Yang-Mills Theory} develops linearised Yang-Mills theory within the formalism of Hack-Schenkel on possibly non-trivial principal bundles, analyses the associated classical phase space of observables and outlines its quantisation within the algebraic formalism. In particular, the notion of Hadamard states is defined therein. In Section~\ref{Sec:Complete Gauge Fixing} we define the \emph{Cauchy radiation gauge}, a complete gauge fixing for linearised Yang-Mills theory generalising the gauge condition first developed for Maxwell's theory in \cite{MurroSchmid}. To achieve this goal, we first establish the twisted Sobolev and Hodge-theoretic framework required to solve the covariant Poisson equation and extend the gauge-fixing construction to non-compact Cauchy hypersurfaces. Moreover, we discuss some analytic properties of the associated projection operator that realises the complete Cauchy radiation gauge. Finally, Section~\ref{Sec:Existence of Hadamard States for Linearised Yang-Mills Theory} is devoted to the construction of Hadamard states for Yang-Mills theory linearised around flat background connections on general globally hyperbolic spacetimes, thereby concluding the quantisation procedure. To achieve this goal, we first construct Hadamard projectors on ultrastatic spacetimes with Cauchy hypersurfaces of bounded geometry by means of covariant parallel transport and pseudodifferential microlocal factorisation. We then use a paracausal deformation argument to extend the construction to arbitrary globally hyperbolic spacetimes and prove the Main Theorem. 

Appendix~\ref{Appendix:Background Material} collects notation and background material on scalar and twisted differential forms, as well as notation and terminology used for Lie groups and connections on principal bundles. Appendix~\ref{Appendix:Poincaré Duality for Flat Vector Bundle-Valued Differential Forms} contains the twisted Poincaré duality used in the analysis of Section \ref{Subsec:Hodge Decomposition in Flat Vector Bundles on Complete Riemannian Manifolds}, while Appendix~\ref{Appendix:Pseudodifferential Calculus on Manifolds of Bounded Geometry} briefly recalls the pseudodifferential calculus on manifolds of bounded geometry employed in Section~\ref{Sec:Existence of Hadamard States for Linearised Yang-Mills Theory}. Appendix~\ref{Appendix:D} contains a list of the most important notation used throughout.

\subsection{Bibliographical Remarks}
As mentioned above, the construction of Hadamard states for linear gauge theories constitutes a much more delicate issue than in the case of ordinary field theories. For Maxwell's theory, the existence of Hadamard states has been investigated under various assumptions on the topology of the underlying spacetime $(\M\cong\mathbb{R}\times\Sigma,\g)$: Fewster-Pfenning~\cite{FewsterPfenning}, for instance, assumed $\Sigma$ to be compact with cohomology $H^{1}(\M)=0$, generalising earlier work of Furlani in \cite{FurlaniMaxwell} for static spacetimes; Finster-Strohmaier~\cite{FinsterStrohmaier}, employing the \textit{Gupta–Bleuler formalism}, considered Cauchy surfaces $\Sigma$ subjected to an ``absence of zero resonance'' condition for the Hodge Laplacian, while Dappiaggi-Siemssen~\cite{DappiaggiSiemssen} worked out the construction on asymptotically flat spacetimes using a \emph{bulk-to-boundary correspondence}, \emph{cf.}~\cite{DMPBook,GerardWrochnaBH1}. The construction of Hadamard states for Maxwell's theory on general globally hyperbolic manifolds has been established by two of us~in\cite{MurroSchmid} (\emph{cf.}, also~\cite{SchmidPhD} for a more detailed treatment) via a \emph{complete gauge-fixing} procedure, which presents the basic framework for the present article on linearised Yang-Mills theory.

The construction of Hadamard states for linearised Yang-Mills theory was first studied by Hollands~\cite{Hollands}, employing the \emph{BRST formalism}. In this work, he considered the theory linearised around the trivial connection $A=0$ of a trivial principal bundle on a spacetime with compact $\Sigma$ and with $H^{1}(\M)=0$. More recently, Gérard-Wrochna~\cite{GerardWrochna} constructed Hadamard states for Yang-Mills theory linearised around non-zero, spacelike-compact background solutions on a trivial $G$-bundle, where $G$ is a linear Lie group with compact Lie algebra. Their result applies to globally hyperbolic spacetimes $\M$ of dimension up to 4, whose Cauchy hypersurface $\Sigma$ is either compact and parallelisable or $\Sigma=\bb{R}^{3}$. In this context, it is also worth mentioning the work of Benini-Capoferri-Dappiaggi~\cite{BCD}, which addresses the construction of Hadamard states for Abelian gauge theories in the presence of a non-trivial topology. Focusing on ultrastatic spacetimes with compact Cauchy surfaces, they constructed Hadamard states by algebraically decoupling the purely dynamical degrees of freedom from the non-trivial topological sectors associated with electric and magnetic fluxes.

Yet, another notable instance of a linear gauge theory is provided by gravity linearised around a globally hyperbolic background spacetime solving the nonlinear Einstein equations. The existence of Hadamard states in the case of globally hyperbolic spacetimes of bounded geometry with compact $\Sigma$ has been established by Gérard \cite{Gerard} via an approach based on \emph{complete gauge fixing}, and recently, this approach has been adapted by Gérard-Molodyk-Wrochna to obtain Hadamard states for linearised gravity on asymptotically flat background spacetimes \cite{GerardMolodykWrochna}. The case of asymptotically flat spacetimes has earlier also been discussed by Benini-Dappiaggi-Murro via an approach based on a bulk-to-boundary correspondence \cite{BeniniMurro}. The quantisation of the latter, however, turns out to be limited to a subspace of observables due to divergences at null infinity. An approach for globally hyperbolic manifolds of bounded geometry that are analytic in Gau{\ss}ian time based on \emph{Wick rotation} has been discussed by Gérard-Murro-Wrochna in \cite{GerardMurroWrochna}. In this approach, the relevant normally hyperbolic operators are transformed into elliptic operators, which are defined in some strip in imaginary Gau{\ss}ian time. Then, one defines a quasi-free state for the Lorentzian theory from \textit{Calderón projectors} of the elliptic theory, \emph{cf.}~\cite{GerardWrochnaCalderon}. For linear gauge theories, this approach inherits a number of difficulties: the Wick rotated elliptic operators need to be invertible and to achieve this, one has to impose suitable boundary conditions. In~\cite{GerardMurroWrochna}, Dirichlet conditions for all the metric components are used, which have the advantage of easily giving invertibility of the Wick-rotated operators and a modified positivity property. However, they are not gauge invariant. Hence, the pseudostates constructed in~\cite{GerardMurroWrochna} are only gauge-invariant up to a smoothing contribution. Motivated by this problem, the spectral properties of more general gauge-invariant boundary conditions satisfying the \textit{Shapiro-Lopatinski conditions} for linearised Euclidean gravity have been discussed by Capoferri-Murro-Schmid~in \cite{SchmidCapoferriMurro}, however, none of them gives invertibility of the relevant elliptic operators on \textit{general} spacetimes. Nevertheless, building on \cite{GerardMurroWrochna}, Hadamard states for linearised gravity on de Sitter space have been constructed by Gérard-Wrochna in \cite{GerardWrochna2}, strongly relying on the fact that the Wick rotation of de Sitter spacetime is a compact Riemannian manifold.

In the specific case of Kerr spacetime, another line of research aims at circumventing the difficulties arising from gauge freedom by expressing the theory in terms of the \textit{Teukolsky fields}, see \emph{e.g.},~\cite{KleinTeu,IulianoTeu}. In particular, H\"afner and Klein~\cite{KleinTeu2} constructed the Unruh state on the physical subalgebra of the Bosonic Teukolsky fields on subextreme Kerr spacetimes and proved its Hadamard property.

The general existence of Hadamard states for linearised gravity on arbitrary background spacetimes therefore remains an important open problem, motivating the development and testing of novel techniques in other gauge theories as a stepping stone towards achieving this goal.

\subsection{General Notation}
In this section, we briefly introduce the \emph{general} notation used throughout the article. Conventions and notation concerning differential forms and Lie groups are collected in Appendix~\ref{Appendix:Background Material}.

\begin{itemize}
    \item[$\bullet$]A \textit{manifold} is a second-countable locally Euclidean Hausdorff space and hence in particular normal, paracompact and metrisable. All our manifolds are assumed to be smooth.
	\item[$\bullet$]For a given vector bundle $\E\xrightarrow{\pi}\M$, we denote by $\Gamma(\E)$ the $C^{\infty}(\sf{M})$-module of \emph{smooth sections} of $\E$ and by $\Gamma_{\mathrm{c}}(\sf{E})$ and $\Gamma_{\mathrm{sc}}(\sf{E})$ the linear subspaces consisting of \emph{compactly-supported} and \emph{spatial compactly-supported}, respectively.
	\item[$\bullet$]Let $\sf{M}$ be a smooth manifold and $(\E,\langle\cdot,\cdot\rangle_{\E})$ a pair consisting of a complex vector bundle $\E\xrightarrow{\pi}\M$ and a non-degenerate Hermitian bundle metric $\langle\cdot,\cdot\rangle_{\E}$. If $\sf{M}$ is equipped with a pseudo-Riemannian metric $\sf{g}$, for any $\psi,\varphi\in\Gamma(\sf{E})$ such that $\mathrm{supp}(\psi)\cap\mathrm{supp}(\varphi)$ is compact, we denote by
	\begin{align}\label{eq:MetricSections}
		(\psi,\varphi)_{\sf{E}}:=\int_{\sf{M}}\,\langle\psi(p),\varphi(p)\rangle_{\sf{E}_{p}}\,\d\mu_{\sf{g}}(p)
	\end{align}
	 the induced non-degenerate sesquilinear form on the level of sections. Here $\mu_{\sf{g}}\:\mathcal{L}(\sf{M})\to[0,\infty]$ denotes the usual (Riemann-Lebesgue) volume measure of $(\sf{M},\sf{g})$ defined on the Lebesgue $\sigma$-algebra $\mathcal{L}(\sf{M})$ of $\sf{M}$.
	\item[$\bullet$]For $\bb{K}\in\{\bb{R},\bb{C}\}$, we denote the \emph{trivial $\bb{K}$-vector bundle} of rank $k\in\mathbb{N}^{\ast} \defeq \mathbb{N} \setminus \{0\}$ by 
	\begin{align*}
		\underline{\mathbb{K}}^{k}_{\M}:=\M\times\mathbb{K}^{k}\quad\xrightarrow{\mathrm{pr}_{1}}\quad\sf{M}\, .
	\end{align*}
	\item[$\bullet$]The $C^{\infty}(\sf{M})$-modules of \emph{smooth vector fields}, \emph{smooth $k$-forms} and \emph{$\sf{E}$-valued} (or \textit{$\E$-twisted}) \textit{smooth $k$-forms} on $\M$ are respectively denoted by
	\begin{align*}
		\mathfrak{X}(\sf{M}):=\Gamma(\sf{T}\sf{M})\,,\quad\Omega^{k}(\sf{M}):=\Gamma\bigg(\bigwedge^{k}\T^{\ast}\M\bigg)\,\quad\text{and}\quad \Omega^{k}(\sf{M},\E):=\Gamma\bigg(\bigwedge^{k}\T^{\ast}\M\otimes \sf{E}\bigg)\,.
	\end{align*}
  Moreover, we shall use the subscript $c$ to denote smooth $k$-forms and $\E$-valued $k$-forms with compact support.
	\item[$\bullet$]\emph{Einstein's summation convention} for repeated indices will be employed throughout.
\end{itemize}

\paragraph{Acknowledgments and Funding:} 
We are grateful for inspiring discussions with Claudio Dappiaggi and Felix Finster on gauge-theoretic aspects of Yang-Mills theory. B.C. and S.M. are partially supported by the INdAM - GNFM Project ``Hadamard states for linearized Yang Mills theories'', CUP E5324001950001. The work of B.C. and of R.D.S. has been supported by a fellowship of the University of Pavia. R.D.S. is grateful to the Department of Mathematics at the University of York for the kind hospitality during the realization of part of this work. The authors acknowledge both the support of INFN, Sezione di Pavia and Sezione di Genova, and of Gruppo Nazionale di Fisica Matematica, part of INdAM. 

\paragraph{AI Disclosure:} During the final stage of the preparation, AI-models were used for linguistic editing, identifying mathematical inaccuracies and potentially missing literature. All mathematical statements and proofs were designed and verified independently by the authors, including the review and assessment of relevant literature.
\section{Geometric Preliminaries on Yang-Mills Theory}
\label{Sec:Geometric Preliminaries on Yang-Mills Theory}
In the following, let $(\M,\g)$ be a $(1+d)$-dimensional globally hyperbolic spacetime, \textit{i.e.}, a Lorentzian manifold of the form
 \begin{align*}
      \M=\R\times\Sigma\,,\qquad\qquad \g=-\beta^2\d t\otimes\d t+\h_{t}\,,
 \end{align*}
where $\beta\in C^{\infty}(\M,(0,\infty))$ is a \emph{lapse function},  each slice $\Sigma_t := \{t\}\times\Sigma$ is a smooth spacelike Cauchy hypersurface, and $(h_t)_{t\in\mathbb R}$ is a smoothly varying family of Riemannian metrics on $\Sigma$.
 Moreover, let $\G$ be a Lie group\footnote{In Appendix~\ref{Appendix:Notation and Terminology for Lie Groups}, the reader can find some basic definitions concerning Lie groups that will be used throughout the work.} with corresponding Lie algebra $(\mathfrak{g},[\cdot,\cdot]_{\mathfrak{g}})$. The Lie group $\G$ serves as the \emph{structure group} of Yang–Mills theory. Henceforth, we do not assume $\G$ to be simple or to admit a faithful matrix representation, but we only require the following:

\begin{assumption}\label{Assumption}
    There exists a \emph{positive definite} inner product $\kappa(\cdot,\cdot)\:\mathfrak{g}\times\mathfrak{g}\to\mathbb{R}$ on $\mathfrak{g}$ that is, in addition, \emph{$\mathrm{Ad}$-invariant}, \textit{i.e.},~$\kappa(\mathrm{Ad}_{g}(X),\mathrm{Ad}_{g}(Y))=\kappa(X,Y)$ for all $g\in\G$ and $X,Y\in\mathfrak{g}$.
\end{assumption}

\begin{remark} Note that $\mathrm{ad}_{Z}$-invariance of $\kappa$ implies that each $\mathrm{ad}_{Z}$ is skew-adjoint with respect to $\kappa$, namely, $\kappa(\mathrm{ad}_{Z}(X),Y)=-\kappa( X,\mathrm{ad}_{Z}(Y))$ for all $X,Y,Z\in\mathfrak{g}$. 
\end{remark}

Observe that connected Lie groups satisfying Assumption~\ref{Assumption} can be completely classified. For the sake of the exposition, we first illustrate a few examples of $\mathrm{Ad}$-invariant inner products on important classes of Lie groups before stating their general classification in Proposition \ref{Prop: Lie groups classification} below.

\begin{examples}\label{Ex:Assumption} Some notable examples of $\mathrm{Ad}$-invariant inner products are listed here: 
    \begin{itemize}
        \item[(i)]Consider the torus $\mathbb{T}^{n}:=\mathbb{R}^{n}/\mathbb{Z}^{n}$ with its canonical Lie group structure. Any positive-definite inner product on its Lie algebra $\mathbb{R}^{n}$  is trivially $\operatorname{Ad}$-invariant since $\mathbb{T}^n$ is Abelian.
        \item[(ii)]Let $\G$ be a compact Lie group and let $\langle\cdot,\cdot\rangle_{\mathfrak{g}}$ be an arbitrary positive-definite inner product on its Lie algebra $\mathfrak{g}$. Then,
        \begin{align*}
            \kappa(X,Y):=\int_{\G}\langle\mathrm{Ad}_{g}(X),\mathrm{Ad}_{g}(Y)\rangle_{\mathfrak{g}}\,\mathrm{d}\mu_{\G}(g)
        \end{align*}
        defines an $\mathrm{Ad}$-invariant inner product on $\mathfrak{g}$, where $\mu_{\G}$ denotes the \emph{Haar measure} of $\G$.
        \item[(iii)]By definition, a Lie algebra $\mathfrak{g}$ is called \emph{semisimple} if it has no non-zero Abelian ideals. By Cartan's criterion, this is equivalent to the fact that the \emph{Killing form} $(X,Y)\mapsto\mathrm{tr}(\mathrm{ad}_X \circ \mathrm{ad}_Y)$ on $\mathfrak{g}$ is non-degenerate. It turns out that the Killing form is \emph{negative-definite} if and only if $\mathfrak{g}$ is a \emph{compact} semisimple Lie algebra, \textit{i.e.},~if it is the Lie algebra of a \emph{compact} semisimple Lie group. Hence, for any compact semisimple\footnote{It should be noted that the term ``(semi)simple'' for Lie \emph{groups} is not used uniformly in the literature. Here, we follow the Lie-theoretic rather than abstract group-theoretic terminology and call a connected Lie group $\G$ (semi)simple if its Lie algebra $\mathfrak{g}$ is (semi)simple. In particular, we allow for non-trivial centers $\mathcal{Z}(\G)$.} Lie group, a negative multiple of the Killing form defines an Ad-invariant inner product on $\mathfrak{g}$. 
        
        If the compact Lie algebra $\mathfrak{g}$ is \emph{simple}, \textit{i.e.}, it is non-Abelian and has no non-trivial ideals, it turns out that \emph{any} Ad-invariant and positive-definite inner product is necessarily a (negative) multiple of the Killing form, see~\cite[Thm.~2.5.3]{Hamilton}. 
        \item[(iv)]The group $\mathrm{U}(n)$ is compact but not semisimple. An Ad-invariant inner product on its Lie algebra is given by $\kappa(X,Y):=-\mathrm{tr}(XY)$ for $X,Y \in \mathfrak{u}(n)$, where $\mathfrak{u}(n)=\{X\in\mathbb{C}^{n\times n}\mid X+X^{\dagger}=0\}$.
    \end{itemize}
\end{examples}

\begin{proposition}
\label{Prop: Lie groups classification}
Let $\G$ be a finite-dimensional \emph{connected} Lie group with Lie algebra $\mathfrak{g}$. Then, the following statements are equivalent:
    \begin{itemize}
        \item[(i)]$\mathfrak{g}$ admits a positive-definite $\mathrm{Ad}$-invariant inner product $\kappa$, \textit{i.e.},~$\G$ satisfies Assumption~\ref{Assumption}.
        \item[(ii)]$\G$ admits a \emph{bi-invariant Riemannian metric}, namely, a Riemannian metric $s\in\Gamma(\T^{\ast}\G^{\otimes_{s}2})$ with the property that $L_{g}^{\ast}s=R_{g}^{\ast}s=s$ for all $g\in\G$. Here $L_{g}$ and $R_{g}$ denote respectively the left and right multiplication of $\G$, see Appendix \ref{Appendix:Notation and Terminology for Lie Groups}.
        \item[(iii)]$\G$ is of the form 
         \begin{align*}
        \G\cong\G_{\mathrm{c}}\times\mathbb{R}^{n}\,,
    \end{align*}
     where $\G_{\mathrm{c}}$ is a \emph{compact} and \emph{connected} Lie group and where $(\mathbb{R}^{n},+)$ for some $n\in\mathbb{N}$ denotes the (Abelian) additive vector group.
    \end{itemize}
\end{proposition}

\begin{proof}
    Claim $(ii) \Rightarrow(i)$ is obvious. Indeed, if $s\colon\T\G\times\T\G\to\mathbb{R}$ is a bi-invariant Riemannian metric on $\G$, then its restriction to $\T_e\G\cong\mathfrak{g}$ defines an $\mathrm{Ad}$-invariant inner product on $\mathfrak{g}$. Claim $(i) \Rightarrow (ii)$ follows from a straightforward computation and we refer to \cite[Thm.~2.2.2]{Hamilton} for further details. 

    For $(iii) \Rightarrow (i)$, we can simply consider the direct sum of an arbitrary inner product on $\mathbb{R}^{n}$ with an $\mathrm{Ad}$-invariant inner product on the Lie algebra of $\G_{\mathrm{c}}$ as defined in Example~\ref{Ex:Assumption}. Claim $(ii) \Rightarrow (iii)$ was established by Milnor in \cite[Lemma~7.5]{Milnor}.
\end{proof}

\begin{remark}
    For completeness, we recall the well-known \emph{classification theorem of compact Lie groups}: every compact and connected Lie group $\G_{\mathrm{c}}$ is of the form
    \begin{align*}
        \G_{\mathrm{c}}\cong(\mathbb{T}^{k}\times\G_{1}\times\dots\times\G_{l})/\mathfrak{Z}\,,
    \end{align*}
    where $\mathfrak{Z}$ is a finite central subgroup, $\mathbb{T}^{k}$ is a torus and each $\G_{i}$ is a \emph{simply-connected compact simple Lie group}, \textit{i.e.},~$\G_{i}$ either isomorphic to one (and only one) of the matrix Lie groups: 
    \begin{align*}
        \mathrm{SU}(n+1),\,\,n\geq 1\qquad \mathrm{Spin}(2n+1),\,\,n\geq 2\qquad \mathrm{Sp}(n),\,\,n\geq 3\qquad \mathrm{Spin}(2n),\,\,n\geq 4\,,
    \end{align*}
    or one of the \emph{exceptional} Lie groups $G_{2}$, $F_{4}$, $E_{6}$, $E_{7}$ and $E_{8}$. We refer to \cite[Sec.~V.8.]{Brocker} for details.

    Now, as explained in item (iii) in Example~\ref{Ex:Assumption}, every $\mathrm{Ad}$-invariant inner product on a simple compact Lie group is necessarily a negative multiple of the Killing form. In particular, this implies that every $\mathrm{Ad}$-invariant inner product on a compact Lie group of the form $\G_{\mathrm{c}}:=\mathbb{T}^{k}\times\G_{1}\times\dots\times\G_{l}$ with $\G_{i}$ simple and compact is of the form (\textit{cf.},~\cite[Thm.~2.5.4]{Hamilton})
    \begin{align*}
        \kappa=\langle\cdot,\cdot\rangle_{\mathfrak{t}^{k}}\oplus c_{1}(-\kappa_{1})\oplus\dots\oplus c_{l}(-\kappa_{l})
    \end{align*}
    where $\langle\cdot,\cdot\rangle_{\mathfrak{t}^{k}}$ is an inner product on $\mathfrak{t}^{k}$, the Lie algebra of $\mathbb{T}^{k}$, $\kappa_{i}$ are the Killing forms on $\mathfrak{g}_{i}$ and $c_{i}\in\mathbb{R}_{>0}$ are positive constants, usually referred to as \emph{coupling constants}.
\end{remark}

\begin{example}[The Standard Model of Particle Physics]
A noteworthy, physically relevant example of a compact Lie group is the gauge group of the celebrated \emph{standard model of particle physics} given by 
\begin{align*}
    \G_{\mathrm{SM}}=(\mathrm{SU}(3)\times\mathrm{SU}(2)\times\mathrm{U}(1))/\mathbb{Z}_{6}.
\end{align*}
By the previous discussion, we note that $\G_{\mathrm{SM}}$ satisfies Assumption~\ref{Assumption}.
\end{example}

It is important to stress that Assumption~\ref{Assumption} generally fails to hold for \emph{arbitrary} non-compact Lie groups. For instance, the Lie algebras of $\mathrm{SO}_0(1,3)$ and $\mathrm{SL}(2,\mathbb{R})$ do not admit positive-definite $\mathrm{Ad}$-invariant inner products, as one can easily see from the characterisation in Proposition~\ref{Prop: Lie groups classification}(iii). We remark that Yang-Mills theories based on such gauge groups do appear in some physical theories, for instance, in \emph{MacDowell-Mansouri gravity} \cite{Mac1,Mac2}, which employs the non-compact gauge groups $\mathrm{SO}(1,4)$ or $\mathrm{SO}(2,3)$ to formulate general relativity as a gauge theory.\footnote{We remark that the compactness of $\G$ also plays a role in ensuring a well-behaved background theory, particularly concerning the well-posedness of the Cauchy problem for non-linear Yang-Mills theory. For instance, for the non-compact Lie group $\G=\mathrm{SU}(2,\mathbb{C})$, there exist finite-time blow-up solutions on Minkowski spacetime \cite{Yang}.}

\subsection{Yang-Mills Theory on Globally Hyperbolic Spacetimes}\label{Subsec:Yang-Mills Theory on Globally Hyperbolic Spacetimes}
In this section, we present the preliminary definitions necessary for the ensuing discussion of Yang-Mills theories on non-trivial principal bundles. We refer the reader to \cite{Baum,Hamilton,RudolphSchmidt2,Taubes} for further details on the geometric structures discussed below. 

Let $\P\xrightarrow{\pi_{\P}}\M$ be a smooth principal $\G$-bundle over a globally hyperbolic spacetime $(\M,\g)$ with Lie group $\G$ satisfying Assumption~\ref{Assumption}. For the sake of notational ease, we shall simply denote by ``$\cdot$'' the free and transitive right-group action $\P_{x}\times\G\to\P_{x}$ at a point $x\in\M$. Moreover, we denote by $r_{g}\:\P\to\P$ the map $r_{g}(p):=p\cdot g$ for all $g\in\G$. Let us further introduce the \emph{adjoint bundle}, \emph{i.e.},~the associated vector bundle 
\begin{align}\label{eq:BundleMetricInduced}
    \mathrm{Ad}(\P):=\P\times_{\mathrm{Ad}}\mathfrak{g}\, .
\end{align}
We recall that the fibre of $\mathrm{Ad}(\P)$ at a point $x\in\M$ is, by definition, the quotient space $\mathrm{Ad}(\P)_{x}:=(\P_{x}\times\mathfrak{g})/\G$ defined with respect to the group action $(p,X)\cdot g = (p\cdot g,\mathrm{Ad}_{g^{-1}}(X))$ for all $p\in\P_{x}$, $g\in\G$ and $X\in\mathfrak{g}$. Such an equivalence class will be denoted by $[p,X]\in\mathrm{Ad}(\P)_{x}$. The adjoint bundle is naturally equipped with a bundle metric $\langle\cdot,\cdot\rangle_{\mathrm{Ad}(\P)}$ induced from $\kappa$ and defined by
\begin{align}\label{eq:InducedBundleMetric}
	\langle [p,X],[p,Y]\rangle_{\mathrm{Ad}(\P)_{x}}:=\kappa(X,Y)
\end{align}
for all $x\in\M$, $p\in\P_{x}$ and $X,Y\in\mathfrak{g}$, which is well-defined by $\mathrm{Ad}$-invariance of $\kappa$. 

Let us next establish the notation and define the relevant inner products for the space of $\mathrm{Ad}(\P)$-twisted differential forms on $\M$.\footnote{We refer to Appendix~\ref{Appendix: Notation and Conventions for Scalar and Twisted Differential Forms} for various details and conventions used for scalar and twisted differential forms.} To start with, we equip the exterior bundles with their usual (Hodge) bundle metrics, \textit{i.e.},~we consider the pseudo-Riemannian vector bundles
\begin{align*}
    \bigwedge^{k}\T^{\ast}\M \,,\qquad \langle\omega,\eta\rangle_{\g}:=\frac{1}{k!}(\g^{-1})^{\otimes k}(\omega,\eta),
\end{align*}
and, following the general notation introduced in Eq.~\eqref{eq:MetricSections}, we denote the induced non-degenerate symmetric bilinear form on the level of sections $\Omega_c^{k}(\M)=\Gamma_c(\bigwedge^{k}\T^{\ast}\M)$ by 
\begin{align*}
    (\alpha,\beta)_{\g}:=\int_{\M}\langle\alpha,\beta\rangle_{\g}\,\d\mu_{\g}=\int_{\M}\alpha\wedge\ast_{\g}\,\beta,
\end{align*}
for all $\alpha,\beta\in\Omega_{\mathrm{c}}^{k}(\M)$. With this notation, we define the pseudo-Riemannian vector bundles
\begin{align}\label{Eq.DefEk1}
    \E_{k}:=\bigwedge^{k}\T^{\ast}\M\otimes\mathrm{Ad}(\P)\,,\qquad \langle\omega\otimes \varphi,\eta\otimes\psi\rangle_{\E_{k}}:=\langle\omega,\eta\rangle_{\g}\langle \varphi,\psi\rangle_{\mathrm{Ad}(\P)}\,,
    \end{align}
whose sections are $\mathrm{Ad}(\P)$-valued differential forms, \emph{i.e.}, $\Gamma(\E_{k})=\Omega^{k}(\M,\mathrm{Ad}(\P))$. Here, we identified $\Omega^{k}(\M,\mathrm{Ad}(\P))\cong\Omega^{k}(\M)\otimes_{C^{\infty}(\M)}\Gamma(\mathrm{Ad}(\P))$ and used the fact that this space is generated by pure tensors of the form $\omega\otimes\varphi$ with $\omega\in\Omega^{k}(\M)$ and $\varphi\in\Gamma(\mathrm{Ad}(\P))$. At the level of sections, the bundle metric $\langle\cdot,\cdot\rangle_{\E_{k}}$ induces a corresponding non-degenerate symmetric bilinear form, in analogy to $(\cdot,\cdot)_{\g}$ above, denoted by
\begin{align}\label{Eq.DefEk2}
    (\omega,\eta)_{\E_{k}}:=\int_{\M}\langle\omega,\eta\rangle_{\E_{k}}\,\d\mu_{\g}=\int_{\M}\langle\omega\wedge\ast_{\g}\,\eta\rangle_{\mathrm{Ad}(\P)},
\end{align}
for $\omega,\eta\in\Omega_{\mathrm{c}}^{k}(\M,\mathrm{Ad}(\P))$, where $\langle\cdot\wedge\cdot\rangle_{\mathrm{Ad}(\P)}\:\Omega^{k}(\M,\mathrm{Ad}(\P))\times\Omega^{l}(\M,\mathrm{Ad}(\P))\to\Omega^{k+l}(\M)$ denotes the wedge-type product induced by $\langle\cdot,\cdot\rangle_{\mathrm{Ad}(\P)}$ (see Example~\ref{Example:WedgeBundleMetric}), while $\ast_{\g}\:\Omega^{k}(\M,\E)\to\Omega^{(1+d)-k}(\M,\E)$ is the Hodge $\ast$-operator that for bundle-valued forms is defined to be the unique linear operator satisfying the relation $\langle\omega\wedge\ast_{\g}\,\eta\rangle_{\mathrm{Ad}(\P)}=\langle\omega,\eta\rangle_{\E_{k}}\mathrm{vol}_{\g}$, generalising the scalar-valued definition. 

Now, let us denote by $\mathcal{C}(\P)\subset\Omega^{1}(\P,\mathfrak{g})$ the space of \emph{connection $1$-forms} on $\P$, whose definition we recall briefly in Appendix~\ref{Appendix:Connections on Principal Bundles}. For a fixed connection $A\in\mathcal{C}(\P)$, we consider the linear differential operators $\mathrm{D}_{A} \:\Omega^{k}(\P,\mathfrak{g})\to\Omega^{k+1}(\P,\mathfrak{g})$ defined by
\begin{align}\label{Eq.Def.OpDA}
    (\mathrm{D}_{A}\omega)(X_{1},\dots,X_{k+1}):=\d\omega(\mathrm{pr}_{A}(X_{1}),\dots,\mathrm{pr}_{A}(X_{k+1}))
\end{align}
for all $X_{i}\in\mathfrak{X}(\P)$, where $\mathrm{pr}_{A}\:\T\P\to\H$ denotes the projection onto the horizontal tangent bundle $\sf{H}:=\mathrm{ker}(A)$ and where $\d\:\Omega^{k}(\P,\mathfrak{g})\to\Omega^{k+1}(\P,\mathfrak{g})$ is the usual exterior derivative of (vector-valued) differential forms, \textit{i.e.}, denoting by $\{t_a\}_{a=1, \ldots, \mathrm{dim}(\mathfrak{g})}$ a basis of the Lie algebra $\mathfrak{g}$, \begin{align*}\d \omega = (\d\omega^a)\otimes t_{a}\,, \qquad \omega=\omega^{a}\otimes t_{a}\quad\text{with}\quad\omega^a \in \Omega^k (\P).\end{align*} Employing this notation, we define the \emph{curvature of $A$} by setting $F^{A}:=\mathrm{D}_{A}A\in\Omega^2(\P,\mathfrak{g})$. It is straightforward to verify that $F^{A}$ can be computed by means of \emph{Cartan's structure equation}
\begin{align*}
    F^{A}=\d A+\frac{1}{2}[A\wedge A]_{\mathfrak{g}}\, ,
\end{align*}
where $[\cdot\wedge\cdot]_{\mathfrak{g}}\:\Omega^{k}(\P,\mathfrak{g})\times\Omega^{l}(\P,\mathfrak{g})\to\Omega^{k+l}(\P,\mathfrak{g})$ denotes the wedge product induced from $[\cdot,\cdot]_{\mathfrak{g}}$, see also Example~\ref{Ex:WedgeLieAlg1} and Remark~\ref{Ex:WedgeLieAlg2} in the appendix for details.

As a next step, we recall the well-known fact that a suitable subspace of $\Omega^{k}(\P,\mathfrak{g})$ can be identified with $\Omega^{k}(\M,\mathrm{Ad}(\P))$. This identification is crucial, as it allows us to define linearised Yang-Mills fields and the curvature of a Yang-Mills connection, which originally live on the principal bundle $\P$, globally on the base manifold $\M$. More precisely, there is an isomorphism\footnote{For the sake of completeness: $(\omega_{\M})_{x}(v_{1},\dots,v_{k}):=[p,\omega_{p}(X_{1},\dots,X_{k})]$ for all $x\in\M$, $p\in\P_{x}$ and $v_{1},\dots,v_{k}\in\T_{x}\M$, where $X_{i}\in\T_{p}\P$ are such that $\d(\pi_{\P})_{p}(X_{i})=v_{i}$, see, \textit{e.g.},~\cite[Satz 3.5]{Baum} or \cite[Thm.~5.13.4]{Hamilton}.}
\begin{align}\label{eq:Iso}
    \Omega^{k}_{\mathrm{hor},\mathrm{Ad}}(\P,\mathfrak{g})\xrightarrow{\cong}\Omega^{k}(\M,\mathrm{Ad}(\P))\,,\qquad \omega\mapsto\omega_{\M}\,,
\end{align}
where $\Omega^{k}_{\mathrm{hor},\mathrm{Ad}}(\P,\mathfrak{g})$ is the linear subspace of $\Omega^{k}(\P,\mathfrak{g})$ consisting of forms $\omega$ that are \emph{horizontal} and of \emph{type $\mathrm{Ad}$}, \textit{i.e.},~the set of all $\omega\in\Omega^{k}(\P,\mathfrak{g})$ for which the following holds:
\begin{align*}
    \text{(i)}&\quad \omega(X_{1},\dots,X_{k})=0\qquad\text{if at least one $X_{i}\in\mathfrak{X}(\P)$ is vertical}\\
    \text{(ii)}&\quad (r_{g}^{\ast}\omega)_{p}=\mathrm{Ad}_{g^{-1}}\circ \omega_{p}, 
\end{align*}
where we recall that $r_g(p):=p\cdot g$ is the right-group action. 
It is easy to see that $F^{A}\in\Omega^{2}_{\mathrm{hor},\mathrm{Ad}}(\P,\mathfrak{g})$ for every $A\in\mathcal{C}(\P)$. Hence, the curvature can be identified with a $2$-form, which we denote by 
\begin{align*}
    F_{\M}^{A}\in\Omega^{2}(\M,\mathrm{Ad}(\P))\, .
\end{align*}

For a fixed connection $A\in\mathcal{C}(\P)$, we introduce, moreover, the following differential operators acting on the $C^{\infty}(\M)$-module of $\mathrm{Ad}(\P)$-valued differential forms:
\begin{itemize}
    \item[$\bullet$]For $\omega\in \Omega^{k}_{\mathrm{hor},\mathrm{Ad}}(\P,\mathfrak{g})$, it is easy to verify that $\mathrm{D}_{A}\omega=\d\omega+[A\wedge\omega]_{\mathfrak{g}}$, which in particular implies that also $\mathrm{D}_{A}\omega\in\Omega^{k+1}_{\mathrm{hor},\mathrm{Ad}}(\P,\mathfrak{g})$. Hence, in view of the isomorphism in Equation~\eqref{eq:Iso}, we obtain a family of differential operators
    \begin{align*}
        \d^{A}\:\Omega^{k}(\M,\mathrm{Ad}(\P))\to \Omega^{k+1}(\M,\mathrm{Ad}(\P))\,,\qquad \d^{A}\omega_{\M}=(\mathrm{D}_{A}\omega)_{\M}\, .
    \end{align*}
    The restriction $\nabla^{A}:=\d^{A}\vert_{\Omega^{0}}\:\Gamma(\mathrm{Ad}(\P))\to\Omega^{1}(\M,\mathrm{Ad}(\P))$ defines a connection on $\mathrm{Ad}(\P)$ and it is easy to check that $\d^{A}$ is the \emph{exterior covariant derivative} induced by $\nabla^{A}$, \emph{i.e.}, $\d^{A}=\d^{\nabla^{A}}$ and hence
    \begin{align*}
		(\d^{A}\omega)(X_{1},\dots,X_{k+1})=&\sum_{i}(-1)^{i-1}\nabla_{X_{i}}^{A}(\omega(X_{1},\dots,\hat{X_{i}},\dots,X_{k+1}))\\&+\sum_{i<j}(-1)^{i+j}\omega([X_{i},X_{j}],X_{1},\dots,\hat{X_{i}},\dots,\hat{X_{j}},\dots,X_{k+1})
\end{align*}
for all $X_{i}\in\mathfrak{X}(\M)$, where the symbol $\hat{X}_{i}$ indicates that $X_i$ is omitted from the sum. 
    \item[$\bullet$]Let us denote the formal adjoint of $\d^{A}$ with respect to $(\cdot,\cdot)_{\E_{\bullet}}$ by $\delta^{A}$. It is explicitly given by
    \begin{align*}
        \delta^{A}\:\Omega^{k+1}(\M,\mathrm{Ad}(\P))\to\Omega^{k}(\M,\mathrm{Ad}(\P))\,,\qquad \delta^{A}=(-1)^{k+1}\ast_{\g}^{-1}\d^{A}\ast_{\g}\, .
    \end{align*}
	If we denote the natural connection on $\E_{k}=\bigwedge^{k}\T^{\ast}\M\otimes\mathrm{Ad}(\P)$ induced from $\nabla^{A}$ and the Levi-Civita connection $\nabla$ on $\M$ by $\nabla^{A,\g}$, it holds that  
	\begin{align}\label{eq:CoDiff}
     (\delta^{A}\omega)(X_{1},\dots,X_{k})=-\g^{\alpha\beta}(\nabla^{A,\g}_{\partial_{\alpha}}\omega)(\partial_{\beta},X_{1},\dots,X_{k})
 \end{align}  
for all $X_{i}\in\mathfrak{X}(\M)$, $i=1, \ldots, k$, where $(\mathcal{U},x^{\alpha})$ is a local chart on some open subset $\mathcal{U}\subset\M$.
    \item[$\bullet$]We denote the $\mathrm{Ad}(\P)$-valued \emph{de Rham-Hodge d'Alembertian} by 
     \begin{align*}
\square^{A}:=\d^{A}\delta^{A}+\delta^{A}\d^{A}\:\Omega^{k}(\M,\mathrm{Ad}(\P))\to\Omega^{k}(\M,\mathrm{Ad}(\P))\, .
     \end{align*} 
 This operator is \emph{normally hyperbolic} and it is related up to a sign to the connection d'Alembertian associated to $\nabla^{A,\g}$ by means of a Weitzenb\"ock identity, \textit{i.e.},
     \begin{align*}
         \square^{A}=-\square^{\nabla^{A,\g}}+\mathcal{R}_{A}+\mathcal{R}_{\g}\,,
     \end{align*}
     where $\mathcal{R}_{A}$ and $\mathcal{R}_{\g}$ are zeroth-order operators involving the curvature of $A$ and Riemann curvature of $(\M,\g)$, respectively, and where
     \begin{align*}
         \square^{\nabla^{A,\g}}:=\mathrm{tr}_{\g}((\nabla^{A,\g})^{2}):=-(\nabla^{A,\g})^{\ast}\nabla^{A,\g}=\g^{\alpha\beta}(\nabla^{A,\g}_{\partial_{\alpha}}\nabla^{A,\g}_{\partial_{\beta}}-\nabla^{A,\g}_{\nabla_{\partial_{\alpha}}\partial_{\beta}})
     \end{align*}
     denotes the \emph{connection} or \emph{rough d'Alembertian}, see \emph{e.g.}~\cite[Cor.~2.7.21]{RudolphSchmidt2} for detail.
\end{itemize}

For direct applications, it is often useful to have local descriptions of the operators $\d^{A}$ and $\delta^{A}$. In the context of gauge theory, we call a section $s\in\Gamma(\mathcal{U},\P)$ on some open subset $\mathcal{U}\subset\M$ a \emph{local gauge}. Local gauges play the role of trivialisations of $\P$, since any such $s$ defines a local trivialisation by \begin{align}\label{eq:TrivialisationsGauges}\Phi_{s}\:\mathcal{U}\times\G\to\pi_{\P}^{-1}(\mathcal{U}), \qquad (x,g)\mapsto s(x)\cdot g,\end{align} and vice versa. Now, fixing a local gauge, we can identify connection $1$-forms and $\mathrm{Ad}(\P)$-valued forms \emph{locally} with $\mathfrak{g}$-valued forms on $\M$, \textit{i.e.}, 
\begin{align*}
    \mathcal{C}(\P\vert_{\mathcal{U}})&\xrightarrow{\cong}\Omega^{1}(\mathcal{U},\mathfrak{g})\,,\qquad A\mapsto A_{s}:=s^{\ast} A\\
    \Omega^{k}(\mathcal{U},\mathrm{Ad}(\P\vert_{\mathcal{U}}))&\xrightarrow{\cong}\Omega^{k}(\mathcal{U},\mathfrak{g})\,,\qquad\alpha\mapsto\alpha_{s}\,,
\end{align*}
where $\alpha_{s}:=s^{\ast}\alpha$ when viewed as a form on $\P$ via the isomorphism in Eq.~\eqref{eq:Iso}.\footnote{Equivalently, we note that the local gauge $s$ together with a basis $(t_{a})_{a=1,\dots,\mathrm{dim}(\mathfrak{g})}$ of $\mathfrak{g}$ induces a local frame $e_{a}(x):=[s(x),t_{a}]$ of $\mathrm{Ad}(\P)$ on $\mathcal{U}$. Then, writing $\alpha=\alpha^{a}\otimes e_{a}$ for $\alpha^{a}\in\Omega^{k}(\mathcal{U})$, we have $\alpha_{s}=\alpha^{a}\otimes t_{a}$.} The forms $\alpha_{s}$ and $A_{s}$ are referred to as the \emph{local representatives of $\alpha$ and $A$ in the gauge $s$}, respectively.

\begin{lemma}
Let $s\in\Gamma(\mathcal{U},\P)$ be a local gauge and let $A\in\mathcal{C}(\P)$ be a connection 1-form. Then, for any $\omega\in\Omega^{k}(\M,\mathrm{Ad}(\P))$, it holds:
    \begin{align*}
        \text{(i)}\qquad & (\d^{A}\omega)_{s}=\d\omega_{s}+[A_{s}\wedge\omega_{s}]_{\mathfrak{g}}\\
        \text{(ii)}\qquad & (\delta^{A}\omega)_{s}=\delta\omega_{s}-[A_{s}^{\sharp}\lrcorner\omega_{s}]_{\mathfrak{g}}\,,
    \end{align*}
    where $\d$ and $\delta:=(-1)^{k}\ast^{-1}_{\g}\d\ast_{\g}$ denote the exterior derivative and codifferential for $\mathfrak{g}$-valued forms and where $[\cdot\lrcorner\cdot]_{\mathfrak{g}}\:\mathfrak{X}(\mathcal{U},\mathfrak{g})\times\Omega^{k}(\mathcal{U},\mathfrak{g})\to\Omega^{k-1}(\mathcal{U},\mathfrak{g})$ is the bilinear pairing that acts as the interior product $\lrcorner$ on the differential form part and as the Lie bracket $[\cdot,\cdot]_{\mathfrak{g}}$ on the Lie algebra part.
\end{lemma}

\begin{proof}
    Claim (i) is standard and follows from a straightforward computation. We refer to \cite[Thm.~5.12.9]{Hamilton} for details. For (ii), we take a basis $(t_{a})_{a=1,\dots,\mathrm{dim}(\mathfrak{g})}$ of $\mathfrak{g}$ and consider the corresponding local frame $e_{a}(x):=[s(x),t_{a}]$ of $\mathrm{Ad}(\P)$. By definition, if we write $\delta^{A}\omega=(\delta^{A}\omega)^{a}\otimes e_{a}$ for local components $(\delta^{A}\omega)^{a}\in\Omega^{k-1}(\mathcal{U})$, then $(\delta^{A}\omega)_{s}=(\delta^{A}\omega)^{a}\otimes t_{a}$. Hence, it remains to compute $(\delta^{A}\omega)^{a}$. Choosing a local chart $(x^{\alpha})_{\alpha=0,\dots,d}$ on $\mathcal{U}$ and writing $\omega=\omega^{a}\otimes e_{a}$ on $\mathcal{U}$, we obtain from Eq.~\eqref{eq:CoDiff},
    \begin{align*}
        \delta^{A}\omega&=-\g^{\alpha\beta}\partial_{\beta}\lrcorner\nabla^{A,\g}_{\partial_{\alpha}}\omega=-\g^{\alpha\beta}\partial_{\beta}\lrcorner\nabla^{A,\g}_{\partial_{\alpha}}(\omega^{a}\otimes e_{a})=\\&=-(\g^{\alpha\beta}\partial_{\beta}\lrcorner\nabla_{\partial_{\alpha}}\omega^{a})\otimes e_{a}-(\g^{\alpha\beta}\partial_{\beta}\lrcorner\omega^{a})\otimes\nabla^{A}_{\partial_{\alpha}}e_{a}=\\&=\delta\omega^{a}\otimes e_{a}-(\g^{\alpha\beta}\partial_{\beta}\lrcorner\omega^{a})\otimes\nabla^{A}_{\partial_{\alpha}}e_{a}\,.
    \end{align*}
    The local gauge representative of the first term is exactly given by $(\delta\omega^{a}\otimes e_{a})_{s}=\delta\omega^{a}\otimes t_{a}=\delta\omega_{s}$. It remains to compute the local gauge representative of $\nabla^{A}_{\partial_{\alpha}}e_{a}$. Noting that $(e_{a})_{s}=t_{a}$, we obtain from item (i) applied to the case of zero-forms, $(\nabla^{A}_{\partial_{\alpha}}e_{a})_{s}=[A_{s}(\partial_{\alpha}),t_{a}]=(A_{s})^{c}_{\alpha}[t_{c},t_{a}]$ and putting everything together, we conclude that $(\delta^{A}\omega)_{s}=\delta\omega_{s}-[A_{s}^{\sharp}\lrcorner\omega_{s}]_{\mathfrak{g}}$, as claimed.
\end{proof}

Using the notation introduced above, we establish the following terminology.

\begin{definition}
    \label{Def:YangMillsCon}
    A \emph{Yang-Mills connection} is a solution $A\in\mathcal{C}(\P)$ to the non-linear Yang-Mills equations $\delta^{A}F_{\M}^{A}=0$, where $F^{A}_{\M}\in\Omega^{2}(\M,\mathrm{Ad}(\P))$ denotes the curvature of $A$.
\end{definition}

The nonlinear Yang-Mills equations $\delta^{A}F_{\M}^{A}=0$ are not hyperbolic. However, they admit a \emph{gauge symmetry}, which allows to formulate a well-posed initial problem. More precisely, a \emph{(global) gauge transformation} is a principal bundle automorphism, namely, a diffeomorphism $f\:\P\to\P$ that preserves the fibres, \textit{i.e.}, $\pi_{\P}\circ f=\pi_{\P}$, and that is $\G$-equivariant, \textit{i.e.}, $f(p\cdot g)=f(p)\cdot g$ for all $p\in\P$ and $g\in\G$. Let us denote the group of all gauge transformations, the \emph{gauge group}, by $\mathcal{G}(\P)$. This group can be identified with the group of $\G$-equivariant smooth maps on $\P$, that is,
\begin{align*}
    \mathcal{G}(\P)\xrightarrow{\,\,\cong\,\,} C^{\infty}(\P,\G)^{\G}:=\{\sigma\in C^{\infty}(\P,\G)\mid \sigma(p\cdot g)=g^{-1}\sigma(p)g\}\,,\qquad f\mapsto\sigma_{f}\,,
\end{align*}
where $\sigma_{f}$ is uniquely determined by $f(p):=p\cdot\sigma_{f}(p)$ for all $p\in\P$. On connection $1$-forms, the gauge group acts via pull-back, \emph{i.e.},
\begin{align*}
	\mathcal{C}(\P)\times \mathcal{G}(\P)\to\mathcal{C}(\P)\,,\qquad (A,f)\mapsto f^{\ast}A=\mathrm{Ad}_{\sigma_{f}^{-1}}\circ A+\sigma_{f}^{\ast}\mu\,,
\end{align*}
where $\mu\in\Omega^{1}(\G,\mathfrak{g})$ denotes the Maurer-Cartan form of $\G$. On the level of the adjoint bundle, the gauge group acts as 
\begin{align*}
	\mathcal{G}(\P)\times\mathrm{Ad}(\P)_{x}\to\mathrm{Ad}(\P)_{x}\,,\qquad (f,[p,X])\mapsto f\cdot [p,X]:=[f(p),X]=[p,\mathrm{Ad}_{\sigma_{f}(p)}(X)]\,,
\end{align*}
which extends to sections of $\Gamma(\mathrm{Ad}(\P))$ and, more generally, differential forms in $\Omega^{k}(\M,\mathrm{Ad}(\P))$ when applied pointwise. In view of the isomorphism in Equation~\eqref{eq:Iso}, it is possible to show that this action is compatible with the pull-back in the sense that
\begin{align*}
	f^{-1}\cdot\omega_{\M}=(f^{\ast}\omega)_{\M}
\end{align*}
for all $f\in\mathcal{G}(\P)$ and $\omega\in\Omega^{k}_{\mathrm{hor},\mathrm{Ad}}(\P,\mathfrak{g})$. The Yang-Mills equations are thus gauge-invariant in the sense that
\begin{align*}
\delta^{f^{\ast}A}F^{f^{\ast}A}_{\M}=f^{-1}\cdot\delta^{A}F^{A}_{\M}\, .
\end{align*}

Employing this invariance, one can rewrite the non-linear Yang-Mills equations as a hyperbolic problem subject to constraints and prove that the corresponding Cauchy problem is globally well-posed on globally hyperbolic spacetimes. For connected Lie groups with compact Lie algebra and sufficiently regular initial data satisfying the Gauss constraint, Chrusciel-Shatah in~\cite{ChruscielShatah} proved global well-posedness in smooth globally hyperbolic spacetimes of dimension 4. Earlier work includes Choquet-Bruhat’s analysis of Yang–Mills–Higgs fields in $2+1$ dimensions in \cite{ChoquetBruhatYangMills} and by various authors with different techniques for the Einstein cylinder $\mathbb{R}\times S^{3}$, \emph{e.g.}, \cite{ChristodoulouEC,Christodoulou,ChoquetBruhatEC,ChoquetBruhatPaneitzSegal}. On Minkowski spacetime, local and global existence results were obtained for various classes of initial data and couplings to other fields, see  \emph{e.g.}, \cite{SegalYangMills,GinibreVelo1,GinibreVelo2}  for local existence, \cite{EardleyM1,EardleyM2,Klainerman,Goganov,Burzlaff} for global existence and also \cite{GlasseyStrauss} for a particular class of global solutions.

\subsection{Flat Yang-Mills Connections}\label{Subsec:Flat Yang-Mills Connections}
In the present work, we focus on a special class of Yang-Mills configurations. Specifically, we consider Yang-Mills theory linearised around a \emph{flat} connection, \emph{i.e.}, a connection $A\in\mathcal{C}(\P)$ satisfying 
\begin{align*}
    F^{A}=0\,.
\end{align*}
Before discussing the linearised theory, let us therefore take a closer look at such configurations. Flat connections are trivially solutions to the nonlinear Yang-Mills equations, namely, $\delta^{\A}F^{\A}=0$. From a physical perspective, this corresponds to the choice of a \emph{vacuum configuration}, \textit{cf.}, the discussion in \cite[Sec.~8.1.2]{Hamilton}. This naturally raises the question of the existence of flat connections on a given principal bundle and underlying spacetime manifold. In this regard, we have to distinguish between two cases.

\paragraph{Trivial Bundles.} First of all, on every smooth manifold $\M$, we can always choose a \emph{trivial principal $\G$-bundle}, \emph{i.e.}, a principal $\G$-bundle $\P\to\M$ that is isomorphic as a principal bundle to the trivial $\G$-bundle $\M\times\G$. It is well-known that a principal bundle $\P$ is trivial if and only if there exists a \emph{global gauge}, \emph{i.e.}, a global section $s\in\Gamma(\M,\P)$. Indeed, in this case, an isomorphism $\Phi_{s}\:\M\times\G\to\P$ can be defined by $\Phi_{s}(x,g):=s(x)\cdot g$, as in Eq.~\eqref{eq:TrivialisationsGauges}, and we can identify the space of connections in this gauge accordingly as
\begin{align*}
    \mathcal{C}(\P)\xrightarrow{\cong} \Omega^{1}(\M,\mathfrak{g})\,,\qquad A\mapsto A_{s}:=s^{\ast}A\,.
\end{align*}
The inverse of this map constructs a connection $A \in \mathcal{C}(\P)$ from a 1-form $\omega\in\Omega^{1}(\M,\mathfrak{g})$ by first defining a connection 1-form $\bar{A}$ on the trivial bundle $\M\times\G$, given by
\begin{align*}
\bar{A}_{(x,g)}(v,w):=\mathrm{Ad}_{g^{-1}}(\omega_{x}(v))+\mu_{g}(w)
\end{align*}
for $(x,g)\in\M\times\G$, $(v,w)\in\T_{(x,g)}(\M\times\G)$ and $\omega\in\Omega^{1}(\M,\mathfrak{g})$, and by then setting $A:=(\Phi_{s}^{-1})^{\ast}\bar{A}$. In this setting, one can always consider the trivial 1-form $\omega\equiv 0$, leaving $\bar{A}$ non-zero exclusively along the $G$-directions, which yields the \emph{trivial flat connection} $\bar{A}:=\mathrm{pr}_{2}^{\ast}\mu$, \textit{i.e.},
\begin{align*}
    A:=(\mathrm{pr}_{2}\circ\Phi_{s}^{-1})^{\ast}\mu\in\mathcal{C}(\P)
\end{align*}
with $\mathrm{pr}_{2}\:\M\times\G\to\G$ being the projection. The fact that this indeed defines a \emph{flat} connection follows easily from the \emph{Maurer-Cartan equation} $\d\mu+\frac{1}{2}[\mu\wedge\mu]_{\mathfrak{g}}=0$. In the Ehresmann picture, the horizontal tangent bundle $\H:=\mathrm{ker}(A)\subset \T\P$ is simply the tangent space to the constant-$g$ slices, that is, $\H_{(x,g)} \cong \T_x\M \oplus \{0\}$. Moreover, the induced vector bundle connection $\nabla^{A}$ on $\mathrm{Ad}(\P)$, which in this case is a trivial vector bundle, \textit{i.e.}, $\mathrm{Ad}(\P)\cong\M\times\mathfrak{g}$, coincides with the \emph{trivial vector bundle connection} $\nabla^{A}\psi=\d\psi$. In other words, its local connection coefficients vanishes in this gauge.

\paragraph{Non-trivial Bundles.} The situation is different in the non-trivial scenario. First and foremost, it is worth recalling that not every manifold admits a non-trivial principal bundle. Indeed, \emph{smooth principal $\G$-bundles} can be classified by the homotopy classes of continuous maps $f\:\M\to\B\G$ with $\B\G$ being the \emph{classifying space of $\G$}, see, \emph{e.g.},~\cite[Thm.~3.6.7]{RudolphSchmidt2}. Hence, the existence of non-trivial principal bundles crucially depends both on the topology of $\M$ and $\G$. 

\begin{examples}
    \begin{itemize}
        \item[(i)]Every principal bundle over a \emph{contractible} manifold $\M$ is trivial. Indeed, the claim follows immediately from the fact that a homotopy equivalence $f\:\M\to\N$ between two smooth manifolds $\M,\N$ induces a bijection between isomorphism classes of principal bundles via pullback,  see, \emph{e.g.},~\cite[Cor.~3.3.5]{RudolphSchmidt2}.
        \item[(ii)]If $\M$ is a smooth connected manifold of dimension $d\leq 3$ and $\G$ is a \emph{simply-connected} Lie group, then any principal $\G$-bundle over $\M$ is trivial, see for instance~\cite[Lemma 2.1]{Freed}. In particular, if $(\M,\g)$ is a $(1+3)$-dimensional globally hyperbolic spacetime and we choose $\G$ to be simply connected, then any principal $\G$-bundle on $\M$ is trivial, since any principal bundle on $\M$ can be obtained as the pullback from a principal bundle of some spatial Cauchy surface (see Lemma~\ref{Lemma:BundleSigma} below for details). Note that this is the case for $\G=\mathrm{SU}(n)$ with $n\geq 2$, which is of particular physical relevance in the \emph{Standard Model of particle physics}.
    \end{itemize}
\end{examples}

Observe that if $\M$ is a given smooth manifold and $\P\to\M$ a non-trivial principal bundle, then there might be further obstructions to the existence of flat connections. Indeed, it is well-known that the non-trivial $S^{1}$-principal \emph{Hopf bundle} $S^{3}\to S^{2}$ does not admit flat connections. More generally, one can show via \emph{Chern-Weil theory} that a necessary (but not sufficient) condition for the existence of a flat connection on a given principal bundle is that all its real Chern–Weil characteristic classes of positive degree vanish, we refer to \cite[Chap.~2]{Morita} for additional details.

Now, to get a better feeling of how the above discussion applies to flat connections, let us consider the space $\mathcal{M}_{\G}(\P):=\mathcal{C}(\P)/\mathcal{G}(\P)$ of gauge-equivalence classes of connections and the subspace 
\begin{align*}
    \mathcal{M}_{\G,\mathrm{flat}}(\P):=\frac{\mathcal{C}_{\mathrm{flat}}(\P)}{\mathcal{G}(\P)}\,,\qquad \mathcal{C}_{\mathrm{flat}}(\P):=\{A\in\mathcal{C}(\P)\mid F^{A}=0\}
\end{align*}
\emph{i.e.}, the \emph{moduli space of flat connections}. As discussed above, $\mathcal{M}_{\G,\mathrm{flat}}(\P)$ may as well be empty. Now, with this definition, we consider the space of equivalence classes of flat bundles, \textit{i.e.}, 
\begin{align*}
    \mathcal{F}_{\G}(\M):=\{(\P,A)\mid \text{$\P$ principal $\G$-bundle over $\M$ and $A\in\mathcal{C}(\P)$\text{ flat }}\}/_{\sim}=\bigcup_{[\P]} \mathcal{M}_{\G,\mathrm{flat}}(\P)\,,
\end{align*}
where we identify $(\P,A)\sim (\P^{\prime},A^{\prime})$ if there is a principal bundle isomorphism $f\:\P\to\P^{\prime}$ with $f^{\ast}A^{\prime}=A$ and where the union on the right-hand side ranges over all equivalence classes of principal $\G$-bundles over $\M$. Assume that $\sf{M}$ is connected.
Then, one can show that there is a bijection
\begin{align}\label{eq:FlatIso}
    \mathcal{F}_{\G}(\M)\cong\frac{\mathrm{Hom}(\pi_{1}(\M),\G)}{\G}\,,
\end{align}
where $\pi_{1}(\M)$ denotes the fundamental group of $\M$ and where $\G$ acts on $\mathrm{Hom}(\pi_{1}(\M),\G)$ by conjugation. The bijection is induced from the map that assigns to a given flat connection $A$ its holonomy representation $\rho\:\pi_{1}(\M)\to\G$. A proof can, for instance, be found in \cite[Sec.~13.9]{Taubes}. 

As a direct consequence of the foregoing discussion, we arrive at the following conclusion: upon restricting attention to flat background connections, the consideration of non-trivial principal bundles necessitates a non-trivial topology of the spatial hypersurface $\Sigma$ of a globally hyperbolic spacetime $(\M,\g)$.

\begin{proposition}
    Let $(\M,\g)$ be a globally hyperbolic spacetime that admits a simply-connected Cauchy surface $\Sigma$. Then, any principal $\G$-bundle over $\M$ that admits a flat connection is trivial.
\end{proposition}

\begin{proof}
    By \cite[Thm.~1.2]{BernalSanchez}, we can find a Cauchy temporal function $t\:\M\to\mathbb{R}$ such that $\Sigma=t^{-1}(0)$ and hence $\M=\mathbb{R}\times\Sigma$. Now, since $\Sigma$ is simply-connected, we have $\pi_{1}(\M)\cong\pi_{1}(\Sigma)=\{0\}$. Hence, using the isomorphism in Eq.~\eqref{eq:FlatIso}, we conclude that $\mathcal{F}_{\G}(\M)$ only contains one element, namely the equivalence class induced from the trivial principal bundle $\M\times\G$ with trivial flat connection.
\end{proof}

As a final remark, we note that any flat connection $A\in\mathcal{C}(\P)$ on some principal $\G$-bundle $\P$ coincides \emph{locally} with the trivial flat connection discussed before, \emph{i.e.}, around every point $x\in\M$, there is a local trivialisation $\psi\:\mathcal{U}\times\G\to\pi^{-1}_{\P}(\mathcal{U})$ such that the restricted connection on $\pi_{\P}^{-1}(\mathcal{U})$ coincides with the trivial flat connection on $\mathcal{U}\times\G$ via pullback by $\psi$, see \cite[Chap.~II, Thm.~9.1]{KobayashiNomizuI}.
\section{The Quantisation of Linearised Yang-Mills Theory}
\label{Sec:The Quantisation of Linearised Yang-Mills Theory}
Having introduced the geometric framework for gauge theories in Section~\ref{Sec:Geometric Preliminaries on Yang-Mills Theory}, we now turn our attention to the quantisation of linearised Yang--Mills theory. In this section, we formulate linearised Yang--Mills theory as a linear gauge theory within the framework of Hack-Schenkel \cite{HackSchenkel}, examine several equivalent formulations of its classical phase space, and finally discuss its quantisation in the framework developed by Hack-Schenkel \cite{HackSchenkel}.  

\subsection{Linearised Yang-Mills Theory in the Hack-Schenkel Formalism}\label{Subsec:Linearised Yang-Mills Theory in the Hack-Schenkel Formalism}
We first discuss the linearisation of the Yang-Mills equations. For convenience and having in mind the quantisation procedure, we work in a \emph{complexified setting} as clarified by the following notational remark. 

\begin{notation} 
For the sake of the ensuing discussion, all objects introduced in the previous sections are understood to be complexified. Thus, for instance, $$ \mathfrak{g} \mapsto \mathfrak{g}_{\mathbb{C}} \defeq \mathfrak{g}\otimes_{\mathbb{R}}\mathbb{C}, \quad \kappa(\cdot, \cdot) \mapsto \kappa_{\mathbb{C}} (\cdot, \cdot), \quad \operatorname{Ad} (\P_\Sigma) \mapsto \operatorname{Ad}(\P_{\Sigma})_\mathbb{C} \defeq \operatorname{Ad}(\P_\Sigma) \otimes_{\mathbb{R}} \mathbb{C},$$ and similarly for the associated constructions, \textit{e.g.}, Lie brackets and wedge products. To avoid unnecessarily burdening the notation, we suppress the subscript $\mathbb C$ throughout the remainder of the paper, with complexification understood unless explicitly stated otherwise. 
\end{notation}

Recalling that the space of connections $\mathcal{C}(\P)$ is an infinite-dimensional affine space modelled over $\Omega^{1}(\M,\mathrm{Ad}(\P))$, we consider a formal perturbation 
\begin{align}\label{eq:FormalPert}
    A\mapsto\hat{A}(\lambda):=A+\lambda\alpha+\mathcal{O}(\lambda^{2})\qquad\text{with}\quad\alpha:=\frac{\d}{\d\lambda}\bigg\vert_{\lambda=0}\hat{A}(\lambda)\in \Omega^{1}(\M,\mathrm{Ad}(\P))
\end{align}
with background $A=\hat{A}(0)\in\mathcal{C}(\P)$. We define the \emph{linearised Yang-Mills operator} $\P\:\Omega^{1}(\M,\mathrm{Ad}(\P))\to \Omega^{1}(\M,\mathrm{Ad}(\P))$ to be the linear differential operator given by
\begin{align*}
    \P\alpha:=\frac{\d}{\d\lambda}\bigg\vert_{\lambda=0}\delta^{\hat{A}(\lambda)}F_{\M}^{\hat{A}(\lambda)}\,\qquad\textit{i.e.}\qquad \delta^{\hat{A}(\lambda)}F_{\M}^{\hat{A}(\lambda)}=\delta^{A}F_{\M}^{A}+\lambda(\P\alpha)+\mathcal{O}(\lambda^{2})\, .
\end{align*}
With the above definition, linearised Yang-Mills theory is an example of a linear gauge theory in the sense of Hack-Schenkel \cite{HackSchenkel} in the specialised setting of Gérard-Wrochna \cite{GerardWrochna}; see also \cite[Chapter 2]{SchmidPhD} for a thorought review, \cite{WrochnaZahn} for a comparison to the (linearised) BRST formalism and \cite{KhavkinePhaseSpace} for a closely related formalism. More precisely, we have the following result. 

\begin{proposition}[\protect{Trivial bundles:\cite[Ex.~3.7]{HackSchenkel}, General case: \cite[Prop.~2.30]{SchmidPhD} and \cite[Sec.~4.2]{WrochnaZahn}}]\label{Prop:LinYangMills}  Let $A\in\mathcal{C}(\P)$ be an arbitrary connection $1$-form, see Appendix \ref{Appendix:Connections on Principal Bundles}. Then, the linearised Yang-Mills operator reads
\begin{align*}
    \P\:\Omega^{1}(\M,\mathrm{Ad}(\P))\to \Omega^{1}(\M,\mathrm{Ad}(\P))\,,\qquad\P=\delta^{A}\d^{A}+\mathcal{F}^{A}\,,
\end{align*}
where $\mathcal{F}^{A}$ is an operator of order zero given by $\mathcal{F}^{A}=-\ast_{\g}[\ast_{\g} F^{A}_{\sf{M}}\wedge\cdot]_{\mathrm{Ad}(\P)}$, with $[\cdot\wedge\cdot]_{\mathrm{Ad}(\P)}$ the natural wedge-product on $\Omega^{\bullet}(\M,\mathrm{Ad}(\P))$ induced from $[\cdot,\cdot]_{\mathfrak{g}}$.\footnote{The Lie bracket $[\cdot,\cdot]_{\mathfrak{g}}$ gives rise to a fibre-wise pairing $[\cdot,\cdot]_{\mathrm{Ad}(\P)_{x}}\:\mathrm{Ad}(\P)_{x}\times\mathrm{Ad}(\P)_{x}\to\mathrm{Ad}(\P)_{x}$ defined by $[[p,X],[p,Y]]_{\mathrm{Ad}(\P)_{x}}:=[p,[X,Y]_{\mathfrak{g}}]_{\mathrm{Ad}(\P)_{x}}$ for $x\in\M$, $p\in\P_{x}$ and $X,Y\in\mathfrak{g}$, which in turn induces the corresponding wedge-type product $[\cdot\wedge\cdot]_{\mathrm{Ad}(\P)}\:\Omega^{k}(\M,\mathrm{Ad}(\P))\times\Omega^{l}(\M,\mathrm{Ad}(\P))\to\Omega^{k+l}(\M,\mathrm{Ad}(\P))$, see~Appendix~\ref{Appendix: Notation and Conventions for Scalar and Twisted Differential Forms}.} Let us further introduce the operator 
\begin{align*}
    \K\:\Omega^{0}(\M,\mathrm{Ad}(\P))\to \Omega^{1}(\M,\mathrm{Ad}(\P))\,,\qquad\K:=\d^{A}\, .
\end{align*}
Then, provided that the background $A$ is a Yang-Mills connection as per Def.~\ref{Def:YangMillsCon}, and denoting by $\K^{\ast}$ the formal adjoint of $\K$ with respect to $(\cdot,\cdot)_{\E_{k}}$, the following hold true:
\begin{itemize}
    \item[(i)]$\P$ is formally self-adjoint with respect to $(\cdot,\cdot)_{\sf{E}_{k}}$ and $\P\circ\K=0$.
    \item[(ii)]The linear differential operators $\mathsf{D}_{0}:=\K^{\ast}\K$ and $\mathsf{D}_{1}:=\P+\K\K^{\ast}$ are explicitly given by
    \begin{align*}
        \mathsf{D}_{0}=\square^{A}\,,\qquad\mathsf{D}_{1}=\square^{A}+\mathcal{F}^{A}
    \end{align*}
    and hence, in particular, they are normally hyperbolic.
\end{itemize}
\end{proposition}

An explicit derivation of the linearised equations can be found in \cite[Sec.~2.2.2]{SchmidPhD}, \textit{cf.},~also \cite{AtiyahBott}.

\begin{remark}
Although the formal perturbation theory used in Eq.~\eqref{eq:FormalPert} suffices to derive the linearised theory, we note that this can also be described in a more geometric fashion. As a matter of fact, the space of connections $\mathcal{C}(\P)$ is an affine space over $\Omega^{1}(\M,\mathrm{Ad}(\P))$ (see Appendix~\ref{Appendix:Connections on Principal Bundles}) and thus carries a natural structure of an $\infty$-dimensional (Fréchet) manifold with tangent space 
    \begin{align*}
        \T_{A}\mathcal{C}(\P) \cong \Omega^{1}(\M, \mathrm{Ad}(\P))
    \end{align*}
    at any $A\in\mathcal{C}(\P)$. The gauge group $\mathcal{G}(\P) \cong C^{\infty}(\P, \G)^{\G}$, on the other hand, carries the structure of an $\infty$-dimensional (Fréchet) Lie group with Lie algebra $\mathfrak{g}(\P)\cong\T_{\mathrm{id}}\mathcal{G}(\P) \cong \Omega^{0}(\M, \mathrm{Ad}(\P))=\Gamma(\mathrm{Ad}(\P))$. Its \emph{exponential map} is given by
    \begin{align*}
        \exp_{\mathcal{G}} \: \mathfrak{g}(\P) \to \mathcal{G}(\P)\,,\qquad (\exp_{\mathcal{G}}(\varepsilon_{\M}))(p) = p \cdot \exp_{\G}(\varepsilon(p))
    \end{align*}
    for $\varepsilon_{\M} \in \Gamma(\mathrm{Ad}(\P))$ and $\varepsilon \in C^{\infty}(\P, \mathfrak{g})$ defined via the isomorphism in Eq.~\eqref{eq:Iso} applied to zero-forms, \textit{i.e.},~$\varepsilon_{\sf{M}}(x)=[x,\varepsilon(x)]$, see for instance~\cite[Sec.~2.4.4]{SchmidGroups}. Choosing a curve $\Phi_{t}:=\mathrm{exp}_{\mathcal{G}}(t\varepsilon_{\M})$ in $\mathcal{G}(\P)$, it is straightforward to verify that $\frac{\d}{\d t}(\Phi^{\ast}_{t}A)_{\M}\vert_{t=0}=\d^{A}\varepsilon_{\M}$, which shows that $\K=\d^{A}$ is indeed the correct operator parametrising gauge transformations at linear order.
\end{remark}

The operator $\K$ encompasses the gauge transformations at the linearised level. In particular, the condition $\P\circ\K=0$ corresponds to the invariance of $\P$ under linear gauge transformations 
\begin{align*}
	\alpha\mapsto\alpha - \K\varepsilon
\end{align*}
for $\alpha\in\Omega^{1}(\M,\mathrm{Ad}(\P))$ and $\varepsilon\in\Omega^{0}(\M,\mathrm{Ad}(\P))$. Furthermore, it encodes the \emph{canonical gauge condition} $\K^{\ast}\alpha=\delta^{A}\alpha=0$, which is called the \emph{Lorenz gauge}. The operator $\mathsf{D}_{1}$ is the gauge-fixed operator with respect to this gauge condition, while $\mathsf{D}_{0}$ describes the remaining gauge freedom. More precisely, the space of equivalence classes of solutions to $\P\alpha=0$ can be identified with the space of solutions to the hyperbolic problem $\mathsf{D}_{1}\alpha=0$ subject to the gauge constraint $\K^{\ast}\alpha=0$, \emph{i.e.},
\begin{align*}
	\cfrac{\mathrm{ker}(\sf{P}\vert_{\Gamma_{\mathrm{sc}}})}{\mathrm{ran}(\sf{K}\vert_{\Gamma_{\mathrm{sc}}})}\cong \cfrac{\mathrm{ker}(\sf{D}_{1}\vert_{\Gamma_{\mathrm{sc}}})\cap\mathrm{ker}(\sf{K}^{\ast}\vert_{\Gamma_{\mathrm{sc}}})}{\sf{K}(\mathrm{ker}(\sf{D}_{0}\vert_{\Gamma_{\mathrm{sc}}}))}\,,
\end{align*}
see, \emph{e.g.}, \cite[Prop.~2.7]{GerardWrochna} or \cite[Prop.~2.5]{SchmidPhD}. The denominator on the right-hand side codifies the fact that a gauge-transformation $\varepsilon\in\Omega^{0}(\M,\mathrm{Ad}(\P))$ preserves the Lorenz gauge if and only if $\mathsf{D}_{0}\varepsilon=0$.

Let us now turn to the analysis of the classical phase space associated with linearised Yang-Mills theory.

\begin{definition}[Classical Phase Space]
\label{Def: classical phase space}
    Let $A\in\mathcal{C}(\P)$ be an arbitrary connection $1$-form satisfying the nonlinear Yang-Mills equations and consider the operators $\P$, $\K$ and $\mathsf{D}_{k}$, $k=0,1$, as defined in Proposition~\ref{Prop:LinYangMills}. Moreover, let us denote by $\G_{k}^{\pm}\; :\Gamma_{\mathrm{c}}(\E_{k})\to\Gamma_{\mathrm{sc}}(\E_{k})$ the retarded $(+)$/advanced $(-)$ Green operators\footnote{We remind the reader that these are the unique linear operators satisfying $\mathsf{D}_{k}\circ\G_{k}^{\pm}=\G_{k}^{\pm}\circ\mathsf{D}_{k}=\mathrm{id}$ on $\Gamma_{\mathrm{c}}(\sf{E}_{k})$ and $\mathrm{supp}(\G^{\pm}_{k}\varphi)\subset\mathcal{J}^{\pm}(\mathrm{supp}(\varphi))$ for all $\varphi\in\Gamma_{\mathrm{c}}(\sf{E}_{k})$. We refer to \cite{BaerGreen} for further details.} of $\mathsf{D}_{k}$ and by
    \begin{align*}
        \G_{k}:=\G_{k}^{+}-\G_{k}^{-}\:\Gamma_{\mathrm{c}}(\E_{k})\to\Gamma_{\mathrm{sc}}(\E_{k})
    \end{align*}
    the corresponding \emph{causal propagators}. Then, the pair $(\mathcal{V}_{\mathrm{P}},\sigma)$ defined by
	\begin{align*}
		 \mathcal{V}_{\mathrm{P}}=\cfrac{\mathrm{ker}(\sf{K}^{\ast}\vert_{\Gamma_{c}})}{\mathrm{ran}(\sf{P}\vert_{\Gamma_{c}})}\,,\qquad \sigma\:\mathcal{V}_{\mathrm{P}}\times\mathcal{V}_{\mathrm{P}}\to\bb{C}\,,\quad \sigma([\psi],[\varphi]):=(\psi,\mathsf{i}\sf{G}_1\varphi)_{\sf{E}_{1}}
	\end{align*}
	is referred to as the \textit{classical phase space}.
\end{definition}
\noindent The space $\mathcal{V}_{\mathrm{P}}$ can be understood as the space of \emph{linear, gauge-invariant, on-shell observables} of the theory. Via the causal propagator, the space $\mathcal{V}_{\mathrm{P}}$ is isomorphic to the space of solutions modulo gauge transformations, as clarified by the following proposition.

\begin{proposition}[\textit{e.g.}~\protect{\cite[Prop.~2.7]{GerardWrochna}}]
\label{Prop:IsoPS}
In the same setting of Definition \ref{Def: classical phase space}, it holds that
    \begin{align*}
        \mathcal{V}_{\mathrm{P}}:=\cfrac{\mathrm{ker}(\sf{K}^{\ast}\vert_{\Gamma_{c}})}{\mathrm{ran}(\sf{P}\vert_{\Gamma_{c}})}\quad\xrightarrow[\cong]{[\G_{1}]}\quad\cfrac{\mathrm{ker}(\sf{P}\vert_{\Gamma_{\mathrm{sc}}})}{\mathrm{ran}(\sf{K}\vert_{\Gamma_{\mathrm{sc}}})}\, .
    \end{align*}
\end{proposition}

We stress that $\sigma$ is in general \emph{degenerate}, as opposed to ordinary field theories without a linear gauge symmetry. This point has been discussed in some details by Hack-Schenkel in \cite[Sec.~5]{HackSchenkel}. In fact, it has been shown therein that $\sigma$ is degenerate whenever $\mathrm{ran}(\sf{K}\vert_{\Gamma_{\mathrm{sc}}})\subset \mathrm{ran}(\sf{K})\cap\Gamma_{\mathrm{sc}}(\sf{E}_{1})$ is a \textit{proper} subspace. Moreover, in \cite[Ex.~5.7]{HackSchenkel}, the authors showed that this is in fact a sufficient condition for Yang-Mills theory \emph{on trivial principal bundles linearised around the trivial connection $A=0$}. Using a generalised Poincaré duality for $\mathrm{Ad}(\P)$-valued forms as discussed in Appendix~\ref{Appendix:Poincaré Duality for Flat Vector Bundle-Valued Differential Forms} of the present article, this can be generalised to the non-trivial case as follows.

\begin{proposition}\label{Prop:DegPS}
    Consider Yang-Mills theory linearised around an arbitrary \emph{flat} background connection. Then, the classical phase space $(\mathcal{V}_{\mathrm{P}},\sigma)$ is weakly\footnote{A Hermitian sesquilinear form $b\:\V\times\V\to\bb{C}$ on some $\mathbb{C}$-vector space is called \emph{weakly non-degenerate} if the induced map $b\:\V\to \overline{\V}^{\ast}$ is injective and \emph{strongly non-degenerate} if it is additionally surjective and, hence, a linear isomorphism. This distinction is clearly only meaningful in the case $\mathrm{dim}_{\mathbb{C}}(\V)=\infty$.} non-degenerate if and only if
    \begin{align*}
        \mathrm{ran}(\sf{K})\cap\Gamma_{\mathrm{sc}}(\sf{E}_{1})=\mathrm{ran}(\sf{K}\vert_{\Gamma_{\mathrm{sc}}})\, .
    \end{align*}
\end{proposition}

\begin{proof}
The direct implication, namely, the fact that non-degeneracy implies $\mathrm{ran}(\sf{K})\cap\Gamma_{\mathrm{sc}}(\sf{E}_{1})=\mathrm{ran}(\sf{K}\vert_{\Gamma_{\mathrm{sc}}})$, is true for general linear gauge theories and the content of \cite[Cor.~5.3]{HackSchenkel}. To establish the reverse implication, suppose instead that $\mathrm{ran}(\sf{K})\cap\Gamma_{\mathrm{sc}}(\sf{E}_{1})=\mathrm{ran}(\sf{K}\vert_{\Gamma_{\mathrm{sc}}})$. Now, let $\psi\in\mathrm{ker}(\sf{K}^{\ast}\vert_{\Gamma_{c}})$ be such that $\sigma([\psi],[\varphi])=0$ for all $[\varphi]\in\mathcal{V}_{\mathrm{P}}$. In other words, $\psi\in\Omega^{1}_{\mathrm{c}}(\M,\mathrm{Ad}(\mathrm{P}))$ is such that $\delta^{A}\psi=0$ and it has the property that
    \begin{align*}
        (\varphi,\sf{G}_1\psi)_{\sf{E}_{1}}=\int_{\sf{M}}\langle\varphi\wedge\ast_{\g}\,\sf{G}_1\psi\rangle_{\mathrm{Ad}(\P)}=0
    \end{align*}
    for all $\varphi\in\Omega^{1}_{\mathrm{c}}(\M,\mathrm{Ad}(\mathrm{P}))$ with $\delta^{\A}\varphi=0$. Using the fact that, as a byproduct of the flatness of $A$, $\d^{A}\circ\d^{A}=0$ and, hence, $\delta^{A}\circ\delta^{A}=0$ as well as $\d^{A}\circ\square^{A}=\square^{A}\circ\d^{A}$ and $\delta^{A}\circ\square^{A}=\square^{A}\circ\delta^{A}$, smoothness implies that
    \begin{align*}
0=(\delta^{A}\omega,\G_{1}\psi)_{\E_{1}}=(\omega,\d^{A}\G_{1}\psi)_{\E_{2}}
    \end{align*}
    for all $\omega\in\Omega^{2}_{\mathrm{c}}(\M,\mathrm{Ad}(\P))$. Non-degeneracy of $(\cdot,\cdot)_{\E_{2}}$ implies that $\d^{A}\sf{G}_1\psi=0$. Hence, by Poincaré duality for $\mathrm{Ad}(\P)$-valued forms, Proposition~\ref{Prop: Poincare duality}, we conclude that there exists $\varepsilon\in\Omega^{0}(\M,\mathrm{Ad}(\P))$ such that $\sf{G}_1\psi=\d^{A}\varepsilon$. By assumption, we can hence find $\varepsilon_{\mathrm{c}}\in\Omega^{0}_{\mathrm{sc}}(\M,\mathrm{Ad}(\P))$ such that $\sf{G}_1\psi=\d^{A}\varepsilon_{\mathrm{c}}$. Injectivity of the map $[\G_{1}]$ in Proposition~\ref{Prop:IsoPS} implies that $\psi=\P\eta$ for some $\eta\in\Omega^{1}_{\mathrm{c}}(\M,\mathrm{Ad}(\P))$, which entails that $[\psi]=0$ in $\mathcal{V}_{\mathrm{P}}$.
\end{proof}

\begin{remark}
    The previous proposition implies that when linearised around flat connections, $(\mathcal{V}_{\mathrm{P}},\sigma)$ is always non-degenerate in the case in which $\Sigma$ is compact. This generalises the well-known analogous fact for Maxwell's theory as well as~\cite[Ex.~5.7]{HackSchenkel} for Yang-Mills theory on trivial principal bundles linearised around the trivial connection $A=0$. We stress that a similar statement, namely, the fact that the classical phase space is non-degenerate in the case of compact $\Sigma$, is also known for linearised gravity \cite[Thm.~4.3]{FewsterHunt}.
    
    In the case in which $\Sigma$ is non-compact, this is not true in general as can be seen from the characterisation in Proposition~\ref{Prop:DegPS} and there are examples of spacetimes on which $\sigma$ is degenerate (see \cite[Ex.~5.7]{HackSchenkel}). Let us also remark that these degeneracies of $\sigma$ are of physical significance. In fact, it has been argued that the degeneracy of $\sigma$ for Maxwell's theory is related to the celebrated \emph{Aharonov-Bohm effect}, see \cite{DappiaggiAharonov}.
\end{remark}

Beyond the issue of degeneracy, yet another question arises in the context of linear gauge theories, namely, whether $(\mathcal{V}_{\mathrm{P}},\sigma)$ indeed represents the ``correct'' space of linear observables for the gauge-equivalence classes of on-shell field configurations. In this context, we define for each element $[\psi]\in\mathcal{V}_{\mathrm{P}}$ of the phase space a linear on-shell observable
\begin{align*}
    \mathcal{O}_{[\psi]}\:\mathrm{ker}(\P)\to\mathbb{C}\,,\qquad \mathcal{O}_{[\psi]}(\varphi):=(\psi,\varphi)_{\E_{1}}=\int_{\M}\langle\psi,\varphi\rangle_{\E_{1}}\,\d\mu_{\g}=\int_{\M}\langle\psi\wedge\ast_{\g}\,\varphi\rangle_{\mathrm{Ad}(\P)}\, .
\end{align*}
The following proposition is yet another application of the Poincaré duality for flat vector bundles discussed in Appendix~\ref{Appendix:Poincaré Duality for Flat Vector Bundle-Valued Differential Forms} and generalises the analogous result for Maxwell's theory in \cite[Thm.~7.6]{BeniniMaxwell}:

\begin{proposition}\label{Prop:SepObs}
    Consider Yang-Mills theory linearised around an arbitrary \emph{flat} background connection $A$ and corresponding phase space $(\mathcal{V}_{\mathrm{P}},\sigma)$. Then, the following holds true:
    \begin{itemize}
        \item[(i)]If $\varphi\in\mathrm{ker}(\P)$ is such that $\mathcal{O}_{[\psi]}(\varphi)=0$ for all $[\psi]\in\mathcal{V}_{\mathrm{P}}$, then $\varphi$ is a pure gauge configuration, \textit{i.e.},~$\varphi=\K\omega$ for some $\omega\in\Gamma(\sf{E}_{0})$.
        \item[(ii)]If $[\psi]\in\mathcal{V}_{\mathrm{P}}$ is such that $\mathcal{O}_{[\psi]}(\varphi)=0$ for all $\varphi\in\mathrm{ker}(\P)$, then $[\psi]=0$, \textit{i.e.},~$\psi=\P\alpha$ for some $\alpha\in\Gamma_{\mathrm{c}}(\E_{1})$.
    \end{itemize}
\end{proposition}

\begin{proof}
The proof follows from similar steps as in~\cite[Thm.~7.6]{BeniniMaxwell} and the discussion preceding it. Note that the assumption of the existence of a ``finite good cover'' in item (ii) of \cite[Thm.~7.6]{BeniniMaxwell} is actually not needed: Poincaré duality in the scalar-valued setting states that, for any smooth $d$-manifold $\M$, the pairing
\begin{align}\label{eq:Dua}
H^{k}(\M)\times H^{d-k}_{\mathrm{c}}(\M)\longrightarrow\mathbb{R},
\qquad
([\alpha],[\beta])\mapsto\int_{\M}\alpha\wedge\beta
\end{align}
induces an isomorphism $H^{k}(\M)\cong\big(H^{d-k}_{\mathrm{c}}(\M)\big)^{\ast}$. If $\M$ admits a ``finite good cover'', \emph{i.e.},~a finite open cover $(\mathcal{U}_{i})_{i\in\mathrm{I}}$ of $\M$ with the property that any intersection of finitely many elements of $(\mathcal{U}_{i})_{i\in\mathrm{I}}$ is diffeomorphic to $\mathbb{R}^{d}$, then both $H^{k}(\M)$ and $H^{k}_{\mathrm{c}}(\M)$ are finite-dimensional, and hence the pairing also induces an isomorphism $H^{k}_{\mathrm{c}}(\M)\cong\big(H^{d-k}(\M)\big)^{\ast}$. Without such a finiteness assumption, the latter map need not be surjective; see, \emph{e.g.}, \cite[Rem.~5.7]{Bott}. Nevertheless, $H^{k}_{\mathrm{c}}(\M)\cong\big(H^{d-k}(\M)\big)^{\ast}$ is actually always injective and hence the pairing in Eq.~\eqref{eq:Dua} \emph{weakly non-degenerate} in \emph{both} entries. Inspecting the proof of~\cite[Thm.~7.6]{BeniniMaxwell}, one sees that this injectivity, rather than surjectivity, is the only property required there.
\end{proof}

Condition (i) entails that $\mathcal{V}_{\mathrm{P}}$ contains sufficiently many observables to distinguish between physical on-shell configurations in the sense that for every $\varphi\in\mathrm{ker}(\P)$, that is not a pure gauge configuration, we can always find observables $[\psi_{1}],[\psi_{2}]\in\mathcal{V}_{\mathrm{P}}$ such that $\mathcal{O}_{[\psi_{1}]}(\varphi)\neq\mathcal{O}_{[\psi_{2}]}(\varphi)$. In other words, $\mathcal{V}_{\mathrm{P}}$ is \emph{separating}. Claim (ii), on the other hand, tells us that there are no redundant elements in the phase space $\mathcal{V}_{\mathrm{P}}$, \emph{i.e.},~for any observable $[\psi]\in\mathcal{V}_{\mathrm{P}}$, we can always find two on-shell $\varphi_{1},\varphi_{2}\in\mathrm{ker}(\P)$ such that $\mathcal{O}_{[\psi]}(\varphi_{1})\neq\mathcal{O}_{[\psi]}(\varphi_{2})$.

\begin{remark}
    In the case of linearised gravity, Fewster-Hunt~\cite{FewsterHunt} showed that the corresponding phase space is non-degenerate and satisfies the two properties of Proposition~\ref{Prop:SepObs} for spatially compact background spacetimes. For non-compact spacetimes, one naturally expects degeneracies in the phase space, analogous to those arising in the Maxwell case. Whether the phase space nevertheless continues to satisfy the properties of Proposition~\ref{Prop:SepObs} is, to the best of the authors' knowledge, an open problem.
\end{remark}

\subsection{Decomposition of Connections and Linearised Yang-Mills Fields}
\label{Subsec:Decomposition of Connections and Linearised Yang-Mills Fields}

In the context of Maxwell's theory on globally hyperbolic spacetime, it is convenient to consider a $(1+d)$-spacetime decomposition of the relevant fields and equations. The goal of this section is to derive a similar decomposition for non-linear and linearised Yang-Mills fields. To start with, let us recall the corresponding formalism in the simpler Maxwell case. We refer to \cite{MurroSchmid} and in particular to \cite[Sec.~3.1]{SchmidPhD} for further details. As before, we consider a globally hyperbolic spacetime
\begin{align*}
     \M=\bb{R}\times\Sigma\,,\qquad\qquad \sf{g}=-\beta^{2}\d t\otimes\d t+\sf{h}_{t}\,.
\end{align*}
Then, denoting by $\pi_{2}\:\M\to\Sigma$ the natural projection map, we consider the $C^{\infty}(\M)$-submodule 
\begin{align}\label{eq:def:TDepFor}
    C^{\infty}(\mathbb{R},\Omega^{k}(\Sigma)):=\Gamma\bigg(\M,\pi_{2}^{\ast}\bigwedge^{k}\T^{\ast}\Sigma\bigg)\subset\Omega^{k}(\M)\,,
\end{align} 
whose elements can be regarded as \emph{time-dependent $k$-forms on $\Sigma$}. If we choose local coordinates $(x^{i})_{i=1,\dots,d}$ of $\Sigma$ defined on some open subset $\mathcal{U}\subset\Sigma$ and the corresponding coordinates $(x^{\mu})_{\mu=0,\dots,d}:=(x^{0}:=t,(x^{i})_{i=1,\dots,d})$ of $\M$ defined on $\mathbb{R}\times\mathcal{U}$, then elements of $C^{\infty}(\mathbb{R},\Omega^{k}(\Sigma))$ are locally given by
\begin{align*}
    \omega(t,\cdot)\vert_{\mathbb{R}\times\mathcal{U}}=\frac{1}{k!}\omega_{i_{1}\dots i_{k}}(t,\cdot)\,\d x^{i_{1}}\wedge\dots\wedge\d x^{i_{k}}
\end{align*}
for \emph{time-dependent} coefficients $\omega_{i_{1}\dots i_{k}}\in C^{\infty}(\mathbb{R}\times\mathcal{U})$. We shall denote by $\partial_{t}\omega:=\mathcal{L}_{\partial_{t}}\omega$ the \emph{time-derivative} of such forms. In local coordinates as above, this implies $(\partial_{t}\omega)_{i_{1}\dots i_{k}}=\partial_{t}\omega_{i_{1}\dots i_{k}}$.

Having fixed the relevant terminology, we consider the \emph{spacetime decomposition}
\begin{align}\label{eq:DecompScalar}
	\Omega^{k}(\M)\,\,\xrightarrow{\cong}\,\, C^{\infty}(\mathbb{R},\Omega^{k-1}(\Sigma))\oplus C^{\infty}(\mathbb{R},\Omega^{k}(\Sigma))\,,\qquad \omega=\d t\wedge\omega_{\mathrm{T}}+\omega_{\Sigma}\,\,\mapsto\,\, (\omega_{\mathrm{T}},\omega_{\Sigma})\,,
\end{align}
for $k\geq 1$, defined by $\omega_{\mathrm{T}}:=\partial_{t}\lrcorner\omega$ and $\omega_{\Sigma}:=\omega-\d t\wedge\omega_{\mathrm{T}}$. In local coordinates adapted to the splitting as above, this implies that the temporal and spatial components $\omega_{\mathrm{T}}$ and $\omega_{\Sigma}$ are locally given by
\begin{align*}
    (\omega_{\mathrm{T}})_{i_{1}\dots i_{k-1}}=\omega_{0i_{1}\dots i_{k-1}}\,,\qquad (\omega_{\Sigma})_{i_{1}\dots i_{k}}=\omega_{i_{1}\dots i_{k}}\, .
\end{align*}

The goal of the ensuing discussion is to generalise the preceding decomposition in Eq.~\eqref{eq:DecompScalar} on general globally hyperbolic spacetimes. Yet, in so doing, we observe that two main complications arise in this setting:
\begin{itemize}
    \item[1.] The background Yang-Mills field is a connection $1$-form defined on the total space $\P$ of a principal $\G$-bundle over $\M$. Consequently, it is a priori unclear how to decompose a connection $A \in \mathcal{C}(\P)$ into a ``temporal'' part $A_{\mathrm{T}}$ and a ``spatial'' part $A_{\Sigma}$, which in turn are necessary to construct operators acting on time-dependent forms on $\Sigma$, such as $\d^{A_{\Sigma}}$.
    \item[2.] If $\E \to\M$ is an arbitrary vector bundle (\textit{e.g.}, $\E = \mathrm{Ad}(\P)$), any $k$-form $\omega \in \Omega^k(\M,\E)$ can be decomposed as
    \begin{align*}
        \omega = \mathrm{d}t \wedge \omega_{\mathrm{T}} + \omega_{\Sigma}\,,
    \end{align*}
    where $\omega_{\mathrm{T}} := \partial_t \lrcorner \, \omega$ and $\wedge : \Omega^1(\M) \times \Omega^{k-1}(\M,\E) \to \Omega^k(\M,\E)$ denotes the wedge product induced by fibrewise scalar multiplication, see Eq.~\eqref{Eq: wedge product induced by a bilinear form}. In this case, $\omega_{\mathrm{T}}$ and $\omega_{\Sigma}$ are sections
    \begin{align*}
        \omega_{\mathrm{T}} \in \Gamma\left(\M,\pi_2^\ast \bigwedge^{k-1}\T^\ast\Sigma\otimes\E\right), \qquad \omega_{\Sigma} \in \Gamma\left(\M,\pi_2^\ast \bigwedge^{k}\T^\ast\Sigma\otimes\E\right).
    \end{align*}
    However, it is a priori non-trivial to establish in what sense these sections can be rigorously identified with genuine ``time-dependent forms'' on $\Sigma$ taking values in a \emph{spatial} vector bundle, which in turn is needed to obtain a good notion of time-derivative replacing $\partial_{t}$ above.
\end{itemize}

Both these issues can be solved on account of the following lemma (\textit{cf.},~\cite{ChruscielShatah}).

\begin{lemma}\label{Lemma:BundleSigma}
    Let $(\M,\g)$ with $\M=\mathbb{R}\times\Sigma$ be a globally hyperbolic spacetime and $\pi_{\P}\:\P\to\M$ be a principal $\G$-bundle over $\M$. Then, there exists a unique (up to isomorphisms) smooth principal $\G$-bundle $\pi_{\P_{\Sigma}}\:\P_{\Sigma}\to\Sigma$ such that $\P\cong\pi_{2}^{\ast}\P_{\Sigma}$. As a consequence, also $\mathrm{Ad}(\P)\cong \pi_{2}^{\ast}\mathrm{Ad}(\P_{\Sigma})$.
\end{lemma}

\begin{proof}
    Let us first recall that $\mathrm{Prin}_{\G}(\M)$, the space of equivalence classes of continuous $\G$-bundles on $\M$, is equivalent to the space $[\M,\B\G]$, which is the space of homotopy classes of continuous maps from $\M$ to the \emph{classifying space} $\B\G$. Now, for a given smooth principal $\G$-bundle $\P\in\mathrm{Prin}_{\G}(\M)$, consider a corresponding classifying map $f\:\M\to\B\G$. Now, since $\R$ is contractible, we note that $\pi_{2}\:\M\to\Sigma$ defines a homotopy equivalence between $\M$ and $\Sigma_{0}:=\{0\}\times\Sigma\cong\Sigma$. Hence, $f$ is homotopic to a map $\tilde{f}\:\Sigma\to\B\G$ that factors through $\Sigma$, \textit{i.e.},~$f\simeq\tilde{f}\circ\pi_{2}$. Pulling back the universal bundle $\E\G\to\B\G$ along $\tilde{f}$ defines a topological $\G$-bundle $\P_{\Sigma}$ over $\Sigma$ such that $\P\cong\pi_{2}^{\ast}\P_{\Sigma}$. Clearly, $\P_{\Sigma}$ inherits a smooth structure from $\P$. Moreover, by construction, it is unique up to isomorphisms.
\end{proof}

\begin{remark}
More explicitly, one may also set $\P_{\Sigma}:=\P\vert_{\Sigma_{0}}$ and establish an isomorphism $\Phi\colon\P\to\pi_{2}^{\ast}\P_{\Sigma}$ by considering the \emph{parallel transport} induced from a fixed but arbitrary connection $A\in\mathcal{C}(\P)$ along $\partial_{t}$ to identify the fibres $\P_{(0,\vec{x})}$ with $\P_{(t,\vec{x})}$ for all $(t,\vec{x})\in\mathbb{R}\times\Sigma$.
\end{remark}

Throughout the following, let us \emph{fix} an isomorphism $\P \cong \pi_2^\ast \P_\Sigma$ and simply identify $\P = \pi_2^\ast \P_\Sigma$. On the level of the total space, recalling that $\pi_{2}^{\ast}\P_{\Sigma}=\{((t,\vec{x}),p)\in\M\times\P_{\Sigma}\mid \pi_{2}(t,\vec{x})=\vec{x}=\pi_{\P_{\Sigma}}(p)\}$, by definition of the pullback bundle, it is useful to note that there is a natural identification
\begin{align*}
    \pi_{2}^*\P_{\Sigma}\xrightarrow{\cong}\mathbb{R}\times\P_{\Sigma}\,,\qquad ((t,\vec{x}),p)\mapsto (t,p)
\end{align*}
with inverse given by $(t,p)\mapsto ((t,\pi_{\P_{\Sigma}}(p)),p)$. In particular, we have identified $\P=\mathbb{R}\times\P_{\Sigma}$. Conceptually, this reduction mirrors the very choice of a spacetime decomposition: just as we select a preferred time direction (\textit{i.e.}, a Cauchy temporal function) to split the (globally hyperbolic) spacetime manifold as $\M= \mathbb{R} \times \Sigma$, we choose a \emph{compatible} temporal framing to trivialise the principal bundle, which describes the internal gauge-theoretic degrees of freedom, over time as $\P=\mathbb{R}\times\P_{\Sigma}$, accordingly. 

Now, fixing an identification $\P=\mathbb{R}\times\P_{\Sigma}$, we obtain a corresponding decomposition of $\mathfrak{g}$-valued forms on $\P$, defined in analogy to Eq.~\eqref{eq:DecompScalar}, \textit{i.e.}, 
    \begin{align}\label{eq:DecompLieAlgVal}
       \Omega^{k}(\P,\mathfrak{g})\xrightarrow{\cong} C^{\infty}(\mathbb{R},\Omega^{k-1}(\P_\Sigma,\mathfrak{g}))\oplus C^{\infty}(\mathbb{R},\Omega^{k}(\P_{\Sigma},\mathfrak{g}))\,,\qquad \omega= \d t\wedge \omega_{\mathrm{T}}+\omega_{\Sigma}\mapsto (\omega_{\mathrm{T}},\omega_{\Sigma})
    \end{align}
    with $\omega_{\mathrm{T}}:=\partial_{t}\lrcorner \omega$, where $\partial_{t}$ is the canonical coordinate vector field associated with the $\bb{R}$-factor of the identification $\P=\mathbb{R}\times\P_{\Sigma}$. As a special case, we obtain the following.

\begin{proposition}[Decomposition of Connections]\label{Prop:DecompConn}
Choose an identification $\P=\mathbb{R}\times\P_{\Sigma}$ and consider the decomposition in Eq.~\eqref{eq:DecompLieAlgVal} in the special case $k=1$, \textit{i.e.}, 
    \begin{align*}
       \Omega^{1}(\P,\mathfrak{g})\xrightarrow{\cong} C^{\infty}(\P,\mathfrak{g})\oplus C^{\infty}(\mathbb{R},\Omega^{1}(\P_{\Sigma},\mathfrak{g}))\,,\qquad \alpha= \alpha_{\mathrm{T}}\d t+\alpha_{\Sigma}\mapsto (\alpha_{\mathrm{T}}:=\partial_{t}\lrcorner \alpha,\alpha_{\Sigma})\,,
    \end{align*}
If $A\in\mathcal{C}(\P)\subset\Omega^{1}(\P,\mathfrak{g})$ is a connection $1$-form on $\P$, the following conditions hold true:
    \begin{itemize}
        \item[(i)]The temporal part $A_{\mathrm{T}}$ is $\G$-equivariant, \textit{i.e.},~$r_{g}^{\ast} A_{\mathrm{T}}=\mathrm{Ad}_{g^{-1}}\circ A_{\mathrm{T}}$, and, hence, in view of the isomorphism in Eq.~\eqref{eq:Iso}, it can be viewed as an element of $\Gamma(\M,\mathrm{Ad}(\P))$.
        \item[(ii)]The spatial part $A_{\Sigma}$ satisfies $A_{\Sigma}(t,\cdot)\in \mathcal{C}(\P_{\Sigma})$ for all $t\in\mathbb{R}$, \textit{i.e.},~$A_{\Sigma}$ can be viewed as a \emph{time-dependent connection of $\P_{\Sigma}$}. We denote the space of all such connections by $C^{\infty}(\mathbb{R},\mathcal{C}(\P_{\Sigma}))$.
    \end{itemize}
In summary, fixing an identification $\P=\mathbb{R}\times\P_{\Sigma}$, we obtain the following decomposition: 
    \begin{align*}
        \mathcal{C}(\P)\xrightarrow{\cong} \Gamma(\M,\mathrm{Ad}(\P))\oplus C^{\infty}(\mathbb{R},\mathcal{C}(\P_{\Sigma})).
    \end{align*}
\end{proposition}

\begin{proof}
    Consider the (fibre-preserving) right-group actions $\cdot\:\P\times\G\to\P$ and $\cdot\:\P_{\Sigma}\times\G\to\P_{\Sigma}$, denoted by the same symbol with a mild abuse of notation. Since $\P=\pi_{2}^{\ast}\P_{\Sigma}$, we note that these actions are related via $(t,p_{\Sigma})\cdot g=(t,p_{\Sigma}\cdot g)$ for all $(t,p_{\Sigma})\in\P$ and $g\in\G$. In particular, we note that
    \begin{align}
    \label{eq.CompatRightAction}
        r_{g}(t,p_{\Sigma})=(t,r_{g}(p_{\Sigma}))\, .
    \end{align}
    Now, consider a connection $1$-form $A\in\mathcal{C}(\P)\subset\Omega^{1}(\P,\mathfrak{g})$ and decompose it accordingly as $A=A_{\mathrm{T}}\d t+A_{\Sigma}$ with $A_{\mathrm{T}}:=\partial_{t}\lrcorner A$. Being a connection $1$-form, it holds that $r_{g}^{\ast}A=\mathrm{Ad}_{g^{-1}}\circ A$ and on account of Eq.~\eqref{eq.CompatRightAction}, we note that $r_{g}^{\ast}A=(r_{g}^{\ast}A_{\mathrm{T}})\d t+(r_{g}^{\ast}A_{\Sigma})$ since $r_{g}^{\ast}\d t=\d t$. In particular, it follows that the temporal and spatial components are individually $\G$-equivariant, \textit{i.e.}, 
    \begin{align*}
        r_{g}^{\ast}A_{\mathrm{T}}=\mathrm{Ad}_{g^{-1}}\circ A_{\mathrm{T}}\,,\qquad r_{g}^{\ast}A_{\Sigma}=\mathrm{Ad}_{g^{-1}}\circ A_{\Sigma}\, .
    \end{align*}
    This proves item (i). For item (ii), we need to show that $A_{\Sigma}(t,\cdot)$ defines a connection on $\P_{\Sigma}$. The above discussion already implies that $A_{\Sigma}(t,\cdot)$ is $\G$-equivariant, namely, $r_{g}^{\ast}A_{\Sigma}(t,\cdot)=\mathrm{Ad}_{g^{-1}}\circ A_{\Sigma}(t,\cdot)$. For the second property, consider $X\in\mathfrak{g}$ and the corresponding fundamental vector fields
    \begin{align*}
        X^{\sharp}_{p}:=\frac{\d}{\d s}\bigg\vert_{s=0}p\cdot\mathrm{exp}_{\G}(sX)\,,\qquad X^{\sharp_{\Sigma}}_{p_{\Sigma}}:=\frac{\d}{\d s}\bigg\vert_{s=0}p_{\Sigma}\cdot\mathrm{exp}_{\G}(sX)\, .
    \end{align*}
   For any point $p=(t,p_{\Sigma})$, we note that Eq.~\eqref{eq.CompatRightAction} implies $X_{p}^{\sharp}=0\cdot\partial_{t}+X^{\sharp_{\Sigma}}_{p_{\Sigma}}=X^{\sharp_{\Sigma}}_{p_{\Sigma}}$. Hence, using that $A$ is a connection, which implies that $A(X^{\sharp})=X$, the above decomposition immediately implies that $A(X^{\sharp_{\Sigma}}) = A_\Sigma(t, \cdot) (X^{\sharp_{\Sigma}})  = X$, which concludes the proof.
\end{proof}

Having established a decomposition of $\mathfrak{g}$-valued differential forms on $\P$ and on the subspace of connection $1$-forms, let us now turn to a decomposition of $\mathrm{Ad}(\P)$-valued forms on $\M$. As above, let $(\M=\bb{R}\times\Sigma,\g)$ be an arbitrary globally hyperbolic spacetime, fix an identification $\P=\pi_{2}^{\ast}\P_{\Sigma}$ and, accordingly, identify $\mathrm{Ad}(\P)\cong\pi_{2}^{\ast}\mathrm{Ad}(\P_{\Sigma})$. In accordance with the notation used for scalar-valued forms in Eq.~\eqref{eq:def:TDepFor}, we denote by
\begin{align*}
    C^{\infty}(\mathbb{R},\Omega^{k}(\Sigma,\mathrm{Ad}(\P_{\Sigma}))):=\Gamma\left(\M,\pi_{2}^{\ast}\left(\bigwedge^{k}\T^{\ast}\Sigma\otimes\mathrm{Ad}(\P_{\Sigma})\right)\right)\subset\Omega^{k}(\M,\mathrm{Ad}(\P)) \, ,
\end{align*}
the space of \emph{time-dependent $\mathrm{Ad}(\P_{\Sigma})$-valued $k$-forms on $\Sigma$}. On account of the above discussion, we obtain the following decomposition of $\mathrm{Ad}(\P)$-valued forms for $k\geq 1$: 
\begin{align}\label{eq:DecompAdValued}
	\Omega^{k}(\M,\mathrm{Ad}(\P))\,\,\xrightarrow{\cong}\,\, C^{\infty}(\mathbb{R},\Omega^{k-1}(\Sigma,\mathrm{Ad}(\P_{\Sigma})))\oplus C^{\infty}(\mathbb{R},\Omega^{k}(\Sigma,\mathrm{Ad}(\P_{\Sigma})))\,,
\end{align}
which is obtained by taking a composition of the identification $\P=\pi_{2}^{\ast}\P_{\Sigma}$ and of the decomposition
\begin{align*}
    \Omega^{k}(\M,\pi_{2}^{\ast}\mathrm{Ad}(\P_{\Sigma}))\ni\omega=\d t\wedge\omega_{\mathrm{T}}+\omega_{\Sigma},\qquad \omega_{\mathrm{T}}:=\partial_t \lrcorner\omega\, .
\end{align*}
Now, let us recall that a suitable subspace of $\Omega^{k}(\P,\mathfrak{g})$ can be identified with $\Omega^{k}(\M,\mathrm{Ad}(\P))$ and similarly for $\Omega^{k}(\P_{\Sigma},\mathfrak{g})$. More precisely, by Eq.~\eqref{eq:Iso}, there exist canonical isomorphisms
\begin{align}\label{eq:Iso2}
    \Omega^{k}_{\mathrm{hor},\mathrm{Ad}}(\P,\mathfrak{g})\cong\Omega^{k}(\M,\mathrm{Ad}(\P))\,,\qquad\text{and}\qquad \Omega^{k}_{\mathrm{hor},\mathrm{Ad}}(\P_{\Sigma},\mathfrak{g})\cong\Omega^{k}(\Sigma,\mathrm{Ad}(\P_{\Sigma}))
\end{align}
It is not too hard to check that the two decompositions in Eq.~\eqref{eq:DecompLieAlgVal} and Eq.~\eqref{eq:DecompAdValued} are compatible in the sense clarified by the following lemma. 

\begin{lemma}
\label{Lem: 2.24}
    Let $\P$ be a principal $\G$-bundle over globally hyperbolic spacetime $(\M,\g)$, and let us fix an identification $\P=\mathbb{R}\times\P_{\Sigma}$. Then, for any $\omega\in\Omega^{k}_{\mathrm{hor},\mathrm{Ad}}(\P,\mathfrak{g})$, decomposed as, $\omega=\d t\wedge\omega_{\mathrm{T}}+\omega_{\Sigma}$, it holds that
    \begin{align*}
        \omega_{\mathrm{T}}(t,\cdot)\in \Omega^{k-1}_{\mathrm{hor},\mathrm{Ad}}(\P_{\Sigma},\mathfrak{g})\,,\qquad \omega_{\Sigma}(t,\cdot)\in \Omega^{k}_{\mathrm{hor},\mathrm{Ad}}(\P_{\Sigma},\mathfrak{g})\, , 
    \end{align*}
    for all $t\in\bb{R}$. In other words, the property of being horizontal and of type $\mathrm{Ad}$ factors through the decomposition in Eq. \eqref{eq:DecompLieAlgVal}:
    \begin{align*}
       \Omega^{k}_{\mathrm{hor},\mathrm{Ad}}(\P,\mathfrak{g})\xrightarrow{\cong} C^{\infty}(\mathbb{R},\Omega^{k-1}_{\mathrm{hor},\mathrm{Ad}}(\P_\Sigma,\mathfrak{g}))\oplus C^{\infty}(\mathbb{R},\Omega^{k}_{\mathrm{hor},\mathrm{Ad}}(\P_{\Sigma},\mathfrak{g}))
    \end{align*}
    Moreover, the decompositions of $\mathfrak{g}$-valued forms on $\P$ and $\mathrm{Ad}(\P)$-valued forms on $\M$ introduced in Eq.~\eqref{eq:DecompLieAlgVal} and Eq.~\eqref{eq:DecompAdValued}, respectively, are compatible with the canonical isomorphisms in Eq.~\eqref{eq:Iso2}, namely, 
    \begin{align*}
        (\omega_{\mathrm{T}})_{\M}=(\omega_{\M})_{\mathrm{T}}\,,\qquad  (\omega_{\Sigma})_{\M}=(\omega_{\M})_{\Sigma}\, .
    \end{align*}
\end{lemma}

\begin{proof}
    Following the discussion in the proof of Proposition~\ref{Prop:DecompConn}, we recall that under the product identification $\P = \mathbb{R} \times \P_{\Sigma}$ the right actions satisfy $r_g(t, p_\Sigma) = (t, r_g(p_\Sigma))$, which implies that the canonical coordinate vector field $\partial_t$ is $G$-invariant, \textit{i.e.},~$(r_{g})_{\ast}\partial_t = \partial_t$. Uniqueness of the corresponding decomposition immediately gives the claimed result.
\end{proof}

\begin{remark}
    Let us remark that the decomposition of $\Omega^{k}(\M,\mathrm{Ad}(\P))$ in Eq.~\eqref{eq:DecompAdValued} is also compatible with taking the decompositions of representatives in local gauges compatible with the foliation $\P\cong\mathbb{R}\times\P_{\Sigma}$. More precisely, let $\omega\in\Omega^{k}(\M,\mathrm{Ad}(\P))$ be arbitrary and decomposed as $\omega=\d t\wedge\omega_{\mathrm{T}}+\omega_{\Sigma}$. Then, consider a local gauge $s\in\Gamma(\mathcal{U},\P)$ that is ``compatible with the identification'' $\P=\mathbb{R}\times\P_{\Sigma}$ in the sense that it is a local gauge defined on $\mathcal{U}:=\mathbb{R}\times\mathcal{U}_{\Sigma}$ obtained as the pullback of some local gauge $s_{\Sigma}\in\Gamma(\mathcal{U}_{\Sigma},\P)$ on $\mathcal{U}_{\Sigma}\subset\Sigma$, \emph{i.e.}, $s(t,\vec{x})=(t,s_{\Sigma}(\vec{x}))$ for all $(t,\vec{x})\in\mathcal{U}$. In this gauge, we can decompose the local gauge representative $\omega_{s}\in\Omega^{k}(\mathcal{U},\mathfrak{g})$ (\textit{cf.},~Section~\ref{Sec:Geometric Preliminaries on Yang-Mills Theory}) as 
    \begin{align*}
        \Omega^{k}(\mathcal{U},\mathfrak{g})\xrightarrow{\cong} C^{\infty}(\bb{R},\Omega^{k-1}(\mathcal{U}_{\Sigma},\mathfrak{g}))\oplus C^{\infty}(\bb{R},\Omega^{k}(\mathcal{U}_{\Sigma},\mathfrak{g}))\,,\qquad \omega_{s}=\d t\wedge \omega_{\mathrm{T}}^{s}+\omega_{\Sigma}^{s}\mapsto (\omega_{\mathrm{T}}^{s},\omega_{\Sigma}^{s})
    \end{align*}
    with $\omega_{\mathrm{T}}^{s}:=\partial_{t}\lrcorner\omega_{s}$. Then, it is not too hard to see that $(\omega_{\mathrm{T}})_{s}=\omega_{\mathrm{T}}^{s}$ as well as $(\omega_{\Sigma})_{s}=\omega_{\Sigma}^{s}$. It is important to stress that this fails in general for \emph{arbitrary} local gauges $s\in\Gamma(\mathcal{U},\P)$, possibly depending on time. Indeed, even if the temporal component $\omega_{\mathrm{T}}$ of $\omega$ is trivial, its local gauge representative $\omega_{s}$ generally has a non-zero temporal component $\omega_{\mathrm{T}}^{s}$, as one can easily see by applying a gauge transformation in the above decomposition to a time-dependent gauge.
\end{remark}

Having derived decompositions for linearised Yang-Mills fields and general $\mathrm{Ad}(\P)$-valued forms on globally hyperbolic spacetimes, the goal of the following discussions is to decompose operators such as $\d^{A}$ and $\delta^{A}$ into their temporal and spatial components, making use of the induced operators $\d^{A_{\Sigma}}$ and $\delta^{A_{\Sigma}}$. To this end, let us restrict ourselves to $(1+d)$-dimensional \emph{ultrastatic} spacetimes, \textit{i.e.}, Lorentzian manifolds $(\M,\g)$ of the form
\begin{align*}
     \M=\R\times\Sigma\,,\qquad\qquad \g=-\d t\otimes\d t+\h\,,
\end{align*}
where $\h$ is a (time-independent) Riemannian metric on $\Sigma$. It is well-known that global hyperbolicity of $(\M,\g)$ in this case is equivalent to requiring $(\Sigma,\h)$ to be a \emph{complete} Riemannian manifold, see, \textit{e.g.},~\cite[Thm.~3.1]{SanchezUltrastatic}. For the following discussions, we shall impose this restriction on the underlying background primarily for computational convenience. However, we remark that for our purposes, it will be sufficient to consider ultrastatic spacetimes, since the construction of states on general globally hyperbolic spacetimes can be obtained by means of a deformation argument. 

As a warm-up, let us start with the scalar case, making use of the decomposition~\eqref{eq:DecompScalar}. We denote by $\d_{\Sigma}\:\Omega^{k}(\Sigma)\to\Omega^{k+1}(\Sigma)$ and $\delta_{\Sigma}:=(-1)^{k+1}\ast_{\h}^{-1}\d\ast_{\h}\:\Omega^{k+1}(\Sigma)\to\Omega^{k}(\Sigma)$ the exterior derivative and codifferential of forms on $\Sigma$, respectively (\textit{cf.},~Appendix~\ref{Appendix: Notation and Conventions for Scalar and Twisted Differential Forms}). Applying them pointwise in time, we obtain corresponding operators defined on  $C^{\infty}(\mathbb{R},\Omega^{\bullet}(\Sigma))$, which we denote by the same symbol. Employing this notation, we can establish the following lemma. 

\begin{lemma}[Special Case of \protect{\cite[Lem.~3.2]{SchmidPhD}}]\label{Lemma:DecScalar}
Let $(\M=\mathbb{R}\times\Sigma,\g=-\d t\otimes\d t+\h)$ be a globally hyperbolic ultrastatic spacetime. Then, the following properties hold true:
\begin{align*}
    \mathrm{(i)}&\qquad (\d f)_{\mathrm{T}}=\partial_{t}f\,,\qquad (\d f)_{\Sigma}=d_{\Sigma}f\qquad &&\text{for}\quad f\in \Omega^{0}(\M)=C^{\infty}(\M)\\
    \mathrm{(ii)}&\qquad (\d\omega)_{\mathrm{T}}=\partial_{t}\omega_{\Sigma}-\d_{\Sigma}\omega_{\mathrm{T}}\,,\qquad (\d \omega)_{\Sigma}=\d_{\Sigma}\omega_{\Sigma}\qquad &&\text{for} \quad\omega\in \Omega^{k\geq 1}(\M)\\
    \mathrm{(iii)}&\qquad  \delta \alpha=\partial_{t} \alpha_{\mathrm{T}}+\delta_{\Sigma}\alpha_{\Sigma}\qquad &&\text{for}\quad \alpha\in \Omega^{1}(\M)\\
    \mathrm{(iv)}&\qquad  (\delta\omega)_{\mathrm{T}}=-\delta_{\Sigma}\omega_{\mathrm{T}}\,,\qquad (\delta\omega)_{\Sigma}=\partial_{t}\omega_{\mathrm{T}}+\delta_{\Sigma}\omega_{\Sigma}\qquad &&\text{for}\quad \omega\in \Omega^{k\geq 2}(\M)
\end{align*}
Moreover, denoting by $\square:=\d\delta+\delta\d$ and $\Delta:=\d_{\Sigma}\delta_{\Sigma}+\delta_{\Sigma}\d_{\Sigma}$ the de Rham-Hodge d'Alembertian acting on $\Omega^{k}(\M)$ and the Laplacian acting on $C^{\infty}(\mathbb{R},\Omega^{k}(\Sigma))$, respectively, we obtain
\begin{align*}
    \mathrm{(v)}&\qquad (\square\omega)_{\mathrm{T}}=(\partial_{t}^{2}+\Delta)\omega_{\mathrm{T}}\,,\qquad (\square\omega)_{\Sigma}=(\partial_{t}^{2}+\Delta)\omega_{\Sigma}\qquad &&\text{for}\quad \omega\in \Omega^{k}(\M). 
\end{align*}
\end{lemma}

\begin{remark}
    We note that the previous discussion can be generalised to non-ultrastatic spacetimes. In this context, it is actually slightly more natural to redefine the decomposition in Eq.~\eqref{eq:DecompScalar} by instead considering $\omega=\eta\wedge\omega_{\mathrm{T}}+\omega_{\Sigma}$ with $\omega_{\mathrm{T}}:=\omega(\nu)$, where $\nu:=\beta^{-1}\partial_{t}$ and $\eta:=\beta\d t$ denote the future-directed timelike normal vector and covector fields, respectively. In the non-ultrastatic case, the decomposition in (i)--(v) becomes more complicated and includes terms involving the lapse function $\beta$ and the second fundamental form $\mathsf{k}:=\partial_{t}\h$ together with its time- and space-derivatives. The generalised formulas (i)--(iv) in this setting can be found in \cite[Lem.~3.2]{SchmidPhD}.
\end{remark}

For later convenience, let us also record the following lemma.

\begin{lemma}\label{Lemma:FlatImpliesFlat} Let $A\in\mathcal{C}(\P)$ be a connection $1$-form, decomposed as $A=A_{\mathrm{T}}\d t+A_{\Sigma}$, see Proposition \ref{Prop:DecompConn}. Then,
\begin{align*}
    A\text{ is flat }\quad\Rightarrow\quad
        A_{\Sigma}(t)\text{ is flat for all $t\in\mathbb{R}$}. 
\end{align*}
\end{lemma}

\begin{proof} 
    This is a direct consequence of Cartan's structure equation. Indeed, denoting by $\d_{\Sigma}$ the exterior derivative acting on $\Omega^{\bullet}(\P_{\Sigma},\mathfrak{g})$, we find analogously to item (ii) in Lemma~\ref{Lemma:DecScalar} the general relation $(\d A)_{\Sigma}=\d_{\Sigma}A_{\Sigma}$, as one can verify by using coordinates adapted to the product $\mathbb{R}\times\P_{\Sigma}$. In particular, denoting by $F^{A}\in\Omega^{2}(\P,\mathfrak{g})$ the curvature of $A$ and by $F^{A_{\Sigma}}\in C^{\infty}(\mathbb{R},\Omega^{2}(\P_{\Sigma},\mathfrak{g}))$ the (time-dependent) curvature of the (time-dependent) connection $A_{\Sigma}\in C^{\infty}(\mathbb{R},\mathcal{C}(\P_{\Sigma}))$, we find
    \begin{align*}
        (F^{A})_{\Sigma}=\bigg(\d A+\frac{1}{2}[A\wedge A]_{\mathfrak{g}}\bigg)_{\Sigma}=\d_{\Sigma}A_{\Sigma}+\frac{1}{2}[A_{\Sigma}\wedge A_{\Sigma}]_{\mathfrak{g}}=F^{A_{\Sigma}}\,.
    \end{align*}
    This entails that $F^{A_{\Sigma}}=0$ if and only if $(F^{A})_{\Sigma}=0$. This is precisely the case whenever $F^{A}=0$.
\end{proof}

Let us now generalise the content of Lemma~\ref{Lemma:DecScalar} to the setting of linearised Yang-Mills fields, \textit{i.e.}, to $\mathrm{Ad}(\P)$-valued forms on $\M$ and to the operators $\d^A$ and $\delta^A$ introduced in Section \ref{Sec:Geometric Preliminaries on Yang-Mills Theory}. In Eq.~\eqref{Eq.DefEk1}, we have introduced the pseudo-Riemannian vector bundles $(\E_{k},\langle\cdot,\cdot\rangle_{\E_{k}})$ and, for a given connection $A\in\mathcal{C}(\P)$, the operators
\begin{align*}
    &\d^{A}\:\Omega^{k}(\M,\mathrm{Ad}(\P))\to\Omega^{k+1}(\M,\mathrm{Ad}(\P))\,,\\
    &\delta^{A}\:\Omega^{k+1}(\M,\mathrm{Ad}(\P))\to\Omega^{k}(\M,\mathrm{Ad}(\P))\,,\qquad \delta^{A}:=(-1)^{k+1}\ast^{-1}_{\g}\d^{A}\ast_{\g}\,,
\end{align*}
where we recall that $\delta^{A}$ is the formal adjoint of $\d^{A}$ with respect to $(\cdot,\cdot)_{\E_{k}}$ defined in Eq.~\eqref{Eq.DefEk2}. In analogy, let us define, on the level of the spatial hypersurface $\Sigma$, the \emph{Riemannian} vector bundle
\begin{align}
\label{Eq: E Sigma k}
    \E_{\Sigma,k}:=\bigwedge^{k}\T^{\ast}\Sigma\otimes\mathrm{Ad}(\P_{\Sigma})\,,\qquad \langle\omega\otimes \varphi,\eta\otimes\psi\rangle_{\E_{\Sigma,k}}:=\langle\omega,\eta\rangle_{\h}\langle \varphi,\psi\rangle_{\mathrm{Ad}(\P_{\Sigma})}
\end{align}
where $\langle\cdot,\cdot\rangle_{\mathrm{Ad}(\P_{\Sigma})}$ is the bundle metric on $\mathrm{Ad}(\P_{\Sigma})$ induced from $\kappa$, defined analogously to $\langle\cdot,\cdot\rangle_{\mathrm{Ad}(\P)}$ in Eq.~\eqref{eq:InducedBundleMetric}, and where $\langle\cdot,\cdot\rangle_{\h}$ is the standard (Hodge) bundle metric on $\bigwedge^{k}\T^{\ast}\Sigma$ defined by
\begin{align*}
    \langle\omega,\eta\rangle_{\h}:=\frac{1}{k!}(\h^{-1})^{\otimes k}(\omega,\eta)\, .
\end{align*}
As usual, we denote the induced inner product at the level of sections by
\begin{align} 
\label{Eq: inner product on E sigma} (\alpha,\beta)_{\E_{\Sigma,k}}:=\int_{\Sigma}\langle\alpha,\beta\rangle_{\E_{\Sigma,k}}\,\d\mu_{\h}=\int_{\Sigma}\langle\alpha\wedge\ast_{\h}\beta\rangle_{\mathrm{Ad}(\P_{\Sigma})}\,,
\end{align}
where $\ast_{\h}\:\Omega^{k}(\Sigma,\mathrm{Ad}(\P_{\Sigma}))\to\Omega^{d-k}(\Sigma,\mathrm{Ad}(\P_{\Sigma}))$ denotes the Hodge $\ast$-operator associated to $(\Sigma,\h)$ and where $\langle\cdot\wedge\cdot\rangle_{\mathrm{Ad}(\P_{\Sigma})}\:\Omega^{k}(\Sigma,\mathrm{Ad}(\P_{\Sigma}))\times\Omega^{l}(\Sigma,\mathrm{Ad}(\P_{\Sigma}))\to\Omega^{k+l}(\Sigma)$ denotes the wedge-type product induced from $\langle\cdot,\cdot\rangle_{\mathrm{Ad}(\P_{\Sigma})}$. Now, following the discussion of Section~\ref{Sec:Geometric Preliminaries on Yang-Mills Theory}, we define the operators 
\begin{align*}
    &\d^{A_{\Sigma}}\:\Omega^{k}(\Sigma,\mathrm{Ad}(\P_{\Sigma}))\to\Omega^{k+1}(\Sigma,\mathrm{Ad}(\P_{\Sigma}))\,,\\ 
    &\delta^{A_{\Sigma}}\:\Omega^{k+1}(\Sigma,\mathrm{Ad}(\P_{\Sigma}))\to\Omega^{k}(\Sigma,\mathrm{Ad}(\P_{\Sigma}))\,,\qquad \delta^{A_{\Sigma}}:=(-1)^{k+1}\ast^{-1}_{\h}\d^{A_{\Sigma}}\ast_{\h}
\end{align*}
for a given connection $A_{\Sigma}\in\mathcal{C}(\P_{\Sigma})$, where we note that $\delta^{A_{\Sigma}}$ is the formal adjoint of $d^{A_{\Sigma}}$ with respect to $(\cdot,\cdot)_{\E_{\Sigma,k}}$. In particular, if we consider a connection $1$-form $A\in\mathcal{C}(\P)$ and decompose it as $A=A_{\mathrm{T}}\d t+A_{\Sigma}$ with $A_{\Sigma}\in C^{\infty}(\mathbb{R},\mathcal{C}(\P_{\Sigma}))$ being a time-dependent connection $1$-form of $\P_{\Sigma}$, in accordance with Proposition~\ref{Prop:DecompConn}, we may consider the exterior covariant derivative $\d^{A_{\Sigma}(t)}$ and corresponding codifferential $\delta^{A_{\Sigma}(t)}$ of the connection $A_{\Sigma}(t)$ of $\P_{\Sigma}$ for every time $t\in\mathbb{R}$. By applying these differential operators pointwise in time, we obtain operators
\begin{align*}
    \d^{A_{\Sigma}}\:C^{\infty}(\mathbb{R},\Omega^{k}(\Sigma,\mathrm{Ad}(\P_{\Sigma})))\to C^{\infty}(\mathbb{R},\Omega^{k+1}(\Sigma,\mathrm{Ad}(\P_{\Sigma})))\,,\qquad (\d^{A_{\Sigma}}\omega)(t):=\d^{A_{\Sigma}(t)}\omega(t)\\
     \delta^{A_{\Sigma}}\:C^{\infty}(\mathbb{R},\Omega^{k+1}(\Sigma,\mathrm{Ad}(\P_{\Sigma})))\to C^{\infty}(\mathbb{R},\Omega^{k}(\Sigma,\mathrm{Ad}(\P_{\Sigma})))\,,\qquad (\delta^{A_{\Sigma}}\omega)(t):=\delta^{A_{\Sigma}(t)}\omega(t) \, . 
\end{align*}

In the same spirit as Lemma~\ref{Lemma:DecScalar}, we aim to decompose forms such as $\d^{A}\omega$ and $\delta^{A}\omega$ into their temporal and spatial components. The operators $\d^{A_{\Sigma}}$ and $\delta^{A_{\Sigma}}$ are the relevant generalisations of $\d_{\Sigma}$ and $\delta_{\Sigma}$ to the $\mathrm{Ad}(\P)$-valued case. It remains to find a suitable replacement for the time-derivative $\partial_{t}$. To this end, we give the following definition.

\begin{definition}
\label{Def: covariant time derivative}
    Let $\P$ be a principal $\G$-bundle over an ultrastatic globally hyperbolic spacetime $(\M,\g)$, fix an identification $\P=\mathbb{R}\times\P_{\Sigma}$ and let $A\in\mathcal{C}(\P)$ be a connection $1$-form decomposed as $A=A_{\mathrm{T}}\d t+A_{\Sigma}$. We define the differential operator
    \begin{align*}
    \widetilde{\mathcal{D}}_{t}^{A_{\mathrm{T}}}\:C^{\infty}(\mathbb{R},\Omega^{k}(\P_{\Sigma},\mathfrak{g}))\to C^{\infty}(\mathbb{R},\Omega^{k}(\P_{\Sigma},\mathfrak{g}))\,,\qquad \widetilde{\mathcal{D}}_{t}^{A_{\mathrm{T}}}\omega:=\partial_{t}\omega+[A_{\mathrm{T}},\omega]_{\mathfrak{g}}\,.
    \end{align*}
    Clearly, $\widetilde{\mathcal{D}}_{t}^{A_{\mathrm{T}}}$ maps $C^{\infty}(\mathbb{R},\Omega^{k}_{\mathrm{hor},\mathrm{Ad}}(\P_{\Sigma},\mathfrak{g}))$ into itself and, hence, in view of the canonical isomorphism in Eq.~\eqref{eq:Iso}, it yields a linear operator
    \begin{align*}
        \mathcal{D}_{t}^{A_{\mathrm{T}}}\:C^{\infty}(\mathbb{R},\Omega^{k}(\Sigma,\mathrm{Ad}(\P_{\Sigma})))\to C^{\infty}(\mathbb{R},\Omega^{k}(\Sigma,\mathrm{Ad}(\P_{\Sigma})))\,.
    \end{align*}   
On account of $C^{\infty}(\mathbb{R},\Omega^{k}(\Sigma,\mathrm{Ad}(\P_{\Sigma})))\cong C^{\infty}(\bb{R},\Omega^{k}(\Sigma))\otimes\Gamma(\mathrm{Ad}(\P))$, this operator acts on pure tensors $\omega\otimes\Phi$ with $\omega\in C^{\infty}(\bb{R},\Omega^{k}(\Sigma))$ and $\Phi\in\Gamma(\mathrm{Ad}(\P))$ as $\mathcal{D}_{t}^{A_{\mathrm{T}}}(\omega\otimes\Phi)=\partial_{t}\omega\otimes\Phi+\omega\otimes\nabla^{A}_{\partial_{t}}\Phi$.
\end{definition}

The operator $\mathcal{D}_{t}^{A_{\mathrm{T}}}$ plays the role of a natural (gauge-covariant) time-derivative for time-dependent $\mathrm{Ad}(\P_{\Sigma})$-valued forms on $\Sigma$, analogously to $\partial_{t}$ on $C^{\infty}(\mathbb{R},\Omega^{\bullet}(\Sigma))$. A direct way to see this is as follows: for forms in $\Omega^{k}(\M,\mathrm{Ad}(\P))=\Gamma(\E_{k})$, a natural time-derivative is provided by considering the directional derivative $\nabla^{A,\g}_{\partial_{t}}\:\Omega^{k}(\M,\mathrm{Ad}(\P))\to\Omega^{k}(\M,\mathrm{Ad}(\P))$, which is the connection on $\E_{k}=\bigwedge^{k}\T^{\ast}\M\otimes\mathrm{Ad}(\P)$ induced from $\nabla^{A}$ and the Levi-Civita connection $\nabla$ of $(\M,\g)$. In the ultrastatic case, it is easy to check that $\nabla^{A,\g}_{\partial_{t}}$ \emph{preserves} the spacetime decomposition $\omega=\d t\wedge \omega_{\mathrm{T}}+\omega_{\Sigma}$ and its two components are exactly the componentwise $\mathcal{D}_{t}^{A_{\mathrm{T}}}$-derivatives. Indeed, we have the following lemma. 

\begin{lemma}
\label{Lem: 2.26}
    Let $(\M=\mathbb{R}\times\Sigma,\g=-\d t\otimes\d t+\h)$ be an ultrastatic globally hyperbolic spacetime and $\P$ be a principal $\G$-bundle over $\M$. Moreover, fixing an identification $\P=\mathbb{R}\times\P_{\Sigma}$, let $A\in\mathcal{C}(\P)$ be a connection $1$-form decomposed as $A=A_{\mathrm{T}}\d t+A_{\Sigma}$.
    \begin{itemize}
        \item[(i)]Let $\omega\in\Omega^{k}(\M,\mathrm{Ad}(\P))$, with $k \ge 1$, be decomposed as $\omega=\d t\wedge\omega_{\mathrm{T}}+\omega_{\Sigma}$ with $\omega_{\mathrm{T}}\in C^{\infty}(\mathbb{R},\Omega^{k-1}(\Sigma,\mathrm{Ad}(\P_{\Sigma})))$ and $\omega_{\Sigma}\in C^{\infty}(\mathbb{R},\Omega^{k}(\Sigma,\mathrm{Ad}(\P_{\Sigma})))$. Then, $\nabla^{A,\g}_{\partial_{t}}\omega\in\Omega^{k}(\M,\mathrm{Ad}(\P))$ decomposes as
    \begin{align*}
        (\nabla^{A,\g}_{\partial_{t}}\omega)_{\mathrm{T}}=\mathcal{D}^{A_{\mathrm{T}}}_{t}\omega_{\mathrm{T}}\,,\qquad (\nabla^{A,\g}_{\partial_{t}}\omega)_{\Sigma}=\mathcal{D}^{A_{\mathrm{T}}}_{t}\omega_{\Sigma}
    \end{align*}
    \item[(ii)]In the case $k=0$, we can identify $C^{\infty}(\mathbb{R},\Omega^{0}(\Sigma,\mathrm{Ad}(\P_{\Sigma})))\cong\Gamma (\M,\mathrm{Ad}(\P))$. On this space, it holds that $\mathcal{D}^{A_{\mathrm{T}}}_{t}=\nabla^{A}_{\partial_{t}}$.
    \item[(iii)]If $A$ is flat, it holds that $[\mathcal{D}_{t}^{A_{\mathrm{T}}},\d^{A_{\Sigma}}]=0$ and $[\mathcal{D}_{t}^{A_{\mathrm{T}}},\delta^{A_{\Sigma}}]=0$.
    \end{itemize}
\end{lemma}

\begin{proof}
    We first prove item (ii): let $\phi\in\Gamma(\M,\mathrm{Ad}(\P)) \equiv \Omega^{0}(\M,\mathrm{Ad}(\P))$. As a special case of the isomorphism in Eq. \eqref{eq:Iso}, $\phi$ can be identified with a $\G$-equivariant element $\varphi\in C^{\infty}(\P,\mathfrak{g})$ such that $\phi = \varphi_\M$. Now, we note that $\nabla^{A}_{\partial_{t}}\phi= \partial_t \lrcorner \d^{A}\phi =(\d^{A}\phi)_{\mathrm{T}}$, since for a $0$-form $\phi \in \Omega^{0}(\M,\mathrm{Ad}(\P))$, it holds that $(\d^A \phi)(\bullet) = \nabla^A_{\bullet} \phi$. By Lemma \ref{Lem: 2.24}, we thus have that
    \begin{align*}
        (\mathrm{D}^{A}\varphi)_{\mathrm{T}}=\partial_{t}\lrcorner (\d\varphi+[A\wedge\varphi]_{\mathfrak{g}}))=\partial_{t}\varphi+[A_{\mathrm{T}},\varphi]_{\mathfrak{g}}=\widetilde{\mathcal{D}}_{t}^{A_{\mathrm{T}}}\varphi\, ,
    \end{align*}
    where $\mathrm{D}^A\: \Omega^k(\P, \mathfrak{g}) \to \Omega^{k+1}(\P, \mathfrak{g})$ is the operator defined in Eq.~\eqref{Eq.Def.OpDA}, which is such that $(\mathrm{D}^A \omega)_\M = \d^A \omega_\M$ for any $\omega \in \Omega_{\mathrm{hor},\mathrm{Ad}}^k(\sf{P}, \mathfrak{g})$, see Eq.~\eqref{eq:Iso}. Since $\widetilde{\mathcal{D}}^{A_\mathrm{T}}_t$ preserves $\G$-equivariant functions, under the identification $C^\infty(P, \mathfrak{g})^\G \simeq \Gamma(\M, \Ad(\P))$, we can conclude that $\nabla^{A}_{\partial_{t}}\phi=\mathcal{D}_{t}^{A_{\mathrm{T}}}\phi$, as claimed. For (i), suppose first that $\omega\in\Omega^{k}(\M,\Ad(\P))$ is a pure tensor of the form $\omega=\alpha\otimes\Phi,$ with $\alpha\in\Omega^{k}(\M)$ and $\Phi\in\Gamma(\Ad(\P)).$ Decomposing $\alpha=\d t\wedge\alpha_{\mathrm T}+\alpha_\Sigma,$ we thus obtain $\omega_{\mathrm T}=\alpha_{\mathrm T}\otimes\Phi,$ and $\omega_\Sigma=\alpha_\Sigma\otimes\Phi.$ Since $\nabla^{A,\g}$ is the tensor-product connection on $\E_k$ induced by the Levi-Civita connection $\nabla^\g$ and by the connection $\nabla^A$ on $\Ad(\P)$, it follows that
$$\nabla^{A,\g}_{\partial_t}\omega = (\nabla^\g_{\partial_t}\alpha)\otimes\Phi + \alpha\otimes\nabla^A_{\partial_t}\Phi.$$
Since the spacetime is ultrastatic, the Levi-Civita connection preserves the splitting into temporal and spatial components, hence we may write $\nabla^\g_{\partial_t}\alpha = \d t\wedge\partial_t\alpha_{\mathrm T} + \partial_t\alpha_\Sigma.$
Consequently,
\begin{align*}
\nabla^{A,\g}_{\partial_t}\omega &= \d t\wedge \Bigl( \partial_t\alpha_{\mathrm T}\otimes\Phi + \alpha_{\mathrm T}\otimes\nabla^A_{\partial_t}\Phi \Bigr) +
\Bigl(\partial_t\alpha_\Sigma\otimes\Phi+\alpha_\Sigma\otimes\nabla^A_{\partial_t}\Phi \Bigr).
\end{align*}
By Definition \ref{Def: covariant time derivative}, we have
$$
\mathcal D_t^{A_{\mathrm T}}
(\alpha_{\mathrm T}\otimes\Phi) = \partial_t\alpha_{\mathrm T}\otimes\Phi + \alpha_{\mathrm T}\otimes\nabla^A_{\partial_t}\Phi, \qquad \mathcal D_t^{A_{\mathrm T}}
(\alpha_\Sigma\otimes\Phi)
=
\partial_t\alpha_\Sigma\otimes\Phi
+
\alpha_\Sigma\otimes\nabla^A_{\partial_t}\Phi.$$
Hence,
$$(\nabla^{A,\g}_{\partial_t}\omega)_{\mathrm T} = \mathcal D_t^{A_{\mathrm T}}\omega_{\mathrm T}, \qquad (\nabla^{A,\g}_{\partial_t}\omega)_\Sigma = \mathcal D_t^{A_{\mathrm T}}\omega_\Sigma.
$$
Since $\Omega^k(\M,\Ad(\P))$ is locally generated, as a $C^\infty(\M)$-module, by such pure tensors, the result follows for arbitrary $\omega\in\Omega^k(\M,\Ad(\P))$. 

To establish item (iii), let $\eta\in C^{\infty}(\bb{R},\Omega_{\mathrm{hor},\mathrm{Ad}}^{k}(\P_{\Sigma},\mathfrak{g}))$ be arbitrary. Then, by definition, we have that $\mathrm{D}^{A_{\Sigma}}\eta=\d_{\Sigma}\eta+[A_\Sigma \wedge \eta]_{\mathfrak{g}}$, while $\widetilde{\mathcal{D}}^{A_{\mathrm{T}}}_{t}\eta=\partial_{t}\eta+[A_{\mathrm{T}},\eta]_{\mathfrak{g}}$. Hence, we find
\begin{equation*}
     [\widetilde{\mathcal{D}}^{A_\mathrm{T}}_t, \mathrm{D}^{A_\Sigma}] \eta = [(\partial_t A_\Sigma - \d_\Sigma A_\mathrm{T} + [A_\mathrm{T}, A_\Sigma]_\mathfrak{g}) \wedge \eta]_{\mathfrak{g}}. 
\end{equation*}
By Cartan's structure equation and the computation in the proof of Lemma~\ref{Lemma:FlatImpliesFlat}, it follows that $F^A = \d t \wedge (\partial_t A_\Sigma - \d_\Sigma A_\mathrm{T} + [A_\mathrm{T}, A_\Sigma]_\mathfrak{g}) + F^{A_\Sigma}$. Hence,
\begin{equation*}
    [\widetilde{\mathcal{D}}^{A_\mathrm{T}}_t, \mathrm{D}^{A_\Sigma}] = [(F^A)_\mathrm{T} \wedge \cdot]_\mathfrak{g} = 0,
\end{equation*}
on account of the fact that the connection $A$ is flat. By the isomorphism~\eqref{eq:Iso} and Lemma~\ref{Lem: 2.24}, we conclude that $[\mathcal{D}^{A_{\mathrm{T}}}_{t},\d^{A_{\Sigma}}]=0$. Similarly, using the fact that $\mathcal{D}^{A_\mathrm{T}}_t$ commutes with $\d^{A_\Sigma}$ by the first part of this proof and with the Hodge-star, it follows that $ [\mathcal{D}^{A_\mathrm{T}}_t, \delta^{A_\Sigma}] = 0$.
\end{proof}

We are finally in the position to generalise Lemma~\ref{Lemma:DecScalar} to the setting of linearised Yang-Mills fields, \textit{i.e.},~$\mathrm{Ad}(\P)$-valued forms on $\M$. 

\begin{proposition}\label{Prop:DecompAdFormsOp}
Let $(\M=\mathbb{R}\times\Sigma,\g=-\d t\otimes\d t+\h)$ be an ultrastatic globally hyperbolic spacetime and let $\P$ be a principal $\G$-bundle over $\M$. Moreover, fixing an identification $\P=\mathbb{R}\times\P_{\Sigma}$, let $A\in\mathcal{C}(\P)$ be a connection $1$-form decomposed as $A=A_{\mathrm{T}}\d t+A_{\Sigma}$. Then, the following properties hold true:
\begin{align*}
    \text{(i)}&\qquad (\d^{A} \varepsilon)_{\mathrm{T}}=\mathcal{D}_{t}^{A_{\mathrm{T}}}\varepsilon\,,\qquad (\d^{A} \varepsilon)_{\Sigma}=\d^{A_{\Sigma}}\varepsilon\qquad &&\text{for}\quad \varepsilon\in \Omega^{0}(\M,\mathrm{Ad}(\P))\\
    \text{(ii)}&\qquad (\d^{A}\omega)_{\mathrm{T}}=\mathcal{D}_{t}^{A_{\mathrm{T}}}\omega_{\Sigma}-\d^{A_\Sigma}\omega_{\mathrm{T}}\,,\qquad (\d^{A} \omega)_{\Sigma}=\d^{A_{\Sigma}}\omega_{\Sigma}\qquad &&\text{for} \quad\omega\in \Omega^{k\geq 1}(\M,\mathrm{Ad}(\P))\\
    \text{(iii)}&\qquad  \delta^{A} \alpha=\mathcal{D}_{t}^{A_{\mathrm{T}}}\alpha_{\mathrm{T}}+\delta^{A_\Sigma}\alpha_{\Sigma}\qquad &&\text{for}\quad\alpha\in \Omega^{1}(\M,\mathrm{Ad}(\P))\\
    \text{(iv)}&\qquad  (\delta^{A}\omega)_{\mathrm{T}}=-\delta^{A_\Sigma}\omega_{\mathrm{T}}\,,\qquad (\delta^{A}\omega)_{\Sigma}=\mathcal{D}_{t}^{A_{\mathrm{T}}}\omega_{\mathrm{T}}+\delta^{A_\Sigma}\omega_{\Sigma}\qquad &&\text{for}\quad \omega\in \Omega^{k\geq 2}(\M,\mathrm{Ad}(\P))
\end{align*}
In addition, assuming $A$ to be flat and denoting by $\square^{A}=\d^{A}\delta^{A}+\delta^{A}\d^{A}$ and $\Delta^{A_{\Sigma}}:=\d^{A_{\Sigma}}\delta^{A_{\Sigma}}+\delta^{A_{\Sigma}}\d^{A_{\Sigma}}$ the twisted de Rham-Hodge d'Alembertian acting on $\Omega^{k}(\M,\mathrm{Ad}(\P))$ and the Laplacian acting on $C^{\infty}(\mathbb{R},\Omega^{k}(\Sigma,\mathrm{Ad}(\P_{\Sigma})))$, respectively, we obtain for all $\omega\in \Omega^{k}(\M,\mathrm{Ad}(\P))$,
\begin{align*}
    \text{(v)}&\qquad (\square^{A}\omega)_{\mathrm{T}}=((\mathcal{D}_{t}^{A_{\mathrm{T}}})^{2}+\Delta^{A_{\Sigma}})\omega_{\mathrm{T}}\,,\qquad (\square^{A}\omega)_{\Sigma}=((\mathcal{D}_{t}^{A_{\mathrm{T}}})^{2}+\Delta^{A_{\Sigma}})\omega_{\Sigma} \, .
\end{align*}
\end{proposition}

\begin{proof}
    Claims (i)-(iv) follow essentially from the definition of the relevant operators and similar steps used in order to obtain Lemma~\ref{Lemma:DecScalar}. More precisely, consider a local coordinate chart $(\mathcal{U},(x^{\alpha})_{\alpha=0,\dots,d})$ of $\sf{M}$ adapted to the product $\sf{M}=\bb{R}\times\Sigma$, \textit{i.e.},~$\mathcal{U}=\bb{R}\times\mathcal{U}_{\Sigma}$ for some open $\mathcal{U}_{\Sigma}\subset\Sigma$, $x^{0}=t$ and $(\mathcal{U}_{\Sigma},x^{i})_{i=1,\dots,d}$ being local coordinates on $\Sigma$. Then, by definition, the operators $\d^{A}$ and $\delta^{A}$ can be written in terms of $\nabla^{A,\g}$ according to
    \begin{align*}
        &(\d^{A}\omega)_{\alpha_{1}\dots\alpha_{k+1}}:=(\d^{A}\omega)(\partial_{\alpha_{1}},\dots,\partial_{\alpha_{k+1}})=\frac{1}{k!}\sum_{\sigma\in\mathfrak{S}^{k+1}}\mathrm{sign}(\sigma)(\nabla^{A,\g}_{\partial_{\alpha_{\sigma(1)}}}\omega)(\partial_{\alpha_{\sigma(2)}},\dots,\partial_{\alpha_{\sigma(k+1)}})\\
        & (\delta^{A}\omega)_{\alpha_{1}\dots\alpha_{k-1}}:=(\delta^{A}\omega)(\partial_{\alpha_{1}},\dots,\partial_{\alpha_{k-1}})=-\g^{\mu\nu}(\nabla^{A,\g}_{\partial_{\mu}}\omega)(\partial_{\nu},\partial_{\alpha_{1}},\dots,\partial_{\alpha_{k-1}})
    \end{align*}
    for all $\omega\in\Omega^{k}(\M,\mathrm{Ad}(\P))$. Now, in the \emph{ultrastatic} case, it is easy to see that $\nabla^{A,\g}_{\partial_{i}}=\nabla^{A_{\Sigma},\h}_{\partial_{i}}$ for \emph{spatial} indices $i=1,\dots,d$. Using Lemma~\ref{Lem: 2.26}(i), a straightforward computation yields the claims. 

    For (v), let $\omega\in\Omega^{k}(\M,\mathrm{Ad}(\P))$. Using (ii) and (iv), we immediately obtain the two equations
    \begin{align*}
        (\d^{A}\delta^{A}\omega)_{\mathrm{T}}&=\mathcal{D}_{t}^{A_{\mathrm{T}}}(\delta^{A}\omega)_{\Sigma}-\d^{A_{\Sigma}}(\delta^{A}\omega)_{\mathrm{T}}=\mathcal{D}_{t}^{A_{\mathrm{T}}}\mathcal{D}^{A_{\mathrm{T}}}_{t}\omega_{\mathrm{T}}+\mathcal{D}_{t}^{A_{\mathrm{T}}}\delta^{A_{\Sigma}}\omega_{\Sigma}+\d^{A_{\Sigma}}\delta^{A_{\Sigma}}\omega_{\mathrm{T}}\\
        (\delta^{A}\d^{A}\omega)_{\mathrm{T}}&=-\delta^{A_{\Sigma}}(\d^{A}\omega)_{\mathrm{T}}=-\delta^{A_{\Sigma}}\mathcal{D}_{t}^{A_{\mathrm{T}}}\omega_{\Sigma}+\delta^{A_{\Sigma}}\d^{A_{\Sigma}}\omega_{\mathrm{T}}
    \end{align*}
    Now, assuming $A$ to be flat, we have that  $[\mathcal{D}_{t}^{A_{\mathrm{T}}},\d^{A_{\Sigma}}]=0$ as well as $[\mathcal{D}_{t}^{A_{\mathrm{T}}},\delta^{A_{\Sigma}}]=0$, by Lemma~\ref{Lem: 2.26}(iii), which proves
    \begin{align*}
        (\square^{A}\omega)_{\mathrm{T}}=((\mathcal{D}_{t}^{A_{\mathrm{T}}})^{2}+\Delta^{A_{\Sigma}})\omega_{\mathrm{T}}\, .
    \end{align*}
    Following similar steps, we obtain $(\square^{A}\omega)_{\Sigma}=((\mathcal{D}_{t}^{A_{\mathrm{T}}})^{2}+\Delta^{A_{\Sigma}})\omega_{\Sigma}$, which concludes the proof.
\end{proof}

\subsection{Phase Spaces on Initial Data and Cauchy Problem}
\label{Subsec:Phase Spaces on Initial Data and Cauchy Problem}
Having discussed both nonlinear and linearised Yang–Mills theory in Sections~\ref{Subsec:Yang-Mills Theory on Globally Hyperbolic Spacetimes} and \ref{Subsec:Linearised Yang-Mills Theory in the Hack-Schenkel Formalism}, respectively, we now turn to the quantisation of the linearised theory. In particular, we begin by defining an equivalent phase space at the level of initial data. Following the general discussion of linearised Yang--Mills theory from
Section~\ref{Subsec:Yang-Mills Theory on Globally Hyperbolic Spacetimes}, let us consider the following general set-up:
\begin{itemize}
    \item[$\bullet$]A globally hyperbolic spacetime $(\M=\mathbb{R}\times\Sigma,\,\g=-\beta^{2}\d t\otimes\d t+\h_{t})$ with corresponding
    foliation by Cauchy hypersurfaces
    $(\Sigma_{t})_{t\in\bb{R}}$, where $\Sigma_t\defeq\{t\}\times\Sigma$.
    \item[$\bullet$] A Lie group $\G$ satisfying Assumption~\ref{Assumption}
    and a principal $\G$-bundle $\P$ over $\M$, for which we fix an
    identification $\P=\mathbb{R}\times\P_\Sigma$, with $\P_\Sigma$ a
    principal $\G$-bundle over $\Sigma$, cf., Section~\ref{Subsec:Decomposition of Connections and Linearised Yang-Mills Fields}.
    \item[$\bullet$]A \emph{flat} background connection
    $\A\in\mathcal C(\P)$.
\end{itemize}

With this set-up in place, we consider \emph{Yang-Mills theory linearised around $A$}, which, according to Proposition~\ref{Prop:LinYangMills}, is characterised by the
Hermitian vector bundles 
\begin{align*}
    \bigg(\E_k:=\bigwedge^k\T^*\M\otimes\Ad(\P)\,,\qquad\langle\cdot,\cdot\rangle_{\E_k}:=\frac{1}{k!}(\sf{g}^{\sharp})^{\otimes k}(\cdot,\cdot)\otimes\langle\cdot,\cdot\rangle_{\mathrm{Ad}(\P)}\bigg)\,,
\end{align*}
\emph{cf.}, Eq.~\eqref{Eq.DefEk1}, together with the linearised Yang--Mills operator
$\P\:\Gamma(\E_1)\to\Gamma(\E_1)$ and linearised gauge transformations
$\K\:\Gamma(\E_0)\to\Gamma(\E_1)$ given respectively by
\begin{align*}
    \P=\delta^\A\d^\A\,,\qquad \K=\d^\A\,.
\end{align*}
Since the background connection is flat, the normally hyperbolic operators
of Proposition~\ref{Prop:LinYangMills} reduce to $\mathsf D_0=\mathsf D_1=\square^\A$. The physical content of the theory is encoded in the \emph{classical phase space}, which, following Definition~\ref{Def: classical phase space}, is the (generally degenerate) Hermitian space given by
\begin{align*}
	\bigg(\mathcal{V}_{\mathrm{P}}=\cfrac{\mathrm{ker}(\sf{K}^{\ast}\vert_{\Gamma_{c}})}{\mathrm{ran}(\sf{P}\vert_{\Gamma_{c}})}\,,\qquad \sigma([\psi],[\varphi]):=(\psi,\mathsf{i}\sf{G}_{1}\varphi)_{\sf{E}_{1}}\bigg)\, .
\end{align*}

Now, as a first step, let us fix the notation for the initial data. Throughout this subsection we keep the reference Cauchy hypersurface $\Sigma_0=\{0\}\times\Sigma$ fixed. When considering the normally hyperbolic operators $\square^\A$
acting on $\Gamma(\E_k)=\Omega^k(\M,\Ad(\P))$, the initial data for the
corresponding Cauchy problems are naturally given by the pair
\begin{align*}
    \left(\omega,\,\,
    \mathrm{i}^{-1}\nabla^{\A,\g}_{\partial_t}\omega\right)^{\top}\big\vert_{t=0},
\end{align*}
where the factor $\mathrm{i}^{-1}$ is purely conventional and where we recall that
$\nabla^{\A,\g}$ is the natural connection acting on sections of $\E_k$,
defined as the tensor product of the gauge covariant derivative
$\nabla^\A$ and the Levi--Civita connection of $(\M,\g)$. Hence, the space of initial data is naturally parametrised as the space of sections of the vector bundles
\begin{align}\label{eq:ERhok}
    \E_{\rho_k}:=\E_{k}\vert_{\Sigma_{0}}\oplus\E_{k}\vert_{\Sigma_{0}}
\end{align}
over $\Sigma$. In view of Eq.~\eqref{eq:DecompAdValued}, it is natural to decompose the initial data accordingly into their temporal and spatial components. We therefore identify
\begin{align*}
    \E_{\rho_{0}}\cong\E_{\Sigma,0}\oplus\E_{\Sigma,0}\,,\qquad \E_{\rho_{k}}\cong \E_{\Sigma,k-1}\oplus \E_{\Sigma,k-1}\oplus \E_{\Sigma,k}\oplus  \E_{\Sigma,k}\,,\qquad k\geq 1\,,
\end{align*}
where the Hermitian bundles $(\E_{\Sigma,k},\langle\cdot,\cdot\rangle_{\E_{\Sigma,k}})$ are defined, in analogy to $(\E_{k},\langle\cdot,\cdot\rangle_{\E_{k}})$, as
\begin{align*}
    \bigg(\E_{\Sigma,k}:=\bigwedge^k\T^*\Sigma\otimes\Ad(\P_{\Sigma})\,,\qquad\langle\cdot,\cdot\rangle_{\E_{\Sigma,k}}:=\frac{1}{k!}(\sf{h}_{0}^{\sharp})^{\otimes k}(\cdot,\cdot)\otimes\langle\cdot,\cdot\rangle_{\mathrm{Ad}(\P_{\Sigma})}\bigg)\,,
\end{align*}
\emph{cf.}, Eq.~\eqref{Eq: E Sigma k}. On the level of sections, this implies that
\begin{align}\label{eq:DecomSecID}
    \Gamma(\E_{\rho_0})&\cong \Omega^0(\Sigma,\Ad(\P_\Sigma))^{\oplus2},\nonumber\\
    \Gamma(\E_{\rho_k})&\cong \Omega^{k-1}(\Sigma,\Ad(\P_\Sigma))^{\oplus2}
    \oplus\Omega^k(\Sigma,\Ad(\P_\Sigma))^{\oplus2},\qquad k\geq1.
\end{align}

In view of the Cauchy problem, we introduce the corresponding \emph{Cauchy data maps} $\rho_k\:\Gamma(\E_k)\to\Gamma(\E_{\rho_k})$ using this parametrisation as
\begin{align}\label{eq:CauchyDataMaps}
    \rho_0(\varepsilon)&\defeq \bigg(\varepsilon,\,\,\mathrm{i}^{-1}\nabla^\A_{\partial_t}\varepsilon\bigg)^{\top}\big\vert_{t=0}\,,\nonumber\\ \rho_k(\omega)&\defeq\bigg(\omega_{\mathrm T},\,\,\mathrm{i}^{-1}(\nabla^{A,\g}_{\partial_{t}}
        \omega)_{\mathrm T},\,\,\omega_\Sigma,\,\,
        \mathrm{i}^{-1}(\nabla^{A,\g}_{\partial_{t}}
        \omega)_{\Sigma}\vert_{t=0}\bigg)^{\top}\big\vert_{t=0},\qquad k\geq1\,.
\end{align}
Since $\square^{A}$ are normally hyperbolic operators, their Cauchy problem is well-posed, meaning that the equation $\square^{A}\omega=0$ with initial data $\rho_{k}(\omega)=\mathfrak{f}$ for some $\mathfrak{f}\in\Gamma(\E_{\rho_{k}})$ admits a unique solution $\omega\in\Gamma(\E_{k})$. In other words, the operators $\rho_{k}$ are invertible when restricted to $\mathrm{ker}(\square^{A})$ and we denote their inverses, the \emph{Cauchy evolution operators}, by $\mathcal{U}_{k}\:\Gamma(\E_{\rho_{k}})\to\Gamma(\E_{k})$. By \emph{finite speed of propagation}, we note that they admit well-defined restrictions $\mathcal{U}_{k}\:\Gamma_{\mathrm{c}}(\E_{\rho_{k}})\to\Gamma_{\mathrm{sc}}(\E_{k})$.

After this preliminary discussion, we now move on to the phase space of linearised Yang--Mills theory on the level of initial data. We follow the general strategy for quantising linear gauge theories developed in \cite{HackSchenkel,GerardWrochna} (cf.~\cite[Chap.~2]{SchmidPhD} for a detailed review). To start with, we encode the gauge transformations and the Lorenz gauge condition on the level of initial data by introducing the linear differential operators
\begin{align*}
    \K_\Sigma\:\Gamma(\E_{\rho_0})\to\Gamma(\E_{\rho_1}),
    &\qquad \K_\Sigma\defeq\rho_1\K\mathcal U_0,
    \nonumber\\
    \K_\Sigma^\dagger\:\Gamma(\E_{\rho_1})\to\Gamma(\E_{\rho_0}),
    &\qquad \K_\Sigma^\dagger\defeq\rho_0\K^*\mathcal U_1.
\end{align*}
We will give explicit expressions of these operators below in the special case of ultrastatic manifolds. Observe that
$\K_\Sigma^\dagger\K_\Sigma=0$. The origin of the notation ``$\dagger$'' will become clear below. With this notation, we introduce the \emph{Cauchy surface phase space} of linearised Yang-Mills theory by 
\begin{align}\label{eq:VSigmaDef}
    \mathcal{V}_{\Sigma}:=\cfrac{\mathrm{ker}(\K_{\Sigma}^{\dagger}\vert_{\Gamma_{\mathrm{c}}})}{\mathrm{ran}(\K_{\Sigma}\vert_{\Gamma_{\mathrm{c}}})}\, .
\end{align}
Let us next equip this space with a suitable Hermitian form defined in analogy to $\sigma$ above. To start with, let us equip the bundles $\E_{\rho_k}$ with their natural (positive-definite) bundle metrics
\begin{align}\label{eq:BundMetrErho}
    \langle\cdot,\cdot\rangle_{\E_{\rho_{0}}}=\langle\cdot,\cdot\rangle_{\E_{\Sigma,0}}^{\oplus 2}\,,\qquad \langle\cdot,\cdot\rangle_{\E_{\rho_{k}}}=\langle\cdot,\cdot\rangle_{\E_{\Sigma,k-1}}^{\oplus 2}\oplus\langle\cdot,\cdot\rangle_{\E_{\Sigma,k}}^{\oplus 2}\,,\qquad k\geq 1\, .
\end{align}
Next, we define linear differential operators
$\G_{\Sigma,k}\:\Gamma(\E_{\rho_k})
\to\Gamma(\E_{\rho_k})$, which are uniquely determined by the
relation
\begin{align}
    \G_k=(\rho_k\G_k)^*\G_{\Sigma,k}(\rho_k\G_k),
    \label{Eq:GSigmaAbstractYM}
\end{align}
where $(\rho_k\G_k)^*$ denotes the adjoint of $\rho_k\G_k$ with respect to
$(\cdot,\cdot)_{\E_k}$ and $(\cdot,\cdot)_{\E_{\rho_k}}$. Using the bundle metrics~\eqref{eq:BundMetrErho}, we define Hermitian forms by setting
\begin{align}
    \sigma_{\Sigma,k}\:\Gamma_{\mathrm{c}}(\E_{\rho_k})\times
    \Gamma_{\mathrm{c}}(\E_{\rho_k})\to\mathbb C,
    \qquad
    \sigma_{\Sigma,k}(\cdot,\cdot)
    \defeq \mathrm{i}(\cdot,\G_{\Sigma,k}\cdot)_{\E_{\rho_k}}.
    \label{Eq:sigmaKSigmaYM}
\end{align}
Note that the choice of the fibre metrics $\langle\cdot,\cdot\rangle_{\sf{E}_{\rho_k}}$ on $\E_{\rho_k}$ is arbitrary and we choose them to be positive-definite for convenience. In particular, also the explicit expression of $\G_{\Sigma,k}$ depends on that choice; however, the Hermitian forms $\sigma_{\Sigma,k}$ are uniquely fixed by the theory, \emph{i.e.}, by
$\G_k$ and the spacetime Hermitian forms on $\E_k$. 

By definition, $\K_\Sigma^\dagger$ is the formal adjoint of $\K_\Sigma$ with respect to
the Hermitian forms $\sigma_{\Sigma,k}$. Hence, $\sigma_{\Sigma,1}$ is well-defined on the quotient space $\mathcal{V}_{\Sigma}$ and we obtain the following identification:

\begin{proposition}[\textit{e.g.}~\protect{\cite[Sec.~2.4]{GerardWrochna}}]
There is a unitary isomorphism
\begin{align*}
   (\mathcal{V}_{\mathrm{P}},\sigma)\xrightarrow{[\rho_{1}\G_1]} (\mathcal{V}_{\Sigma},\sigma_{\Sigma,1})\, .
\end{align*}
\end{proposition}

For the remainder of this section, the goal is to derive explicit expressions of the operators $\K_{\Sigma}$, $\K_{\Sigma}^{\dagger}$ and $\G_{\Sigma,k}$ in the special case of linearised Yang-Mills theory on \emph{ultrastatic spacetimes}. Therefore, throughout the following discussion, we restrict ourselves to an ultrastatic globally hyperbolic spacetime of the form
\begin{align*}
    \M=\mathbb{R}\times\Sigma\,,\qquad \g=-\d t\otimes\d t+\h\,,
\end{align*}
where $(\Sigma,\h)$ is a complete Riemannian manifold. Following Proposition~\ref{Prop:DecompConn}, we decompose our flat background connection $\A\in\mathcal C(\P)$ as $\A=\A_{\mathrm T}\d t+\A_\Sigma$ with $\A_\Sigma\in C^\infty(\mathbb R,\mathcal C(\P_\Sigma))$. Now, in the ultrastatic case, let us recall that the time-dependent forms $(\nabla^{A,\g}_{\partial_{t}}\omega)_{\mathrm{T}}$ and $(\nabla^{A,\g}_{\partial_{t}}\omega)_{\Sigma}$ appearing in the definition of the Cauchy data maps $\rho_{k}$ introduced in Eq.~\eqref{eq:CauchyDataMaps} above only depend on the temporal component $A_{\mathrm{T}}$ of the given background connection. In particular, recalling the covariant time-derivative operators $\mathcal{D}_{t}^{A_{\mathrm{T}}}$ defined in Definition~\ref{Def: covariant time derivative}, we have that
\begin{align*}
     \rho_k(\omega)=\bigg(\omega_{\mathrm T},\,\,\mathrm{i}^{-1}\mathcal{D}_{t}^{A_{\mathrm{T}}}\omega_{\mathrm T},\,\,\omega_\Sigma,\,\,
        \mathrm{i}^{-1}\mathcal{D}_{t}^{A_{\mathrm{T}}}
        \omega_{\Sigma}\bigg)^{\top}\big\vert_{t=0}
\end{align*}
in this case. First and foremost, we compute explicitly the operators $\K_\Sigma$ and $\K_\Sigma^\dagger$ in this setting. As in the Maxwell case, it is convenient to use a slightly different parametrisation of initial data for the operators $\mathcal{D}_{i}$ due to Furlani \cite{FurlaniCauchy} (\emph{cf}, also, \cite{GerardWrochna,Pfenning}) as an intermediate computational tool, namely
\begin{align} \label{Eq:FurlaniYM}
    \rho_{0}^{\mathrm{F}}:=\rho_{0}\,,\qquad\rho_{1}^{\mathrm{F}}(\alpha):=\bigg(\alpha_{\mathrm T},\,\,\mathrm{i}^{-1}\delta^\A\alpha,\,\,\alpha_\Sigma,\,\,\mathrm{i}^{-1}(\d^\A\alpha)_{\mathrm T}\bigg)^{\top}\big\vert_{t=0}\, .
\end{align}
It is quite easy to see that $\delta^{A}\alpha\vert_{t=0}$ serves as a replacement for the initial datum $\partial_{t}\alpha_{\mathrm{T}}\vert_{t=0}$, while $(\d^\A\alpha)_{\mathrm T}$ is meant to replace $\partial_{t}\alpha_{\Sigma}\vert_{t=0}$. Using Proposition~\ref{Prop:DecompAdFormsOp}, we observe that the initial data map $\rho_{1}$ introduced in Eq.~\eqref{eq:CauchyDataMaps} and the Furlani initial data map $\rho_{1}^{\mathrm{F}}$ in Eq.~\eqref{Eq:FurlaniYM} are related by $\rho_1^{\mathrm F}=T_{\mathrm F}\rho_1$ with bijective transition operator $T_{\mathrm F}\:\Gamma(\E_{\rho_{1}})\to\Gamma(\E_{\rho_{1}})$ given by
\begin{align}
    T_{\mathrm F}:=
    \begin{pmatrix}
        \mathds 1&0&0&0\\
        0&\mathds 1&\mathrm{i}^{-1}\delta^{\A_\Sigma^0}&0\\
        0&0&\mathds 1&0\\
        -\mathrm{i}^{-1}\d^{\A_\Sigma^0}&0&0&\mathds 1
    \end{pmatrix},\qquad T_{\mathrm F}^{-1}=
    \begin{pmatrix}
        \mathds 1&0&0&0\\
        0&\mathds 1&-\mathrm{i}^{-1}\delta^{\A_\Sigma^0}&0\\
        0&0&\mathds 1&0\\
        \mathrm{i}^{-1}\d^{\A_\Sigma^0}&0&0&\mathds 1
    \end{pmatrix}\,,
    \label{Eq:RFYM}
\end{align}
where we set $\A_\Sigma^0\defeq \A_\Sigma\vert_{t=0}\in\mathcal{C}(\P_{\Sigma})$. The advantage of using Furlani initial data in treating the Cauchy problem for the Maxwell and linearised Yang–Mills equations is that they are perfectly well adapted to the setting of differential forms. This becomes particularly transparent when computing the operators $\K_{\Sigma}$ and $\K_{\Sigma}^{\dagger}$ in this setting. For this, we introduce the notation
\begin{align*}
    \K_\Sigma^{\mathrm F}\defeq\rho_1^{\mathrm F}\K\mathcal U_0,
    \qquad
    \K_\Sigma^{\mathrm{F},\dagger}
    \defeq\rho_0\K^*\mathcal U_1^{\mathrm F}\,,
\end{align*}
where $\mathcal U_1^{\mathrm F}\defeq\mathcal U_1T_{\mathrm F}^{-1}$ is the Cauchy evolution operator with respect to $\rho_{1}^{\mathrm{F}}$.

\begin{lemma}
\label{Lem:FurlaniGaugeMapsYM}
On ultrastatic spacetimes and $A\in\mathcal{C}(\P)$ flat, it holds that 
\begin{align*}
    \K_\Sigma^{\mathrm F}
    =
    \begin{pmatrix}
        0&\mathrm{i}\mathds 1\\
        0&0\\
        \d^{\A_\Sigma^0}&0\\
        0&0
    \end{pmatrix},
    \qquad\qquad
     \K_\Sigma^{\mathrm{F},\dagger}
    =
    \begin{pmatrix}
        0&\mathrm{i}\mathds 1&0&0\\
        0&0&0&\delta^{\A_\Sigma^0}
    \end{pmatrix}\,,
\end{align*}
where we identified $\Gamma(\E_{\rho_{0}})$ and $\Gamma(\E_{\rho_{1}})$ as in Eq.~\eqref{eq:DecomSecID}.
\end{lemma}

\begin{proof}
Let $\mathfrak{f}=(\mathfrak{a},\pi)\in\Gamma(\E_{\rho_0})$ be $0$-form initial data and set
$\varepsilon\defeq\mathcal U_0(\mathfrak{a},\pi)\in\Gamma(\E_{0})$. By definition, $\varepsilon$ is the unique solution to the equation $\mathcal{D}_{0}\varepsilon=\square^{A}\varepsilon=\delta^{A}\d^{A}\varepsilon=0$ with initial data $\rho_{0}(\varepsilon)=(\mathfrak{a},\pi)$, \emph{i.e.}, $\varepsilon\vert_{t=0}=\mathfrak{a}$ and $\mathcal D_t^{\A_{\mathrm T}}\varepsilon\vert_{t=0}=i\pi$. Using Proposition~\ref{Prop:DecompAdFormsOp} and the fact that $\d^{A}\circ\d^{A}=0$, since $A$ is assumed to be flat, we hence find
\begin{align*}
    \K_\Sigma^{\mathrm F}\mathfrak{f}=\rho_1^{\mathrm F}(\d^\A\varepsilon)&=\bigg((\d^{A}\varepsilon)_{\mathrm T},\,\,\mathrm{i}^{-1}\delta^\A\d^{A}\varepsilon,\,\,(\d^{A}\varepsilon)_\Sigma,\,\,\mathrm{i}^{-1}(\d^\A\d^{A}\varepsilon)_{\mathrm T}\bigg)^{\top}\big\vert_{t=0}=\\&=\bigg(\mathcal{D}_t^{\A_{\mathrm T}}\varepsilon,\,\,0,\,\,\d^{\A_\Sigma}\varepsilon,\,\,0\bigg)^{\top}\big\vert_{t=0}=(i\pi,0,\d^{A_{\Sigma}^{0}}\mathfrak{a},0)^{\top}\,,
\end{align*}
which proves the claim for $\K_{\Sigma}^{\mathrm{F}}$.

Let now $\mathfrak{g}=(\mathfrak{a}_{\mathrm T},\pi_{\mathrm T},\mathfrak{a}_\Sigma,\pi_\Sigma)^\top\in\Gamma(\E_{\rho_{1}})$
be $1$-form initial data and set $\alpha\defeq\mathcal U_1^{\mathrm F}\mathfrak{g}$. By definition, $\mathcal{D}_{1}\alpha=\square^{A}\alpha=(\d^{A}\delta^{A}+\delta^{A}\d^{A})\alpha=0$ and $\rho_{1}^{\mathrm{F}}(\alpha)=\mathfrak{g}$, \emph{i.e.}, $\alpha_{\mathrm{T}}\vert_{t=0}=\mathfrak{a}_{\mathrm{T}}$, $\delta^{A}\alpha\vert_{t=0}=i\pi_{\mathrm{T}}$, $\alpha_{\Sigma}\vert_{t=0}=\mathfrak{a}_{\Sigma}$ and $(\d^{A}\alpha)_{\mathrm{T}}\vert_{t=0}=i\pi_{\Sigma}$. With Proposition~\ref{Prop:DecompAdFormsOp}, we then obtain
\begin{align*}
    \K_\Sigma^{\mathrm{F},\dagger}\mathfrak{g}=\rho_{0}(\delta^{A}\alpha)=\bigg(\delta^{A}\alpha,\,\,i^{-1}\mathcal{D}^{A_{\mathrm{T}}}_{t}(\delta^{A}\alpha)\bigg)^{\top}\big\vert_{t=0}=(i\pi_{\mathrm{T}},\,\,\delta^{A_{\Sigma}^{0}}\pi_{\Sigma})^{\top}\, .
\end{align*}
For the second component, we used Proposition~\ref{Prop:DecompAdFormsOp} and $\square^{A}\alpha=0$, which implies 
\begin{align*}
    \mathcal{D}^{A_{\mathrm{T}}}_{t}(\delta^{A}\alpha)=(\d^{A}\delta^{A}\alpha)_{\mathrm{T}}=-(\delta^{A}\d^{A}\alpha)_{\mathrm{T}}=\delta^{A_{\Sigma}}(\d^{A}\alpha)_{\mathrm{T}}\, .
\end{align*}
In summary, we obtain the claimed expression for $\K_\Sigma^{\mathrm{F},\dagger}$.
\end{proof}

Using the transformation operator $T_{\mathrm{F}}$ introduced in Eq.~\eqref{Eq:RFYM}, we obtain the expressions of $\K_{\Sigma}$ and $\K_{\Sigma}^{\dagger}$ with respect to the standard initial data maps $\rho_{i}$.

\begin{proposition}
\label{Prop:ExplicitKSigmaYM}
On ultrastatic spacetimes and $A\in\mathcal{C}(\P)$ flat, it holds that 
\begin{align*}
    \K_\Sigma =
    \begin{pmatrix}
        0&\mathrm{i}\mathds 1\\
        \mathrm{i}\Delta^{\A_\Sigma^0}&0\\
        \d^{\A_\Sigma^0}&0\\
        0&\d^{\A_\Sigma^0}
    \end{pmatrix},
    \qquad\qquad
     \K_\Sigma^\dagger
    =
    \begin{pmatrix}
        0&\mathrm{i}\mathds 1&\delta^{\A_\Sigma^0}&0\\
        \mathrm{i}\Delta^{\A_\Sigma^0}&0&0&\delta^{\A_\Sigma^0}
    \end{pmatrix}\,,
\end{align*}
where we identified $\Gamma(\E_{\rho_{0}})$ and $\Gamma(\E_{\rho_{1}})$ as in Eq.~\eqref{eq:DecomSecID}.
\end{proposition}

\begin{proof}
    By definition, we have the relations
\begin{align*}
    \K_\Sigma=T_{\mathrm F}^{-1}\K_\Sigma^{\mathrm F},
    \qquad
    \K_\Sigma^\dagger=\K_\Sigma^{\mathrm F,\dagger} T_{\mathrm F}\, .
\end{align*}
Using Lemma~\ref{Lem:FurlaniGaugeMapsYM} and the explicit expression of $T_{\mathrm{F}}$ in Equations~\eqref{Eq:RFYM}, we obtain the result. 
\end{proof}

Last but not least, let us compute the operators $\G_{\Sigma,0}$ and $\G_{\Sigma,1}$ appearing in the Hermitian forms $\sigma_{\Sigma,i}$ introduced in Eq.~\eqref{Eq:sigmaKSigmaYM} in the setting of linearised Yang-Mills theory on ultrastatic spacetimes.

\begin{proposition}\label{Prop:GSigma}
    On ultrastatic spacetimes and $A\in\mathcal{C}(\P)$ flat, it holds that
\begin{align*}
    \G_{\Sigma,0}
    =\frac1{\mathrm i}
    \begin{pmatrix}
        0&\mathds 1\\
        \mathds 1&0
    \end{pmatrix}, \qquad 
    \G_{\Sigma,1}
    =\frac1{\mathrm i}
    \begin{pmatrix}
        0&-\mathds 1&0&0\\
        -\mathds 1&0&0&0\\
        0&0&0&\mathds 1\\
        0&0&\mathds 1&0
    \end{pmatrix},
\end{align*}
where we identified $\Gamma(\E_{\rho_{0}})$ and $\Gamma(\E_{\rho_{1}})$ as in Eq.~\eqref{eq:DecomSecID}.
\end{proposition}

\begin{proof}
    This follows from a straightforward computation using \emph{Green's identity}. In general, if $\N=-\square^{\nabla^{\E}}+\sf{N}_{0}$ is a formally self-adjoint normally hyperbolic operator on some Hermitian vector bundle $(\E,\langle\cdot,\cdot\rangle_{\E})$, then it holds that
    \begin{align*}
		(\psi,\sf{G}\varphi)_{\sf{E}}=\int_{\Sigma_{0}}\bigg(\langle\nabla^{\E}_{\nu}\mathfrak{G}_{\psi},\mathfrak{G}_{\varphi}\rangle_{\sf{E}}-\langle\mathfrak{G}_{\psi},\nabla^{\E}_{\nu}\mathfrak{G}_{\varphi}\rangle_{\sf{E}}\bigg)\bigg\vert_{t=0}\,\d\mu_{\sf{h}_{0}}
	\end{align*}
    for all $\psi,\varphi\in\Gamma_c(\E)$, where $\mathfrak{G}_{\psi}:=\sf{G}\psi$ and $\mathfrak{G}_{\varphi}:=\sf{G}\varphi$ with $\G$ denoting the causal propagator of $\N$ and with $\nu$ denoting the future-directed timelike unit normal to $\Sigma_{0}$, which in the ultrastatic case is given by $\nu=\partial_{t}\vert_{t=0}$. For a proof, we refer, \emph{e.g.}, to \cite[Prop.~3.7]{SchmidPhD} or \cite[Lem.~2.2]{GerardMurroWrochna}. Applying this to the operators $\square^{A}$ on $\Gamma(\E_{i})=\Omega^{i}(\M,\mathrm{Ad}(\P))$ for $i=0,1$ and using the definitions of $\langle\cdot,\cdot\rangle_{\E_{k}}$, $\langle\cdot,\cdot\rangle_{\E_{\rho_k}}$ and $\G_{\Sigma,k}$ in Eq.~\eqref{Eq.DefEk2}, \eqref{eq:BundMetrErho} and \eqref{Eq:GSigmaAbstractYM}, respectively, a straightforward computation gives the claimed result (\emph{cf.}, \cite[Prop.~3.8]{SchmidPhD} for the analogous result in Maxwell's theory).
\end{proof}

\subsection{Quantisation and Hadamard States}
\label{Subsec:Quantisation and Hadamard States}
Let us now outline the quantisation of linearised Yang-Mills theory in this setting: the quantisation of a linear, free field gauge theory is a two-step procedure. First, one assigns to the classical phase space $(\mathcal V_{\mathrm P},\sigma)$ a unital $*$-algebra $\mathrm{CCR}(\mathcal V_{\mathrm P},\sigma)$, generated by $\mathds 1$, $\Phi(v)$, $\Phi^*(w)$ for all $v,w\in\mathcal V_{\mathrm P}$, with $v\mapsto\Phi(v)$ $\mathbb C$-anti-linear, $w\mapsto\Phi^*(w)$ $\mathbb C$-linear, $\Phi(v)^*=\Phi^*(v)$, and fulfilling the canonical commutation relations
    \begin{align*}
        [\Phi(v),\Phi(w)]=[\Phi^*(v),\Phi^*(w)]=0\,,\qquad
        [\Phi(v),\Phi^*(w)]=\sigma(v,w)\mathds 1\,.
    \end{align*}
Then, one determines the admissible physical states of the system by identifying a suitable subclass of the linear, positive and normalised functionals $\omega\:\mathrm{CCR}(\mathcal V_{\mathrm P},\sigma)\to\mathbb C$. Once a state is specified, the Gelfand--Naimark--Segal (GNS) construction guarantees the existence of a representation of the quantum field algebra as operators defined on a common dense subspace of some Hilbert space. In this paper, we will not be concerned with the explicit construction of such a representation and we restrict our attention to \emph{quasifree} states, which are fully determined by a pair of spacetime covariances, \textit{i.e.},~Hermitian forms $\Lambda^\pm\:\mathcal V_{\mathrm P}\times\mathcal V_{\mathrm P}\to\mathbb C$ defined by
\begin{align*}
    \Lambda^+(v,w)\defeq\omega(\Phi(v)\Phi^*(w))\,,\qquad
    \Lambda^-(v,w)\defeq\omega(\Phi^*(w)\Phi(v))\,.
\end{align*}
Indeed, a pair of non-negative Hermitian forms $\Lambda^\pm$ defines a quasifree state on $\mathcal V_{\mathrm P}$ precisely when $\Lambda^+-\Lambda^-=\sigma$. We refer to \cite[Sec.~2]{GerardWrochna}, \cite{GerardBook} and \cite[Sec.~2.3]{SchmidPhD} for further details. In practise, we will consider states coming from a pair of \emph{pseudo-covariances}:

\begin{proposition}[cf.~\protect{\cite[Lem.~2.9]{GerardMurroWrochna}}]
\label{Prop:SpacetimePseudoCovariancesYM}
Consider a pair of pseudo-covariances\footnote{The term ``pseudo-covariance'' reflects that $\lambda^\pm$ need only to be positive w.r.t. $(\cdot,\cdot)_{\E_1}$ on $\ker(\K^*\vert_{\Gamma_{\c}})$, rather than on all $\Gamma(\E_1)$ as in the case of ordinary field theories such as for Klein-Gordon fields.}, namely continuous linear maps $\lambda^\pm\:\Gamma_{\c}(\E_1)\to\Gamma(\E_1)$ satisfying:
\begin{itemize}
    \item[$\mathrm(i)$] $(\lambda^\pm)^*=\lambda^\pm$ with respect to $(\cdot,\cdot)_{\E_1}$ and
    $\lambda^\pm(\ran(\K\vert_{\Gamma_{\c}}))
    \subset \ran(\K)$;
    \item[$\mathrm(ii)$] $\mathsf D_1\lambda^\pm=\lambda^\pm\mathsf D_1=0$ and
    $(\lambda^+-\lambda^-) \zeta = \mathrm{i}\G_1 \zeta$ modulo
    $\mathrm{ran}(\K)$ for every
    $\zeta \in\ker(\K^*\vert_{\Gamma_{\c}})$;
    \item[$\mathrm(iii)$] $(\zeta,\lambda^\pm \zeta)_{\E_1}\geq0$ for every
    $\zeta \in\ker(\K^*\vert_{\Gamma_{\c}})$.
\end{itemize}
Then, the Hermitian forms $\Lambda^\pm\:\mathcal V_{\mathrm P}\times\mathcal V_{\mathrm P}\to\mathbb C$ defined by
\begin{align*}
    \Lambda^\pm([\zeta],[\zeta'])\defeq(\zeta,\lambda^\pm \zeta')_{\E_1}
\end{align*}
for $\zeta, \zeta' \in\ker(\K^*\vert_{\Gamma_{\c}})$ are the spacetime covariances of a quasifree state on $\mathrm{CCR}(\mathcal V_{\mathrm P},\sigma)$.
\end{proposition}

Physically admissible quasifree states are required to satisfy the \emph{Hadamard condition}, which is naturally formulated in microlocal terms.

\begin{definition}[Hadamard States]
A quasifree state $\omega$ on $\mathrm{CCR}(\mathcal V_{\mathrm P},\sigma)$ is called a Hadamard state if it admits a pair of pseudo-covariances $\lambda^\pm$ as in Proposition \ref{Prop:SpacetimePseudoCovariancesYM} satisfying
\begin{align*}
    \operatorname{WF}'(\lambda^\pm)\subset\mathcal N^\pm\times\mathcal N^\pm\,,
\end{align*}
where $\operatorname{WF}'(\lambda^\pm)$ denotes the primed wavefront set of the Schwartz kernels $\lambda^{\pm}\in\Gamma_{\mathrm{c}}^{\prime}(\M\times\M,\E_{1}\boxtimes\E_{1}^{\ast})$ and where
\begin{align*}
    \mathcal N^\pm\defeq\left\{(p,\xi)\in \T^*\M\setminus\{0\}\;:\;
    \g^{-1}(\xi,\xi)=0\ \text{and}\ \pm\xi(v)>0\ \forall\text{future-directed timelike }v\in \T_p\M\right\}.
\end{align*}
\end{definition}

This definition is the standard Hadamard condition for linear gauge theories, see~\cite{GerardWrochna,GerardMurroWrochna}, cf., also \cite{Radzikowski1,Radzikowski2} and \cite{SahlmannVerch} for general considerations on the Hadamard condition for scalar and vector-valued quantum fields. The construction of Hadamard states is in general a hard task and, in order to accomplish it, it is often preferable to work on the level of initial data.

\begin{definition}[Cauchy Pseudo-Covariances.]\label{Def:CauchyPseudoCov}
Let $\omega$ be a quasifree state on $\mathrm{CCR}(\mathcal V_{\mathrm P},\sigma)$ induced by pseudo-covariances $\lambda^\pm$ as per Proposition \ref{Prop:SpacetimePseudoCovariancesYM}. The continuous linear maps $\lambda_\Sigma^\pm\:\Gamma_{\mathrm{c}}(\E_{\rho_1})\to\Gamma(\E_{\rho_1})$ defined by
\begin{align*}
    \lambda_\Sigma^\pm\defeq
    (\rho_1^*\G_{\Sigma,1})^*\lambda^\pm(\rho_1^*\G_{\Sigma,1}),
\end{align*}
where the adjoints are taken w.r.t.~$(\cdot,\cdot)_{\E_1}$ and $(\cdot,\cdot)_{\E_{\rho_1}}$, are called \emph{Cauchy pseudo-covariances}. 
\end{definition}

Since the bundle metric on $\E_{\rho_1}$ can be chosen arbitrarily, it is more natural to work with the physical Hermitian form $\sigma_{\Sigma,1}$. Following~\cite[Prop.~2.11]{GerardMurroWrochna} (\emph{cf.}, also \cite[Prop.~11.1.1]{GerardBook}), we obtain the following characterisation.

\begin{proposition}
\label{Prop:CauchyPseudoCovariancesYM}
Suppose $c^\pm\:\Gamma_\c(\E_{\rho_1})\to\Gamma(\E_{\rho_1})$ are continuous linear operators satisfying:
\begin{itemize}
    \item[$\mathrm(i)$] $(c^\pm)^\dagger=c^\pm$ with respect to $\sigma_{\Sigma,1}$ and
    $c^\pm(\ran(\K_\Sigma))
    \subset\ran(\K_\Sigma)$;
    \item[$\mathrm(ii)$] $(c^++c^-)\mathfrak{f}=\mathfrak{f}$ modulo
    $\ran(\K_\Sigma)$ for every
    $\mathfrak{f}\in\ker(\K_\Sigma^\dagger\vert_{\Gamma_\c})$;
    \item[$\mathrm(iii)$] $\pm\sigma_{\Sigma,1}(\mathfrak{f},c^\pm \mathfrak{f})\geq0$ for every
    $\mathfrak{f}\in\ker(\K_\Sigma^\dagger\vert_{\Gamma_\c})$.
\end{itemize}
Then, $\lambda_\Sigma^\pm\defeq\pm \mathrm{i}\G_{\Sigma,1}c^\pm$ are the Cauchy pseudo-covariances of a quasifree state on $\mathrm{CCR}(\mathcal V_{\mathrm P},\sigma)$, or in other words, $\lambda^\pm=(\rho_1\G_1)^*\lambda_\Sigma^\pm(\rho_1\G_1)=\pm \mathrm{i} \, \mathcal U_1c^\pm(\rho_1\G_1)$ are pseudo-covariances of a quasifree state on $\mathrm{CCR}(\mathcal V_{\mathrm P},\sigma)$. Furthermore, if for some neighbourhood $\mathcal O$ of $\Sigma_0$ in $\M$ one has
\begin{align*}
    \operatorname{WF}'(\mathcal U_1c^\pm)
    \subset(\mathcal N^\pm\cup F)\times \T^*\Sigma
    \qquad\text{over }\quad\mathcal O\times\Sigma,
\end{align*}
where $F\subset \T^*\M$ is conic and $F\cap\mathcal N=\varnothing$, then the associated state is Hadamard.
\end{proposition}

We omit the proof since it can be straightforwardly adapted from that of~\cite[Prop.~2.11]{GerardMurroWrochna}, with $\d$, $\delta$ and $\square_k$ replaced by $\d^\A$, $\delta^\A$ and $\mathsf D_k$, respectively. The argument uses only the abstract linear gauge theory identities and propagation of singularities for the normally hyperbolic operator $\mathsf D_1$. Compared to \cite[Prop.~2.11]{GerardMurroWrochna}, notice that in our formulation, we allow the identity in (ii) to hold modulo $\ran(\K_\Sigma)$. This does not make any difference, since in the end this identity needs be true on the level of sesquilinear forms, which are defined $\mathrm{ker}(\sf{K}_{\Sigma}^{\dagger}\vert_{\Gamma_{\mathrm{c}}})$, see also \cite[Sec.~2.9]{Gerard}.

\begin{remark}
As in~\cite[Lem.~3.7]{GerardWrochna}, $\rho_1\G_1$ admits a continuous extension to distributional sections.
\end{remark}
\section{Complete Gauge Fixing}
\label{Sec:Complete Gauge Fixing}

To complete the quantisation procedure, the main remaining task is to construct Hadamard states for linearised Yang-Mills theory, \textit{i.e.}, a pair of pseudo-covariances $c^{\pm}$ satisfying the properties of Proposition~\ref{Prop:CauchyPseudoCovariancesYM}. As indicated in the introduction, our approach is based on a \emph{complete gauge-fixing}. In this section, we therefore address the problem of achieving a complete gauge fixing for linearised Yang-Mills theory on (not necessarily spatially compact) manifolds by means of the \emph{Cauchy radiation gauge}, first introduced for Maxwell's theory in \cite{MurroSchmid,SchmidPhD}. We shall restrict our discussion to \emph{ultrastatic spacetimes}, as the construction will be carried out exclusively on such backgrounds in Section~\ref{Subsec:Construction of Hadamard States on Ultrastatic Spacetimes}, while their general existence is established via a deformation argument in Section~\ref{Subsec:Construction of Hadamard States on Globally Hyperbolic Spacetimes}. To achieve the Cauchy radiation gauge, we generalise the Hodge decomposition to Sobolev spaces of $\Ad(\P_\Sigma)$-valued differential forms on Riemannian manifolds that are not necessarily compact but are assumed to be complete. This allows us to solve the Poisson equation with a divergence-type source and thereby to impose the Cauchy radiation gauge for Yang–Mills fields on globally hyperbolic (ultrastatic) spacetimes with non-compact Cauchy surfaces. Finally, we establish several relevant properties of the resulting gauge projection.

\subsection{Cauchy Radiation Gauge}
\label{Subsec:Cauchy Radiation Gauge}

The Lorenz gauge, $\delta^{A}\alpha=0$, while yielding hyperbolic equations of motion, leaves residual gauge freedom and it does not eliminate the negative contribution of the temporal component to the indefinite fibre metric on one-forms, which is one the main sources of complications in constructing states for linear gauge theories, as discussed in the introduction of the present article. Therefore, the idea is to impose a further condition with the aim of getting rid of the negative degrees of freedom. To this end, we consider the \emph{Cauchy radiation gauge}, which, for Maxwell's theory, has been introduced in \cite{MurroSchmid}, see \cite[Sec.~3.1.4]{SchmidPhD} for an extended discussion.

\begin{definition}[Cauchy Radiation Gauge] \label{Def: Cauchy radiation gauge} Let $(\M=\mathbb{R}\times\Sigma,\g=-\d t^{2}+\h)$ be a globally hyperbolic ultrastatic spacetime, $A=A_{\mathrm{T}}\d t+A_{\Sigma}$ be a flat connection and let $\alpha\in\Omega^{1}(\M,\mathrm{Ad}(\P))$ be a linearised Yang-Mills field. We say that it satisfies the \emph{Cauchy radiation gauge} if the following holds:
\begin{align*}
    \text{(i)}&\qquad \K^{\ast}\alpha=\delta^{A}\alpha=0 && \text{(Lorenz gauge)}\\
    \text{(ii)}&\qquad \alpha_{\mathrm{T}}\vert_{t=0}=\mathcal{D}_{t}^{A_{\mathrm{T}}}\alpha_{\mathrm{T}}\vert_{t=0}=0 && \text{(Cauchy temporal gauge)}
\end{align*}
\end{definition}

On ultrastatic spacetimes, the equivalence between the Cauchy radiation gauge and the \emph{(full) radiation gauge}, on solutions to the equations of motion, is readily established.

\begin{lemma}\label{Lem:RadEquiv}
    Let $(\M=\mathbb{R}\times\Sigma,\g=-\d t^{2}+\h)$ be a globally hyperbolic ultrastatic spacetime, $A=A_{\mathrm{T}}\d t+A_{\Sigma}$ be a flat connection and $\alpha\in\Omega^{1}_{\mathrm{sc}}(\sf{M},\mathrm{Ad}(\sf{P}))\cap\mathrm{ker}(\sf{P})$ be a solution to the linearised Yang-Mills equations. Then, $\alpha$ satisfies the Cauchy radiation gauge if and only if it satisfies 
    \begin{align*}
        \delta^{A}\alpha=\alpha_{\mathrm{T}}=0 && \emph{(radiation gauge)}\, .
    \end{align*}
\end{lemma}

\begin{proof}
    Since $\sf{P}\alpha=\delta^{A}\alpha=0$, we have $\square^{A}\alpha=0$. By Proposition~\ref{Prop:DecompAdFormsOp}, this equation decoupled into two separate hyperbolic equations for $\alpha_{\mathrm{T}}$ and $\alpha_{\Sigma}$. In particular, $(\square^{A}\alpha)_{\mathrm{T}}=((\mathcal{D}_{t}^{A_{\mathrm{T}}})^{2}+\Delta^{A_{\Sigma}})\alpha_{\mathrm{T}}=0$. Hence, $\alpha_{\mathrm{T}}=0$ if and only if its initial data are zero, \textit{i.e.},  $\alpha_{\mathrm{T}}\vert_{t=0}=\mathcal{D}_{t}^{A_{\mathrm{T}}}\alpha_{\mathrm{T}}\vert_{t=0}=0$
\end{proof}

\noindent As in the Maxwell case \cite{MurroSchmid}, the existence of a gauge transformation bringing the field into the Cauchy radiation gauge can be reduced to the solvability of a Poisson equation on the Cauchy hypersurface. More precisely, the following proposition establishes both the attainability and the completeness of this gauge on spatially compact ultrastatic spacetimes. 

\begin{proposition}
\label{Prop:AchCauchyRadGauge}
    Let $(\M=\mathbb{R}\times\Sigma,\g=-\d t^{2}+\h)$ be a globally hyperbolic ultrastatic spacetime with $\Sigma$ compact and let $\K \defeq \d^A$ with $A$ being a flat connection, \textit{cf.} Proposition \ref{Prop:LinYangMills}. Then, for every $\alpha\in\Omega^{1}(\M,\mathrm{Ad}(\P))$ there exists a gauge transformation $\varepsilon\in\Omega^{0}(\M,\mathrm{Ad}(\P))$, unique modulo $\mathrm{ker}(\K)$, such that $\alpha^{\prime}=\alpha-\K\varepsilon$ is in the Cauchy radiation gauge, see Definition \ref{Def: Cauchy radiation gauge}.
\end{proposition}

\begin{proof}
Let $\alpha\in\Omega^{1}(\M,\mathrm{Ad}(\P))$ be an arbitrary $\mathrm{Ad}(\P)$-valued one-form on $\M$ and decompose it as $\alpha=\alpha_{\mathrm{T}}\d t+\alpha_{\Sigma}$, where $\alpha_{\mathrm{T}} \in C^\infty(\mathbb{R}, \Omega^0(\Sigma, \Ad(\P_{\Sigma})))$, while $\alpha_\Sigma \in C^\infty(\mathbb{R}, \Omega^1(\Sigma, \Ad(\P_{\Sigma})))$. By item (iii) in Proposition~\ref{Prop:DecompAdFormsOp}, we note that $\delta^{A}\alpha=\mathcal{D}^{A_\mathrm{T}}_{\partial_{t}}\alpha_{\mathrm{T}}+\delta^{A_{\Sigma}}\alpha_{\Sigma}$. On account also of item (ii) in Lemma \ref{Lem: 2.26}, we observe that the Cauchy radiation gauge is equivalent to the conditions
\begin{equation*}
        \delta^{A}\alpha=0\,,\qquad \alpha_{\mathrm{T}}\vert_{t=0}=\delta^{A_{\Sigma}}\alpha_{\Sigma}\vert_{t=0}=0\, .
\end{equation*}
Now, consider a gauge transformation $\varepsilon\in\Omega^{0}(\M,\mathrm{Ad}(\P))$. Using again Proposition~\ref{Prop:DecompAdFormsOp}, we note that the transformed field $\alpha^{\prime}=\alpha-\K\varepsilon$ satisfies the Cauchy radiation gauge if and only if
 \begin{equation}
 \label{eq:CauchyProof}
        \begin{cases}
            \square^{A}\varepsilon &=\delta^{A}\alpha\\
            (\mathcal{D}^{A_\mathrm{T}}_{t}\varepsilon)\vert_{t=0} &=\alpha_{\mathrm{T}}\vert_{t=0}\\
            (\Delta^{A_{\Sigma}}\varepsilon)\vert_{t=0} &=(\delta^{A_{\Sigma}}\alpha_{\Sigma})\vert_{t=0}
        \end{cases} \, . 
\end{equation}
For notational brevity, let us write $A_{\Sigma}^{0} \defeq A_{\Sigma} \vert_{t=0}$. By assumption, we note that $A_{\Sigma}^{0}$ defines a flat connection. This implies that $\d^{A_{\Sigma}^{0}}\circ\d^{A_{\Sigma}^{0}}=0$, \textit{i.e.},~we obtain a cochain complex $(\Omega^{\bullet}(\Sigma,\mathrm{Ad}(\P_{\Sigma})),\d^{A_{\Sigma}^{0}})$. Moreover, since $\Sigma$ is assumed to be compact, we have the well-known Hodge-decomposition theorem, which states that, for any $k \in \mathbb{N}$,
\begin{align*}
        \Omega^{k}(\Sigma,\mathrm{Ad}(\P_{\Sigma}))=\d^{A_{\Sigma}^{0}}\Omega^{k-1}(\Sigma,\mathrm{Ad}(\P_{\Sigma}))\oplus\delta^{A_{\Sigma}^{0}}\Omega^{k+1}(\Sigma,\mathrm{Ad}(\P_{\Sigma}))\oplus\mathrm{ker}(\Delta^{A^{0}_{\Sigma}})
  \end{align*}
For $k = 1$, this implies that we can decompose $\alpha_{\Sigma}\vert_{t=0}\in\Omega^{1}(\Sigma,\mathrm{Ad}(\P_{\Sigma}))$ as $\alpha_{\Sigma}\vert_{t=0}=\d^{A_{\Sigma}^{0}}\mathfrak{f}+\omega$ for some $\mathfrak{f}\in\Omega^{0}(\Sigma,\Ad(\P_{\Sigma}))$ and $\omega\in\Omega^{1}(\Sigma,\Ad(\P_{\Sigma}))$ with $\delta^{A_{\Sigma}^{0}}\omega=0$. Observe that $\mathfrak{f}$ is a solution to the Poisson equation 
\begin{align*}
    \Delta^{A_\Sigma^0} \mathfrak{f} =\delta^{A_{\Sigma}^0}\alpha_{\Sigma}\vert_{t=0}
\end{align*} 
and, thus, we may set $\varepsilon\vert_{t=0}=\mathfrak{f}$. More precisely, a solution $\varepsilon$ to the system in Eq.~\eqref{eq:CauchyProof} is provided by solving the following Cauchy problem 
 \begin{equation*}
        \begin{cases}
            \square^{A}\varepsilon &=\delta^{A}\alpha\\
            (\mathcal{D}^{A_\mathrm{T}}_{t}\varepsilon)\vert_{t=0} &=\alpha_{\mathrm{T}}\vert_{t=0}\\
             \varepsilon\vert_{t=0} &=\mathfrak{f}
         \end{cases}\, .
\end{equation*}
Since $\square^A$ is a normally hyperbolic operator, for each choice of $\mathfrak f$ solving the Poisson equation, the corresponding Cauchy problem admits a unique smooth solution $\varepsilon$. In general, however, the solution $\mathfrak f$ of the Poisson equation is determined only up to an element of $\ker(\Delta^{A_\Sigma^0}\vert_{\Omega^{0}}) \equiv \ker(\d^{A_\Sigma^0}\vert_{\Omega^{0}}).$ We claim that $\varepsilon$ is uniquely determined modulo $\mathrm{ker}(\K)$. To prove such claim, suppose that $\tilde{\varepsilon}$ is any other gauge transformation such that $\tilde{\alpha}:=\alpha-\K\tilde{\varepsilon}$ still satisfies the Cauchy radiation gauge and set $\varepsilon' \defeq \tilde{\varepsilon} - \varepsilon$. We want to show that $\K \varepsilon' = 0$. On account of the fact that both $\varepsilon$ and $\tilde{\varepsilon}$ abide by Eq. \eqref{eq:CauchyProof}, we have that $\varepsilon'$ satisfies 
\begin{equation}\label{eq:CauchyProof1}
       \begin{cases}
            \square^{A}\varepsilon^{\prime}&=0\\
             (\mathcal{D}^{A_\mathrm{T}}_{t}\varepsilon^{\prime})\vert_{t=0}&=0\\
             (\Delta^{A}\varepsilon^{\prime})\vert_{t=0}&=0
        \end{cases} \, .
    \end{equation}
    The condition $(\Delta^{A}\varepsilon^{\prime})\vert_{t=0}=0$, in particular, implies that $\d^{A_{\Sigma}^{0}}\varepsilon' \vert_{t=0}$. Now, let us set $\eta:=\sf{K}\varepsilon^{\prime}=\d^{A}\varepsilon^{\prime}$.Flatness of $A$ entails that $\square^A \d^A = \d^A \square^A$ and, therefore, $\square^A \eta = 0$. We argue that the initial data for $\eta$, i.e., $\eta_{\mathrm{T}}\vert_{t=0}$, $\eta_{\Sigma}\vert_{t=0}$, $\mathcal{D}^{A_{\mathrm{T}}}_{t}\eta_{\mathrm{T}}\vert_{t=0}$ and $\mathcal{D}^{A_{\mathrm{T}}}_{t}\eta_{\Sigma}\vert_{t=0}$ are zero. Item (i) in Proposition \ref{Prop:DecompAdFormsOp} yields $\eta_\mathrm{T} \vert_{t=0} = \mathcal{D}_t^{A_\mathrm{T}} \varepsilon' \vert_{t=0} = 0$ as well as $\eta_\Sigma \vert_{t=0} = \d^{A_\Sigma^0} \varepsilon' \vert_{t=0} = 0$. Item (v) in Proposition \ref{Prop:DecompAdFormsOp}, on the other hand, together with Eq.~\eqref{eq:CauchyProof1} entails that $\mathcal{D}_t^{A_\mathrm{T}} \eta_\mathrm{T} \vert_{t=0} = (\mathcal{D}_t^{A_\mathrm{T}})^2 \varepsilon' \vert_{t=0} = 0$. Item (iii) in Lemma \ref{Lem: 2.26} and flatness of the background connection finally imply also that $\mathcal{D}_t^{A_\mathrm{T}} \eta_\Sigma \vert_{t=0} = \mathcal{D}_t^{A_\mathrm{T}} \K \varepsilon' \vert_{t=0} = \K \mathcal{D}_t^{A_\mathrm{T}} \varepsilon' \vert_{t=0} = 0$. We conclude that $\eta=0$, i.e., $\sf{K}\varepsilon^{\prime}=0$.
\end{proof}

\begin{remark}
    The Cauchy radiation gauge can also be defined on general globally hyperbolic spacetimes, by replacing $\mathcal{D}_{t}^{A_{\mathrm{T}}}\alpha_{\mathrm{T}}$ appropriately by $\nabla^{A}_{\partial_{t}}(\alpha_{\mathrm{T}})$. In this setting, it is no longer equivalent to the full radiation gauge in the non-ultrastatic case, \textit{i.e.}, Lemma~\ref{Lem:RadEquiv} fails in general. 
    
    In fact, the attainability of the full radiation gauge on general curved spacetimes is not guaranteed without additional geometric assumptions, \textit{cf.}~\cite{Tolksdorf1992,Tolksdorf1993}, which discusses a generalised radiation gauge for Maxwell's theory in the case on static backgrounds and on simply-connected spacetimes, with possibly non-trivial principal $\mathrm{U}(1)$-bundles. The achievability of the Cauchy radiation gauge, on the other hand, can be established for many spacetimes. In particular,  Proposition~\ref{Prop:AchCauchyRadGauge} can easily be extended to the non-ultrastatic case, see, \textit{e.g.}, \cite[Prop.~3.17]{SchmidPhD} for a proof in the Abelian case. In summary, while the Cauchy radiation gauge and full radiation gauge are equivalent on ultrastatic spacetimes, the former is generally weaker and less restrictive.
\end{remark}

The above proposition shows that the obstruction to achieving the Cauchy radiation gauge is reduced to the solvability of the covariant Poisson equation on the compact Riemannian manifold $(\Sigma, \h)$. In order to drop the compactness hypothesis on the Riemannian manifold, we need to enlarge our functional setting. In particular, this motivates the study of a Hodge decomposition for $\operatorname{Ad}(\P_\Sigma)$-valued differential forms on complete (but not necessarily compact) Riemannian manifolds. This will allow us to solve the Poisson equation with a divergence source and to achieve the Cauchy radiation gauge for Yang-Mills fields also on globally hyperbolic manifold, which are spatially non-compact.

\subsection{Hodge Decomposition in Flat Bundles on Complete Riemannian Manifolds}
\label{Subsec:Hodge Decomposition in Flat Vector Bundles on Complete Riemannian Manifolds}

Motivated by the previous discussion, the goal of the following section is establish a Hodge decomposition for $\operatorname{Ad}(\P_\Sigma)$-valued differential forms in Sobolev spaces on complete Riemannian manifolds, with the aim of solving the Poisson equation on a suitable space, which, in turn, is needed to generalise Proposition~\ref{Prop:AchCauchyRadGauge} to non-spatially compact spacetimes. Since the Hodge decomposition and its application to the Poisson equation on non-compact manifold is of independent interest in considerably more general settings than that required for our applications, we shall formulate and prove it in a more abstract geometric framework, \textit{i.e.}, by working with generic \emph{flat vector bundles}. Readers interested primarily in the applications to gauge theory may take the resulting Hodge decomposition and solvability statement of the Poisson equation as a black box applied to the setting relevant to linearised Yang-Mills theory outlined in Example~\ref{ExampleHodge} below.

Throughout this section, let $(\Sigma, \h)$ be an oriented, connected and \emph{complete} $d$-dimensional Riemannian manifold with empty boundary $\partial \Sigma = \emptyset$. Moreover, let $\E_{\Sigma}\to\Sigma$ be a complex vector bundle endowed with a \emph{positive-definite} bundle metric $\langle\cdot,\cdot\rangle_{\E_{\Sigma}}$ and a connection $\nabla^{\E_{\Sigma}}$ with the following two properties:
\begin{itemize}
    \item[(i)]$\nabla^{\E_{\Sigma}}$ is \emph{compatible} with the metric $\langle\cdot,\cdot\rangle_{\E_{\Sigma}}$, \textit{i.e.},
    \begin{align*}
\d\langle\psi,\phi\rangle_{\E_{\Sigma}}=\langle\nabla^{\E_{\Sigma}}_{\bullet}\psi,\phi\rangle_{\E_{\Sigma}}+\langle\psi,\nabla^{\E_{\Sigma}}_{\bullet}\phi\rangle_{\E_{\Sigma}}\,,\qquad\qquad\forall\psi,\phi\in\Gamma(\E_{\Sigma});
    \end{align*}
    \item[(ii)]$\nabla^{\E_{\Sigma}}$ is \emph{flat},~\textit{i.e.},~its curvature $2$-form $F^{\nabla^{\E_{\Sigma}}}\in\Omega^{2}(\Sigma,\mathrm{End}(\E_{\Sigma}))$ is identically zero:
    \begin{align*}
        F^{\nabla^{\E_{\Sigma}}}(X,Y):=[\nabla^{\E_{\Sigma}}_{X},\nabla^{\E_{\Sigma}}_{Y}]-\nabla^{\E_{\Sigma}}_{[X,Y]}\equiv 0\,,\qquad\qquad\forall X,Y\in\mathfrak{X}(\Sigma)\, .
    \end{align*}
\end{itemize}

Herein, we work intrinsically on the Riemannian manifold $(\Sigma,\h)$, without assuming that it arises as a Cauchy hypersurface of an \emph{ambient} globally hyperbolic spacetime. In the applications considered later, however, $(\Sigma,\h)$ will be identified with a Cauchy hypersurface of a $(d+1)$-dimensional globally hyperbolic spacetime $(\M,\g)$. More precisely, the primary situation to keep in mind is the following:

\begin{example}\label{ExampleHodge}
    Let $(\M=\mathbb{R}\times\Sigma,\sf{g}=-\d t\otimes\d t+\sf{h})$ be a globally hyperbolic \emph{ultrastatic} spacetime and $\P\to\M$ be a principal $\G$-bundle admitting a flat connection $A\in\mathcal{C}(\P)$. As before, $\P$ may be identified with the pullback of a principal bundle $\P_\Sigma\to\Sigma$ (\textit{cf.}~Lemma \ref{Lemma:BundleSigma}). After fixing such an identification, the background connection can be decomposed as $A=A_{\mathrm T}\,\d t+A_\Sigma(t)$ (\textit{cf.}~Proposition~\ref{Prop:DecompConn}), where $A_\Sigma(t)$ is a time-dependent family of \emph{flat} connections on $\P_{\Sigma}$, each of which induces a \emph{flat} connection $\nabla^{A_{\Sigma}(t)}$ on the adjoint bundle $\mathrm{Ad}(\P_{\Sigma})$. Now, global hyperbolicity of $(\M,\g)$ implies that $(\Sigma,\h)$ is a \emph{complete} Riemannian manifold. Moreover, if $\G$ satisfies Assumption~\ref{Assumption}, there is a \emph{positive-definite} $\mathrm{Ad}$-invariant inner product $\kappa$ on $\mathfrak{g}$, which in turn induces a \emph{positive-definite} bundle metric $\langle\cdot,\cdot\rangle_{\mathrm{Ad}(\P_{\Sigma})}$ on $\mathrm{Ad}(\P_{\Sigma})$ that is compatible with the connection $\nabla^{A_{\Sigma}(t)}$ for each fixed $t\in\bb{R}$. Hence, for a fixed (initial) time $t_{0}\in\bb{R}$, we may set $\E_{\Sigma}=\mathrm{Ad}(\P_{\Sigma})$ equipped with the metric $\langle\cdot,\cdot\rangle_{\mathrm{Ad}(\P_{\Sigma})}$ and with a compatible flat connection $\nabla^{A_{\Sigma}(t_{0})}$. 
\end{example}

Similar to the setting of Section~\ref{Sec:Geometric Preliminaries on Yang-Mills Theory}, the bundle metric $\langle\cdot,\cdot\rangle_{\E_{\Sigma}}$ together with the Riemannian metric $\h$ induces bundle metrics $\langle\cdot,\cdot\rangle_{\E_{\Sigma,k}}$ on $\E_{\Sigma,k}:=\bigwedge^{k}\T^{\ast}\Sigma\otimes\E_{\Sigma}$ (\textit{cf}.~Appendix~\ref{Appendix: Notation and Conventions for Scalar and Twisted Differential Forms}), which in turn give rise to the $\L^2$-space of $\E_{\Sigma}$-valued $k$-forms by 
\begin{equation*}
    \L^2_k(\Sigma, \E_{\Sigma}) \defeq \overline{\Omega^{k}_{\c}(\Sigma,  \E_{\Sigma})}^{\L^{2}}, \qquad \langle \alpha, \beta \rangle_{\mathrm{L}^2} \defeq \int_\Sigma \langle \alpha, \beta \rangle_{\E_{\Sigma,k}} \, \d\mu_\h=\int_{\Sigma}\langle\alpha\wedge\ast_{\h}\,\beta\rangle_{\E_{\Sigma}}, 
\end{equation*}
for any $\alpha, \beta \in \Omega^k_c (\Sigma, \E_{\Sigma})$. In the setting relevant to the subject of this paper, see Example~\ref{ExampleHodge}, we note that $\langle \cdot,\cdot \rangle_{\mathrm{L}^2}$ agrees with what has previously been denoted by $(\cdot,\cdot)_{\E_{\Sigma,k}}$ in Eq.~\eqref{Eq: inner product on E sigma}. As usual, we denote by
\begin{align*}
    \d^{\nabla^{\E_{\Sigma}}} &:\Omega^{k}(\Sigma, \E_{\Sigma})\to\Omega^{k+1}(\Sigma, \E_{\Sigma}), \\
    \delta^{\nabla^{\E_{\Sigma}}} &: \Omega^{k+1}(\Sigma, \E_{\Sigma})\to\Omega^{k}(\Sigma, \E_{\Sigma}), \qquad \delta^{\nabla^{\E_{\Sigma}}} \defeq (-1)^{k+1} \ast_\h^{-1} \d^{\nabla^{\E_{\Sigma}}} \ast_\h,
\end{align*}
the (covariant) exterior derivative and codifferential respectively, and let us recall that $\delta^{\nabla^{\E_{\Sigma}}}$ is the formal adjoint of $\d^{\nabla^{\E_{\Sigma}}}$ w.r.t.~the inner product defined above, \textit{i.e.}, for any $\alpha, \beta \in \Omega_{\mathrm{c}}^\bullet(\Sigma, \E_{\Sigma})$, it holds that 
\begin{align*}
    \langle \alpha, \d^{\nabla^{\E_{\Sigma}}} \beta \rangle_{\mathrm{L}^2} = \langle \delta^{\nabla^{\E_{\Sigma}}} \alpha, \beta \rangle_{\mathrm{L}^2}
\end{align*}
as a consequence of the compatibility of $\nabla^{\E_{\Sigma}}$ and $\langle\cdot,\cdot\rangle_{\E_{\Sigma}}$ (for more details on bundle-valued forms, we refer to Appendix~\ref{Appendix: Notation and Conventions for Scalar and Twisted Differential Forms}). Moreover, let us view the twisted de Rham-Hodge Laplacian as a densely-defined operator of the form 
\begin{equation}
  \label{Eq: DeltaA}  
    \Delta^{\nabla^{\E_{\Sigma}}} \defeq \d^{\nabla^{\E_{\Sigma}}}\delta^{\nabla^{\E_{\Sigma}}}+\delta^{\nabla^{\E_{\Sigma}}}\d^{\nabla^{\E_{\Sigma}}}\:\mathrm{dom}(\Delta^{\nabla^{\E_{\Sigma}}})\to \L^{2}_k(\Sigma, \E_{\Sigma}) \, , 
\end{equation}
with domain $\mathrm{dom}(\Delta^{\nabla^{\E_{\Sigma}}}):=\Omega^k_{\mathrm{c}}(\Sigma, \E_{\Sigma})$. Flatness of $\nabla^{\E_{\Sigma}}$ entails that $\d^{\nabla^{\E_{\Sigma}}}\circ \d^{\nabla^{\E_{\Sigma}}}=0$ and hence also $\delta^{\nabla^{\E_{\Sigma}}}\circ \delta^{\nabla^{\E_{\Sigma}}}=0$ as well as
\begin{align*}
    \d^{\nabla^{\E_{\Sigma}}}\circ \Delta^{\nabla^{\E_{\Sigma}}}=\Delta^{\nabla^{\E_{\Sigma}}}\circ\d^{\nabla^{\E_{\Sigma}}}\,,\qquad \delta^{\nabla^{\E_{\Sigma}}}\circ \Delta^{\nabla^{\E_{\Sigma}}}=\Delta^{\nabla^{\E_{\Sigma}}}\circ\delta^{\nabla^{\E_{\Sigma}}}\, .
\end{align*}

\begin{lemma}
 \label{Lem: Chernoff thm}
The twisted de Rham-Hodge Laplacian in Eq.~\eqref{Eq: DeltaA} is non-negative and essentially self-adjoint on a complete Riemannian manifold $(\Sigma, \h)$. 
\end{lemma}
 
\begin{proof}
The non-negativity of $\Delta^{\E_\Sigma}$ is a direct consequence of the fact that $\delta^{A_{\Sigma}}$ is the formal adjoint of $\d^{A_{\Sigma}}$ w.r.t. the $\mathrm{L}^2$-inner product defined above. Indeed, for every $\alpha \in \Omega^{k}(\Sigma,\E_{\Sigma})$, 
\begin{equation*}
    \langle \alpha, \Delta^{\nabla^{\E_{\Sigma}}} \alpha \rangle_{\L^2} = \lVert \d^{\nabla^{\E_{\Sigma}}} \alpha \rVert^2_{\L^2} + \lVert \delta^{\nabla^{\E_{\Sigma}}} \alpha \rVert^2_{\L^2} \ge 0. 
\end{equation*}
To prove essential self-adjointness, consider the formally self-adjoint Dirac-type operator 
\begin{align*}
    \mathcal{D}^{\nabla^{\E_{\Sigma}}}:= \d^{\nabla^{\E_{\Sigma}}}+\delta^{\nabla^{\E_{\Sigma}}}\, ,
\end{align*}
acting on sections of the total bundle $\E_{\Sigma, \bullet} \defeq \bigoplus_{k=0}^d \E_{\Sigma, k}$. Its principal symbol, up to conventional factor $i$, is given by
    \begin{align*}
        \sigma_{\mathcal{D}^{\nabla^{\E_{\Sigma}}}}(x,\xi)\omega=\xi\wedge\omega+\xi\lrcorner\omega\,,
    \end{align*}
and, therefore, $\lVert  \sigma_{\mathcal{D}^{\nabla^{\E_{\Sigma}}}}(x,\xi) \rVert = |\xi|_\h$. Thus, in the terminology of \cite{Chernoff}, the \emph{maximal propagation speed} of the corresponding symmetric hyperbolic system 
\begin{align*}
    \partial_{t}+i \mathcal{D}^{\nabla^{\E_{\Sigma}}}
\end{align*} is uniformly bounded by $1$. Since $(\Sigma, \h)$ is complete per hypothesis, Chernoff's theorem \cite[Thm.~2.2]{Chernoff} applied to the formally skew-adjoint operator $\mathrm{i} \mathcal{D}^{\nabla^{\E_{\Sigma}}}$ implies that $\mathcal{D}^{\nabla^{\E_{\Sigma}}}$ and all its integer powers are essentially self-adjoint, which proves the claim.
\end{proof}

\begin{definition}
Let $E \defeq (\mathds{1}+\overline{\Delta^{\nabla^{\E_{\Sigma}}}}): \operatorname{dom}(\overline{\Delta^{\nabla^{\E_{\Sigma}}}}) \to \L^2_k(\Sigma, \E_{\Sigma}).$ Then, for all $\mathfrak{s}\in\mathbb{R}$, we define $E^{\mathfrak{s}} \:\mathrm{dom}(E^{\mathfrak{s}})\to\L^{2}_{k}(\Sigma, \E_{\Sigma}),$ by means of the spectral theorem, \textit{i.e.},
	\begin{align*}
		\mathrm{dom}(E^{\mathfrak{s}})\defeq\{\omega\in\L^{2}_{k}(\Sigma, \E_{\Sigma})\mid (\lambda\mapsto\lambda^{\mathfrak{s}})\in \L^{2}(\sigma(E),\d\mu^{E}_{\omega})\},\quad E^{\mathfrak{s}}\defeq \int_{\sigma(E)}\,\lambda^{\mathfrak{s}}\,\mathrm{d}\mu^{E}(\lambda)\, ,
	\end{align*}
	where $\mu^{E}$ denotes the spectral measure of the self-adjoint operator $E$ and  $\mu^{E}_{\omega}(\cdot)\defeq \langle \mu^{E}(\cdot)\omega,\omega\rangle_{\L^{2}}$ is the corresponding complex measure.
\end{definition}

Note that $E^\mathfrak{s}$ is well-defined for all $\mathfrak s \in \mathbb{R}$ since $\sigma(E)\subset [1,\infty)$ by Lemma ~\ref{Lem: Chernoff thm}, while for $\mathfrak{s}<0$, the operator $E^{\mathfrak{s}}$ is bounded. It follows that, for any $\mathfrak s \in [0, \infty)$, $E^{\mathfrak{s}}: \mathrm{dom}(E^{\mathfrak{s}}) \to \L^2_k(\Sigma, \E_{\Sigma})$ is invertible with bounded inverse $E^{-\mathfrak{s}}$, see \cite[Prop. 3.24]{SchmidPhD} for further details. This motivates the following definition.

\begin{definition}
For $\mathfrak{s}\in[0,\infty)$ and $k\in \mathbb{N}_0$, we define the \emph{Sobolev space of order $\mathfrak{s}$} of $\E_{\Sigma}$-valued $k$-forms by
\begin{equation*}
    \H^{\mathfrak{s}}_k(\Sigma,\E_{\Sigma})
    \defeq
    \operatorname{dom} (E^{\frac{\mathfrak{s}}{2}}), \qquad 
    \langle \cdot ,\cdot \rangle_{\H^{\mathfrak{s}}}
    \defeq \langle E^{\frac{\mathfrak s}{2}}\cdot, E^{\frac{\mathfrak s}{2}}\cdot \rangle_{\L^2}.
\end{equation*}
\end{definition}
\noindent Note that, since $E^{\frac{\mathfrak{s}}{2}}\:\H^{\mathfrak{s}}_{k}(\Sigma, \E_{\Sigma})\to\L^{2}_{k}(\Sigma, \E_{\Sigma})$ is unitary, $$(\H^{\mathfrak{s}}_{k}(\Sigma, \E_{\Sigma}),\langle\cdot,\cdot\rangle_{\H^{\mathfrak{s}}})\, ,$$ is a Hilbert space. Furthermore, by the core property for fractional powers of elliptic Laplace-type operators, $\Omega^{k}_{\c}(\Sigma, \E_{\Sigma})$ is dense in $\H^{\mathfrak{s}}_{k}(\Sigma, \E_{\Sigma})$ w.r.t.~$\H^\mathfrak{s}$-norm (see \cite[Lem.~5.1]{Bandara} and \cite[Thm.~4.3]{Strichartz}). 

\begin{remark}
\label{Rmk: bounded geometry}
Completeness of the Cauchy hypersurface $(\Sigma, \h)$ suffices\footnote{We remark, however, that completeness is by no means a necessary condition, see~\textit{e.g.}, \cite{Gaffney1951,Gaffney1954}.} for the essential self-adjointness of the relevant Hodge-de Rham operators and, consequently, for defining the associated Sobolev spaces by spectral calculus. For the microlocal factorisation of the covariant Cauchy evolution considered later, however, we shall impose the stronger assumption that $(\Sigma,\h)$ is of \emph{bounded geometry} (see \textit{e.g.}, \cite{Shubin1992} and \cite[Chap.~10]{GerardBook}), namely that its injectivity radius is positive and that $\sup_{x \in \Sigma} |\nabla^j_{\h} \operatorname{Riem}(\h) (x)|_{\h} < \infty$ for all $j \in \mathbb{N}_0$, where $\operatorname{Riem}(\h)$ is the Riemann curvature tensor, while $\nabla_{\h}$ is the Levi-Civita connection induced by $\h$. Recall that every compact Riemannian manifold has bounded geometry and that this property, in particular, implies completeness. A further standard example is provided by the Euclidean space $(\mathbb{R}^d,\delta)$. Moreover, if $(\Sigma,\h)$ is of bounded geometry, then the Hermitian vector bundle $\E_\Sigma \to \Sigma$, equipped with the flat connection $\nabla^{\E_\Sigma}$, is also of bounded geometry, since its curvature vanishes identically.

In particular, if $(\Sigma, \h)$ is a Riemannian manifold of bounded geometry, we have the following properties:
\begin{itemize}
    \item[(i)] The \emph{Sobolev embedding theorem} holds and yields continuous embeddings (see \textit{e.g.},~\cite{AubinBook,EichhornBook,HebeyBook})
\begin{equation*}
        \H^m_k(\Sigma,\E_{\Sigma})
    \hookrightarrow
    \Gamma^\ell(\Sigma,\E_{\Sigma,k}),
    \qquad
    m-\ell>\frac d2, \; m,\ell\in\mathbb N_0.
\end{equation*}
\item[(ii)] For $m\in\mathbb N_0$, the spectral Sobolev space $\H^m_k(\Sigma,\Ad(\P_\Sigma))$ agrees, with equivalent norms, with the standard covariant Sobolev space $W^{m,2}_k(\Sigma,\Ad(\P_\Sigma))$ defined via weak/distributional covariant derivatives. A proof of this fact can be found in \cite{Aubin1976,AubinBook} (for functions), \cite{Dodziuk1981} (for real-valued differential forms) and \cite{Eichhorn} (for general bundle-valued forms). 
\end{itemize}
\end{remark}

The next lemma shows that there exists a continuous embedding between Sobolev spaces also on manifolds that are not of bounded geometry. The proof is well-known and can, for instance, be obtained by a straightforward adaptation of \cite[Prop.~3.28(ii)]{SchmidPhD}; see also \cite[Lem.~3.3]{MurroSchmid}. 

\begin{lemma}[see, \textit{e.g.}, \protect{\cite[Prop.~3.28, (ii)]{SchmidPhD} or \cite[Lem.~3.3]{MurroSchmid}}] \label{Lemma:WeakCovergence}
	For any  $0 \le \mathfrak s \le \mathfrak r$, the Sobolev space $\H^{\mathfrak r}_{\bullet}(\Sigma, \E_{\Sigma})$ embeds continuously into $\H^{\mathfrak s}_{\bullet}(\Sigma, \E_{\Sigma})$. 
\end{lemma}

On account of Lemma \ref{Lemma:WeakCovergence}, we can also define the \emph{Sobolev space of order $\infty$} by 
\begin{equation*}
    \H^\infty_k(\Sigma,\E_{\Sigma}) \defeq \bigcap_{\mathfrak{s}\geq 0}
    \H^{\mathfrak{s}}_k(\Sigma,\E_{\Sigma})\, ,
\end{equation*}
endowed with the natural projective limit topology, which equips $\H^\infty_k(\Sigma,\E_{\Sigma})$ with the structure of a Fréchet space. If $(\Sigma,\h)$ is a manifold of bounded geometry (Remark~\ref{Rmk: bounded geometry}), then clearly $\H^\infty_k(\Sigma,\E_{\Sigma})\subset\Omega^{k}(\Sigma,\E_{\Sigma})$ as a consequence of the Sobolev embedding theorems.\\

Having introduced Sobolev spaces of differential forms in flat vector bundles, let us now turn to the Hodge decomposition. To start with, let us introduce a special notation for \emph{smooth} differential forms cotained in a given Sobolev space.

\begin{definition}
For any $\mathfrak s \in [0,\infty]$ and $k \in \mathbb{N}$, we denote the linear subspace of \emph{smooth} $\sf{E}_{\Sigma}$-valued $k$-forms in the Sobolev space $\H^\mathfrak{s}_k (\Sigma, \sf{E}_{\Sigma})$ by 
\begin{equation*}
  \Omega^{k}_{\mathfrak s}(\Sigma, \sf{E}_{\Sigma})\defeq \Omega^{k}(\Sigma, \sf{E}_{\Sigma})\cap\H^{\mathfrak s}_{k}(\Sigma, \sf{E}_{\Sigma})\,.
\end{equation*}
For notational ease and whenever no confusion can arise, we will write $\Omega^k_\mathfrak{s}$ in place of $\Omega^{k}_{\mathfrak s}(\Sigma,\sf{E}_{\Sigma})$. 
\end{definition}

The following theorem establishes a (strong) Hodge-type decomposition for the spaces $\Omega^{k}_{\mathfrak s}(\Sigma,\sf{E}_{\Sigma})$, which is needed to obtain solutions to the Poisson equation on non-compact (but complete) manifolds.

\begin{theorem}[Hodge Decomposition for Sobolev Spaces in Flat Vector Bundles]\label{Theorem.HodgeDecom} Let $(\Sigma,\h)$ be a complete Riemannian manifold and $(\E_{\Sigma},\langle\cdot,\cdot\rangle_{\E_{\Sigma}})$ be a Riemannian vector bundle equipped with a compatible flat connection $\nabla^{\E_{\Sigma}}$. The space $\Omega^{k}_{\mathfrak{s}}(\Sigma,\sf{E}_{\Sigma})$ admits the $\sf{H}^{\mathfrak{s}}$-orthogonal decompositions
	\begin{align*}
		\Omega^{k}_{\mathfrak{s}}(\Sigma,\sf{E}_{\Sigma})\cong \Omega^{k}_{\mathfrak{s},\d}(\Sigma,\sf{E}_{\Sigma})\oplus\Omega^{k}_{\mathfrak{s},\delta}(\Sigma,\sf{E}_{\Sigma})\oplus\mathrm{ker}(\Delta^{\nabla^{\sf{E}_{\Sigma}}}\vert_{\Omega^{k}_{\mathfrak{s}}})
	\end{align*}
	for all $s\in [0,\infty]$ and $k\in\bb{N}_{0}$, where $\Omega_{\mathfrak{s},\d}^{k}(\Sigma)$ and $\Omega_{\mathfrak{s},\delta}^{k}(\Sigma)$ are the subspaces defined by
	\begin{align*}
		\Omega^{k}_{\mathfrak{s},\d}(\Sigma,\sf{E}_{\Sigma}):=\Omega^{k}(\Sigma,\sf{E}_{\Sigma})\cap\overline{\d^{\nabla^{\E_\Sigma}}\Omega^{k-1}_{\infty}(\Sigma,\sf{E}_{\Sigma})}^{\sf{H}^{\mathfrak{s}}}\,,\quad \Omega^{k}_{\mathfrak{s},\delta}(\Sigma,\sf{E}_{\Sigma}):=\Omega^{k}(\Sigma,\sf{E}_{\Sigma})\cap\overline{\delta^{\nabla^{\E_\Sigma}}\Omega^{k+1}_{\infty}(\Sigma,\sf{E}_{\Sigma})}^{\sf{H}^{\mathfrak{s}}}\, .
	\end{align*}
	Moreover, if we decompose $\omega\in\Omega^{k}_{\mathfrak{s}}(\Sigma,\sf{E}_{\Sigma})$ accordingly as $\omega=\alpha+\beta+\gamma$ for (unique) $\alpha\in\Omega^{k}_{\mathfrak{s},\d}(\Sigma,\sf{E}_{\Sigma})$, $\beta\in\Omega^{k}_{\mathfrak{s},\delta}(\Sigma,\sf{E}_{\Sigma})$ and $\gamma\in\mathrm{ker}(\Delta^{\nabla^{\E_\Sigma}}\vert_{\Omega^{k}_{\mathfrak{s}}})$, the following holds true:
	\begin{itemize}
		\item[\emph{(i)}]$\Omega^{k}_{\mathfrak{s},\d}(\Sigma,\sf{E}_{\Sigma})\subset\mathrm{ran}(\d^{\nabla^{\E_\Sigma}}\vert_{\Omega^{k-1}})$, i.e.~$\alpha=\d^{\nabla^{\E_\Sigma}}\psi$ for some smooth $\psi\in\Omega^{k-1}(\Sigma,\sf{E}_{\Sigma})$.
		\item[\emph{(ii)}]$\Omega^{k}_{\mathfrak{s},\delta}(\Sigma,\sf{E}_{\Sigma})\subset\mathrm{ran}(\delta^{\nabla^{\E_\Sigma}}\vert_{\Omega^{k+1}})$, i.e.~$\beta=\delta^{\nabla^{\E_\Sigma}}\phi$ for some smooth $\phi\in\Omega^{k+1}(\Sigma,\sf{E}_{\Sigma})$.
		\item[\emph{(iii)}]$\mathrm{ker}(\d^{\nabla^{\E_\Sigma}}\vert_{\Omega^{k}_{s}})=\Omega^{k}_{\mathfrak{s},\d}(\Sigma,\sf{E}_{\Sigma})\oplus\mathrm{ker}(\Delta^{\nabla^{\E_\Sigma}}\vert_{\Omega^{k}_{\mathfrak{s}}})$, i.e.~$\d^{\nabla^{\E_\Sigma}}\omega=0$ if and only if $\beta=0$.
		\item[\emph{(iv)}]$\mathrm{ker}(\delta_{\nabla^{\E_\Sigma}}\vert_{\Omega^{k}_{\mathfrak{s}}})=\Omega^{k}_{\mathfrak{s},\delta}(\Sigma,\sf{E}_{\Sigma})\oplus\mathrm{ker}(\Delta^{\nabla^{\E_\Sigma}}\vert_{\Omega^{k}_{\mathfrak{s}}})$, i.e.~$\delta^{\nabla^{\E_\Sigma}}\omega=0$ if and only if $\alpha=0$.
	\end{itemize}
\end{theorem}

\begin{remark}
In the case $\mathfrak{s}=\infty$, the respective closures are taken in the relevant Fréchet topology on $\sf{H}^{\infty}_{k}$ and orthogonality holds with respect to every $\sf{H}^{\mathfrak{s}}_{k}$-inner product. We remark that the closure in this topology satisfies
\begin{align*}
    \overline{\sf{S}}^{\sf{H}^{\infty}}=\bigcap_{s\in [0,\infty)}\overline{\sf{S}}^{\sf{H}^{\mathfrak{s}}}
\end{align*}
for an arbitrary subset $\sf{S}\subset\sf{H}^{\infty}_{k}(\Sigma,\sf{E}_{\Sigma})$, which easily follows from the definition of the projective limit topology and a diagonalisation argument, see \textit{e.g.}~\cite[Lemma~3.47]{SchmidPhD} for details.
\end{remark}

Theorem~\ref{Theorem.HodgeDecom} extends Theorem 1.2 in \cite{MurroSchmid} (\textit{cf.}~\cite[Thm.~3.44]{SchmidPhD}) to bundle-valued differential forms. Because the proof follows naturally from this prior work, we outline only the essential steps here, supplemented by appropriate references.

\begin{proof}[Proof of Theorem~\ref{Theorem.HodgeDecom} (sketch).] For notational convenience, throughout the proof, we shall adopt the short-hand notation $\d:=\d^{\nabla^{\E_\Sigma}}$, $\delta:=\delta^{\nabla^{\E_\Sigma}}$ and
$\Delta_k:=\d\delta+\delta \d$ on $\Omega^k(\Sigma,\E_\Sigma)$.
We limit ourselves to outlining only the main steps, following the strategy of
\cite[Sec.~3]{MurroSchmid}: 
\begin{itemize}
    \item[1.] \emph{Sobolev adjoints.} Fix $\mathfrak{s}\in[0,\infty)$. Flatness of $\nabla^{\E_\Sigma}$ implies $\d^2=\delta^2=0$ and, hence, it yields the intertwining relations
    \begin{equation*}
\d\Delta_k=\Delta_{k+1} \d,\qquad
\delta\Delta_k=\Delta_{k-1}\delta.
\end{equation*}
By spectral calculus, these relations extend to the fractional powers of $1+\overline{\Delta}_k$ on their natural common domains, as in \cite[Lem.~5.8]{MorettiMurroVolpe}. Consequently, $\d$ and $\delta$, initially defined on $\Omega^\bullet_\infty(\Sigma,\E_\Sigma)$, are formal adjoints with respect to the $\H^\mathfrak{s}$-inner products. Using suitable regularisations and the density of smooth
sections in the relevant graph norms (\textit{cf.}
\cite{EichhornBook,BaerBallmann2016}), one concludes
that the closures of $d$ and $\delta$ are mutually
adjoint with respect to the Sobolev inner products.
The argument follows the strategy of \cite[Lemmas~3.4--3.7]{MurroSchmid}, adapted to the flat Hermitian bundle setting.

\item[2.] \emph{Weak Hodge Decomposition.} Since $\d^2=0$, the closures of the degree components of $\d$ define a Hilbert complex. The abstract Hodge decomposition for Hilbert complexes \cite[Lem.~2.1]{BruningLesch1992} yields
\begin{equation*}
\H^\mathfrak{s}_k = \overline{\d\Omega^{k-1}_\infty}^{\H^\mathfrak{s}} \oplus \overline{\delta\Omega^{k+1}_\infty}^{\H^\mathfrak{s}} \oplus
\mathrm{Har}^\mathfrak{s}_k,
\end{equation*}
where $\mathrm{Har}^\mathfrak{s}_k \defeq \ker \overline{\d}\cap\ker \overline{\delta} = \ker(\Delta_k|_{\Omega^k_\mathfrak{s}}).$ The last identification follows from the non-negativity of $\Delta_k$ and elliptic regularity.

\item[3.] \emph{Smoothness and Exactness.} Let $\omega\in\Omega^k_\mathfrak{s}$ and write $\omega=\alpha+\beta+\gamma$ according to the preceding decomposition. Since convergence in $\H^\mathfrak{s}$ implies convergence in $\L^2$, flatness gives $\d\alpha=0$ and $\delta\beta=0$ distributionally. As $\gamma$ is harmonic, we obtain
\begin{equation*}
(\d+\delta)\alpha=\delta\omega,\qquad
(\d+\delta)\beta=\d\omega
\end{equation*}
in the distributional sense. The operator $\d+\delta$ is elliptic and $\omega$ is smooth. Elliptic regularity thus entails that $\alpha$ and $\beta$ are smooth. The same holds for $\gamma$. 

To prove (i), let $\eta\in\Omega^k_\c(\Sigma,\E_\Sigma)$
satisfy $\delta\eta=0$. By the definition of $\alpha$,
there exists a sequence $\alpha_n\in\Omega^{k-1}_\infty(\Sigma,\E_\Sigma)$ such that $\d\alpha_n\to\alpha$ in $\L^2$. Thus,
\begin{equation*}
\langle\alpha,\eta\rangle_{\L^2} =
\lim_{n\to\infty}
\langle \d\alpha_n,\eta\rangle_{\L^2} =
\lim_{n\to\infty} \langle\alpha_n,\delta\eta\rangle_{\L^2} =0.
\end{equation*}
Since $\alpha$ is smooth and closed, twisted Poincaré duality (Proposition \ref{Prop: Poincare duality}) implies that
$\alpha=\d\psi$ for some $\psi\in\Omega^{k-1}(\Sigma,\E_\Sigma)$.
Claim (ii) follows analogously, using the Hodge star operator.

Finally, if $\d\omega=0$, then $\d\beta=0$. Together with $\delta\beta=0$, this implies $\beta\in\mathrm{Har}^\mathfrak{s}_k$ and hence $\beta=0$ by orthogonality. This proves (iii); item (iv) follows simply by exchanging the r\^ole of $\d$ and $\delta$.

\item[4.] \emph{The case $\mathfrak{s}=\infty$.} Since $\H^\infty_k=\bigcap_{\mathfrak{s}\geq0} \H^\mathfrak{s}_k$, the uniqueness of the orthogonal decomposition and the continuous embeddings $\H^\mathfrak{r}_k\hookrightarrow\H^\mathfrak{s}_k$ for $\mathfrak r\geq\mathfrak{s}$ imply that the components of $\omega\in\Omega^k_\infty$ are independent of $\mathfrak{s}$. They therefore belong to $\H^\infty_k$, yielding the claimed decomposition in the projective limit topology.
\end{itemize}
\end{proof}

As an immediate consequence of Theorem~\ref{Theorem.HodgeDecom}, we get existence of solutions to the Poisson equation, which is closely related to the achievability of the Cauchy radiation gauge as discussed in Section \ref{Subsec:Cauchy Radiation Gauge}.

\begin{proposition}\label{prop:Poisson}
Let $\mathfrak s\in [0,\infty]$ and $\omega\in\Omega_{\mathfrak s}^{1}$. Then, the Poisson equation
\begin{align*}
	\Delta^{\nabla^{\E_\Sigma}} \mathfrak f=\delta^{\nabla^{\E_\Sigma}} \omega\, ,
\end{align*}
has a unique solution, modulo $\ker (\d^{\nabla^{\E_\Sigma}})$, on the space $\{\mathfrak f\in  \Omega^{0}(\Sigma, \E_\Sigma)\mid \d^{\nabla^{\E_\Sigma}} \mathfrak f\in\Omega_{\mathfrak s,\d}^{1}(\Sigma, \E_\Sigma)\}$.
\end{proposition}

\begin{proof}
By item (iv) of Theorem~\ref{Theorem.HodgeDecom}, $\omega$ can be written as $\omega=\alpha+\beta$, where $\alpha\in\Omega_{\mathfrak s,\d}^{1}(\Sigma, \E_\Sigma)$ and $\beta\in\ker(\delta^{\nabla^{\E_\Sigma}} \vert_{\Omega^{1}_{\mathfrak s}})$. Furthermore, $\alpha=\d^{\nabla^{\E_\Sigma}} \mathfrak f$ for some $\mathfrak f\in \Omega^0(\Sigma, \E_\Sigma)$. It follows that $\delta^{\nabla^{\E_\Sigma}} \omega=\delta^{\nabla^{\E_\Sigma}} \alpha=\Delta^{\nabla^{\E_\Sigma}} \mathfrak f$. For uniqueness, suppose that $\mathfrak f\in \Omega^0(\Sigma, \E_\Sigma)$ is such that $\d^{\nabla^{\E_\Sigma}} \mathfrak f\in\Omega^{1}_{\mathfrak s,\d}(\Sigma, \E_\Sigma)$ and $\Delta_{0}^{\nabla^{\E_\Sigma}} \mathfrak f=0$. Then, the $1$-form $\eta\defeq \d^{\nabla^{\E_\Sigma}} \mathfrak f$ satisfies $\d^{\nabla^{\E_\Sigma}} \eta=\delta^{\nabla^{\E_\Sigma}} \eta=0$ and, hence, in particular $\eta\in\ker(\Delta^{\nabla^{\E_\Sigma}} \vert_{\Omega^{1}_{\mathfrak s}})$. Since also $\eta\in\Omega^{1}_{\mathfrak s,\d}(\Sigma, \E_\Sigma)$ by assumption, we conclude that $\eta= \d^{\nabla^{\E_\Sigma}} \mathfrak f = 0$, by the fact that the decomposition in Theorem \ref{Theorem.HodgeDecom} is a direct sum decomposition. Consequently, $\mathfrak f$ as in the statement of the proposition is unique modulo elements lying in  $\ker (\d^{\nabla^{\E_\Sigma}})$. 
\end{proof}

\subsection{Cauchy Problem and Complete Gauge Fixing in the Non-Compact Case}
\label{Subsec:Cauchy Problem and Complete Gauge Fixing}

As already indicated at the beginning of Section~\ref{Subsec:Cauchy Radiation Gauge}, we shall now employ the Hodge decomposition developed in Section~\ref{Subsec:Hodge Decomposition in Flat Vector Bundles on Complete Riemannian Manifolds} in order to achieve the Cauchy radiation gauge also in the setting of non-spatially compact spacetimes. Throughout the following, we consider a $(1+d)$-dimensional globally hyperbolic and \emph{ultrastatic} spacetime
\begin{align*}
     \M=\R\times\Sigma\,,\qquad\qquad \g=-\d t\otimes\d t+\h
\end{align*}
and a flat background connection $A=A_{\mathrm{T}}\d t+A_{\Sigma}\in\mathcal{C}(\P)$. By global hyperbolicity, $(\Sigma,\h)$ is complete and, following the discussion in Section~\ref{Subsec:Hodge Decomposition in Flat Vector Bundles on Complete Riemannian Manifolds}, we denote by $\sf{H}^{\mathfrak{s}}_{k}(\Sigma_{t},\Ad(\P_\Sigma))$ for $\mathfrak{s}\in [0,\infty]$ and $t\in\bb{R}$ the Sobolev space of $\mathrm{Ad}(\P_{\Sigma})$-valued $k$-forms on $\Sigma$ associated with the spatial connection $\A_\Sigma(t)$, where $\Sigma_t \defeq \{t\} \times \Sigma$.\footnote{The Hermitian form $(\cdot,\cdot)_{\sf{E}_{\Sigma,k}}$ on $\Gamma(\E_{\Sigma,k})=\Omega^{k}(\Sigma,\mathrm{Ad}(\P_{\Sigma}))$ induced from the natural bundle metric $\langle\cdot,\cdot\rangle_{\sf{E}_{\Sigma,k}}$  on $\sf{E}_{\Sigma,k}$, \emph{cf.}, Eq.~\eqref{Eq: E Sigma k}, coincides with what we called $\langle\cdot,\cdot\rangle_{\sf{L}^{2}}$ in Section~\ref{Subsec:Hodge Decomposition in Flat Vector Bundles on Complete Riemannian Manifolds}, see also Example~\ref{ExampleHodge}.} Note that these spaces explicitly depend on time and we hence obtain the $\infty$-dimensional bundles $(\sf{H}^{\mathfrak{s}}_{k}(\Sigma_{t},\Ad(\P_\Sigma)))_{t\in\bb{R}}$. 
For $s<\infty$, we denote the space of smooth sections of this Hilbert bundle by $C^\infty(\mathbb R,\sf{H}_k^s(\Sigma_\bullet,\operatorname{Ad}(P_\Sigma)))$, where regularity in time has to be understood with respect to the gauge-covariant time-derivative $\mathcal{D}^{A_{\mathrm{T}}}_{t}$ introduced in Definition~\ref{Def: covariant time derivative}. By taking the projective limit, we obtain the space $C^\infty(\mathbb R,\sf{H}_k^\infty(\Sigma_\bullet,\operatorname{Ad}(P_\Sigma)))$.
Recalling the decomposition of $\mathrm{Ad}(\P)$-valued forms on $\M$ introduced in Section~\ref{Subsec:Decomposition of Connections and Linearised Yang-Mills Fields} and the bundles of initial data $\E_{\rho_{k}}$ introduced in Section~\ref{Subsec:Phase Spaces on Initial Data and Cauchy Problem}, Eq.~\eqref{eq:ERhok}, we denote the spaces of smooth spacelike Sobolev
fields of degree $\mathfrak s=\infty$ and the corresponding spaces of initial data respectively by
\begin{align*}
    \Gamma_\infty(\E_0)
    &\defeq \Gamma(\E_{0})\cap C^{\infty}(\bb{R},\sf{H}^{\infty}_{0}(\Sigma_{\bullet},\mathrm{Ad}(\sf{P}_{\Sigma}))),
    \nonumber\\
    \Gamma_\infty(\E_k)
    &\defeq \Gamma(\E_{k})\cap\bigg(C^{\infty}(\bb{R},\sf{H}^{\infty}_{k-1}(\Sigma_{\bullet},\mathrm{Ad}(\sf{P}_{\Sigma})))\oplus C^{\infty}(\bb{R},\sf{H}^{\infty}_{k}(\Sigma_{\bullet},\mathrm{Ad}(\sf{P}_{\Sigma})))\bigg),\qquad k\geq1,
    \nonumber\\
    \Gamma_\infty(\E_{\rho_0})
    &\defeq \Omega^0_\infty(\Sigma_0,\Ad(\P_\Sigma))^{\oplus2},
    \nonumber\\
    \Gamma_\infty(\E_{\rho_k})
    &\defeq \Omega^{k-1}_\infty(\Sigma_0,\Ad(\P_\Sigma))^{\oplus2}
    \oplus\Omega^k_\infty(\Sigma_0,\Ad(\P_\Sigma))^{\oplus2},\qquad k\geq1.
\end{align*}
where we recall that $\Gamma(\E_{k})=\Omega^{k}(\M,\mathrm{Ad}(\P))$ and where we choose $\Sigma_{0}\cong\{0\}\times\Sigma$ as the initial slice. 
In particular, every $\omega\in\Gamma_\infty(E_k)$ with $k\geq 1$ satisfies
$$
\omega_{\mathrm{T}}(t,\cdot)\in\Omega_\infty^{k-1}(\Sigma_t,\operatorname{Ad}(P_\Sigma)),\qquad
\omega_\Sigma(t,\cdot)\in\Omega_\infty^k(\Sigma_t,\operatorname{Ad}(P_\Sigma))
$$
for all $t\in\mathbb R$. for which 
\begin{align*}
    \omega_{\mathrm{T}}(t,\cdot)\in\Omega^{k-1}_{\infty}(\Sigma_{t},\mathrm{Ad}(\P_{\Sigma}))\,,\qquad \omega_{\Sigma}(t,\cdot)\in\Omega^{k}_{\infty}(\Sigma_{t},\mathrm{Ad}(\P_{\Sigma}))
\end{align*} 
for all $t\in\bb{R}$. Last but not least, we note that the Cauchy data maps are well-defined when regarded as operators of the form $\rho_k\:\Gamma_\infty(\E_k)\to\Gamma_\infty(\E_{\rho_k})$.
We further set
\begin{align*}
    \Gamma_{\tc,\infty}(\E_k)
    \defeq \Gamma_{\tc}(\E_k)\cap\Gamma_\infty(\E_k)\,,
\end{align*}
where $\Gamma_{\tc}(\E_k)$ denotes the space of smooth sections, whose support is contained in the region between two Cauchy surfaces of our foliation.

As a first step, we show that the Cauchy problem of $\square^{A}$ is well-posed in the setting of smooth $\sf{H}^{\infty}$-forms. This fact, together with the corresponding properties of the Green operators, is summarised in the following theorem, extending the results of \cite[Thm.~4.2 and Prop.~4.7]{MurroSchmid} to the present twisted setting.

\begin{theorem}[The Cauchy Problem for Smooth Spatial $\H^\infty$-Sections]\label{Thm:CauchyProblemHInfty}
Let $(\M,\g)$ be an ultrastatic globally hyperbolic spacetime. Then, the following statements hold:
\begin{itemize}
    \item[(i)] For any $\varphi\in\Gamma_\infty(\E_k)$ and $f\in\Gamma_\infty(\E_{\rho_k})$, the Cauchy problem for the twisted de Rham-Hodge d'Alembertian $\square^A$, \textit{i.e.}, 
\begin{equation*}
    \begin{cases}
        \square^A \omega = \varphi\\
        \rho_k(\omega) = \mathfrak f,
    \end{cases}
\end{equation*}
admits a unique solution $\omega\in\Gamma_\infty(\E_k)$. In particular, the Cauchy data maps restrict to well-defined isomorphism $\rho_k\:\ker(\square^{A} \vert_{\Gamma_\infty})\to\Gamma_\infty(\E_{\rho_k})$ with inverse given by the Cauchy evolution operators $\mathcal U_k\:\Gamma_\infty(\E_{\rho_k})
    \longrightarrow
    \ker(\mathsf \square^{A}\vert_{\Gamma_\infty})$.
\item[(ii)] The retarded and advanced Green operators are well-defined as linear operators of the form $\G_k^\pm:\Gamma_{\mathrm{tc}, \infty}(\E_k) \to \Gamma_\infty(\E_k)$ and they satisfy: 
\begin{itemize}
    \item[$\mathrm{(iia)}$] $\G_k^\pm \circ \D_k \vert_{\Gamma_{\mathrm{tc}, \infty}} = \D_k  \circ \G_k^\pm = \operatorname{id}_{\Gamma_{\mathrm{tc}, \infty}}$; 
    \item[$\mathrm{(iib)}$] $\supp(\G_k^\pm \psi) \subset \mathcal{J}^\pm(\supp (\psi))$, for all $\psi \in \Gamma_{\mathrm{tc}, \infty}(\E_k)$.
\end{itemize}
Furthermore, the corresponding causal propagator 
$\G_k:=\G_k^+-\G_k^-\:\Gamma_{\tc,\infty}(\E_k)
    \to\Gamma_\infty(\E_k)$ defines an exact sequence
        \begin{align*}
		0\xrightarrow{}\Gamma_{\mathrm{tc},\infty}(\E_{k})\xrightarrow{\square^{A}}\Gamma_{\mathrm{tc},\infty}(\E_{k})\xrightarrow{\sf{G}_{k}}\Gamma_{\infty}(\E_{k})\xrightarrow{\square^{A}}\Gamma_{\infty}(\E_{k})\xrightarrow{}0\, .
	\end{align*}
\end{itemize}
\end{theorem}

\begin{proof}[Proof (sketch).]
Since the background connection $A$ is flat and $(\M,\g)$ ultrastatic, Proposition~\ref{Prop:DecompAdFormsOp}(v) implies that the equation $\square^{A}\omega=0$ decouples into independent normally hyperbolic equations for the components $\omega_{\mathrm{T}}$ and $\omega_{\Sigma}$, i.e., 
\begin{align*}
    \begin{cases}
    \big((\mathcal{D}^{A_{\mathrm{T}}}_{t})^{2}+\Delta^{A_{\Sigma}}\big)\omega_{\mathrm{T}}&=\varphi_{\mathrm{T}}\\
    \omega_{\mathrm{T}}\vert_{t=0}&=\mathfrak{a}_{\mathrm{T}}\\
    \mathcal{D}_{t}^{A_{\mathrm{T}}}\omega_{\mathrm{T}}\vert_{t=0}&=i\pi_{\mathrm{T}}
    \end{cases}\,,\qquad\text{and}\qquad \begin{cases}
    \big((\mathcal{D}^{A_{\mathrm{T}}}_{t})^{2}+\Delta^{A_{\Sigma}}\big)\omega_{\Sigma}&=\varphi_{\Sigma}\\
    \omega_{\Sigma}\vert_{t=0}&=\mathfrak{a}_{\Sigma}\\
    \mathcal{D}_{t}^{A_{\Sigma}}\omega_{\Sigma}\vert_{t=0}&=i\pi_{\Sigma}
    \end{cases}
\end{align*} where we wrote $\mathfrak{f}=(\mathfrak{a}_{\mathrm{T}},\pi_{\mathrm{T}},\mathfrak{a}_{\Sigma},\pi_{\Sigma})$. Now, for $\eta\in\{\omega_{\mathrm{T}},\omega_{\Sigma}\}$, we consider the \emph{energy} functional $\mathcal{E}_{s}(\eta,t):=\Vert\eta\vert_{\Sigma_{t}}\Vert_{\sf{H}^{\mathfrak{s}}}^{2}+\Vert\mathcal{D}^{A_{\mathrm{T}}}_{t}\eta\vert_{\Sigma_{t}}\Vert_{\sf{H}^{\mathfrak{s}-1}}^{2}$ for $\mathfrak{s}\geq 1$. Note that the $\sf{H}^{\mathfrak{s}}$-norms explicitly depend on time, although we suppressed this dependence in the notation. Since $[\mathcal{D}_{t}^{A_{\mathrm{T}}},\Delta^{A_{\Sigma}}]=0$, by flatness and Lemma~\ref{Lem: 2.26}(iii), we also obtain $[\mathcal{D}_{t}^{A_{\mathrm{T}}},E^{\mathfrak{s}}]=0$, with $E:=\mathds{1}+\overline{\Delta^{A_{\Sigma}^{t}}}$, as in Section~\ref{Subsec:Hodge Decomposition in Flat Vector Bundles on Complete Riemannian Manifolds}, by spectral calculus, on a suitably chosen domain. In particular, the standard energy estimates go through without major difficulties and for each $t_{0},t_{1}\in\bb{R}$ with $t_{0}<t_{1}$, we can find a constant $C=C(t_{0},t_{1})>0$, such that
\begin{align*}
		\mathcal{E}_{s}(\eta,t_1)\leq \mathcal{E}_{s}(\eta,t_{0})e^{C(t_{1}-t_{0})}+\int_{t_0}^{t_1}e^{C(t_1-\tau)}\Vert\big((\mathcal{D}^{A_{\mathrm{T}}}_{t})^{2}+\Delta^{A_{\Sigma}}\big)\eta\vert_{\Sigma_\tau}\Vert_{\sf{H}^{s-1}}^{2}\,\d\tau\,.
\end{align*}
For more details we refer to \cite[Thm.~8]{BaerWafo}. In particular, it follows from standard density arguments that the Cauchy problem
\begin{align*}
    \begin{cases}
    \big((\mathcal{D}^{A_{\mathrm{T}}}_{t})^{2}+\Delta^{A_{\Sigma}}\big)\eta&=\varphi\\
    \eta\vert_{t=0}&=\mathfrak{a}\\
    \mathcal{D}_{t}^{A_{\mathrm{T}}}\eta\vert_{t=0}&=\pi
    \end{cases}
\end{align*}
for Cauchy data $(\mathfrak{a},\pi)\in\sf{H}^{s}_{k}(\Sigma_{0},\mathrm{Ad}(\P_{\Sigma}))\times\sf{H}^{s-1}_{k}(\Sigma_{0},\mathrm{Ad}(\P_{\Sigma}))$ and $\varphi\in C^{0}(\bb{R},\sf{H}^{s-1}_{k}(\Sigma_{\bullet},\mathrm{Ad}(\P_{\Sigma})))$ admits a unique solution $\eta\in C^{0}(\bb{R},\sf{H}^{s}_{k}(\Sigma_{\bullet},\mathrm{Ad}(\P_{\Sigma})))\cap C^{1}(\bb{R},\sf{H}^{s-1}_{k}(\Sigma_{\bullet},\mathrm{Ad}(\P_{\Sigma})))$. In our setting with $\sf{H}^{\infty}$-Cauchy data, we hence obtain, inductively,
\begin{align*}
    \omega_{\mathrm{T}}\in C^{\infty}(\bb{R},\sf{H}^{\infty}_{k-1}(\Sigma_{\bullet},\mathrm{Ad}(\P_{\Sigma})))\,,\qquad \omega_{\Sigma}\in C^{\infty}(\bb{R},\sf{H}^{\infty}_{k}(\Sigma_{\bullet},\mathrm{Ad}(\P_{\Sigma})))
\end{align*}
and since our Cauchy data is additionally assumed to be smooth, we conclude that $\omega\in\Gamma_{\infty}(\E_{k})$, cf., \cite[Thm.~3.2.11]{BaerBook}. The first assertion thus follows. For more details on the Cauchy problem in the Maxwell setting we refer to \cite[Sec.~4]{MurroSchmid} and \cite[Sec.~3.3.1]{SchmidPhD}.

The Green operators are obtained by solving the inhomogeneous Cauchy problem with vanishing data to the past or future of the temporally compact source. Their support properties follow from finite propagation speed, while the remaining identities and exactness of the sequence follow from uniqueness and the standard time-cutoff argument, see e.g., \cite[Thm.~3.5]{GinouxBaer}.
\end{proof}

Having discussed the Cauchy problem for linearised Yang-Mills fields in the $\sf{H}^{\infty}$-setting, let us now turn to the achievability of the \emph{Cauchy radiation gauge}, as per Definition~\ref{Def: Cauchy radiation gauge}, in the non-compact case. To this end, we follow the same steps as in the compact case, Proposition~\ref{Prop:AchCauchyRadGauge}, with solvability discussion concerning the Poisson equation replaced accordingly with Proposition~\ref{prop:Poisson}. In this setting, we introduce the following subspaces of $\Gamma_\infty(\E_1)$ and
$\Gamma_\infty(\E_{\rho_1})$:
\begin{align*}
    \Gamma_{\infty,\d}(\E_1)
    &\defeq
    \left\{\alpha=\alpha_{\mathrm T}\d t+\alpha_\Sigma
    \in\Gamma_\infty(\E_1)\ \middle|\
    \alpha_\Sigma\vert_{\Sigma_0},\,
    \mathcal D_t^{\A_{\mathrm T}}\alpha_\Sigma\vert_{\Sigma_0}
    \in\Omega^1_{\infty,\d}(\Sigma_0,\Ad(\P_\Sigma))\right\},
    \nonumber\\
    \Gamma_{\infty,\d}(\E_{\rho_1})
    &\defeq
    \Omega^0_\infty(\Sigma_0,\Ad(\P_\Sigma))^{\oplus2}
    \oplus
    \Omega^1_{\infty,\d}(\Sigma_0,\Ad(\P_\Sigma))^{\oplus2},
\end{align*}
with the notation introduced in Theorem \ref{Theorem.HodgeDecom}. Clearly,
$\rho_1\:\Gamma_{\infty,\d}(\E_1) \to\Gamma_{\infty,\d}(\E_{\rho_1})$.

\begin{remark}
Note that the definition of $\Gamma_{\infty,\d}(\E_1)$ depends on the
choice of the initial hypersurface $\Sigma_0$, since the condition
$\alpha_\Sigma\vert_{\Sigma_0}\in
\Omega^1_{\infty,\d}(\Sigma_0,\Ad(\P_\Sigma))$ does not need to propagate.
\end{remark}

Using the discussion of Section~\ref{Subsec:Hodge Decomposition in Flat Vector Bundles on Complete Riemannian Manifolds}, we obtain the following
generalisation of the complete gauge fixing to non-compact Cauchy
hypersurfaces.

\begin{corollary}
\label{Cor:CompleteGaugeFixingYM}
Let $(\M,\g)$ be an ultrastatic and globally hyperbolic spacetime and $A\in\mathcal{C}(\P)$ a flat background connection. For any $\alpha\in\Gamma_\infty(\E_1)$, there exists a gauge transformation
$\varepsilon\in\Gamma(\E_0)$ with the property $\K\varepsilon\in\Gamma_{\infty,\d}(\E_1)$, unique modulo $\ker(\K)$, such that
$\alpha'\defeq\alpha-\K\varepsilon$ satisfies the Cauchy radiation gauge
condition of Definition~\ref{Def: Cauchy radiation gauge}.
\end{corollary}

\begin{proof}
    The proof is the covariant counterpart of the discussion of Maxwell's theory in \cite[Cor.~5.3]{MurroSchmid}. Existence of $\varepsilon\in\Gamma(\E_0)$ such that $\alpha'\defeq\alpha-\K\varepsilon$ satisfies the Cauchy radiation gauge follows from the same steps as in Proposition~\ref{Prop:AchCauchyRadGauge}, when solving the Poisson equation as in Proposition~\ref{prop:Poisson}.

    Moreover, setting $\eta:=\sf{K}\varepsilon$, flatness of $\sf{A}$ implies that $\square^{A}\eta=\d^{A}\delta^{A}\alpha$. It is easy to check that its initial data satisfy 
    \begin{align*}
        \eta_{\mathrm{T}}\vert_{t=0}\in \Omega_{\infty}^{0}\,,\qquad \mathrm{D}^{A_{\mathrm{T}}}_{t}\eta_{\mathrm{T}}\vert_{t=0}\in\Omega^{0}_{\infty}\,,\qquad\eta_{\Sigma}\vert_{t=0}\in\Omega^1_{\infty,\d}\,,\qquad \mathrm{D}^{A_{\mathrm{T}}}_{t}\eta_{\Sigma}\vert_{t=0}\in\Omega^1_{\infty,\d}\,,
    \end{align*}
    by using the assumptions, the properties of the solution to the Poisson equation and the equations $\square^{A}\eta=\d^{A}\delta^{A}\alpha$. In particular, this implies $\eta=\K\varepsilon\in\Gamma_{\infty,\d}(\E_{1})$ by Theorem~\ref{Thm:CauchyProblemHInfty}(i). Uniqueness modulo $\mathrm{ker}(\K)$ then follows from similar steps as in Proposition~\ref{prop:Poisson}. For more details we refer to the analogous statement for Maxwells theory in \cite[Cor.~5.3]{MurroSchmid} and \cite[Prop.~3.55]{SchmidPhD}.
\end{proof}

Let us now come back to the phase space of linearised Yang-Mills theory. Following the discussion in Section~\ref{Subsec:Linearised Yang-Mills Theory in the Hack-Schenkel Formalism} and Section~\ref{Subsec:Phase Spaces on Initial Data and Cauchy Problem}, we recall the two phases spaces
\begin{align*}
   \bigg(\mathcal{V}_{\mathrm{P}}=\cfrac{\mathrm{ker}(\sf{K}^{\ast}\vert_{\Gamma_{c}})}{\mathrm{ran}(\sf{P}\vert_{\Gamma_{c}})},\, \sigma([\cdot],[\cdot])=(\cdot,\mathsf{i}\sf{G}_{1}\cdot)_{\sf{E}_{1}}\bigg)\xrightarrow[\cong]{[\rho_{1}\G_{1}]} \bigg(\mathcal{V}_{\Sigma}=\cfrac{\mathrm{ker}(\K_{\Sigma}^{\dagger}\vert_{\Gamma_{\mathrm{c}}})}{\mathrm{ran}(\K_{\Sigma}\vert_{\Gamma_{\mathrm{c}}})},\,\sigma_{\Sigma,1}([\cdot],[\cdot])=(\cdot,\mathsf{i}\sf{G}_{\Sigma,1}\cdot)_{\sf{E}_{\rho_1}}\bigg)\, .
\end{align*}

In the following, we shall adapt these phase spaces to the setting of smooth fields that are spatially in the $\sf{H}^{\infty}$-Sobolev space. Moreover, we discuss the complete gauge fixing in this context. To start with, we encode the Cauchy temporal gauge both globally and on the
level of initial data by introducing the operators
\begin{align*}
    \mathsf{R}_\Sigma\:\Gamma(\E_{\rho_1})\to\Gamma(\E_{\rho_1}),
    &\qquad
    \mathsf{R}_\Sigma\defeq
    \begin{pmatrix}
        \mathds1&0&0&0\\
        0&\mathds1&0&0\\
        0&0&0&0\\
        0&0&0&0
    \end{pmatrix},
    \nonumber\\
    \mathsf{R}\:\Gamma(\E_1)\to\Gamma(\E_1),
    &\qquad \mathsf{R}\defeq\mathcal U_1\mathsf{R}_\Sigma\rho_1.
\end{align*}
In other words, $\alpha\in\ker(\mathsf{R})$ if and only if it satisfies the Cauchy temporal gauge, \emph{i.e.}, 
\begin{align*}
    \alpha_{\mathrm T}\vert_{t=0}=0,
    \qquad
    \mathcal D_t^{\A_{\mathrm T}}\alpha_{\mathrm T}\vert_{t=0}=0\, .
\end{align*}

With this notation, we define the phase space of initial data in the Cauchy radiation gauge, denoted by $\mathcal{V}_{\sf{R}}^{\infty}$, and the phase space of $\sf{H}^{\infty}$-initial data, which we denote by $\mathcal{V}_{\Sigma}^{\infty}$ in order avoid confusion between the smooth Sobolev phase space $\mathcal{V}_{\Sigma}$ recalled above. More precisely, we set
\begin{align}
    \mathcal V^\infty_\Sigma
    &\defeq
    \frac{\ker(\K_\Sigma^\dagger\vert_{\Gamma_\infty})}
    {\ran(\K_\Sigma)\cap\Gamma_{\infty,\d}(\E_{\rho_1})},
    \label{Eq:VSigmaAbstractYM}\\[1mm]
    \mathcal V^\infty_{\mathrm R}
    &\defeq
    \ker(\K_\Sigma^\dagger\vert_{\Gamma_\infty})
    \cap\ker(\sf{R}_\Sigma\vert_{\Gamma_\infty}).\label{Eq:VSigmaAbstractYM2}
\end{align}
Here and below, restrictions to $\Gamma_\infty$ are understood with the appropriate bundle as domain. The identity clearly induces a linear map $[\mathrm{id}]\:\mathcal{V}_{\Sigma}\to\mathcal{V}_{\Sigma}^{\infty}$, however, this map is in general neither injective nor surjective. In other words, there is no simple relation between the phase spaces  $\mathcal{V}_{\Sigma}^{\infty}$ and $\mathcal{V}_{\Sigma}$. The space $\mathcal{V}_{\Sigma}^{\infty}$ can be seen as the class of observables, on the level of initial data, allowing for the Cauchy radiation gauge to be imposed. In the end, this space will only play an auxiliary role in our construction and Hadamard states will be defined on $\mathcal{V}_{\Sigma}$, as in the quantisation procedure outlined in Section~\ref{Subsec:Quantisation and Hadamard States}. The space $\mathcal{V}_{\sf{R}}^{\infty}$, on the other hand, contains initial data $\mathfrak{f}$ for $1$-forms whose solution $\alpha$ of $\square^{A}\alpha=0$ with $\rho_{1}(\alpha)=\mathfrak{f}$ satisfies the Cauchy radiation gauge.

Since the Cauchy radiation gauge can be achieved on the level of $\Gamma_{\infty}$-fields by Corollary~\ref{Cor:CompleteGaugeFixingYM}, we obtain the following commutative diagram of phases spaces.

\begin{proposition}[cf.~\protect{\cite[Prop.~5.4]{MurroSchmid}}]\label{Prop:CommDiaGP}
There exists a map $\mathsf{T}_\Sigma:\Gamma_\infty(\E_{\rho_1})\to\Gamma_\infty(\E_{\rho_1})$ such that the following diagram is commutative and every map is an isomorphism:
\begin{equation*}
\begin{tikzcd}[column sep=large,row sep=large]
\displaystyle
\mathcal V^\infty_\Sigma
\arrow[r,"{[\mathcal U_1]}"',"\cong"]
\arrow[d,"{[\mathsf{T}_\Sigma]}","\cong"']
&
\displaystyle
\frac{\ker(\mathsf P\vert_{\Gamma_\infty})}
{\ran(\K)\cap\Gamma_{\infty,\d}(\E_1)}
\\
\displaystyle
\mathcal V^\infty_{\mathrm R}
\arrow[r,"\mathcal U_1","\cong"']
&
\displaystyle
\ker(\mathsf D_1\vert_{\Gamma_\infty})
\cap\ker(\K^*\vert_{\Gamma_\infty})
\cap\ker(\sf{R}\vert_{\Gamma_\infty}) 
\arrow[u,hook,"\cong", "{[\mathsf{id}]}"'] .
\end{tikzcd}
\end{equation*}
The maps denoted by hooked arrows are induced by the inclusions, and
$[\mathsf{T}_\Sigma]$ assigns to a Cauchy-data gauge class its unique representative in
the Cauchy radiation gauge.
\end{proposition}

\begin{proof}
    The isomorphisms $[\mathcal{U}_{1}]$ and $[\rho_{1}]$ follow directly from the Cauchy problem in the $\sf{H}^{\infty}$-setting, Theorem~\ref{Thm:CauchyProblemHInfty}, and the achievability of the Cauchy radiation gauge established in Corollary~\ref{Cor:CompleteGaugeFixingYM}. The map $\sf{T}_{\Sigma}$ is the \emph{gauge projector}, that maps a given equivalence class $[\mathfrak{f}]\in\mathcal{V}_{\Sigma}^{\infty}$ into its unique representative $\mathfrak{f}_{\mathrm{CR}}$ satisfying the Cauchy radiation gauge on the level of initial data, i.e., $\sf{K}_{\Sigma}^{\dagger}\mathfrak{f}_{\mathrm{CR}}=\sf{R}_{\Sigma}\mathfrak{f}_{\mathrm{CR}}=0$. The inverse of the map $[\mathrm{id}]$ is the corresponding gauge projector on the level of spacetime. By construction, it is clear that the diagram commutes. For more details in the context of Maxwell's theory, see \cite[Prop.~5.4]{MurroSchmid} and \cite[Cor.~3.56]{SchmidPhD}. 
\end{proof}

As a next step, let us derive an explicit expression of the gauge projector $\sf{T}_{\Sigma}$ defined abstractly in Proposition~\ref{Prop:CommDiaGP}. To start with, we note that the explicit form of $\K_\Sigma^\dagger$ in Prop.~\ref{Prop:ExplicitKSigmaYM} gives the following
characterisation of the Cauchy radiation-gauge initial data:
\begin{align}
    \mathcal V^\infty_{\mathrm R}
    =\left\{
    \begin{pmatrix}0\\0\\a_\Sigma\\\pi_\Sigma\end{pmatrix}
    \in\Gamma_\infty(\E_{\rho_1})\;:\;
    \delta^{\A_\Sigma^0}a_\Sigma=0,
    \ \delta^{\A_\Sigma^0}\pi_\Sigma=0
    \right\}.
    \label{Eq:ExplicitVRYM}
\end{align}
Moreover, by the smooth Hodge decomposition of Section~\ref{Subsec:Hodge Decomposition in Flat Vector Bundles on Complete Riemannian Manifolds}, we have that 
\begin{align}
    \Omega^1_\infty(\Sigma_0,\Ad(\P_\Sigma))
    =\Omega^1_{\infty,\d}(\Sigma_0,\Ad(\P_\Sigma))
    \oplus
    \ker\bigl(\delta^{\A_\Sigma^0}\vert_{\Omega^1_\infty}\bigr).
    \label{Eq:HodgeSplitSection4YM}
\end{align}
Let us denote by 
\begin{align*}
    \Pi:\Omega^1_\infty(\Sigma_0,\Ad(\P_\Sigma))
    \longrightarrow
    \ker\bigl(\delta^{\A_\Sigma^0}\vert_{\Omega^1_\infty}\bigr)
\end{align*}
the corresponding orthogonal projector. Being a projector, it has the following obvious properties:

\begin{lemma}
\label{Lem:PiYM}
Let $(\Sigma_0,\h)$ be a complete Riemannian manifold and let $\Pi$ be the orthogonal projector associated with~\eqref{Eq:HodgeSplitSection4YM}. Then, the following holds true:
\begin{itemize}
    \item[$\mathrm(i)$] $\Pi^2=\Pi$, $\ran(\Pi)=\ker(\delta^{\A_\Sigma^0}\vert_{\Omega^1_\infty})$ and $\ker(\Pi)=\Omega^1_{\infty,\d}(\Sigma_0,\Ad(\P_\Sigma))$;
    \item[$\mathrm(ii)$] $\Pi$ is orthogonal with respect to $\langle\cdot,\cdot\rangle_{\sf{H}^{\mathfrak{s}}}$ and any $\mathfrak{s}\in [0,\infty)$;
    \item[$\mathrm(iii)$] $\Pi\Delta^{\A_\Sigma^0}=\Delta^{\A_\Sigma^0}\Pi$ on $\Omega^1_\infty(\Sigma_0,\Ad(\P_\Sigma))$;
    \item[$\mathrm(iv)$]The operator $\d^{\A_\Sigma^0}(\Delta^{\A_\Sigma^0})^{-1}\delta^{\A_\Sigma^0}$ is well-defined on $\Omega^{1}_{\infty}(\Sigma_0,\mathrm{Ad}(\P_{\Sigma}))$ and it holds that
    \begin{equation}
        \Pi=\mathds 1-\d^{\A_\Sigma^0}(\Delta^{\A_\Sigma^0})^{-1}\delta^{\A_\Sigma^0}\,,
        \label{Eq:PiYM}
    \end{equation}
\end{itemize}
\end{lemma}

\begin{proof}
Items (i) and (ii) follow directly from its definition in Eq.~\eqref{Eq:HodgeSplitSection4YM}. For (iii), flatness gives
$\Delta^{\A_\Sigma^0}\d^{\A_\Sigma^0}=\d^{\A_\Sigma^0}\Delta^{\A_\Sigma^0}$ and
$\delta^{\A_\Sigma^0}\Delta^{\A_\Sigma^0}=\Delta^{\A_\Sigma^0}\delta^{\A_\Sigma^0}$, so $\Delta^{\A_\Sigma^0}$ preserves both summands in Eq.~\eqref{Eq:HodgeSplitSection4YM}. Finally, writing $\omega=\d^{\A_\Sigma^0}f+\beta$ with $\delta^{\A_\Sigma^0}\beta=0$, one has $\Delta^{\A_\Sigma^0}f=\delta^{\A_\Sigma^0}\omega$. Moreover, by Proposition~\ref{prop:Poisson}, we note that $\Delta^{A_{\Sigma}^{0}}$ is bijective as an operator $\Delta^{A_{\Sigma}^{0}}\:\{f\in\Omega^{0}\mid \d^{A_{\Sigma}^{0}}f\in\Omega^{1}_{\infty,\d}\}/\mathrm{ker}(\d^{A^{0}_{\Sigma}})\to\ran(\delta^{\A_\Sigma^0}\vert_{\Omega^1_\infty})$, which shows that $\d^{\A_\Sigma^0}(\Delta^{\A_\Sigma^0})^{-1}\delta^{\A_\Sigma^0}\:\Omega^{1}_{\infty}(\Sigma_0,\mathrm{Ad}(\P_{\Sigma}))\to\Omega^{1}_{\infty}(\Sigma_0,\mathrm{Ad}(\P_{\Sigma}))$ is well-defined. Claim (iv) then follows immediately.
\end{proof}

\begin{remark}
    The projection operators and $\mathds{1}-\d\Delta^{-1}\delta$ and $\d\Delta^{-1}\delta$ acting on real-valued form can also understood as \emph{pseudodifferential operators} and as such have been constructed explicitly in \cite[Sec.~5]{CapoferriCurl} on spatially compact spacetimes. Generalisations to flat vector bundles are straightforward.
\end{remark}

We now possess all the ingredients to realise the complete gauge fixing explicitly.

\begin{proposition}
\label{Prop:TSigmaYM}
Let $(\M,\g)$ be a globally hyperbolic and ultrastatic spacetime with complete Cauchy surface $(\Sigma,\h)$. Then, the operator
\begin{align*}
    \mathsf{T}_\Sigma\defeq
    \begin{pmatrix}
        0&0&0&0\\
        0&0&0&0\\
        0&0&\Pi&0\\
        0&0&0&\Pi
    \end{pmatrix}
    :\Gamma_\infty(\E_{\rho_1})\longrightarrow\Gamma_\infty(\E_{\rho_1})
\end{align*}
has the following properties:
\begin{itemize}
    \item[$\mathrm(i)$] $\mathsf{T}_\Sigma^2=\mathsf{T}_\Sigma$ and $\mathsf{T}_\Sigma\vert_{\mathcal V_{\mathrm R}}=\mathds 1$;
    \item[$\mathrm(ii)$] $\mathsf{T}_\Sigma=\mathds 1-\K_\Sigma(\sf{R}_\Sigma\K_\Sigma)^{-1}\sf{R}_\Sigma$ on $\ker(\K_\Sigma^\dagger\vert_{\Gamma_\infty(\E_{\rho_1})})$;
    \item[$\mathrm(iia)$] $\ker(\mathsf{T}_\Sigma\vert_{\ker(\K_\Sigma^\dagger\vert_{\Gamma_\infty})})=\ran(\K_\Sigma)\cap\Gamma_{\infty,\d}(\E_{\rho_1})$;
    \item[$\mathrm(iib)$] $\ran(\mathsf{T}_\Sigma\vert_{\ker(\K_\Sigma^\dagger\vert_{\Gamma_\infty})})=\mathcal V_{\mathrm R}$.
\end{itemize}
Hence, $\mathsf{T}_\Sigma$ induces a bijection $[\mathsf{T}_\Sigma]\:\mathcal V_\Sigma^{\infty}\xrightarrow{\cong}\mathcal V_{\mathrm R}^{\infty}$, \textit{i.e.}, the gauge projector defined implicitly in the commutative diagram of Proposition~\ref{Prop:CommDiaGP}.
\end{proposition}

\begin{proof}
The first statement follows from Lemma~\ref{Lem:PiYM} together with
Eq.~\eqref{Eq:ExplicitVRYM}. By Proposition~\ref{Prop:ExplicitKSigmaYM},
\begin{align*}
    \sf{R}_\Sigma\K_\Sigma
    =
    \begin{pmatrix}
        0& \mathrm{i}\mathds 1\\
        \mathrm{i}\Delta_0^{\A_\Sigma^0}&0\\
        0&0\\
        0&0
    \end{pmatrix}.
\end{align*}
The covariant Poisson result of Section~\ref{Subsec:Hodge Decomposition in Flat Vector Bundles on Complete Riemannian Manifolds}, Proposition~\ref{prop:Poisson}, then implies that this
operator is bijective on the gauge-data space
\begin{align*}
    \left(
    \frac{\{a\in\Omega^0(\Sigma_0,\Ad(\P_\Sigma)):\
    \d^{\A_\Sigma^0}a\in\Omega^1_{\infty,\d}\}}
    {\ker(\d^{\A_\Sigma^0})}
    \right)
    \oplus\Omega^0_\infty(\Sigma_0,\Ad(\P_\Sigma))
\end{align*}
onto the temporal data compatible with the constraint
$\K_\Sigma^\dagger f=0$.  For
$f=(a_{\mathrm T},\pi_{\mathrm T},a_\Sigma,\pi_\Sigma)^\top$ in that
constraint space, Proposition~\ref{Prop:ExplicitKSigmaYM} gives
\begin{align*}
    \mathrm{i}\pi_{\mathrm T}
    &=-\delta^{\A_\Sigma^0}a_\Sigma,
    &
    \mathrm{i}\Delta_0^{\A_\Sigma^0}a_{\mathrm T}
    &=-\delta^{\A_\Sigma^0}\pi_\Sigma.
\end{align*}
Using~\eqref{Eq:PiYM}, a direct computation then yields
\begin{align*}
    \left(\mathds 1-\K_\Sigma(\sf{R}_\Sigma\K_\Sigma)^{-1}\sf{R}_\Sigma\right)f
    =
    \begin{pmatrix}
        0\\0\\\Pi a_\Sigma\\\Pi\pi_\Sigma
    \end{pmatrix}
    =\mathsf{T}_\Sigma f.
\end{align*}
The statements on kernel and range follow from
$\ker(\Pi)=\Omega^1_{\infty,\d}$ and
$\ran(\Pi)=\ker\delta^{\A_\Sigma^0}$.
\end{proof}

By the previous proposition, we hence have an explicit description of the gauge projector $[\mathsf{T}_\Sigma]\:\mathcal V^{\infty}_\Sigma\xrightarrow{\cong}\mathcal V^{\infty}_{\mathrm R}$ introduced in Proposition~\ref{Prop:ExplicitKSigmaYM}. Let us now equip these phase spaces with suitable Hermitian forms. Following the same line of reasoning as in \cite[Prop.~2.21]{SchmidPhD}, we first notice that the Hermitian form $\sigma_{\Sigma,1}$, defined in Eq.~\eqref{Eq:sigmaKSigmaYM}, is not only well-defined on $\mathcal{V}_{\Sigma}$, but also induces a well-defined sesquilinear form on the space $\mathcal{V}_{\Sigma}^\infty$, which is defined using smooth Sobolev data.

\begin{lemma}[Smooth Sobolev Phase Space]\label{Lem:WellDefPS}
    The map $\sigma_{\Sigma,1}$ as per Eq.~\eqref{Eq:sigmaKSigmaYM} is a well-defined sesquilinear form on the space $\mathcal{V}^\infty_\Sigma$ introduced in Eq.~\eqref{Eq:VSigmaAbstractYM}.
\end{lemma}

\begin{proof}[Proof (sketch).]
    Let $\mathfrak{f}\in\mathrm{ker}(\sf{K}_{\Sigma}^{\dagger}\vert_{\Gamma_{\infty}})$ be arbitrary and $\mathfrak{g}\in\Gamma(\E_{\rho_{0}})$ be such that $\K_{\Sigma}\mathfrak{g}\in\Gamma_{\infty,\d}(\E_{\rho_{1}})$. Naively, we would like to compute
    \begin{align*}
        \sigma_{\Sigma,1}(\mathfrak{f},\K_{\Sigma}\mathfrak{g})=\sigma_{\Sigma,0}(\K_{\Sigma}^{\dagger}\mathfrak{f},\mathfrak{g})=0\,,
    \end{align*}
    however, this is a priori unjustified since $\mathfrak{g}$ is just smooth and does not necessarily have any decay, which implies that $\sigma_{\Sigma,0}(\K_{\Sigma}^{\dagger}\mathfrak{f},\mathfrak{g})$ is undefined, despite the fact that $\K_{\Sigma}^{\dagger}\mathfrak{f}=0$. However, by assumption, we have that $\K_{\Sigma}\mathfrak{g}\in\Gamma_{\infty,\d}(\E_{\rho_{1}})$. Hence, we know that the $1$-form components $\d^{A_{\Sigma}^{0}}\mathfrak{a}$ and $\d^{A_{\Sigma}^{0}}\pi$ of $\K_{\Sigma}\mathfrak{g}=(i\pi,i\Delta^{A_{\Sigma}^{0}}\mathfrak{a},\d^{A_{\Sigma}^{0}}\mathfrak{a},\d^{A_{\Sigma}^{0}}\pi)$, where we wrote $\mathfrak{g}=(\mathfrak{a},\pi)$, can be approximated by sequences $\d^{A_{\Sigma}^{0}}\mathfrak{a}_{n}$ and $\d^{A_{\Sigma}^{0}}\pi_{n}$ with $\mathfrak{a}_{n},\pi_{n}\in\Omega_{\infty}^{0}$. In particular, one can find a sequence $\mathfrak{g}_{n}\in\Gamma_{\infty}(\E_{\rho_{0}})$ for which $\K_{\Sigma}\mathfrak{g}_{n}\to\K_{\Sigma}\mathfrak{g}$ in a suitably sense, which allows the above computation to be carried out in the limit. For more details, we refer to the discussion for Maxwell's theory in \cite[Lem.~3.59]{SchmidPhD}.
\end{proof}

We recall that the Hermitian form $\sigma_{\Sigma,1}$ is given by $\sigma_{\Sigma,1}([\cdot],[\cdot])=i(\cdot,\G_{\Sigma,1}\cdot)_{\E_{\rho_{1}}}=\langle\cdot,\G_{\Sigma,k}\cdot\rangle_{\sf{L}^{2}}$, where the explicit form of $\G_{\Sigma,1}$ has been derived in Proposition~\ref{Prop:GSigma}. We conclude this section by giving an explicit expression of the Hermitian form on $\mathcal V^{\infty}_{\mathrm R}$ induced by the isomorphism $[\T_{\Sigma}]$. To this end, we set
\begin{align}
    \sigma_{\Sigma,\mathrm R}(f,g)
    \defeq \mathrm{i}(f,\G_{\Sigma,\mathrm R}g)_{\E_{\rho_1}},
    \qquad
    \G_{\Sigma,\mathrm R}
    \defeq\frac1{\mathrm i}
    \begin{pmatrix}
        0&0&0&0\\
        0&0&0&0\\
        0&0&0&\mathds 1\\
        0&0&\mathds 1&0
    \end{pmatrix}.
    \label{Eq:SigmaRYM}
\end{align}
Since $\Pi$ is orthogonal, $\mathsf{T}_\Sigma$ is formally self-adjoint with respect to $\sigma_{\Sigma,1}$ and it holds that
\begin{align}
    \sigma_{\Sigma,1}(\mathfrak{f},\mathsf{T}_\Sigma \mathfrak{g})
    =\sigma_{\Sigma,\mathrm R}(\mathsf{T}_\Sigma \mathfrak{f},\mathsf{T}_\Sigma \mathfrak{g})
    \label{Eq:TSigmaIsometryYM}
\end{align}
for constrained smooth Sobolev data $\mathfrak{f},\mathfrak{g}$. In particular, we obtain the following:

\begin{lemma}\label{Lemma:VR}
    The map $\mathsf{T}_\Sigma:(\mathcal V^{\infty}_\Sigma,\sigma_{\Sigma,1})\xrightarrow{\cong}
(\mathcal V^{\infty}_{\mathrm R},\sigma_{\Sigma,\mathrm R})$ is unitary.
\end{lemma}
\section{Existence of Hadamard States for Linearised Yang-Mills Theory}
\label{Sec:Existence of Hadamard States for Linearised Yang-Mills Theory}

Goal of this section is to construct Hadamard states for Yang--Mills theory linearised around flat connections on globally hyperbolic spacetimes. First, the construction is carried out on ultrastatic spacetimes whose Cauchy surface is of bounded geometry, \emph{cf.}, Remark~\ref{Rmk: bounded geometry}, where we follow the strategy developed for Maxwell theory in~\cite{MurroSchmid}. The main additional feature in the present setting is that, although the spacetime metric is ultrastatic, the spatial part $\A_\Sigma(t)$ of the flat background connection need not be independent of time. This problem will be overcome by means of covariant parallel transport, which identifies the time-dependent spatial problem with a family of operators acting on a fixed reference
Cauchy hypersurface. In the second part of this section, we remove the restriction to the ultrastatic setting by means of a paracausal deformation argument, thereby completing the proof of the Main Theorem stated in the introduction.

\subsection{Construction of Hadamard States on Ultrastatic Spacetimes} \label{Subsec:Construction of Hadamard States on Ultrastatic Spacetimes}
Throughout the following discussion, let $(\M=\bb{R}\times\Sigma,\g=-\d t\otimes\d t+\h)$ be a globally hyperbolic and ultrastatic manifold, where we assume that the complete Riemannian manifold $(\Sigma,\h)$ is additionally of bounded geometry, see~Remark~\ref{Rmk: bounded geometry}. Moreover, we fix a flat background connection $A\in\mathcal{C}(\P)$, decomposed as $A=A_{\mathrm{T}}\d t+A_{\Sigma}$, as usual.

The main difference from the microlocal factorisation for Maxwell theory (see \cite[Sec.~5]{MurroSchmid}) is that the background connection $\A_\Sigma(t)$, which enters linearised Yang–Mills theory as a coefficient of the relevant PDE, may depend on time. Flatness allows us to remove this time dependence before applying the pseudodifferential construction. For $k=0,1$, let
\begin{align*}
    W_k(t,s):\Gamma(\E_{\Sigma,k})\longrightarrow
    \Gamma(\E_{\Sigma,k})
\end{align*}
be the parallel transport operator induced by the gauge covariant time derivative $\mathcal D_t^{\A_{\mathrm T}}$ defined in Definition~\ref{Def: covariant time derivative}, \textit{i.e.}, 
\begin{align*}
    \mathcal D_t^{\A_{\mathrm T}}W_k(t,s)=0,
    \qquad W_k(s,s)=\mathds 1.
\end{align*}

\begin{lemma}
\label{Lem:ParallelTransportMicrolocalYM}
For every $k$ and every $t,s\in\mathbb R$,
\begin{align}
    \d^{\A_\Sigma(t)}W_k(t,s)
    &=W_{k+1}(t,s)\d^{\A_\Sigma(s)},
    \nonumber\\
    \delta^{\A_\Sigma(t)}W_{k+1}(t,s)
    &=W_k(t,s)\delta^{\A_\Sigma(s)},
    \label{Eq:IntertwiningWYM}
\end{align}
and, consequently,
\begin{align}
    \Delta_k^{\A_\Sigma(t)}W_k(t,s)
    =W_k(t,s)\Delta_k^{\A_\Sigma(s)}.
    \label{Eq:IntertwiningDeltaWYM}
\end{align}
Moreover, $W_k(t,s)$ extends to a unitary map between the Sobolev spaces associated with the de Rham-Hodge d'Alembertians $\Delta_k^{\A_\Sigma(s)}$ and
$\Delta_k^{\A_\Sigma(t)}$.
\end{lemma}

\begin{proof}
The first identity in Eq.~\eqref{Eq:IntertwiningWYM} is the integrated form of
the commutation relation
$[\mathcal D_t^{\A_{\mathrm T}},\d^{\A_\Sigma}]=0$ from
Lemma~\ref{Lem: 2.26}, which holds true because the mixed curvature component of
the background connection vanishes. Since $W_k(t,s)$ is unitary for the
bundle metric and $\h$ is time-independent, taking adjoints gives the
identity for $\delta^{\A_\Sigma}$.  Equation~\eqref{Eq:IntertwiningDeltaWYM}
follows.  Finally, spectral calculus yields
\begin{align*}
    (\mathds 1+\Delta_k^{\A_\Sigma(t)})^{\mathfrak s/2}W_k(t,s)
    =W_k(t,s)(\mathds 1+\Delta_k^{\A_\Sigma(s)})^{\mathfrak s/2},
\end{align*}
which proves the last statement.
\end{proof}

We observe that, by Proposition~\ref{Prop:DecompAdFormsOp}, the temporal and spatial components of a solution of $\square^\A\omega=0$ satisfy equations of the form
\begin{align}
    \left((\mathcal D_t^{\A_{\mathrm T}})^2
    +\Delta_k^{\A_\Sigma(t)}\right)u=0,
    \qquad k=0,1\,,
    \label{Eq:ComponentWaveYM}
\end{align}
where $A^{0}_{\Sigma}:=A_{\Sigma}(0)$. If $v(t)\defeq W_k(0,t)u(t),$ then Lemma~\ref{Lem:ParallelTransportMicrolocalYM} gives the time-independent equation
\begin{align}
    (\partial_t^2+\Delta_k^{\A_\Sigma^0})v=0.
    \label{Eq:TransportedWaveYM}
\end{align}
Thus, all pseudodifferential operators needed in the factorisation can be constructed on the single reference hypersurface $\Sigma_0$.

As next step, we construct approximate square roots of the covariant Hodge Laplacians, based on the discussion in \cite[Lem.~5.1]{GerardMurroWrochna} and in particular \cite[Lem.~5.18]{MurroSchmid}.

\begin{lemma}
\label{Lem:EpsilonYM}
For $k=0,1$ there exist formally self-adjoint, strictly positive and
invertible operators
\begin{align*}
    \epsilon_k\in\Psi_{\mathrm b}^1(\Sigma,\E_{\Sigma,k})
\end{align*}
and
$r_{k,-\infty}\in W^{-\infty}(\Sigma,\E_{\Sigma,k})$ such that
\begin{itemize}
    \item[(i)]
    $
        \epsilon_k^2
        =\Delta_k^{\A_\Sigma^0}+r_{k,-\infty};
    $
    \item[(ii)]
    $
        \sigma_{\mathrm{pr}}(\epsilon_k)(x,\xi)
        =|\xi|_{\h}\,\mathds 1_{\E_{\Sigma,k}};
    $
    \item[(iii)]
    $
        [\epsilon_1,\Pi]
        \in W^{-\infty}(\Sigma;\E_{\Sigma,1}).
    $
\end{itemize}
In particular,
$\epsilon_k^{-1}\in\Psi_{\mathrm b}^{-1}(\Sigma;\E_{\Sigma,k})$.
\end{lemma}

\begin{proof}
The construction is the covariant analogue of
\cite[Lem.~5.18]{MurroSchmid}.  Since
$\Delta_k^{\A_\Sigma^0}$ is a non-negative Laplace-type operator on a
bundle of bounded geometry, the bounded pseudodifferential calculus gives a
formally self-adjoint elliptic operator $\epsilon_k$ with principal symbol
$|\xi|_{\h}\mathds 1$ and
\begin{align*}
    \epsilon_k^2-\Delta_k^{\A_\Sigma^0}
    \in W^{-\infty}(\Sigma, \E_{\Sigma,k}).
\end{align*}
It can be chosen strictly positive and invertible without imposing a mass
gap on $\Delta_k^{\A_\Sigma^0}$.

To conclude the proof, it remains to show that $\varepsilon_1$ commutes up to smoothing with $\Pi$. \\
For $k=1$, Lemma \ref{Lem:PiYM} gives $[\Delta_1^{A_\Sigma^0},\Pi]=0$ and, by duality, $\Pi$ is bounded on $\H^{s}_{1}(\Sigma, \E_{\Sigma})$ for every $s\in\mathbb R$. Hence $R:=[\varepsilon_1^2,\Pi]=[r_{1,-\infty},\Pi]\in W^{-\infty}(\Sigma;\mathsf E_{\Sigma,1})$. \\
Now set $S_\lambda:=(\varepsilon_1^2+\lambda)^{-1}$ for $\lambda>0$. The integral representation of the positive square root gives
$$
\varepsilon_1u=\frac1\pi\int_0^\infty \lambda^{-1/2}\varepsilon_1^2S_\lambda u\,d\lambda
$$
for any $u\in\H^{2}_{1}(\Sigma, \E_{\Sigma})$.
 Combining the commutator identity for the resolvent
$$[S_\lambda,\Pi]=-S_\lambda[\varepsilon^2_1+\lambda^2,\Pi]S_\lambda = -S_\lambda R S_\lambda $$
 with the identity $\varepsilon_1^2S_\lambda=1-\lambda S_\lambda$, we obtain $[\varepsilon_1^2S_\lambda,\Pi]=\lambda S_\lambda R S_\lambda$. Therefore, it follows
\begin{equation*}
\begin{aligned}
[\varepsilon_1,\Pi]u
&=\varepsilon_1(\Pi u)-\Pi(\varepsilon_1u)\\
&=\frac1\pi\int_0^\infty \lambda^{-1/2}
\big(\varepsilon_1^2 S_\lambda\Pi u-\Pi\varepsilon_1^2 S_\lambda u\big)\,\d\lambda\\
&=\frac1\pi\int_0^\infty \lambda^{-1/2}
[\varepsilon_1^2S_\lambda,\Pi]u\,\d\lambda\\
&= \frac1\pi\int_0^\infty\lambda^{1/2}S_\lambda R S_\lambda u\,\d\lambda\,.
\end{aligned}
\end{equation*}
For each $\lambda>0$, the operator $S_\lambda R S_\lambda$ is smoothing, since $R$ is smoothing and $S_\lambda$ has order $-2$. It remains to check convergence of the integral.
Choose $c>0$ such that $\varepsilon_1\geq c$.
The spectral theorem gives
$$\|S_\lambda\|_{\mathsf{H}_1^0\to\mathsf{H}_1^0}\leq(c^2+\lambda)^{-1}\,.$$
Since $S_\lambda$ commutes with $\varepsilon_1^s$, we have
$$
\|\varepsilon_1^s S_\lambda u\|_{\mathsf{H}_1^0}
= \|S_\lambda\varepsilon_1^s u\|_{\mathsf{H}_1^0}
\leq (c^2+\lambda)^{-1}
     \|\varepsilon_1^s u\|_{\mathsf{H}_1^0}.
$$
By ellipticity, $\|\varepsilon_1^s u\|_{\mathsf{H}_1^0}$ is an
equivalent norm on $\mathsf{H}_1^s$.
Applying the estimate to the right resolvent at order $-m$ and to
the left resolvent at order $m$, we see that the
$\mathsf{H}_1^{-m}\to\mathsf{H}_1^m$ norm of the integrand is bounded
by a constant times $\lambda^{1/2}(c^2+\lambda)^{-2}$.
This bound is integrable for every $m\geq0$, proving
$[\varepsilon_1,\Pi]\in W^{-\infty}(\Sigma;\mathsf{E}_{\Sigma,1})$.
\end{proof}

Let $\mathcal U_{\Sigma,k}(t,s)$ denote the Cauchy evolution of Eq.~\eqref{Eq:ComponentWaveYM} on $\Gamma(\E_{\Sigma,k})^{\oplus2}$, with Cauchy vector
\begin{align*}
    \Psi_k(t)
    \defeq
    \begin{pmatrix}
        u(t)\\[1mm]
        \mathrm{i}^{-1}\mathcal D_t^{\A_{\mathrm T}}u(t)
    \end{pmatrix}.
\end{align*}
Set
\begin{align*}
    \mathbf W_k(t,s)
    \defeq
    \begin{pmatrix}
        W_k(t,s)&0\\
        0&W_k(t,s)
    \end{pmatrix}
\end{align*}
and
\begin{align*}
    S_{\epsilon_k}
    \defeq
    \mathrm{i}^{-1}
    \begin{pmatrix}
        \mathds 1&-\mathds 1\\
        \epsilon_k&\epsilon_k
    \end{pmatrix}
    (2\epsilon_k)^{-1/2}.
\end{align*}
Exactly as in the Maxwell case, Eq.~\eqref{Eq:TransportedWaveYM} can be diagonalised modulo smoothing terms.

\begin{proposition}
\label{Prop:MicrolocalFactorizationYM}
For $k=0,1$ one has
\begin{align}
    \mathcal U_{\Sigma,k}(t,0)
    =\mathbf W_k(t,0) \left(S_{\epsilon_k}
    \begin{pmatrix}
        e^{\mathrm{i}t\epsilon_k}&0\\
        0&e^{-\mathrm{i}t\epsilon_k}
    \end{pmatrix}
    S_{\epsilon_k}^{-1}
    +R_{k,-\infty}(t)\right),
    \label{Eq:MicrolocalFactorizationYM}
\end{align}
where $R_{k,-\infty}(t)$ is a smooth family of operators in $W^{-\infty}(\Sigma,\E_{\Sigma,k}^{\oplus2})$.
\end{proposition}

\begin{proof}
Conjugating by $\mathbf W_k(0,t)$ reduces the Cauchy problem exactly to
Eq.~\eqref{Eq:TransportedWaveYM}. Lemma~\ref{Lem:EpsilonYM} then gives
the same first-order diagonalisation as in
\cite[Prop.~5.19]{MurroSchmid}. The error is smoothing and remains
smoothing under the corresponding evolution by Duhamel's formula. Conjugating back yields precisely Eq.~\eqref{Eq:MicrolocalFactorizationYM}.
\end{proof}

We define the positive and negative frequency projectors on the Cauchy data
of Eq.~\eqref{Eq:TransportedWaveYM} by
\begin{align*}
    \pi_k^\pm
    \defeq
    \frac12
    \begin{pmatrix}
        \mathds 1&\pm\epsilon_k^{-1}\\
        \pm\epsilon_k&\mathds 1
    \end{pmatrix},
    \qquad k=0,1,
\end{align*}
and, on one-form Cauchy data,
\begin{align*}
    \pi^\pm\defeq\pi_0^\pm\oplus\pi_1^\pm.
\end{align*}
They satisfy
\begin{align*}
    (\pi_k^\pm)^2=\pi_k^\pm,
    \qquad
    \pi_k^+\pi_k^-=0,
    \qquad
    \pi_k^++\pi_k^-=\mathds 1.
\end{align*}
Equivalently,
$\pi_k^\pm=S_{\epsilon_k}\Pi_k^\pm S_{\epsilon_k}^{-1}$ with
$\Pi_k^+=\mathrm{diag}(\mathds1,0)$ and
$\Pi_k^-=\mathrm{diag}(0,\mathds1)$. Let
\begin{align*}
    F\defeq\{(t,x;\tau,\xi)\in \T^*\M\setminus\{0\}:\xi=0\}
\end{align*}
and denote by $\mathcal N^\pm$ the two connected components of the
characteristic set of $\square^\A$.  The factorisation above yields the
following microlocal splitting.

\begin{proposition}
For $k=0,1$,
\begin{align}
    \WF'\!\left(\mathcal U_{\Sigma,k}\pi_k^\pm\right)
    \subset
    (\mathcal N^\pm\cup F)\times \T^*\Sigma.
    \label{Eq:WavefrontFrequencyYM}
\end{align}
Consequently,
\begin{align}
    \WF'\!\left(\mathcal U_1\pi^\pm\right)
    \subset
    (\mathcal N^\pm\cup F)\times \T^*\Sigma.
    \label{Eq:WavefrontFullFrequencyYM}
\end{align}
\end{proposition}

\begin{proof}
After conjugation by the smooth bundle isomorphism $W_k(0,t)$, the problem
is reduced to the static one.  Let
\begin{align*}
    Q_k^\pm\defeq\partial_t\mp \mathrm{i}\epsilon_k.
\end{align*}
Modulo a smooth kernel, Proposition~\ref{Prop:MicrolocalFactorizationYM}
shows that $Q_k^\pm\mathcal U_{\Sigma,k}\pi_k^\pm$ is smooth.  Away from
$F$, $Q_k^\pm$ is microlocally a classical pseudodifferential operator on
$\M$ with principal symbol
\begin{align*}
    \mathrm{i}(\tau\mp|\xi|_{\h})\mathds 1.
\end{align*}
It is elliptic away from $\mathcal N^\pm\cup F$.  Microlocal elliptic
regularity therefore gives Eq.~\eqref{Eq:WavefrontFrequencyYM}.  Multiplication
by the parallel transport $W_k(t,0)$ does not enlarge the wavefront set.
The full one-form evolution is the direct sum of the degree-zero temporal
and degree-one spatial evolutions, which gives
Eq.~\eqref{Eq:WavefrontFullFrequencyYM}.
\end{proof}

We are finally in the position to define the Cauchy pseudo-covariances. Set
\begin{align}
    c^\pm\defeq \sf{T}_\Sigma\pi^\pm \sf{T}_\Sigma.
    \label{Eq:HadamardProjectorsYM}
\end{align}

\begin{theorem}
\label{Thm:HadamardProjectorsYM}
Let $(\M=\bb{R}\times\Sigma,\g=-\d t\otimes\d t+\h)$ be a globally hyperbolic and ultrastatic manifold, such that $(\Sigma,\h)$ is of bounded geometry, and let $\A$ be a flat background connection. The operators $c^\pm$ defined in Eq.~\eqref{Eq:HadamardProjectorsYM} satisfy:
\begin{itemize}
    \item[(i)] $(c^\pm)^\dagger=c^\pm$ with respect to
    $\sigma_{\Sigma,1}$ and
    $
        c^\pm\bigl(\ran(\K_\Sigma)\cap
        \Gamma_{\infty,\d}(\E_{\rho_1})\bigr)
        \subset
        \ran(\K_\Sigma)\cap\Gamma_{\infty,\d}(\E_{\rho_1});
   $
    \item[(ii)] for every
    $f\in\ker(\K_\Sigma^\dagger\vert_{\Gamma_\infty(\E_{\rho_1})})$, it holds $
        (c^++c^-)f
        =f
        \quad\mathrm{mod}\ 
        \ran(\K_\Sigma)\cap\Gamma_{\infty,\d}(\E_{\rho_1});
    $
    \item[(iii)] for every such $f$, it holds
   $
        \pm\sigma_{\Sigma,1}(f,c^\pm f)\geq0;
  $
    \item[(iv)]
    $
        \WF'\!\left(\mathcal U_1c^\pm\right)
        \subset
        (\mathcal N^\pm\cup F)\times T^*\Sigma.
    $
\end{itemize}
After restriction to compact Cauchy data, the corresponding spacetime
pseudo-covariances
\begin{align*}
    \lambda^\pm\defeq\pm \mathrm{i}\,\mathcal U_1c^\pm(\rho_1\G_1)
\end{align*}
define a quasifree Hadamard state on the physical phase space
$\mathcal V_{\mathrm P}$ of Definition~\ref{Def: classical phase space}, by the
standard Cauchy-surface characterisation for linear gauge theories, cf., Section~\ref{Subsec:Quantisation and Hadamard States}.
\end{theorem}

\begin{proof}
Since $\epsilon_k$ is formally self-adjoint,
$\pi^\pm$ is formally self-adjoint with respect to
$\sigma_{\Sigma,1}$.  The same is true for $\sf{T}_{\Sigma}$ by
Lemma~\ref{Lem:PiYM}.  Hence $(c^\pm)^\dagger=c^\pm$.  Moreover,
$\sf{T}_{\Sigma}\K_\Sigma=0$ on the exact smooth Sobolev gauge directions, which
proves (i). For (ii), using $\pi^++\pi^-=\mathds1$ and $\sf{T}_{\Sigma}^2=\sf{T}_{\Sigma}$, we have
\begin{align*}
    (c^++c^-)f=\sf{T}_{\Sigma} f.
\end{align*}
Proposition~\ref{Prop:TSigmaYM} gives
$\sf{T}_{\Sigma} f=f$ modulo
$\ran(\K_\Sigma)\cap\Gamma_{\infty,\d}(\E_{\rho_1})$. For (iii), write
\begin{align*}
    \sf{T}_{\Sigma} f=
    \begin{pmatrix}0\\0\\a_\Sigma\\\pi_\Sigma\end{pmatrix}
    \in\mathcal V_{\mathrm R}.
\end{align*}
Using~\eqref{Eq:TSigmaIsometryYM},
Equation~\eqref{Eq:SigmaRYM} and the explicit expression of $\pi_1^\pm$,
we obtain
\begin{align*}
    \pm\sigma_{\Sigma,1}(f,c^\pm f)
    =\frac12
    \left\|
        \epsilon_1^{1/2}a_\Sigma
        \pm\epsilon_1^{-1/2}\pi_\Sigma
    \right\|_{\E_{\Sigma,1}}^2
    \geq0.
\end{align*}
Finally, Lemma~\ref{Lem:EpsilonYM} gives
$[\epsilon_1,\Pi]\in W^{-\infty}$ and therefore also
$[\epsilon_1^{-1},\Pi]\in W^{-\infty}$.  Consequently,
\begin{align*}
    [\sf{T}_{\Sigma},\pi^\pm]\in W^{-\infty}.
\end{align*}
Using $\sf{T}_{\Sigma}^2=\sf{T}_{\Sigma}$,
\begin{align*}
    \mathcal U_1c^\pm
    =\mathcal U_1\pi^\pm \sf{T}_{\Sigma}
    +\mathcal U_1[\sf{T}_{\Sigma},\pi^\pm]\sf{T}_{\Sigma}.
\end{align*}
An elliptic parametrix for the twisted Hodge Laplacian shows that $\Pi$ is a properly supported pseudodifferential operator of order zero modulo a smoothing remainder.
The pseudodifferential part of $\sf{T}_\Sigma$ therefore preserves the wavefront bound in~\eqref{Eq:WavefrontFullFrequencyYM} under right composition, while the smoothing terms remain smooth under Cauchy evolution, proving (iv).
\end{proof}

\begin{remark}
The construction does not require a mass gap for
$\Delta_k^{\A_\Sigma^0}$.  The invertible operators $\epsilon_k$ are
constructed directly in the bounded pseudodifferential calculus and differ
from a square root of the covariant Hodge Laplacian only by smoothing terms.
Furthermore, no temporal gauge is imposed on the background connection.
The possible time dependence of $\A_\Sigma(t)$ is removed, for the purpose
of the microlocal factorisation, by the parallel transport introduced in
Lemma~\ref{Lem:ParallelTransportMicrolocalYM}.
\end{remark}

\subsection{Construction of Hadamard States on Globally Hyperbolic Spacetimes} 
\label{Subsec:Construction of Hadamard States on Globally Hyperbolic Spacetimes}

We are finally in the position to prove the existence of Hadamard states on globally hyperbolic spacetimes for linearised Yang--Mills theory around a flat background connection. The argument follows the deformation argument used for Maxwell theory in~\cite[Sec.~6.3]{MurroSchmid}. Before applying it, we record the simple observation that flatness allows us to keep the Yang--Mills background fixed while deforming the spacetime metric.

\begin{lemma}
\label{Lem:FlatBackgroundMetricDeformation}
Let $\P\to\M$ be a principal $\G$-bundle over a globally hyperbolic spacetime $(\sf{M},\sf{g})$ and let $\A\in\mathcal C(\P)$ be flat.  For every Lorentzian metric $\widetilde\g$ on $\M$, the connection $\A$ is a Yang--Mills connection with respect to $\widetilde\g$.  Moreover, denoting explicitly the metric dependence of the formal adjoint by $\delta^{\A,\widetilde\g}$, the corresponding linear gauge theory is given by
\begin{align*}
    \K&=\d^\A, &
    \K_{\widetilde\g}^*&=\delta^{\A,\widetilde\g}, &
    \P_{\widetilde\g}&=\delta^{\A,\widetilde\g}\d^\A,
\end{align*}
and the gauge-fixed operators are
\begin{align*}
    \mathsf D_{0,\widetilde\g}
    &=\K_{\widetilde\g}^*\K
      =\delta^{\A,\widetilde\g}\d^\A,
    &
    \mathsf D_{1,\widetilde\g}
    &=\P_{\widetilde\g}+\K\K_{\widetilde\g}^*
      =\delta^{\A,\widetilde\g}\d^\A+\d^\A\delta^{\A,\widetilde\g}.
\end{align*}
In particular, whenever $\widetilde\g$ is globally hyperbolic, $\mathsf D_{0,\widetilde\g}$ and $\mathsf D_{1,\widetilde\g}$ are normally hyperbolic and
\begin{align}
    \mathsf D_{1,\widetilde\g}\K
    =\K\mathsf D_{0,\widetilde\g},
    \qquad
    \K_{\widetilde\g}^*\mathsf D_{1,\widetilde\g}
    =\mathsf D_{0,\widetilde\g}\K_{\widetilde\g}^*.
    \label{Eq:MetricDeformationIntertwiningYM}
\end{align}
\end{lemma}

\begin{proof}
Since $F^\A=0$, one has $\delta^{\A,\widetilde\g}F^\A=0$ for every Lorentzian metric $\widetilde\g$, hence $\A$ is a Yang--Mills connection independently of the metric.  The displayed formulas are then the metric-dependent versions of Proposition~\ref{Prop:LinYangMills}.  Finally, flatness gives $(\d^\A)^2=0$, and the intertwining identities follow exactly as in the proof of Proposition~\ref{Prop:LinYangMills}.
\end{proof}

The previous lemma is the only additional ingredient needed with respect to the Maxwell case: while $\K=\d^\A$ and the background connection are kept fixed throughout the deformation, the operators $\K^*$, $\P$ and $\mathsf D_k$ vary with the metric. We now possess all the ingredients to prove the Main Theorem.

\begin{proof}[Proof of Main Theorem.]
We follow~\cite[Sec.~6.3]{MurroSchmid} and split the proof into three steps.

\smallskip
\noindent\textit{Step 1: The ultrastatic case.}
Let $(\M,\g_0)$ be ultrastatic with $\g_0=-\d t\otimes\d t+\h_0$, where $(\Sigma,\h_0)$ is of bounded geometry.  By Theorem~\ref{Thm:HadamardProjectorsYM}, the operators $c_0^\pm$ constructed in Subsection~\ref{Subsec:Construction of Hadamard States on Ultrastatic Spacetimes}, after restriction to compact Cauchy data, satisfy the compact-support counterpart of Proposition~\ref{Prop:CauchyPseudoCovariancesYM}.  Hence they determine a quasifree Hadamard state on the CCR-algebra associated with $(\mathcal V_{\mathrm P},\sigma)$.

\smallskip
\noindent\textit{Step 2: Deformation between comparable metrics.}
Consider two globally hyperbolic metrics $\g_0$ and $\g_1$ on $\M$ such that $\g_0\preceq\g_1$.  Let $\chi\in C^\infty(\M,[0,1])$ and set
\begin{align*}
    \g_\chi\defeq(1-\chi)\g_0+\chi\g_1.
\end{align*}
Choosing $\chi$ as in~\cite[Sec.~6.3]{MurroSchmid}, the metric $\g_\chi$ is globally hyperbolic, $\g_0\preceq\g_\chi\preceq\g_1$, and there exist Cauchy hypersurfaces $\Sigma_0,\Sigma_1\subset(\M,\g_\chi)$, with $\Sigma_1\subset I^+_{\g_\chi}(\Sigma_0)$, such that $\g_\chi=\g_0$ in a neighbourhood of $J^-_{\g_\chi}(\Sigma_0)$ and $\g_\chi=\g_1$ in a neighbourhood of $J^+_{\g_\chi}(\Sigma_1)$; see also~\cite[Thm.~2.18]{MorettiMurroVolpeParacausal}.

By Lemma~\ref{Lem:FlatBackgroundMetricDeformation}, the same flat connection $\A$ is a Yang--Mills background for all three metrics.  Let $\mathcal U_{1,\chi}^{\Sigma_0}$ denote the Cauchy evolution for $\mathsf D_{1,\g_\chi}$ from data on $\Sigma_0$, and let $\rho_{1,\Sigma_1}$ be the Cauchy data map on $\Sigma_1$.  We set
\begin{align*}
    \mathcal R_{10}\defeq\rho_{1,\Sigma_1}\mathcal U_{1,\chi}^{\Sigma_0}.
\end{align*}
The operator $\mathcal R_{10}$ descends to a unitary isomorphism between the Cauchy phase spaces associated with $\g_0$ on $\Sigma_0$ and $\g_1$ on $\Sigma_1$.  Indeed, if $u$ is a solution of $\mathsf D_{1,\g_\chi}u=0$ whose initial data satisfy the constraint, then by Eq.~\eqref{Eq:MetricDeformationIntertwiningYM} the section $\K_{\g_\chi}^*u$ solves $\mathsf D_{0,\g_\chi}\K_{\g_\chi}^*u=0$ with vanishing Cauchy data on $\Sigma_0$, and therefore vanishes identically.  Similarly, the second identity in~\eqref{Eq:MetricDeformationIntertwiningYM} shows that gauge data are mapped to gauge data.  The preservation of the Hermitian form follows from Green's identity for the normally hyperbolic operator $\mathsf D_{1,\g_\chi}$, exactly as in~\cite[Sec.~6.3]{MurroSchmid}.

Suppose now that $c_0^\pm$ are Cauchy pseudo-covariances on $\Sigma_0$ satisfying the conditions of the compact-support version of Proposition~\ref{Prop:CauchyPseudoCovariancesYM}.  Define
\begin{align*}
    c_1^\pm\defeq(\mathcal R_{10}^{-1})^\dagger c_0^\pm\mathcal R_{10}^{-1}.
\end{align*}
Green's identity with one compactly supported argument gives $(\mathcal R_{10}^{-1})^\dagger = \mathcal R_{10}$, so
$c_1^\pm = \mathcal R_{10} c_0^\pm \mathcal R_{10}^{-1}$ is well defined by Cauchy evolution on smooth data.
Moreover since $\mathcal R_{10}$ is unitary on the quotient phase spaces, properties (i)--(iii) of Proposition~\ref{Prop:CauchyPseudoCovariancesYM} are preserved.  Property (iv) follows from propagation of singularities for $\mathsf D_{1,\g_\chi}$, using that $\g_\chi$ agrees with $\g_0$ near $\Sigma_0$ and with $\g_1$ near $\Sigma_1$.  Thus $c_1^\pm$ determine a quasifree Hadamard state for the theory associated with $(\M,\g_1,\A)$.  The case $\g_1\preceq\g_0$ is obtained by reversing the construction.  Apart from the metric-independence of the flat background established in Lemma~\ref{Lem:FlatBackgroundMetricDeformation}, this is \emph{verbatim} the argument of~\cite[Step~2 in the proof of Thm.~1.1]{MurroSchmid}.

\smallskip
\noindent\textit{Step 3: The general globally hyperbolic case.}
Let now $\g$ be arbitrary. By~\cite[Cor.~2.23]{MorettiMurroVolpeParacausal}, $\g$ is paracausally related to an ultrastatic globally hyperbolic metric.  Moreover, the spatial metric of the latter may be chosen of bounded geometry without changing its paracausal class, as explained in~\cite[Sec.~5]{MorettiMurroVolpe}. Hence there exists a finite sequence of globally hyperbolic metrics
\begin{align*}
    \g_0,\g_1,\ldots,\g_N=\g,
\end{align*}
where $\g_0$ is ultrastatic with Cauchy surface of bounded geometry and, for every $j$, either $\g_j\preceq\g_{j+1}$ or $\g_{j+1}\preceq\g_j$.  By Step~1 there exists a quasifree Hadamard state for $(\M,\g_0,\A)$.  Iterating Step~2 along the finite paracausal chain transports it to a quasifree Hadamard state for $(\M,\g,\A)$.
\end{proof}
\appendix
\section{Background Material}\label{Appendix:Background Material}
\subsection{Notation and Conventions for Scalar and Twisted Differential Forms}\label{Appendix: Notation and Conventions for Scalar and Twisted Differential Forms}
This section provides a succinct review of the definitions from the theory of differential forms that are relevant to our analysis, with the aim of fixing the notation and sign conventions used throughout the manuscript. 

\paragraph{Scalar-Valued Forms: Wedge Product and Exterior Derivative --} Let $\M$ be a smooth $d$-dimensional manifold and consider the space of smooth $k$-forms $\Omega^{k}(\M) \defeq \Gamma(\bigwedge^{k}\T^{\ast}\M)$. For the wedge product, we use the \emph{determinant normalisation convention}, {\it cf.},~\cite[Chap.~14]{LeeSmooth}. In other words, for any $\alpha\in\Omega^{k}(\M)$, $\beta\in\Omega^{l}(\M)$ and for all $v_{1},\dots,v_{k+l}\in\T_{p}\M$ with $p\in\M$, we set
\begin{align}
\label{Eq: wedge for scalar-valued forms}
    (\alpha\wedge \beta)_{p}(v_{1},\dots,v_{k+l}):=\frac{1}{k!l!}\sum_{\sigma\in\mathfrak{S}^{k+l}}\operatorname{sgn}(\sigma) \, \alpha_{p}(v_{\sigma(1)},\dots,v_{\sigma(k)}) \, \beta_{p}(v_{\sigma(k+1)},\dots,v_{\sigma(k+l)}),
\end{align}
where $\mathfrak{S}^{k}$ denotes the \emph{symmetric group of degree $k$} and $\mathrm{sgn}(\sigma)$ is the \emph{sign} of a permutation $\sigma\in\mathfrak{S}^{k}$. This normalisation convention has two important consequences: First, for a given local chart $(\mathcal{U},(x^{i})_{i=1,\dots,d})$ of $\M$, we can locally write
    \begin{align*}
        \alpha\vert_{\mathcal{U}}=\frac{1}{k!}\alpha_{\mu_{1}\dots\mu_{k}}\,\d x^{\mu_{1}}\wedge\dots\wedge\d x^{\mu_{k}}=\sum_{1\leq\mu_{1}<\dots<\mu_{k}\leq d}\alpha_{\mu_{1}\dots\mu_{k}}\,\d x^{\mu_{1}}\wedge\dots\wedge\d x^{\mu_{k}}
    \end{align*}
    for components $\alpha_{\mu_{1}\dots\mu_{k}}:=\alpha(\partial_{\mu_{1}},\dots,\partial_{\mu_{k}})\in C^{\infty}(\mathcal{U})$. Moreover, we obtain the local formula
    \begin{align*}
        (\alpha\wedge\beta)_{\mu_{1}\dots\mu_{k+l}}=\frac{(k+l)!}{k!l!}\alpha_{[\mu_{1}\dots\mu_{k}}\beta_{\mu_{k+1}\dots\mu_{k+l}]}\,,
    \end{align*}
  where $[\ldots]$ denotes the \emph{antisymmetrisation} over the enclosed indices, normalised so as to define an idempotent operation, {\it e.g.},~$\T_{[\mu_{1}\mu_{2}]}:=\frac{1}{2!}(\T_{\mu_{1}\mu_{2}}-\T_{\mu_{2}\mu_{1}})$ for a $(0,2)$-tensor $\T\in\Gamma(\T^{\ast}\M^{\otimes 2})$. The \emph{exterior derivative} $\d\:\Omega^{k}(\M)\to\Omega^{k+1}(\M)$ is defined in the usual way, {\it i.e.}, for all $\alpha\in\Omega^{k}(\M)$ and $X_{1},\dots,X_{k+1}\in\mathfrak{X}(\M)$,
  \begin{equation}
    \begin{aligned}
    \label{Eq: scalar ext derivative}
        (\d\alpha)(X_{1},\dots,X_{k+1}):=&\sum_{i=1}^{k+1}(-1)^{i-1}X_{i}(\alpha(X_{1},\dots,\hat{X_{i}},\dots,X_{k+1}))\\&+\sum_{i<j}(-1)^{i+j}\alpha([X_{i},X_{j}],X_{1},\dots,\hat{X_{i}},\dots,\hat{X_{j}},\dots,X_{k+1}),
    \end{aligned}
    \end{equation}
  where the symbol $\hat{\cdot}$ indicates that the corresponding element is omitted from the sum. In a local coordinate chart $(\mathcal{U},(x^{i})_{i=1,\dots,d})$ of $\M$ as above, this implies that locally
    \begin{align*}
	   (\d\alpha)_{\mu_{1}\dots\mu_{k+1}}=(1+k)\partial_{[\mu_{1}}\alpha_{\mu_{2}\dots\mu_{k+1}]}=(1+k)\nabla_{[\mu_{1}}\alpha_{\mu_{2}\dots\mu_{k+1}]}\, ,
    \end{align*}
    where $\nabla$ denotes the connection induced on the exterior bundles $\bigwedge^{k}\sf{T}^{\ast}\M$ by any \emph{torsion-free} connection on $\T\M$, {\it e.g.}, the Levi--Civita connection associated with a pseudo-Riemannian metric.

    \paragraph{Scalar-Valued Forms: Hodge Operator and Codifferential --} Assume now that $(\M,\g)$ is a smooth, \emph{oriented} pseudo-Riemannian $d$-manifold. We denote by $\mathrm{ind}(\g)$ the \emph{index} of $\g$, {\it i.e.},~the dimension of the largest subspace of $\T\M$ on which $\g$ is negative-definite. In particular, $\mathrm{ind}(\g)=0$ for a Riemannian metric $\g$, while $\mathrm{ind}(\g)=(d-1)$ for a Lorentzian metric $\g$ with signature convention $(+,-,\dots,-)$. Using the metric $\g$, we equip $\bigwedge^{k}\T^{\ast}\M$ with the bundle metric: 
    \begin{align*}
        \langle\alpha,\beta\rangle_{\g}:=\frac{1}{k!}(\g^{-1})^{\otimes k}(\alpha,\beta), \qquad \text{for all} \; \alpha, \beta \in \Omega^k(\M). 
    \end{align*}
    The \emph{Hodge $\ast$-operator} $\ast_{\g}\:\Omega^{k}(\M)\to\Omega^{d-k}(\M)$ is then defined by 
    \begin{align*}
        \alpha\wedge\ast_{\g}\,\beta=\langle\alpha,\beta\rangle_{\g}\mathrm{vol}_{\g}\,,
    \end{align*}
    where $\mathrm{vol}_{\g}$ denotes the volume form induced by $\g$ and the chosen orientation of $\M$. With this sign convention, it holds that $\ast_{\g}\mathrm{vol}_{\g}=s_{\g}$ and
    \begin{align*}
        \ast_{\g}\ast_{\g}=(-1)^{k(d-k)}s_{\g}\mathrm{id}_{\Omega^{k}}\,,\qquad\qquad \ast_{\g}^{-1}=(-1)^{k(d-k)}s_{\g}\ast_{\g}\qquad\text{on}\quad\Omega^{k}(\M)\,,
    \end{align*}
    where $s_{\g}:=(-1)^{\mathrm{ind}(\g)}$ is the \emph{signature scalar}, {\it i.e.},~the sign of the determinant of $\g$. With the notation and conventions introduced above, we can thus define the \emph{codifferential} $\delta\:\Omega^{k}(\M)\to\Omega^{k-1}(\M)$ by setting
    \begin{align*}
        \delta:=(-1)^{d(k-1)+1}s_{\g}\ast_{\g}\d\ast_{\g}=(-1)^{k}\ast_{\g}^{-1}\d\ast_{\g}\, ,
    \end{align*}
    which entails that $\d$ and $\delta$ are formal adjoints of each other with respect to the Hodge bilinear form, {\it i.e.},~the corresponding bilinear form on the level of sections induced from $\langle\cdot,\cdot\rangle_{\g}$, see also~Eq.~\eqref{eq:BundleMetricInduced},
    \begin{align*}
        (\alpha,\beta)_{\g}:=\int_{\M}\langle\alpha,\beta\rangle_{\g}\,\d\mu_{\g}=\int_{\M}\alpha\wedge\ast_{\g}\,\beta, \qquad \text{for all} \; \alpha,\beta\in\Omega_{\mathrm{c}}^{k}(\M).
    \end{align*}
    In addition, on a local coordinate chart $(\mathcal{U},(x^{i})_{i=1,\dots,d})$ of $\M$, we have that 
    \begin{align*}
        (\delta\alpha)_{\mu_{1}\dots\mu_{k-1}}=-\g^{\rho\sigma}\nabla_{\rho}\alpha_{\sigma\mu_{1}\dots\mu_{k-1}}\,,
    \end{align*}
    where $\nabla$ denotes the Levi-Civita connection on $(\M,\g)$ acting on $k$-forms.

\paragraph{Bundle-Valued Forms:~Wedge-Product and Exterior Derivative --} Let us now turn our attention to \emph{twisted} differential forms, namely, forms taking values in some vector space or vector bundle. To start with, we consider the following general concept of wedge-type product used throughout: Consider $\bb{K}$-vector spaces $\sf{V}$, $\sf{W}$ and $\sf{U}$. Every bilinear map $\mathcal{B}\:\sf{V}\times\sf{W}\to\sf{U}$ yields a \emph{wedge-type product} $\wedge_{\mathcal{B}}\:\Omega^{k}(\sf{M},\sf{V})\times\Omega^{l}(\sf{M},\sf{W})\to\Omega^{k+l}(\sf{M},\sf{U})$ defined by 
    \begin{align}
    \label{Eq: wedge for bundle-valued forms}
        (\omega\wedge_{\mathcal{B}}\eta)_{p}(v_{1},\dots,v_{k+l}):=\frac{1}{k!l!}\sum_{\sigma\in\mathfrak{S}^{k+l}}\operatorname{sgn}(\sigma)\mathcal{B}(\omega_{p}(v_{\sigma(1)},\dots,v_{\sigma(k)}),\eta_{p}(v_{\sigma(k+1)},\dots,v_{\sigma(k+l)})), 
    \end{align}
    for all $\omega\in\Omega^{k}(\sf{M},\sf{V})$, $\eta\in\Omega^{l}(\sf{M}, \sf{W})$ with $k, l \in \mathbb{N}$, $p\in\sf{M}$ and $v_{1},\dots,v_{k+l}\in \sf{T}_{p}\sf{M}$. Equivalently, choosing bases $(e_{a})_{a=1,\dots,\mathrm{dim}(\sf{V})}$ and $(f_{b})_{b=1,\dots,\mathrm{dim}(\sf{W})}$ of $\sf{V}$ and $\sf{W}$, respectively, and writing $\omega=\omega^{a}\otimes e_{a}$ and $\eta=\eta^{b}\otimes f_{b}$ for coefficient forms $\omega^{a}\in\Omega^{k}(\sf{M})$ and $\eta^{b}\in\Omega^{l}(\sf{M})$, this is equivalent to
    \begin{align*}
    		\omega\wedge_{\mathcal{B}}\eta=(\omega^{a}\wedge\eta^{b})\otimes \mathcal{B}(e_{a},f_{b})\, .
    \end{align*}

    \begin{example}\label{Ex:WedgeLieAlg1}
       For our purposes, a particularly relevant example is the wedge product induced by the Lie bracket of a Lie algebra $(\mathfrak{g},[\cdot,\cdot]_{\mathfrak{g}})$, which we shall denote by
        \begin{align*}
        [\cdot\wedge\cdot]_{\mathfrak{g}}\:\Omega^{k}(\M, \mathfrak{g})\times\Omega^{l}(\M,\mathfrak{g})\to\Omega^{k+l}(\M,\mathfrak{g}),
    \end{align*}
    in order to avoid ambiguity with the wedge products defined in Eqs.~\eqref{Eq: wedge for scalar-valued forms} and \eqref{Eq: wedge for bundle-valued forms}. This product equips $\bigoplus_{k\geq 0}\Omega^{k}(\M,\mathfrak{g})$ with the structure of an \emph{$\mathbb{N}$-graded Lie superalgebra}, \textit{i.e.}, it enjoys the following two properties: 
 \begin{align*}
    \text{(i)} \; &[\omega\wedge\eta]_{\mathfrak{g}}=-(-1)^{\vert\omega\vert\cdot\vert\eta\vert}[\eta\wedge\omega]_{\mathfrak{g}} \\
    \text{(ii)} \; &(-1)^{\vert\omega\vert\cdot\vert\sigma\vert}[\omega\wedge[\eta\wedge\sigma]_{\mathfrak{g}}]_{\mathfrak{g}}+(-1)^{\vert\sigma\vert\cdot\vert\eta\vert}[\sigma\wedge[\omega\wedge\eta]_{\mathfrak{g}}]_{\mathfrak{g}} +(-1)^{\vert\eta\vert\cdot\vert\omega\vert}[\eta\wedge[\sigma\wedge\omega]_{\mathfrak{g}}]_{\mathfrak{g}}=0\,,
\end{align*}
    where $\vert\omega\vert\in\mathbb{N}$ denotes the \emph{degree} of a $\mathfrak{g}$-valued differential form $\omega$.
    \end{example}

\begin{remark}\label{Ex:WedgeLieAlg2}
	   Suppose that $\mathfrak{g}$ is a \emph{matrix} Lie algebra, namely, it is a Lie subalgebra of $\mathfrak{gl}(n,\mathbb{C})$. In this case, there are \emph{two} natural bilinear operations to take into account: the matrix multiplication $X\cdot Y$ and the Lie algebra bracket, which is inherited from  $\mathfrak{gl}(n,\mathbb{C})$, {\it i.e.},~$[X,Y]_{\mathfrak{g}}:=X\cdot Y-Y\cdot X$. Both of them induce natural wedge products as per Eq.~\eqref{Eq: wedge for bundle-valued forms}. As highlighted in Example~\ref{Ex:WedgeLieAlg1}, the wedge product induced from $[\cdot,\cdot]_{\mathfrak{g}}$ will always be denoted by $[\cdot\wedge\cdot]_{\mathfrak{g}}$. To avoid possible confusion, we denote the one induced from matrix multiplication by $\dot{\wedge}\:\Omega^{k}(\sf{M}, \mathfrak{g})\times\Omega^{l}(\sf{M}, \mathfrak{g})\to\Omega^{k+l}(\sf{M}, \mathfrak{g})$. It is then easy to check that
	\begin{align*}
		[\alpha\wedge\beta]_{\mathfrak{g}}=\alpha\,\dot{\wedge}\,\beta-(-1)^{kl}\beta\dot{\wedge}\alpha, 
	\end{align*}
	for all $\alpha\in\Omega^{k}(\sf{M},\mathfrak{g})$ and $\beta\in\Omega^{l}(\sf{M},\mathfrak{g})$. In particular, if $A \in\mathcal{C}(\P)\subset\Omega^{1}(\P,\mathfrak{g})$ is the connection $1$-form of some principal $\G$-bundle $\P$ over $\M$, {\it cf.}, Section \ref{Appendix:Connections on Principal Bundles}, we have that $[A\wedge A]_{\mathfrak{g}}=2(A\,\dot{\wedge}\,A)$, which implies that Cartan's structure equation can also be expressed in the form $F^{A}=\d A+A\,\dot{\wedge}\,A$.
    \end{remark}

    More generally, given smooth finite-rank $\bb{K}$-vector bundles $\sf{E}$, $\sf{F}$ and $\sf{J}$ over $\sf{M}$ and a fibre-wise bilinear map $\mathcal{B}_{p}\:\sf{E}_{p}\times\sf{F}_{p}\to\sf{J}_{p}$ depending smoothly on $p\in\sf{M}$, {\it i.e.},~a section $\mathcal{B}\in\Gamma(\sf{E}^{\ast}\otimes\sf{F}^{\ast}\otimes\sf{J})$, we call the induced bilinear map $\wedge_{\mathcal{B}}\:\Omega^{k}(\sf{M},\sf{E})\times\Omega^{l}(\sf{M},\sf{F})\to\Omega^{k+l}(\sf{M},\sf{J})$ defined by 
    \begin{align}
    \label{Eq: wedge product induced by a bilinear form}
        (\omega\wedge_{\mathcal{B}}\eta)_{p}(v_{1},\dots,v_{k+l}):=\frac{1}{k!l!}\sum_{\sigma\in\mathfrak{S}^{k+l}}\operatorname{sgn}(\sigma) \, \mathcal{B}_{p}(\omega_{p}(v_{\sigma(1)},\dots,v_{\sigma(k)}),\eta_{p}(v_{\sigma(k+1)},\dots,v_{\sigma(k+l)}))
    \end{align}
    for all $\omega\in\Omega^{k}(\sf{M},\sf{E})$, $\eta\in\Omega^{l}(\sf{M},\sf{F})$ with $k, l \in \mathbb{N}$, $p\in\sf{M}$ and $v_{1},\dots,v_{k+l}\in \sf{T}_{p}\sf{M}$, the \emph{wedge product induced from $\mathcal{B}$}. Equivalently, choosing local frames $(e_{a})_{a=1,\dots,\mathrm{rank}(\sf{E})}$ and $(f_{b})_{b=1,\dots,\mathrm{rank}(\sf{F})}$ of $\sf{E}$ and $\sf{F}$ defined on some open subset $\mathcal{U}\subset\sf{M}$, respectively, and writing $\omega\vert_{\mathcal{U}}=\omega^{a}\otimes e_{a}$ and $\eta\vert_{\mathcal{U}}=\eta^{b}\otimes f_{b}$ for local coefficient forms $\omega^{a}\in\Omega^{k}(\mathcal{U})$ and $\eta^{b}\in\Omega^{l}(\mathcal{U})$, we can locally write
    \begin{align*}
    		(\omega\wedge_{\mathcal{B}}\eta)\vert_{\mathcal{U}}=(\omega^{a}\wedge\eta^{b})\otimes \mathcal{B}(e_{a},f_{b})
    \end{align*}
    where $\mathcal{B}(e_{a},f_{b})\in \Gamma(\mathcal{U},\sf{J})$ is defined pointwise, that is $\mathcal{B}(e_{a},f_{b})(p):=\mu_{p}(e_{a}(p),f_{b}(p))$.

\begin{example}
\label{Example:WedgeBundleMetric}
    Let $\mathbb{K} \in \{\mathbb{R}, \mathbb{C}\}$ and let $\E$ be a finite-rank $\mathbb{K}$-vector bundle equipped with a bundle metric $\langle\cdot,\cdot\rangle_{\E}$. Then, we can introduce a wedge product induced by such metric 
    \begin{align*}
        \langle\cdot\wedge\cdot\rangle_{\E}\:\Omega^{k}(\M,\E)\times\Omega^{l}(\M,\E)\to\Omega^{k+l}(\M)\,,
    \end{align*} 
    which will be used below to obtain a suitable generalisation of the Hodge $\ast$-operator for $\E$-valued differential forms. 
\end{example}

In analogy to Eq.~\eqref{Eq: scalar ext derivative}, given a connection $\nabla^{\E}\:\Gamma(\E)\to\Omega^{1}(\M,\E)$ on a vector bundle $\E$, we define the \emph{exterior covariant derivative} $\d^{\nabla^{\E}}\:\Omega^{k}(\M,\E)\to\Omega^{k+1}(\M,\E)$ by 
    \begin{align}
    \label{Eq: ext derivative}
        (\d^{\nabla^{\E}}\omega)(X_{1},\dots,X_{k+1}):=&\sum_{i=1}^{k+1}(-1)^{i-1}\nabla^{\E}_{X_{i}}(\omega(X_{1},\dots,\hat{X_{i}},\dots,X_{k+1})) \notag \\&+\sum_{i<j}(-1)^{i+j}\omega([X_{i},X_{j}],X_{1},\dots,\hat{X_{i}},\dots,\hat{X_{j}},\dots,X_{k+1})\, .
    \end{align}
    for all $\omega\in\Omega^{k}(\M,\E)$ and $X_{1},\dots,X_{k+1}\in\mathfrak{X}(\M)$. Note that, in general, $\d^{\nabla^{\E}}\circ\d^{\nabla^{\E}}\neq 0$. As a matter of fact, if we define the \emph{curvature} $F^{\nabla^{\E}}\in\Omega^{2}(\M,\mathrm{End}(\E))$ of $(\E,\nabla^{\E})$ by 
    \begin{align}\label{eq:CurvBundle}
        F^{\nabla^{\E}}(X,Y)\psi:=\nabla^{\E}_{X}\nabla^{\E}_{Y}\psi-\nabla^{\E}_{Y}\nabla^{\E}_{X}\psi-\nabla^{\E}_{[X,Y]}\psi
    \end{align}
    for all $X,Y\in\mathfrak{X}(\M)$, then, for all $\omega\in\Omega^{k}(\M,\E)$, it holds that
    \begin{align*}
        \d^{\nabla^{\E}}\d^{\nabla^{\E}}\omega=F^{\nabla^{\E}}\wedge_{\Phi}\omega\,,
    \end{align*}
    where we denote by $\wedge_{\Phi}\:\Omega^{k}(\sf{M},\mathrm{End}(\sf{E}))\times\Omega^{l}(\sf{M},\sf{E})\to\Omega^{k+l}(\sf{M},\sf{E})$ the wedge-type product induced from the natural pairing $\Phi_{p}\:\mathrm{End}(\sf{E}_{p})\times\sf{E}_{p}\to\sf{E}_{p}$. If $\nabla^{\E}$ is a \emph{flat connection}, {\it i.e.},~$F^{\nabla^{\E}}=0$, we hence obtain a \emph{generalised $\E$-twisted de Rham complex} $(\Omega^{\bullet}(\M,\E),\d^{\nabla^{\E}})$. 

    \paragraph{Bundle-Valued Forms: Hodge Operator and Codifferential --} Assume now that $(\M,\g)$ is a smooth, \emph{oriented} pseudo-Riemannian $d$-manifold and let $\E$ be a smooth vector bundle equipped with a bundle metric $\langle\cdot,\cdot\rangle_{\E}$ and a connection $\nabla^{\E}$, that is compatible with $\langle\cdot,\cdot\rangle_{\E}$, \textit{i.e.}, 
    \begin{align}\label{comment}
        \mathcal{L}_{X}\langle\psi,\phi\rangle_{\E} \equiv \d\langle\psi,\phi\rangle_{\E}(X)=\langle\nabla^{\E}_{X}\psi,\phi\rangle_{\E}+\langle\psi,\nabla^{\E}_{X}\phi\rangle_{\E}, \qquad \text{for all} \; \psi, \varphi \in \Gamma(\E),\,X\in\mathfrak{X}(\M)
    \end{align}
    or, equivalently, $\nabla^{\E^{\ast}\otimes\E^{\ast}}\langle\cdot,\cdot\rangle_{\E}=0$. Using the metric $\g$ and the bundle metric $\langle\cdot,\cdot\rangle_{\E}$, we equip $\E_{k}:=\bigwedge^{k}\T^{\ast}\M\otimes\E$ with the induced bundle metric
    \begin{align*}\langle\alpha\otimes\psi,\beta\otimes\varphi\rangle_{\E_{k}}:=\langle\alpha,\beta\rangle_{\g}\langle\psi,\varphi\rangle_{\E}\, .
    \end{align*}
    The \emph{Hodge $\ast$-operator} $\ast_{\g}\:\Omega^{k}(\M,\E)\to\Omega^{d-k}(\M,\E)$, still denoted by the same symbol with a mild abuse of notation, is then defined by 
    \begin{align*}  \langle\omega\wedge\ast_{\g}\,\eta\rangle_{\E}=\langle\omega,\eta\rangle_{\E_{k}}\mathrm{vol}_{\g}\,,
    \end{align*}
    where $\langle\cdot\wedge\cdot\rangle_{\E}\:\Omega^{k}(\M,\E)\times\Omega^{l}(\M,\E)\to\Omega^{k+l}(\M)$ denotes the wedge product induced from $\langle\cdot,\cdot\rangle_{\E}$ as in Example~\ref{Example:WedgeBundleMetric}. From its definition, it is easy to see that $\ast_{\g}$ can be computed componentwise, {\it i.e.}, if we choose a local frame $(e_{a})_{a = 1,\dots,\mathrm{rank}(\E)}\subset\Gamma(\mathcal{U},\E)$ of $\E$ defined on some open subset $\mathcal{U}\subset\M$ and locally write $\omega\vert_{\mathcal{U}}=\omega^{a}\otimes e_{a}$ for components $\omega^{a}\in\Omega^{k}(\mathcal{U})$, then $(\ast_{\g}\omega)\vert_{\mathcal{U}}=(\ast_{\g}\omega^{a})\otimes e_{a}$. 
    
    With this notation, we define the \emph{exterior codifferential} $\delta^{\nabla^{\E}}\:\Omega^{k}(\M,\E)\to\Omega^{k-1}(\M,\E)$ induced by the connection $\nabla^{\E}$ by
    \begin{align*}
        \delta^{\nabla^{\E}}:=(-1)^{d(k-1)+1}s_{\g}\ast_{\g}\d^{\nabla^{\E}}\ast_{\g}=(-1)^{k}\ast_{\g}^{-1}\d^{\nabla^{\E}}\ast_{\g}\, ,
    \end{align*}
    which implies that $\d^{\nabla^{\E}}$ and $\delta^{\nabla^{\E}}$ are formal adjoints of each other with respect to the Hodge bilinear form, that is, the bilinear form on the level of sections induced from the pairing $\langle\cdot,\cdot\rangle_{\E_{k}}$, {\it cf.},~Eq.~\eqref{eq:BundleMetricInduced}:
    \begin{align*}
        (\omega,\eta)_{\E_k}:=\int_{\M}\langle\omega,\eta\rangle_{\E_{k}}\,\d\mu_{\g}=\int_{\M}\langle\omega\wedge\ast_{\g}\,\eta\rangle_{\E}
    \end{align*}
    for all $\omega,\eta\in\Omega^{k}_{\mathrm{c}}(\M,\E)$.

\subsection{Notation and Terminology for Lie Groups}\label{Appendix:Notation and Terminology for Lie Groups}
The term \emph{Lie group} will always mean a \emph{smooth, real} and \emph{finite-dimensional Lie group}. If $\G$ is a Lie group in this sense, we define the \emph{left-multiplication} $L_{g}\colon\G\to\G$, \emph{right-multiplication} $R_{g}\colon\G\to\G$ and \emph{conjugation map} $C_{g}\colon\G\to\G$, respectively, by setting
    \begin{align*}
        L_{g}(h):=gh\,,\qquad R_{g}(h):=hg\,,\qquad C_{g}:=ghg^{-1}
    \end{align*}
    for all $g,h\in\G$. If $(\mathfrak{g},[\cdot,\cdot]_{\mathfrak{g}})$ is the Lie algebra of $\G$, we denote the \emph{adjoint representation} by
    \begin{align*}
        \mathrm{Ad}\:\G\to\mathrm{Aut}(\mathfrak{g})\,,\quad g\mapsto\mathrm{Ad}_{g}:=(\mathrm{d}C_{g})_{e}\:\mathfrak{g}\to\mathfrak{g}\,,
    \end{align*}
    where $e\in\G$ denotes the neutral element and where we identified $\T_{e}\G\cong\mathfrak{g}$. In the special case in which $\G$ is a \emph{matrix Lie group}, \textit{i.e.}, a closed subgroup of $\mathrm{Gl}(n,\mathbb{C})$, we have $\mathrm{Ad}_{g}(X)=gXg^{-1}$ for any $X \in \mathfrak{g}$. Furthermore, we denote the induced Lie-algebra representation by 
    \begin{align*}
        \mathrm{ad}\:\mathfrak{g}\to\mathrm{End}(\mathfrak{g})\,,\qquad X\mapsto \mathrm{ad}_{X}:=(\mathrm{d}\mathrm{Ad}_{e})_{X}\:\mathfrak{g}\to\mathfrak{g}\,.
    \end{align*}
    It holds that $\mathrm{ad}_{X}(Y)=[X,Y]_{\mathfrak{g}}$ for arbitrary (not necessarily matrix) Lie groups $\G$. Last but not least, if $\G$ is a Lie group with Lie algebra $(\mathfrak{g},[\cdot,\cdot]_{\mathfrak{g}})$, we denote its \emph{Maurer-Cartan form} by 
    \begin{align*}
        \mu\in\Omega^{1}(\G,\mathfrak{g})\,,\qquad \mu_{g}(X):=(\d L_{g^{-1}})_{g}(X)
    \end{align*}
    for all $g\in\G$, $X\in\T_{g}\G$. By definition, $\mu$ satisfies the \emph{Maurer--Cartan equations}, $\d\mu+\frac{1}{2}[\mu\wedge\mu]_{\mathfrak{g}}=0$.

\subsection{Connections on Principal Bundles}
\label{Appendix:Connections on Principal Bundles}
Let $\P\to\M$ be a smooth principal $\G$-bundle over a smooth manifold $\M$ and let $(\mathfrak{g}, [\cdot, \cdot]_{\mathfrak{g}})$ be the Lie algebra associated to $\G$, see \cite[Sec.~10]{Taubes}. A \emph{connection $1$-form} on $\P$ is a $1$-form $\A \in\Omega^{1}(\P,\mathfrak{g})$ with the property that for all $g\in\G$, $p\in\P$ and $X\in\mathfrak{g}$:
\begin{itemize}
\item[(i)]$(r_{g}^{\ast}\A)_{p}=\mathrm{Ad}_{g^{-1}}\circ \A_{p}$, where $r_{g}\:\P\to\P$ is defined by $r_{g}(p):=p\cdot g$.
\item[(ii)]$\A(X^{\#})=X$, where $X^{\#}\in\mathfrak{X}(\P)$ is the \textit{fundamental vector field} of $X$, \textit{i.e.}, \begin{align*}
	\bullet^{\#}\:\mathfrak{g}\xrightarrow{\cong} \V_{p}\subset\mathfrak{X}(\P)\,,\qquad\X^{\#}_{p}:=\frac{\d}{\d t}\bigg\vert_{t=0}(p\cdot\mathrm{exp}_{\G}(tX))\,,
\end{align*}
with $\V_{p}:=\mathrm{ker}(\d_{p}\pi)\subset\T_{p}\P$ being the vertical tangent space at $p\in\P$ and $\mathrm{exp}_{\G}\:\mathfrak{g}\to\sf{G}$ denoting the \textit{exponential map} of $\G$.
\end{itemize} 
Every connection $1$-form $A\in\mathcal{C}(\P)$ induces a \emph{(principal) Ehresmann connection}, namely, a vector sub-bundle (or \emph{distribution}) $\H\subset\T\P$, such that: 
\begin{itemize}
    \item $\T\P=\V\oplus\H$
    \item $(r_{g}^{\ast}\H)_{p}=\H_{p\cdot g}$ for all $p\in\P$ and $g\in\G$, where $\H_{p}:=\mathrm{ker}(\A_{p})$. 
\end{itemize}
The converse also holds true, that is, every Ehresmann connection $\H\subset\T\P$ induces a connection by setting $\A(X_p^{\#}+Y_p):=X$ for all $X\in\mathfrak{g}$ and $Y\in\H$. Tangent vectors in $\H$ are called \emph{horizontal}. It is easy to see that, for any two connections $\A_{1},\A_{2}\in\mathcal{C}(\P)$, their difference satisfies
\begin{align*}
    \A_{1}-\A_{2}\in\Omega^{1}_{\mathrm{hor},\mathrm{Ad}}(\P,\mathfrak{g})\cong\Omega^{1}(\M,\mathrm{Ad}(\P))\,,
\end{align*}
which implies that $\mathcal{C}(\P)$ is an infinite-dimensional affine space modelled over $\Omega^{1}(\M,\mathrm{Ad}(\P))$. 

As discussed in the main text, we define the curvature of a connection $A\in\mathcal{C}(\P)$ to be the 2-form $F^{\A}:=\mathrm{D}_{\A}\A=\d\A+\frac{1}{2}[\A\wedge\A]_{\mathfrak{g}}$. However, we may also consider the connection $\nabla^{A}$ on $\mathrm{Ad}(\P)$ and define the associated bundle-curvature $F^{\nabla^{\A}}\in\Omega^{2}(\M,\mathrm{End}(\mathrm{Ad}(\P)))$ defined as in Eq.~\eqref{eq:CurvBundle}. Note that these two notions of curvature are compatible in the sense that for any local gauge $s\in\Gamma(\mathcal{U},\P)$, it holds that
\begin{align}\label{FlatBnudleP}
    (F^{\nabla^{\A}}(X,Y))\Phi=[s,[(F^{\A})_{s}(X,Y),\varphi]_{\mathfrak{g}}]
\end{align}
for all $X,Y\in\mathfrak{X}(\M)$ and $\Phi\in\Gamma(\mathcal{U},\mathrm{Ad}(\P))$ written as $\Phi(x)=[s(x),\varphi(x)]$, where $\varphi:=\Phi_{s}\in C^{\infty}(\mathcal{U},\mathfrak{g})$ is $\mathfrak g$-valued local representative of $\Phi$ with respect to $s$, see \textit{e.g.}, \cite[Satz~3.21]{Baum}.

\section{Poincaré Duality for Flat Vector Bundle-Valued Differential Forms}
\label{Appendix:Poincaré Duality for Flat Vector Bundle-Valued Differential Forms}
In this section, we succinctly review a version of \emph{Poincaré duality} for differential forms with values in a \emph{flat} (pseudo-Riemannian) vector bundle. As a preliminary step, we fix the general set-up used throughout this section: let $\M$ be a smooth and oriented $d$-dimensional manifold without boundary, and let $\E \to \M$ be a finite-rank $\bb{R}$-vector bundle equipped with a non-degenerate bundle metric $\langle \cdot,\cdot\rangle_\E$ and a connection $\nabla^\E$ that is \emph{compatible} with the metric, \textit{i.e.}, satisfies Eq.~\eqref{comment}, and that is \emph{flat}, \textit{i.e.}, $F^{\nabla^\E}=0$. In the following, we shall refer to such a triple $(\E,\langle\cdot,\cdot\rangle_{\E},\nabla^{\E})$ as a \emph{flat pseudo-Riemannian vector bundle}. Note that this is a similar set-up as in Sec.~\ref{Subsec:Hodge Decomposition in Flat Vector Bundles on Complete Riemannian Manifolds}, with the difference that here we allow for non-positive fibre metrics.

The following two examples will serve as the main motivation for the analysis that follows.

\begin{examples}\label{Ex: AdP}
\begin{itemize}
    \item[(i)]Consider the trivial line bundle $\E=\underline{\mathbb{R}}_\M$ whose fibres $\E_p \cong \mathbb{R}$ carry the standard inner product on $\bb{R}$. Identifying sections $\Gamma(\E) \cong C^\infty(\M)$, as usual, the canonical flat connection defined by $\nabla^\E_X f = \mathrm{d}f(X)$ for $f \in C^\infty(\M)$ and $X \in \mathfrak{X}(\M)$ is clearly compatible with this metric. Hence, $(\underline{\mathbb{R}}_\M,\cdot,\d)$ defines a flat pseudo-Riemannian vector bundle.
    \item[(ii)]Let $\G$ be a Lie group with corresponding Lie algebra $(\mathfrak{g},[\cdot,\cdot]_{\mathfrak{g}})$, \textit{cf.}, Appendix \ref{Appendix:Notation and Terminology for Lie Groups}, and let $\P\to\M$ be a smooth principal $\G$-bundle over $\M$. Suppose that $\G$ satisfies Assumption~\ref{Assumption}, which entails the existence of an \emph{$\mathrm{Ad}$-invariant} \emph{positive definite} inner product $\kappa\:\mathfrak{g}\times\mathfrak{g}\to\mathbb{R}$, which in turn defines a non-degenerate bundle metric $\langle\cdot,\cdot\rangle_{\mathrm{Ad}(\P)}$ on the \emph{adjoint bundle} $\mathrm{Ad}(\P):=\P\times_{\mathrm{Ad}}\mathfrak{g}$ as per Eq.~\eqref{eq:InducedBundleMetric}. Now, if we fix a connection $1$-form $A\in\mathcal{C}(\P)$, then the induced connection $\nabla^{A}$ as defined in Section~\ref{Sec:Geometric Preliminaries on Yang-Mills Theory} is compatible with $\langle\cdot,\cdot\rangle_{\mathrm{Ad}(\P)}$, as a consequence of $\mathrm{Ad}$-invariance of $\kappa$. Hence, assuming $A$ to be flat, which implies that $\nabla^{A}$ is flat as well on account of Eq.~\eqref{FlatBnudleP}, we conclude that $(\mathrm{Ad}(\P),\langle\cdot,\cdot\rangle_{\mathrm{Ad}(\P)},\nabla^{A})$ defines a flat pseudo-Riemannian vector bundle.
\end{itemize}
\end{examples}

Now, for a given flat pseudo-Riemannian vector bundle $(\E,\langle\cdot,\cdot\rangle_{\E},\nabla^{\E})$, consider the exterior covariant derivative $\d^{\nabla^\E}$ induced by the connection $\nabla^{\E}$ on $\E$, \textit{cf.}, Eq. \eqref{Eq: ext derivative}. By assumption, we have
\begin{align*}
    \d^{\nabla^{\E}}\circ\d^{\nabla^{\E}}=F^{\nabla^{\E}}\wedge_{\Phi}\bullet=0\,,
\end{align*}
where $\wedge_{\Phi}$ is as in Appendix~\ref{Appendix: Notation and Conventions for Scalar and Twisted Differential Forms}, which implies that we obtain $\E$-twisted de Rham complexes $(\Omega^{\bullet}(\M,\E),\d^{\nabla^{\E}})$ and $(\Omega^{\bullet}_{\mathrm{c}}(\M,\E),\d^{\nabla^{\E}})$. We denote by 
\begin{flalign*}
     H^k(\M,\,\E) \;\coloneqq\;
  \frac{\ker\bigl(\d^{\nabla^\E}\vert_{\Omega^k(\M, \E)}\bigr)}%
       {\im\bigl(\d^{\nabla^{\E}}\vert_{\Omega^{k-1}(\M,\E)}\bigr)},\qquad H_{\mathrm{c}}^k(\M,\,\E) \;\coloneqq\;
  \frac{\ker\bigl(\d^{\nabla^\E}\vert_{\Omega_{\mathrm{c}}^k(\M, \E)} \bigr)}%
       {\im\bigl(\d^{\nabla^{\E}}\vert_{\Omega_{\mathrm{c}}^{k-1}(\M,\E)}\bigr)}
\end{flalign*}
the corresponding \emph{$\E$-twisted de Rham cohomology groups} and their compactly-supported counterparts. In the special case of the trivial line bundle $(\underline{\mathbb{R}}_\M,\cdot,\d)$ from Example~\ref{Ex: AdP}(i), we recover the usual de Rham cohomology groups $H^k(\M)$ and $H^k_{\mathrm{c}}(\M)$.

Now, Poincaré's duality in its usual differential geometric formulation entails that the bilinear pairing defined by
\begin{align*}
    \langle\cdot,\cdot\rangle_{H^{\bullet}}\:H^{k}(\M)\times H^{d-k}_{\mathrm{c}}(\M)\to\bb{R}\,,\qquad \langle [\alpha],[\beta]\rangle_{H^{\bullet}}:=\int_{\M}\alpha\wedge\beta
\end{align*}
gives rise to a well-defined linear isomorphism $H^{k}(\M)\cong (H^{d-k}_{\mathrm{c}}(\M))^{\ast}$ defined by $[\alpha]\mapsto \langle[\alpha],-\rangle_{H^{\bullet}}$. A proof can, for instance, be found in \cite[Thm.~10.26]{LeeManifolds}. Standard approaches from a differential-geometric perspective typically assume that the manifold $\M$ admits a \emph{finite good cover}, \emph{i.e.}, a finite open cover $(\mathcal{U}_{i})_{i\in\mathrm{I}}$ of $\M$ such that any non-empty finite intersection of elements of $(\mathcal{U}_{i})_{i\in\mathrm{I}}$ is diffeomorphic to $\mathbb{R}^{d}$. Nevertheless, a more detailed topological analysis shows that this assumption can be omitted, see, \emph{e.g.}, \cite[Sec.~5.12]{Greub}. It is important to stress, however, that for non-compact manifolds, in the absence of such covers, the relevant cohomology spaces may be infinite-dimensional, and hence, in general, $H^{d-k}_{\mathrm{c}}(\M) \not\cong (H^{k}(\M))^{\ast}$. Indeed, the induced map $[\beta]\mapsto \langle -,[\beta]\rangle_{H^{\bullet}}$ may fail to be surjective, see, \textit{e.g.}, \cite[Rem.~5.7]{Bott}.

In the case of flat pseudo-Riemannian vector bundles, we obtain the following generalisation: 

\begin{proposition}[$\E$-Twisted Poincaré Duality]\label{Prop: Poincare duality}
Let $\M$ be an oriented smooth $d$-dimensional manifold and $(\E,\langle\cdot,\cdot\rangle_{\E},\nabla^{\E})$ be a flat pseudo-Riemannian vector bundle. Then,
\begin{equation}
\label{eq:main_pairing}
  \langle -,- \rangle_{H^\bullet_\E}\colon
  H^k(\M,\,\E) \;\times\; H_c^{d-k}(\M, \,\E)
  \;\to\; \mathbb{R}, \qquad  \langle [\alpha], [\beta] \rangle_{H^\bullet_\E}
  \;\coloneqq\;\int_\M  \langle \alpha \wedge \beta \rangle_\E,
\end{equation}
where $\langle \cdot \wedge \cdot \rangle_\E: \Omega^k(\M, \E) \times \Omega^l(\M, \E) \to \Omega^{k+l}(\M)$ denotes the induced wedge product as defined in Example \ref{Example:WedgeBundleMetric},
is a well-defined bilinear form that gives rise to a linear isomorphism
\[
  H^k(\M,\,\E) \;\xrightarrow{\sim} (H_c^{d-k}(\M,\,\E))^{\ast}\,,\qquad [\alpha]\mapsto \langle [\alpha],-\rangle_{H^{\bullet}_{\E}} \, .
\]
\end{proposition}

Before proving the preceding proposition, let us observe that the Poincaré duality for $\E$-valued forms does not follow directly from the scalar case. Indeed, although there is a natural $C^{\infty}(\M)$-module isomorphism
$\Omega^{k}(\M,\E)\cong\Omega^{k}(\M)\otimes_{C^{\infty}(\M)}\Gamma(\E)$,
the corresponding cohomology groups do not admit such a straightforward decomposition, since the exterior covariant derivative mixes the components. More precisely, if $(e_{a})_{a=1,\dots,\mathrm{rank}(\E)}$ is a local frame for $\E$ over some open subset $\mathcal{U}\subset\M$, then locally
\begin{align*}
(\d^{\nabla^{\E}}\omega)\vert_{\mathcal{U}}
= \d\omega^{a}\otimes e_{a}
+ (-1)^{\lvert\omega\rvert}\omega^{a}\wedge\nabla^{\E}e_{a},
\end{align*}
for every $\omega\in\Omega^{k}(\M,\E)$ written as $\omega\vert_{\mathcal{U}}=\omega^{a}\otimes e_{a}$ with local coefficients $\omega^{a}\in\Omega^{k}(\mathcal{U})$. Nevertheless, Proposition~\ref{Prop: Poincare duality} can be proven in similar way as its scalar counterpart. Since the proof is standard, we shall only sketch the main steps in the following.

\begin{proof}[Proof of Proposition~\ref{Prop: Poincare duality} (sketch).]
    To start with, let us recall that the exterior covariant derivative satisfies a Leibniz rule of the form 
    \begin{align*}
        \d \langle \alpha\wedge\beta \rangle_{\E}
  = \langle \d^{\nabla^\E} \alpha\wedge\beta \rangle_{\E}
    + (-1)^{\vert\alpha\vert}\, \langle \alpha\wedge \d^{\nabla^\E} \beta \rangle_{\E}
    \end{align*}
    for all $\alpha,\beta\in\Omega^{\bullet}(\M,\E)$, which follows easily from a computation in local coordinates by using the standard Leibniz rule for scalar-valued forms and the compatibility condition of $\nabla^{\E}$ and $\langle\cdot \wedge \cdot\rangle_{\E}$ (a proof can be found, for instance, in \cite[Lem.~A.2]{SchmidMSc}). In particular, Stokes' theorem together with the fact that $\d^{\nabla^{\E}}\circ\d^{\nabla^{\E}}=0$ by assumption, immediately implies that $\langle -,- \rangle_{H^\bullet_\E}$ is independent of representatives and hence well-defined.

    Now, suppose that $\mathcal{U}\subset\sf{M}$ is a subset that is diffeomorphic to $\bb{R}^{d}$. Since $\mathcal{U}$ trivialises the bundle $\E$ and moreover, since $\nabla^{\E}$ is assumed to be flat, we can find a local frame $(e_{a})_{a=1,\dots,\mathrm{rank}(\E)}$ consisting of \emph{parallel} sections, \textit{i.e.}, $\nabla^{\E}e_{a}=0$. In particular, it follows that $(\d^{\nabla^{\E}}\omega)\vert_{\mathcal{U}}=\d\omega^{a}\otimes e_{a}$ for all $\omega\in\Omega^{k}(\M,\E)$ in this chart. This implies that over $\mathcal{U}$, the $\E$-twisted de Rham complex decouples into a direct sum of $\mathrm{rank}(\E)$-copies of the standard scalar-valued de Rham complex. Consequently, the twisted cohomology groups over $\mathcal{U}$ decompose as $$H^{k}(\mathcal{U},\E\vert_{\mathcal{U}}) \cong \bigoplus_{a=1}^{\mathrm{rank}(\E)} H^{k}(\mathcal{U}),$$ and similarly for the compactly supported cohomology groups $H_{\mathrm{c}}^{d-k}(\mathcal{U},\E\vert_{\mathcal{U}})$. Since Poincaré duality holds for $\mathbb{R}^{d}$ (and hence for $\mathcal{U}$), the pairing in Eq. \eqref{eq:main_pairing} induces a linear isomorphism $H^{k}(\mathcal{U},\E\vert_{\mathcal{U}}) \xrightarrow{\sim} (H_{\mathrm{c}}^{d-k}(\mathcal{U},\E\vert_{\mathcal{U}}))^{\ast}$. Thus, the proposition holds for any coordinate chart diffeomorphic to $\mathbb{R}^{d}$.

   Now, if $\mathcal{U},\mathcal{V}\subset\sf{M}$ are two open subsets that are diffeomorphic to $\bb{R}^{d}$ such that also $\mathcal{U}\cap\mathcal{V}$ is diffeomorphic to $\bb{R}^{d}$ as well, the standard arguments used for the scalar Poincaré duality based on the \emph{Meyer-Vietoris sequence} and the \emph{five-lemma} shows that the proposition also holds for the union $\mathcal{U}\cup\mathcal{V}$, \textit{cf.}, \cite[Thm.~10.26]{LeeManifolds} or also \cite[Prop.~6]{KhavkineGrav}. In particular, an inductive argument yields the claim for manifolds that admit a finite good cover. For more general manifolds, we can follow steps similar to those outlined in~\cite[Sec.~5.12]{Greub}.
\end{proof}

\begin{remark}
    Proposition~\ref{Prop: Poincare duality} can also be rephrased without relying on a bundle metric: consider a flat vector bundle $(\E,\nabla^{\E})$ and denote by $\nabla^{\E^{\ast}}$ the induced connection on the dual bundle $\E^{\ast}$, which is clearly flat as well. Then, the bilinear map
\begin{equation}\label{eq:22}
  H^k(\M,\,\E^{\ast}) \;\times\; H_c^{d-k}(\M, \,\E)
  \;\to\; \mathbb{R}, \qquad  \langle [\alpha], [\beta] \rangle_{H^\bullet_\E}
  \;\coloneqq\;\int_\M  \alpha \wedge_{\Phi} \beta ,
\end{equation}
where $\wedge_{\Phi}\: \Omega^k(\M, \E^{\ast}) \times \Omega^l(\M, \E) \to \Omega^{k+l}(\M)$ denotes wedge product induced from the duality pairing $\Phi_{p}\:\E^{\ast}_{p}\times\E_{p}\to\bb{R}$,
is well-defined and gives rise to a linear isomorphism
\begin{align*}
    H^k(\M,\,\E^{\ast}) \;\xrightarrow{\sim} (H_c^{d-k}(\M,\,\E))^{\ast}\, .
\end{align*}

Now, if $\langle\cdot,\cdot\rangle_{\E}$ is a (non-degenerate) bundle metric that is compatible with $\nabla^{\E}$, we obtain a linear isomorphism $\Psi_{\E}\:\E\xrightarrow{\sim}\E^{\ast}$ given by $\psi\mapsto\langle\psi,-\rangle_{\E}$. With this notation, it is easy to see that the two wedge-type products $\langle\cdot,\cdot\rangle_{\E}$ and $\wedge_{\Phi}$ are related as $\langle\alpha,\beta\rangle_{\E}=\Psi_{\E}(\alpha)\wedge_{\Phi}\beta$ for all $\alpha,\beta\in\Omega^{\bullet}(\M,\E)$. Moreover, it is easy to see that the bilinear pairing~in Eq.~\eqref{eq:main_pairing} and the pairing in Eq.~\eqref{eq:22} are canonically identified under this isomorphism: more precisely, since the metric $\langle\cdot,\cdot\rangle_\E$ is parallel with respect to $\nabla^\E$, the identification $\Psi_{\E}$ intertwines the exterior covariant derivatives, that is, $\d^{\nabla^{\E^*}}(\Psi_{\E}(\alpha)) =\Psi_{\E} (\d^{\nabla^\E}\alpha)$ for all $\alpha\in\Omega^{k}(\M,\E)$, and hence induces an isomorphism on the level of cohomology groups, 
\begin{align*}
    H^k(\M, \E) \xrightarrow{\sim} H^k(\M, \E^{\ast})\, .
\end{align*}
In this formulation, a proof of Proposition~\ref{Prop: Poincare duality} can also be found in \cite[Prop.~6]{KhavkineGrav} following similar steps as mentioned above. Moreover, we note that the Poincaré duality in Proposition~\ref{Prop: Poincare duality} is a particular case of the \emph{Verdier duality} applied to the locally constant sheaf associated to a flat vector bundle $(\E,\nabla^{\E})$. For further details, we refer to \cite{Bredon97,KS90}.
\end{remark}

While the preceding discussion treated the case of \emph{real} vector bundles, the complex case follows analogously, with the relevant constructions adapted to sesquilinear pairings and complex-linear or antilinear maps, as appropriate.

\section{Pseudodifferential Calculus on Manifolds of Bounded Geometry}
\label{Appendix:Pseudodifferential Calculus on Manifolds of Bounded Geometry}

We briefly recall the bounded pseudodifferential calculus employed in Section \ref{Subsec:Construction of Hadamard States on Ultrastatic Spacetimes}, in order to fix notation.  We refer to \cite{GerardStoskopf,GerardMurroWrochna,GerardOulghaziWrochna}, \cite[Chap.~10]{GerardBook} and \cite[App.~1]{Shubin1992} for details on that subject.  Let $(\Sigma,\h)$ be of bounded geometry
and let $\E\to\Sigma$ be a Hermitian vector bundle of bounded geometry.  In
our application $\E=\E_{\Sigma,k}$ with $k=0,1$; Remark~\ref{Rmk: bounded geometry}
and the flatness of $\A_\Sigma^0$ ensure that these bundles are of bounded
geometry. We denote by
\begin{align*}
    \BS^m(\T^*\Sigma,\operatorname{End}(\E))
\end{align*}
the space of classical bounded symbols of order $m$ and by
$\operatorname{Op}$ a bounded quantisation associated with a bounded atlas and a bounded
partition of unity. Different choices of bounded quantisation differ by an
operator in the smoothing ideal
\begin{align*}
    W^{-\infty}(\Sigma;\E)
    \defeq
    \bigcap_{N\in\mathbb N}
    \mathcal{B}\bigl(\H^{-N}(\Sigma;\E),\H^N(\Sigma;\E)\bigr)\,,
\end{align*}
where $\mathcal{B}(\sf{X},\sf{Y})$ denotes the space of linear and bounded operators between two topological vector spaces $\sf{X}$ and $\sf{Y}$. We then set
\begin{align*}
    \Psi_{\mathrm b}^m(\Sigma;\E)
    \defeq
    \operatorname{Op}\bigl(\BS^m(\T^*\Sigma,\operatorname{End}(\E))\bigr)
    +W^{-\infty}(\Sigma;\E).
\end{align*}
The classes $\Psi_{\mathrm b}^m$ are closed under composition and taking
adjoints, and satisfy the usual elliptic parametrix construction modulo
$W^{-\infty}$.  We shall use these facts throughout the manuscript without further comment.

\section{List of Notation}
\label{Appendix:D}
\begin{small}
\begin{tabularx}{0.88\textwidth}{
    >{\centering\arraybackslash}p{0.12\textwidth}
    X
}
\label{tab:notation}\\

\toprule
\textbf{Symbol} & \textbf{Meaning} \\
\midrule
\endfirsthead

\toprule
\textbf{Symbol} & \textbf{Meaning} \\
\midrule
\endhead

\renewcommand{\arraystretch}{1.5}
		$(\sf{M},\g)$ & globally hyperbolic spacetime, $\M\cong\bb{R}\times\Sigma$, $\g=-\beta^{2}\d t\otimes\d t+\h_{t}$\\
		$\G$ & Lie group \\
		$(\mathfrak{g},[\cdot,\cdot]_{\mathfrak{g}})$ & Lie algebra of a Lie group $\sf{G}$\\
		&\\
        $\sf{E}_{k}$ & bundles $\sf{E}_{k}:=\bigwedge^{k}\sf{T}^{\ast}\sf{M}\otimes\mathrm{Ad}(\sf{P})$, Eq.~\eqref{Eq.DefEk1} \\
        $\sf{E}_{\Sigma,k}$ & bundles $\sf{E}_{\Sigma,k}:=\bigwedge^{k}\sf{T}^{\ast}\Sigma\otimes\mathrm{Ad}(\sf{P}_{\Sigma})$, Eq.~\eqref{Eq: E Sigma k}\\
        $\sf{E}_{\rho_{k}}$ & bundles of Initial Data, $\sf{E}_{\rho_{k}}=\sf{E}_{k}\vert_{\Sigma_{0}}\oplus\sf{E}_{k}\vert_{\Sigma_{0}}$, Eq.~\eqref{eq:ERhok}\\
        $\P$ & principal $\G$-bundle over $\sf{M}$, Section~\ref{Subsec:Yang-Mills Theory on Globally Hyperbolic Spacetimes}\\
        $\P_{\Sigma}$ & principal $\G$-bundle over $\Sigma$, Lemma~\ref{Lemma:BundleSigma}\\
        $\mathrm{Ad}(\P)$ & $:=\sf{P}\times_{\mathrm{Ad}}\mathfrak{g}$, adjoint bundle, Section~\ref{Subsec:Yang-Mills Theory on Globally Hyperbolic Spacetimes}\\
        $\mathrm{Ad}(\P_{\Sigma})$ & $:=\sf{P}_{\Sigma}\times_{\mathrm{Ad}}\mathfrak{g}$, adjoint bundle, Section~\ref{Subsec:Yang-Mills Theory on Globally Hyperbolic Spacetimes}\\
        &\\
       
        $\langle\cdot,\cdot\rangle_{\sf{g}}$ & Lorentzian Hodge metric on $\bigwedge^{k}\sf{T}^{\ast}\sf{M}$, \emph{i.e.}, $\langle\cdot,\cdot\rangle_{\g}:=\frac{1}{k!}(\g^{-1})^{\otimes k}(\cdot,\cdot)$\\
        $\langle\cdot,\cdot\rangle_{\sf{h}}$ & Riemannian Hodge metric on $\bigwedge^{k}\sf{T}^{\ast}\Sigma$, \emph{i.e.}, $\langle\cdot,\cdot\rangle_{\h}:=\frac{1}{k!}(\h^{-1})^{\otimes k}(\cdot,\cdot)$\\
        $\kappa(\cdot,\cdot)$ & Ad-invariant inner product on $\mathfrak{g}$, Assumption~\ref{Assumption}\\
        $\langle\cdot,\cdot\rangle_{\mathrm{Ad}(\P)}$ & bundle metric on $\mathrm{Ad}(\P)=\P\times_{\mathrm{Ad}}\mathfrak{g}$ induced from $\kappa$, Eq.~\eqref{eq:InducedBundleMetric}\\
        $\langle\cdot,\cdot\rangle_{\mathrm{Ad}(\P_{\Sigma})}$ & bundle metric on $\mathrm{Ad}(\P_{\Sigma})=\P_{\Sigma}\times_{\mathrm{Ad}}\mathfrak{g}$ induced from $\kappa$\\
        $\langle\cdot,\cdot\rangle_{\sf{E}_{k}}$ & bundle metric on $\sf{E}_{k}$ induced from  $\langle\cdot,\cdot\rangle_{\mathrm{Ad}(\P)}$ and $\langle\cdot,\cdot\rangle_{\sf{g}}$, Eq.~\eqref{Eq.DefEk1}\\
        $\langle\cdot,\cdot\rangle_{\sf{E}_{\Sigma,k}}$ & bundle metric on $\sf{E}_{\Sigma,k}$ induced from  $\langle\cdot,\cdot\rangle_{\mathrm{Ad}(\P_{\Sigma})}$ and $\langle\cdot,\cdot\rangle_{\sf{h}}$, Eq.~\eqref{Eq: E Sigma k}\\
        $\langle\cdot,\cdot\rangle_{\sf{E}_{\rho_{k}}}$ & bundle metric on $\E_{\rho_{k}}$ induced from $\langle\cdot,\cdot\rangle_{\sf{E}_{\Sigma,k}}$, Eq.~\eqref{eq:BundMetrErho}\\ 
        $(\cdot,\cdot)_{\sf{F}}$ & for a vector bundle $\sf{F}$ over pseudo-Riemannian manifold $(\sf{N},\sf{k})$ with bundle metric $\langle\cdot,\cdot\rangle_{\sf{F}}$, round brackets denote $(\cdot,\cdot)_{\sf{F}}:=\int_{\sf{N}}\langle\cdot,\cdot\rangle_{\F}\,\d\mu_{\sf{k}}$, \emph{i.e.}, the integrated Hermitian form  on sections $\Gamma_{\mathrm{c}}(\sf{F})$; Eq.~\eqref{eq:MetricSections}\\
        $\langle\cdot,\cdot\rangle_{\sf{L}^{2}}$ & $=(\cdot,\cdot)_{\sf{E}_{\Sigma,k}}$, the $\sf{L}^{2}$-inner product on $\sf{L}^{2}_{k}(\Sigma,\mathrm{Ad}(\sf{P}_{\Sigma}))$, Section~\ref{Subsec:Hodge Decomposition in Flat Vector Bundles on Complete Riemannian Manifolds}\\
        $\langle\cdot,\cdot\rangle_{\sf{H}^{\mathfrak{s}}}$
       & $:=\langle E^{\frac{\mathfrak{s}}{2}}\cdot,E^{\frac{\mathfrak{s}}{2}}\cdot\rangle_{\sf{L}^{2}}$, inner product of Sobolev spaces $\sf{H}^{\mathfrak{s}}_{k}(\Sigma,\mathrm{Ad}(\sf{P}_{\Sigma}))$\\
        &\\
        $[\cdot\wedge\cdot]_{\mathfrak{g}}$ & wedge product induced from  $[\cdot,\cdot]_{\mathfrak{g}}$, Section~\ref{Subsec:Yang-Mills Theory on Globally Hyperbolic Spacetimes} and Appendix~\ref{Appendix: Notation and Conventions for Scalar and Twisted Differential Forms}\\
        $\langle\cdot\wedge\cdot\rangle_{\sf{E_{k}}}$ & wedge product induced from $\langle\cdot,\cdot\rangle_{\sf{E}_{k}}$, Section~\ref{Subsec:Yang-Mills Theory on Globally Hyperbolic Spacetimes} and Appendix~\ref{Appendix: Notation and Conventions for Scalar and Twisted Differential Forms}\\
        $\langle\cdot\wedge\cdot\rangle_{\sf{E_{\Sigma,k}}}$ & wedge product induced from $\langle\cdot,\cdot\rangle_{\sf{E}_{\Sigma,k}}$, Section~\ref{Subsec:Yang-Mills Theory on Globally Hyperbolic Spacetimes} and Appendix~\ref{Appendix: Notation and Conventions for Scalar and Twisted Differential Forms}\\
       &\\
       $\mathcal{V}_{\sf{P}}$ & physical phase space for Yang-Mills theory, Definition~\ref{Def: classical phase space}\\
       $\mathcal{V}_{\Sigma}$ & initial data phase space, Eq.~\eqref{eq:VSigmaDef}\\
       $\mathcal{V}_{\Sigma}^{\infty}$ & Initial data phase space for smooth $\sf{H}^{\infty}$-data, Eq.~\eqref{Eq:VSigmaAbstractYM}\\
       $\mathcal{V}_{\sf{R}}^{\infty}$ & phase space for Cauchy radiation gauge initial data, Eq.~\eqref{Eq:VSigmaAbstractYM2}\\
       &\\
		$\sigma([\cdot],[\cdot])$ & Hermitian form on $\mathcal{V}_{\sf{P}}$, $\sigma([\cdot],[\cdot])=(\cdot,\mathsf{i}\sf{G}_{1}\cdot)_{\sf{E}_{1}}$, Def.~\ref{Def: classical phase space} \\
		$\sigma_{\Sigma,k}(\cdot,\cdot)$ & Hermitian forms on $\Gamma(\sf{E}_{\rho_k})$, $\sigma_{\Sigma,k}(\cdot,\cdot)
    \defeq \mathrm{i}(\cdot,\G_{\Sigma,k}\cdot)_{\E_{\rho_k}}$, Eq.~\eqref{Eq:sigmaKSigmaYM}, which induce Hermitian forms on $\mathcal{V}_{\Sigma}$ and $\mathcal{V}_{\Sigma}^{\infty}$, Sec.~\ref{Subsec:Phase Spaces on Initial Data and Cauchy Problem} and Sec.~\ref{Subsec:Cauchy Problem and Complete Gauge Fixing} \\
    $\sigma_{\Sigma,\sf{R}}(\cdot,\cdot)$ & Hermitian form on $\mathcal{V}_{\sf{R}}$, Eq.~\eqref{Eq:SigmaRYM}\\
       &\\
       $\nabla$ & Levi-Civita connection of $(\M,\g)$\\
       $\nabla^{A}$ & covariant derivative induced from $A\in\mathcal{C}(\sf{P})$ on $\mathrm{Ad}(\P)$, Section~\ref{Subsec:Yang-Mills Theory on Globally Hyperbolic Spacetimes}\\
       $\nabla^{A,\g}$ & connection on $\E_{k}$ induced from $\nabla^{A}$ and Levi-Civita con., Section~\ref{Subsec:Yang-Mills Theory on Globally Hyperbolic Spacetimes}\\
       $\mathrm{D}_{A}$ & exterior covariant derivative, Section~\ref{Subsec:Yang-Mills Theory on Globally Hyperbolic Spacetimes}\\
       $\d^{A}$ & exterior covariant derivative, Section~\ref{Subsec:Yang-Mills Theory on Globally Hyperbolic Spacetimes}\\
       $\delta^{A}$ & exterior covariant codifferential, Section~\ref{Subsec:Yang-Mills Theory on Globally Hyperbolic Spacetimes}\\
       $\square^{A}$ & $\mathrm{Ad}(\P)$-valued de Rham-Hodge d'Alembertian, Section~\ref{Subsec:Yang-Mills Theory on Globally Hyperbolic Spacetimes}\\
       $\d^{A_{\Sigma}}$ & exterior covariant derivative, Section~\ref{Subsec:Decomposition of Connections and Linearised Yang-Mills Fields}\\
       $\delta^{A_{\Sigma}}$ & exterior covariant codifferential, Section~\ref{Subsec:Decomposition of Connections and Linearised Yang-Mills Fields}\\
       $\Delta^{A_{\Sigma}}$ & $\mathrm{Ad}(\P_{\Sigma})$-valued de Rham-Hodge Laplacian, Section~\ref{Subsec:Decomposition of Connections and Linearised Yang-Mills Fields}\\
       $\mathcal{D}_{t}^{A_{\mathrm{T}}}$ & covariant time-derivative, Definition~\ref{Def: covariant time derivative}\\
       $E$ & $:=\mathds{1}+\overline{\Delta^{A_{\Sigma}}}$, used to define the Bessel potentials $E^{\frac{\mathfrak{s}}{2}}$, Section~\ref{Subsec:Hodge Decomposition in Flat Vector Bundles on Complete Riemannian Manifolds}\\
       $\P$ & linearised Yang-Mills operator, $\P=\delta^{A}\d^{A}$ for flat $A$; Proposition~\ref{Prop:LinYangMills}\\
       $\K$ & gauge transformations, $\K=\d^{A}$; Proposition~\ref{Prop:LinYangMills}\\
       $\sf{R}$ & operator realising Cauchy radiation gauge, Section~\ref{Subsec:Cauchy Problem and Complete Gauge Fixing}\\
       $\sf{R}_{\Sigma}$ & operator realising Cauchy radiation gauge on initial data, Section~\ref{Subsec:Cauchy Problem and Complete Gauge Fixing}\\
       $\sf{D}_{0}$, $\sf{D}_{1}$ & $\sf{D}_{0}=\sf{K}^{\ast}\sf{K}$ and $\sf{D}_{1}=\sf{P}+\sf{K}\sf{K}^{\ast}$; for flat $A$: $\sf{D}_{0}=\sf{D}_{1}=\square^{A}$; Prop.~\ref{Prop:LinYangMills}\\
       $\rho_{k}$ & $\rho_{k}\:\Gamma(\E_{k})\to\Gamma(\E_{\rho_{k}})$ initial data maps for $\square^{A}$, Sec.~\ref{Subsec:Phase Spaces on Initial Data and Cauchy Problem}\\
       $\mathcal{U}_{k}$ & $\mathcal{U}_{k}:=(\rho_{k}\vert_{\mathrm{ker}(\square^{A})})^{-1}$, Cauchy evolution operators for $\square^{A}$, Sec.~\ref{Subsec:Phase Spaces on Initial Data and Cauchy Problem}\\
       $\K_{\Sigma}$ & Gauge transformations on the level of initial data, Sec.~\ref{Subsec:Phase Spaces on Initial Data and Cauchy Problem}\\
       $\K_{\Sigma}^{\dagger}$ & Lorentz gauge on the level of initial data, Sec.~\ref{Subsec:Phase Spaces on Initial Data and Cauchy Problem}\\
       $\G_{k}$ & causal propagators of $\square^{A}$, Definition~\ref{Def: classical phase space}\\
       $\G_{\Sigma,k}$ & $\G_k=(\rho_k\G_k)^*\G_{\Sigma,k}(\rho_k\G_k)$, Eq.~\eqref{Eq:GSigmaAbstractYM} and Prop.~\ref{Prop:GSigma}\\
       $\G_{\Sigma,\sf{R}}$ & Eq.~\eqref{Eq:TSigmaIsometryYM}\\
       $\Pi$ & projector $\Pi:\Omega^1_\infty(\Sigma_0,\Ad(\P_\Sigma))
    \longrightarrow
    \ker\bigl(\delta^{\A_\Sigma^0}\vert_{\Omega^1_\infty}\bigr)$, Lemma~\ref{Lem:PiYM}\\
       $\T_{\Sigma}$ & projector of complete gauge fixing, Prop.~\ref{Prop:CommDiaGP} and Prop.~\ref{Prop:TSigmaYM}\\
       &\\
    $\lambda^{\pm}$ & pseudo-covariances of a (Hadamard) state, Prop.~\ref{Prop:SpacetimePseudoCovariancesYM}\\
    $\lambda^{\pm}_{\Sigma}$  & Cauchy pseudo-covariances of a (Hadamard) state, Definition~\ref{Def:CauchyPseudoCov}\\
    $c^{\pm}$ & $\lambda_\Sigma^\pm\defeq\pm \mathrm{i}\G_{\Sigma,1}c^\pm$, Proposition~\ref{Prop:CauchyPseudoCovariancesYM}\\
    &\\
    $A$ & connection $1$-form; background Yang-Mills field, Appendix~\ref{Appendix:Connections on Principal Bundles}\\
    $A_{\Sigma}$ & $A=A_{\mathrm{T}}\d t+A_{\Sigma}$ with $A_{\Sigma}\in C^{\infty}(\bb{R},\mathcal{C}(\P_{\Sigma}))$, Proposition~\ref{Prop:DecompConn}\\
    $F^{A}$ & curvature of $A$, Section~\ref{Subsec:Yang-Mills Theory on Globally Hyperbolic Spacetimes}\\
    $F^{A}_{\sf{M}}$ & curvature of $A$ as a $\mathrm{Ad}(\P)$-valued $2$-form, Section~\ref{Subsec:Yang-Mills Theory on Globally Hyperbolic Spacetimes}\\
    $\mu$ & Maurer-Cartan form on $\G$, Appendix~\ref{Appendix:Notation and Terminology for Lie Groups}\\
       &\\
$\mathcal{C}(\P)$ & \hspace*{1.8cm}space of connections on $\P$, Appendix~\ref{Appendix:Connections on Principal Bundles}\\
$\mathcal{C}(\P_{\Sigma})$ & \hspace*{1.8cm}space of connections on $\P_{\Sigma}$, Appendix~\ref{Appendix:Connections on Principal Bundles}\\
$\mathcal{G}(\P)$ & \hspace*{1.8cm}gauge group, Section~\ref{Subsec:Yang-Mills Theory on Globally Hyperbolic Spacetimes}\\
 $C^{\infty}(\bb{R},\mathcal{C}(\P_{\Sigma}))$ & \hspace*{1.8cm}time-dependent connections on $\P_{\Sigma}$, Sec.~\ref{Subsec:Decomposition of Connections and Linearised Yang-Mills Fields}\\
 $\Omega^{k}_{\mathrm{hor},\mathrm{Ad}}(\P,\mathfrak{g})$ & \hspace*{1.8cm}subspace of $\Omega^{k}(\P,\mathfrak{g})$, see the isomorphism~\eqref{eq:Iso}\\
$\Omega^{k}(\sf{M},\mathrm{Ad}(\P))$ & \hspace*{1.8cm}$\Omega^{k}(\M,\mathrm{Ad}(\P))=\Gamma(\E_{k})$, $\mathrm{Ad}(\P)$-valued $k$-forms on $\sf{M}$\\
       $\Omega^{k}(\Sigma,\mathrm{Ad}(\P_{\Sigma}))$ &  \hspace*{1.8cm}$\Omega^{k}(\Sigma,\mathrm{Ad}(\P_{\Sigma}))=\Gamma(\E_{\Sigma,k})$, $\mathrm{Ad}(\P_{\Sigma})$-valued $k$-forms on $\Sigma$\\
       $C^{\infty}(\bb{R},\Omega^{k}(\Sigma,\mathrm{Ad}(\P_{\Sigma})))$ & \hspace*{1.8cm}time-dependent $\mathrm{Ad}(\P_{\Sigma})$-valued $k$-forms on $\Sigma$, Sec.~\ref{Subsec:Decomposition of Connections and Linearised Yang-Mills Fields}\\
       $\sf{L}^{2}_{k}(\Sigma,\mathrm{Ad}(\P_{\Sigma}))$ & \hspace*{1.8cm}$\sf{L}^{2}$-space of $\mathrm{Ad}(\P_{\Sigma})$-valued $k$-form on $\Sigma$, Sec.~\ref{Subsec:Hodge Decomposition in Flat Vector Bundles on Complete Riemannian Manifolds}\\
       $\sf{H}^{\mathfrak{s}}_{k}(\Sigma,\mathrm{Ad}(\P_{\Sigma}))$ & \hspace*{1.8cm}$\sf{H}^{\mathfrak{s}}$-Sobolev space of $\mathrm{Ad}(\P_{\Sigma})$-valued $k$-form on $\Sigma$, Sec.~\ref{Subsec:Hodge Decomposition in Flat Vector Bundles on Complete Riemannian Manifolds}\\
       $C^{\infty}(\bb{R},\sf{H}^{\mathfrak{s}}_{k}(\Sigma_{\bullet},\mathrm{Ad}(\sf{P}_{\Sigma})))$ & \hspace*{1.8cm}time-dependent $\sf{H}^{\mathfrak{s}}$-forms on $\Sigma$, Sec.~\ref{Subsec:Cauchy Problem and Complete Gauge Fixing}\\
       $\Omega_{\mathfrak{s}}^{k}(\Sigma,\mathrm{Ad}(\P_{\Sigma}))$ & \hspace*{1.8cm}subspace of $\mathsf{H}^{\mathfrak{s}}_{k}(\Sigma,\mathrm{Ad}(\P_{\Sigma}))$ of smooth forms, Sec.~\ref{Subsec:Hodge Decomposition in Flat Vector Bundles on Complete Riemannian Manifolds}\\
       $\Omega^{k}_{\mathfrak{s},\d}(\Sigma,\mathrm{Ad}(\P_{\Sigma}))$ & \hspace*{1.8cm}$\Omega^{k}_{\mathfrak{s},\d}(\Sigma,\mathrm{Ad}(\P_{\Sigma}))=\Omega^{k}(\Sigma,\mathrm{Ad}(\P_{\Sigma}))\cap\overline{\d^{A_{\Sigma}}\Omega^{k-1}_{\infty}(\Sigma,\mathrm{Ad}(\P_{\Sigma}))}^{\sf{H}^{s}}$\\
       $\Omega^{k}_{\mathfrak{s},\delta}(\Sigma,\mathrm{Ad}(\P_{\Sigma}))$ & \hspace*{1.8cm}$\Omega^{k}_{\mathfrak{s},\delta}(\Sigma,\mathrm{Ad}(\P_{\Sigma}))=\Omega^{k}(\Sigma,\mathrm{Ad}(\P_{\Sigma}))\cap\overline{\delta^{A_{\Sigma}}\Omega^{k+1}_{\infty}(\Sigma,\mathrm{Ad}(\P_{\Sigma}))}^{\sf{H}^{s}}$\\
       $\Gamma_{\infty}(\E_{k})$ & \hspace*{1.8cm}subspace of $\Gamma(\E_{k})$ with $\omega_{\mathrm{T}}(t,\cdot)\in\Omega^{k-1}_{\infty}(\Sigma_{t},\mathrm{Ad}(\P_{\Sigma}))$ and \hspace*{1.8cm}$ \omega_{\Sigma}(t,\cdot)\in\Omega^{k}_{\infty}(\Sigma_{t},\mathrm{Ad}(\P_{\Sigma}))$, Section~\ref{Subsec:Cauchy Problem and Complete Gauge Fixing}\\
       $\Gamma_{\infty}(\E_{\rho_{k}})$ & \hspace*{1.8cm}$:=\Omega^{k-1}_\infty(\Sigma_0,\Ad(\P_\Sigma))^{\oplus2}
    \oplus\Omega^k_\infty(\Sigma_0,\Ad(\P_\Sigma))^{\oplus2}$, Sec.~\ref{Subsec:Cauchy Problem and Complete Gauge Fixing}\\
       $\Gamma_{\infty,\d}(\E_{1})$ & \hspace*{1.8cm}subspace consisting of forms $\alpha\in\Gamma_{\infty}(\E_{k})$ with\\&\hspace*{1.8cm}$\alpha_\Sigma\vert_{\Sigma_0},\,
    \mathcal D_t^{\A_{\mathrm T}}\alpha_\Sigma\vert_{\Sigma_0}
    \in\Omega^1_{\infty,\d}(\Sigma_0,\Ad(\P_\Sigma))$, Section~\ref{Subsec:Cauchy Problem and Complete Gauge Fixing}\\
       $\Gamma_{\infty,\d}(\E_{\rho_1})$ & \hspace*{1.8cm}$:=\Omega^0_\infty(\Sigma_0,\Ad(\P_\Sigma))^{\oplus2}
    \oplus
    \Omega^1_{\infty,\d}(\Sigma_0,\Ad(\P_\Sigma))^{\oplus2}$, Sec.~\ref{Subsec:Cauchy Problem and Complete Gauge Fixing}\\
    &\\
    $\Psi^{m}_{\mathrm{b}}(\Sigma,\mathrm{Ad}(\P_{\Sigma}))$ & \hspace*{1.8cm}space of pseudodifferential operators, Appendix~\ref{Appendix:Pseudodifferential Calculus on Manifolds of Bounded Geometry}\\
    $W^{-\infty}(\Sigma,\mathrm{Ad}(\P_{\Sigma}))$ & \hspace*{1.8cm}ideal of smoothing operators on $\Sigma$, Appendix~\ref{Appendix:Pseudodifferential Calculus on Manifolds of Bounded Geometry}\\
       \bottomrule
\end{tabularx}
\end{small}
\bibliographystyle{abbrv} 

\end{document}